\documentstyle[12pt,twoside,epsfig,psfig,uibthesis]{report}
\author{Anders Strand Vestb{\o}}
\title{Pattern Recognition and Data Compression for the ALICE High
Level Trigger}
\department{Institutt for fysikk og teknologi}
\degreemonth{May}
\degreeyear{2004}

\newcommand{\bfig}{\begin{figure}}
\newcommand{\efig}{\end{figure}}
\newcommand{\insertplot}[4]{
  \begin{center}
     \centering
     \epsfig{file=#1,width=#2}
     \end{center} 
     \caption[#3]
             {#4}
}	
\newcommand{\pt}{$p_t$\ }
\newcommand{\dndy}{dN$_{\mathrm{ch}}$/d$\eta$\ }
\newcommand{\beq}{\begin{equation}}
\newcommand{\eeq}{\end{equation}}

\begin{document}
\newlength{\tmptextheight} \setlength{\tmptextheight}{\textheight}
\newlength{\tmptextwidth}  \setlength{\tmptextwidth}{\textwidth}
\setlength{\textheight}{\tmptextheight}
\setlength{\textwidth}{\tmptextwidth}

\maketitle
\thispagestyle{empty}
\newpage

\chapter*{Acknowledgements}
\thispagestyle{empty}

There are several people who deserve acknowledgement for
their contribution one way or another to the work compiled in this thesis. 

Most of all, my sincerest thanks goes to my supervisor,
Prof.\ Dieter R{\"o}hrich, for excellent guidance and
support. His relaxed attitude and
detailed insight in a wide range of topics has provided me with an
ideal working environment.
Furthermore, I would like to thank Constantin Albrecht Loizides for
all the academic and social interactions during the last two years.
I appreciate all the nice discussions -- both fundamental and
shallow, enlightening questions and answers, great parties
and valuable comments to the thesis.
I would also like to thank Dr.\ Ulrich
Frankenfeld for a great time shared working together during his
Post.\ Doc.\ period in Bergen, and later in various
pubs around the world discussing the crew on German warships etc.

I am grateful to all the people in the Experimental Nuclear Physics Group in
Bergen for maintaining a good environment for both research and
friendship. In particular,
I would like to mention Jens Ivar J{\o}rdre, Zhongbao Yin, J{\o}rgen
Lien, Are Severin Martinsen, Gaute {\O}vrebekk, Kenneth Aamodt and
former students Bj{\o}rn Tore Knudsen and Espen Vorland.

Thanks also to Timm Morten Steinbeck and Arne Wiebalck for being
excellent hosts during
my visits to Heidelberg, and for making the HLT data-challenge in
Paderborn such an interesting experience.
I am also grateful to the STAR L3 group under direction of Dr.\ Jens
S{\"o}ren Lange, for giving me a boost into the world of High Level
Triggers during my stay at BNL, spring 2000.

I also wish to thank Prof.\ Bernhard Skaali, the project leader of the
Norwegian ALICE Group, for hiring me as a Dr.\ Scient.\ student at the University
of Oslo, and for giving me the opportunity to attend a number of
various international conferences and workshops.

Finally, I am deeply in debt to Renate, for encouragement,
improving my thesis and for having absolute confidence in me. 

\vspace{2cm}

\noindent Bergen, March 2004\\
Anders Strand Vestb{\o}

\vspace{1cm}
\hspace{11.1cm} ``{\it In principle it's easy.}''

\thispagestyle{empty}
\cleardoublepage

\pagenumbering{roman}
\sloppy   

\tableofcontents
\listoffigures
\listoftables
\cleardoublepage

\rm
\bibliographystyle{physsort}
\pagenumbering{arabic}

\widowpenalty=10000

\chapter*{Introduction}
\addcontentsline{toc}{chapter}{Introduction}

The primary objective of high energy physics is to study the
fundamental forces and symmetries which exist in nature and their
macroscopic manifestations. Over the
last decades, a detailed theory of elementary particles and their
fundamental interactions has been established in the {\it Standard
Model}. Still, very little is known about the properties of nuclear or
hadronic matter, i.e. matter that is composed of quarks and
bound by the strong force -- one of the fundamental forces in
nature. Under normal conditions the quarks are confined in protons and
neutrons, interacting via the nuclear force. At low energy densities
these hadronic bound states are the degrees of freedom of nuclear matter. At
higher energy densities the degrees of freedom are quarks and gluons
interacting via the strong force.

The focus of heavy ion physics is to study and understand the
properties of the different phases of nuclear matter.
At very high densities and temperatures the nucleons are expected to
dissolve into their constituents and form a plasma consisting of
quarks and gluons, the so-called quark-gluon plasma. 
According to Big Bang cosmology such a phase transition from the
quark-gluon plasma into hadronic matter took place during the first microsecond
after the Big Bang. By colliding heavy ions
at very high energies similar conditions can be generated in
the laboratory. This creates instantaneously a partonic phase
which quickly equilibrates into a quark-gluon plasma.

The study of such a phase transition, and the physics of the
quark-gluon plasma state,
requires numerous systematic measurements of nuclear collisions with
varying initial conditions.
The main challenge of heavy-ion physics is to
record and analyze the large number of particles which emerge from these
collisions. The ALICE experiment at the upcoming Large Hadron Collider
(LHC) at CERN will be dedicated to the study of heavy ion
collisions at energies which go far beyond the critical
energy density for a phase transition. At these energies,
up to 20\,000 particles will be detected in every central collision,
generating a wealth of information which has to be recorded for
subsequent analysis. In order to accumulate enough statistics for a
coherent measurement of the wide range of predicted observables, the
experiment has to collect as many events as possible within the given
runtime. The allowed event rate, however, will produce about one order
of magnitude more data than the foreseen data rate to mass storage.
This inconsistency between the available data rate and the limited mass
storage bandwidth can be overcome by introducing a
layer in the readout-system which is able to efficiently reduce the data rate
by online event selection and data compression. Such a {\it High
Level Trigger system} will have to perform real-time analysis of the
detector information, requiring fast pattern recognition in order to
reconstruct the particle tracks.

The ALICE High Level Trigger system is designed to accomplish this task.
The system entails a large scale generic processing farm of
the order of several hundred separate nodes. The overall architecture of the
system follows a hierarchical structure, driven by the intrinsic
parallelism of the data flow from the detectors and the demand
for a full event reconstruction. The system components will be based on
commercially available PCs connected with a high bandwidth, low latency
network. A number of nodes will be equipped with FPGA co-processors
for designated pre-processing tasks.

The main processing task of the system is fast parallel
detector specific pattern recognition. Given the large uncertainty of
the anticipated particle multiplicity, different approaches to the
pattern recognition problem need to be considered.
Once the particle tracks have been reconstructed event selections can
be performed on the basis of various physics analysis algorithms.
Such applications may include event rate
reduction by complete event selection/rejection, or event size
reduction by region-of-interest readout or data compression.

\chapter{Ultrarelativistic Heavy Ion Collisions}
\label{HIC}

\section{Quarks and gluons}
One of the fundamental assumptions in modern elementary particle physics is the
quark model defined by Gell-Mann and Zweig~\cite{gellmann,zweig}. It states
that all hadrons consist of a multiple of quarks in a bound
state. Most common are the baryons and mesons, with
three quarks ($qqq$) and a quark and a anti-quark ($q\bar{q}$)
respectively. In addition,
recent experimental evidence for the so-called {\it pentaquark} state
($qqqq\bar{q}$) has been reported~\cite{nakano,barmin,step,barth,alt}.
It is possible
to reconstruct and explain all the properties of the hadrons 
(charge, mass, magnetic moment, isospin etc.) 
from the quantum numbers of the
quarks. For instance, to build a single nucleon one needs at least two different
types of quarks, which are designated by {\it up} ($u$)$\ $ and {\it down
(d)} and have charge 2/3 and -1/3 charge respectively.
The proton consists of three quarks ($uud$), resulting in a total charge of
+1, while the neutron contains the combination ($udd$). 
Quarks are identified by the quantum property {\it flavor}. There are
in total 6 different quarks, here listed with increasing mass:
{\it up (u), down (d), strange (s), charm (c),
bottom (b) and top (t)}. 

The interaction binding the quarks into hadrons is called the strong
interaction, and is described by the theory of {\it Quantum
Chromodynamics} (QCD). Such a fundamental interaction is, according to
the standard model, always connected with virtual particle
exchanges. Analogous to the electromagnetic interaction in which
photons are exchanged between electrically charged particles, 
gluons are the mediators of the strong force and couple to a quantum
number called color charge. This quantum number can assume three
values (red, blue and green), and each quark of a given
flavor carries a color quantum number. In contrast to the
photons which have no charge, the gluons carry simultaneous color and
anti-color. This has the effect that they do not only couple to
quarks, but also to other gluons. As a consequence, the
strong coupling constant shows a strong dependence on the quark-quark
separation. For large distances the coupling constant grows towards
infinity, which implies that an infinite amount of energy would be
required to separate two color charges. Consequently, no quark
or gluon may exist as ``free'' particle. This is reflected through the
fact the the quarks are always arranged
in such a way that all particles which exist in physical vacuum are
colorless. This phenomenon is commonly referred to as {\it
confinement}.
However, for very small distances the coupling decreases
asymptotically. In the high energy limit quarks can be considered to be ``free'', and
this is called {\it asymptotic freedom}.

\section{Hot and dense nuclear matter}
The asymptotic behavior of QCD at high densities has been predicted to
be a phase transition in nuclear matter~\cite{cabbibo}. 
Such a phase transition is expected to occur at extreme temperatures
and energy densities, forcing the nuclear matter to undergo a transition into
a deconfined state of quarks and gluons. The new phase is known as
the Quark-Gluon Plasma (QGP), and unlike ordinary 
nuclear matter where quarks and gluons are confined in
bound states as hadrons, they are now considered as almost ``free'' particles. Such a
phase transition will consequently lead to a dramatic jump in the energy density of
the state, due to the sudden increase of the number of degrees of
freedom. There will be more spin and color states available to the quarks and gluons
when moving freely compared to the number of states available within
the hadrons.

In addition, QCD predicts that in a high temperature phase transition
a fundamental symmetry of the QCD theory, which are valid only at
high energy densities, is restored. This {\it chiral symmetry} is
spontaneously broken at normal nuclear density, and the current quark masses
originate as a direct consequence of this symmetry breaking mechanism.
During a phase transition into QGP the chiral symmetry is
approximately restored and the quark masses are reduced from the large
effective values in hadronic matter to their small bare ones.


\subsubsection{Lattice QCD and the phase diagram}
Phase transitions are related to large
distance phenomena in a thermal
medium, and go along with long range collective phenomena and
the spontaneous breaking of global symmetries. 
Thus in order to study such a phase transition within the theory of
QCD, a numerical
approach that is capable of dealing with the equilibrium
thermodynamics of the strong interactions is needed. 
Lattice QCD~\cite{lattice}
provides a first principle approach that allows to study large
distance, non-perturbative aspects of the strong interaction. This is
done by introducing a discrete space-time lattice, which makes it 
suited for numerical calculations. 


\bfig[htb]
\insertplot{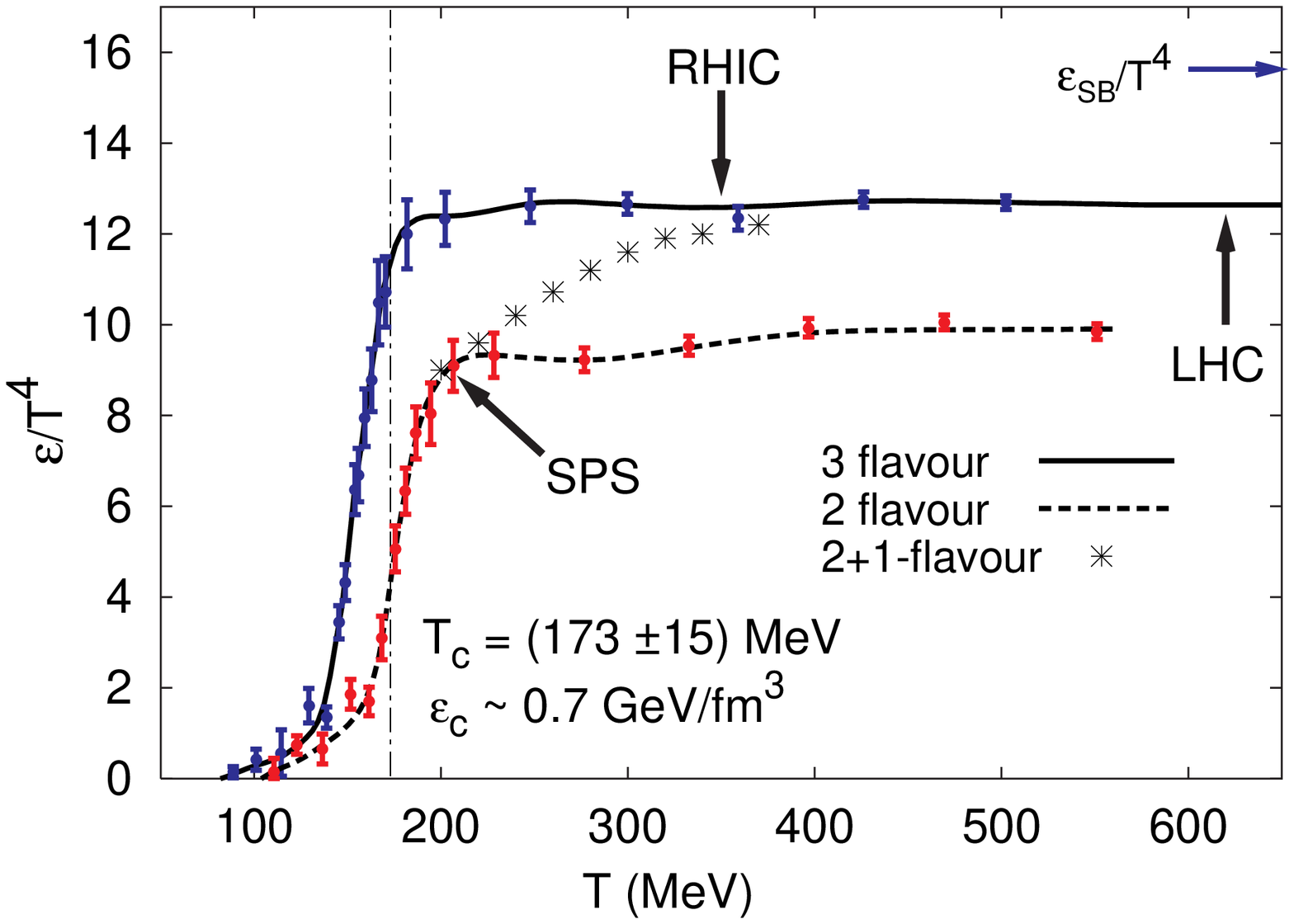}{10cm}
	{Energy density as a function of temperature calculated within lattice QCD.}
        {Energy density as a function of temperature calculated in lattice QCD at
zero baryon chemical potential with various numbers of degenerate
quark flavors~\cite{karsch1}.} 
\label{HIC_energyvstemp}
\efig

In the lattice calculations a discontinuity in the energy density as a
function of temperature is found at a critical temperature of the
order of $T_c\approx$170\,MeV, corresponding to an energy density of
$\epsilon\approx$1\,GeV/fm$^3$~\cite{karsch1}, Figure~\ref{HIC_energyvstemp}. 
There are however many uncertainties involved
regarding the actual value of this temperature, and the order of the
phase transition. The reason is that both depend on the number
of flavors and the bare quark masses being used in the calculations. In
the high temperature and zero baryon density limit, the phase
transition is fully described by the chiral symmetry of the QCD
Lagrangian~\cite{pisark}. This symmetry is a global intrinsic symmetry of
the theory which is exact only in the limit of vanishing quark
masses. However, the quarks in nature are not massless, 
and in particular the heavy quarks (charm, bottom and
top) are too heavy to play a role in the thermodynamics in the
vicinity of the phase transition. However, the strange
quark, whose mass is of the order of $T_c$, plays a crucial role in
deciding about the nature of the transition at vanishing baryon density.
In the massless limit a three flavor QCD shows a first order phase transition. 
Recent lattice calculations indicate that the phase transition for realistic
values of the up, down and strange quark masses may even be a {\it
rapid crossover} taking place in a narrow temperature interval
around $T_c\sim$170\,MeV~\cite{karsch}.

\bfig[htb]
\insertplot{hic/lattice.eps}{9cm}
	{Lattice calculations at finite baryon chemical potential.}
	{Lattice calculations at finite baryon chemical potential
from~\cite{scikor} and references therein. The lines indicate a rapid
crossover transition at low $\mu_B$ which becomes first order above
the tri-critical point. The position of this point (indicated by the
rectangle on the figure) is subject to significant uncertainties.}
\label{HIC_finitemu}
\efig

At finite baryon chemical potential ($\mu_B\neq$0) the standard
Monte-Carlo sampling techniques used in lattice
calculations, Figure~\ref{HIC_energyvstemp}, are no longer
applicable. However, recent
theoretical progress has overcome this problem, and consequently extends the lattice
simulations of the QCD phase transition for values up to
$\mu_b$=0.5-0.8\,GeV~\cite{scikor}, Figure~\ref{HIC_finitemu}. The results
show a slight decrease of $T_c$ with increasing $\mu_B$.

\bfig[htb]
\insertplot{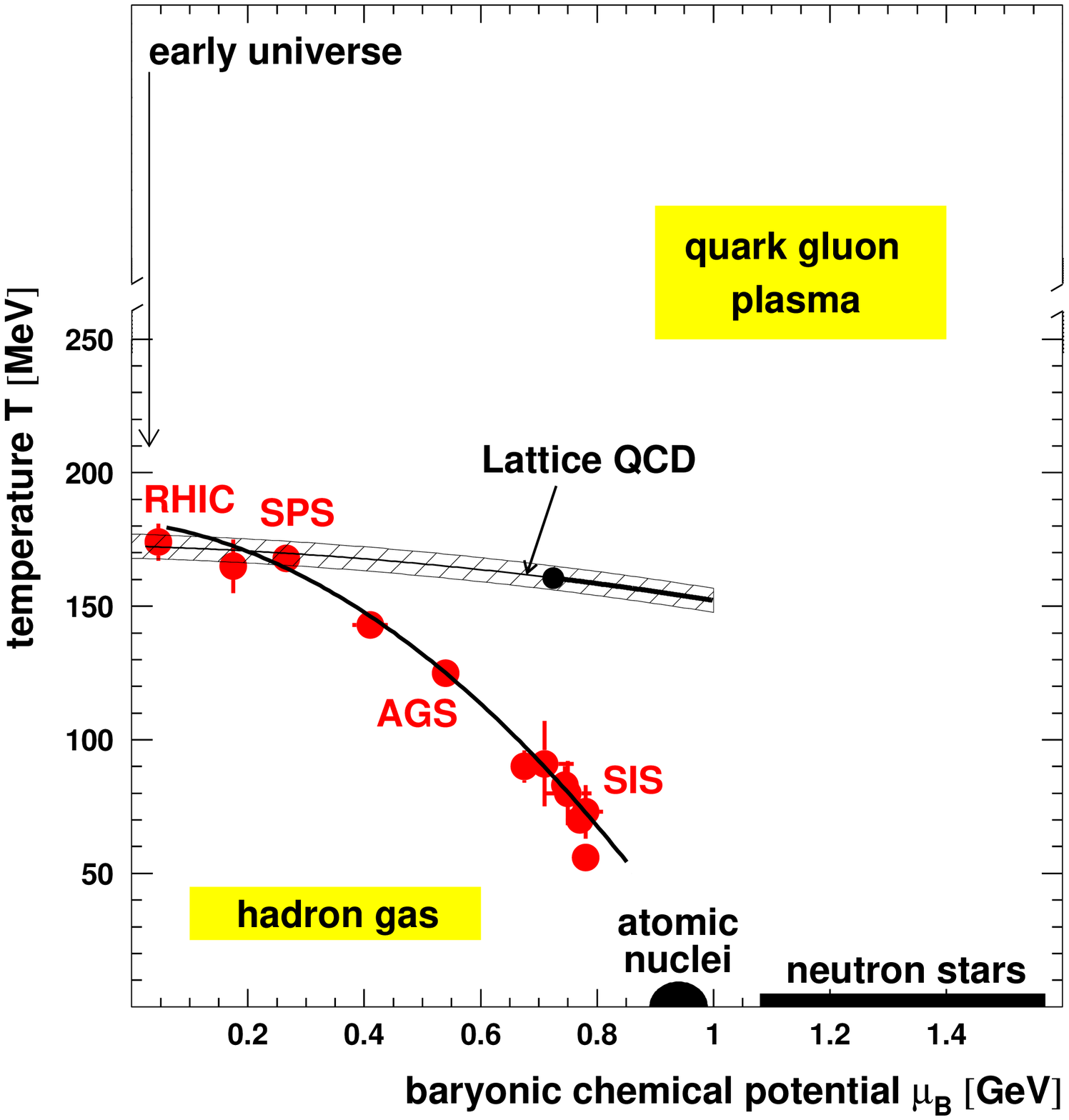}{9cm}
	{QCD phase diagram.}
	{QCD phase diagram summarizing the present understanding about
the structure of nuclear matter at different densities and
temperatures. The points marks illustrates the results achieved by the
different ultrarelativistic collider experiments, and the dashed line
represent the lattice QCD calculations.}
\label{HIC_phase}
\efig

The present experimental and theoretical knowledge about the different
phases of strongly interacting
matter can be summarized in a generic QCD phase diagram, Figure~\ref{HIC_phase}.
In addition to the phase transition at high temperatures,
deconfinement is expected at sufficiently large density (several
times normal nuclear matter density) and low temperature. However, in
this case the evidence is less compelling due to the lack of
lattice results for very high values of $\mu_B$.



\subsubsection{Relativistic heavy ion collisions}
Relativistic heavy ion collisions offer a unique tool to probe hot
and dense nuclear matter in the laboratory.
During the last decades a great number of experiments have
been carried out in order to explore the nuclear state of matter as a
function of temperature and energy density. The main motivation is the
search for a QGP phase. At the CERN SPS accelerator a series of fixed
target experiments has collected a wealth of information about nuclear
matter at center-of-mass energies of $\sqrt{s}$=5-20\,A\,GeV. However,
the results have not firmly established the existence of the QGP phase
yet, as the energy density obtained only slightly exceeds the critical
temperature, $T_c$.

Current heavy-ion experiments at the Relativistic Heavy Ion
Collider (RHIC) at Brookhaven National Laboratories (USA) and
scheduled experiments at the Large
Hadron Collider (LHC) at CERN (Switzerland), will generate
sufficiently high energy densities to form a baryon-free plasma. At
the RHIC accelerator, four experiments are dedicated to the study of
Au--Au collisions at center-of-mass energies up to
$\sqrt{s}$=200\,A\,GeV. Furthermore, the LHC will make Pb--Pb nuclei
collide at $\sqrt{s}$=5.5\,A\,TeV which will be studied by the ALICE
experiment. At these energies, nuclear matter is predicted to be
transparent enough to form baryon-free matter, heated well beyond
the expected phase transition temperature.

\section{The dynamics of heavy ion collisions}
Even though relativistic heavy ion experiments in the laboratory may recreate the
conditions for a
phase transition, direct comparison to lattice QCD calculations is generally
very difficult.
The reason is that lattice QCD exclusively describes
matter in a thermodynamical equilibrium, while the
outcome of
a heavy ion collision is a finite, highly excited and dynamical
non-equilibrated system. The correct theoretical treatment of such a system is
therefore not a trivial task, and involves concepts which go far
beyond the capabilities of simple statistical thermodynamics. 
These models can however be valuable
as they can provide first order (quasi-)analytic solutions that
can be compared directly with measured quantities. 


The evolution of a heavy ion collision in space and time depends
extensively on the initial conditions of the system. 
Consequently, heavy ion collisions are generally divided in two energy
domains: Lower energies where the stopping power is sufficient
to stop the colliding nuclear matter, and higher energies, where
the colliding baryons initially penetrate each other. The former case is
applicable to the energy range of the AGS and
SPS experiments, and is commonly described within Landau's
fluid-dynamical model.
For the energies which will be obtained at the LHC, the
latter scenario is most likely to be the case, and is often described with
the scaling hydrodynamical model of Bjorken~\cite{bjorken},
Figure~\ref{HIC_bjorken}. In this picture, the
Lorentz contracted nuclei become almost completely transparent to each
other, and the valence quarks maintain their initial rapidities. At their
inter-penetration, however, the partons interact creating a
high energy density chromo-electric field between the two nuclei.
Within the chromo-electric field a system of non-equilibrated deconfined quarks and
gluons is created. This matter constitutes the so-called {\it pre-equilibrium
phase}, which after a certain formation time might lead to a local
thermal equilibrium provided that there are enough interactions among
the constituents. The initial conditions at which an equilibrium is
reached is defined by the proper time, $\tau$.
The proper time is defined as the local time in the rest frame of any fluid
element. If all the particles originate from one point in space-time
the proper time can be expressed as
\begin{eqnarray*}
\tau=t/\gamma=\sqrt{t^2-z^2}.
\end{eqnarray*}	
After a formation time $\tau_0$ which is likely to be 0.5 to 2\,fm/c the system reaches
thermal equilibrium which is characterized by a uniform energy density
and temperature. From that time on the system is
treated using 1+1 dimensional (spatial + time) ideal relativistic fluid
dynamics, where the system expands in the longitudinal direction. 

As the system expands, the equilibrated plasma of deconfined quarks
and gluons quickly cools down to the temperature where a phase
transition into a hadron gas takes place. Depending on the type of
the transition, the system may spend some time in a mixed phase where
the QGP coexists with the hadron gas. Finally, the size of the system
becomes larger than the mean free path of hadrons, in which they
undergo a {\it freeze-out} and stream freely towards the
detectors. This freeze-out process is most usually treated as a sudden
freeze-out, implying that at a given instant in the space-time all
constituents within the fluid become independent, and final
interactions and collisions are neglected. 
\bfig
\insertplot{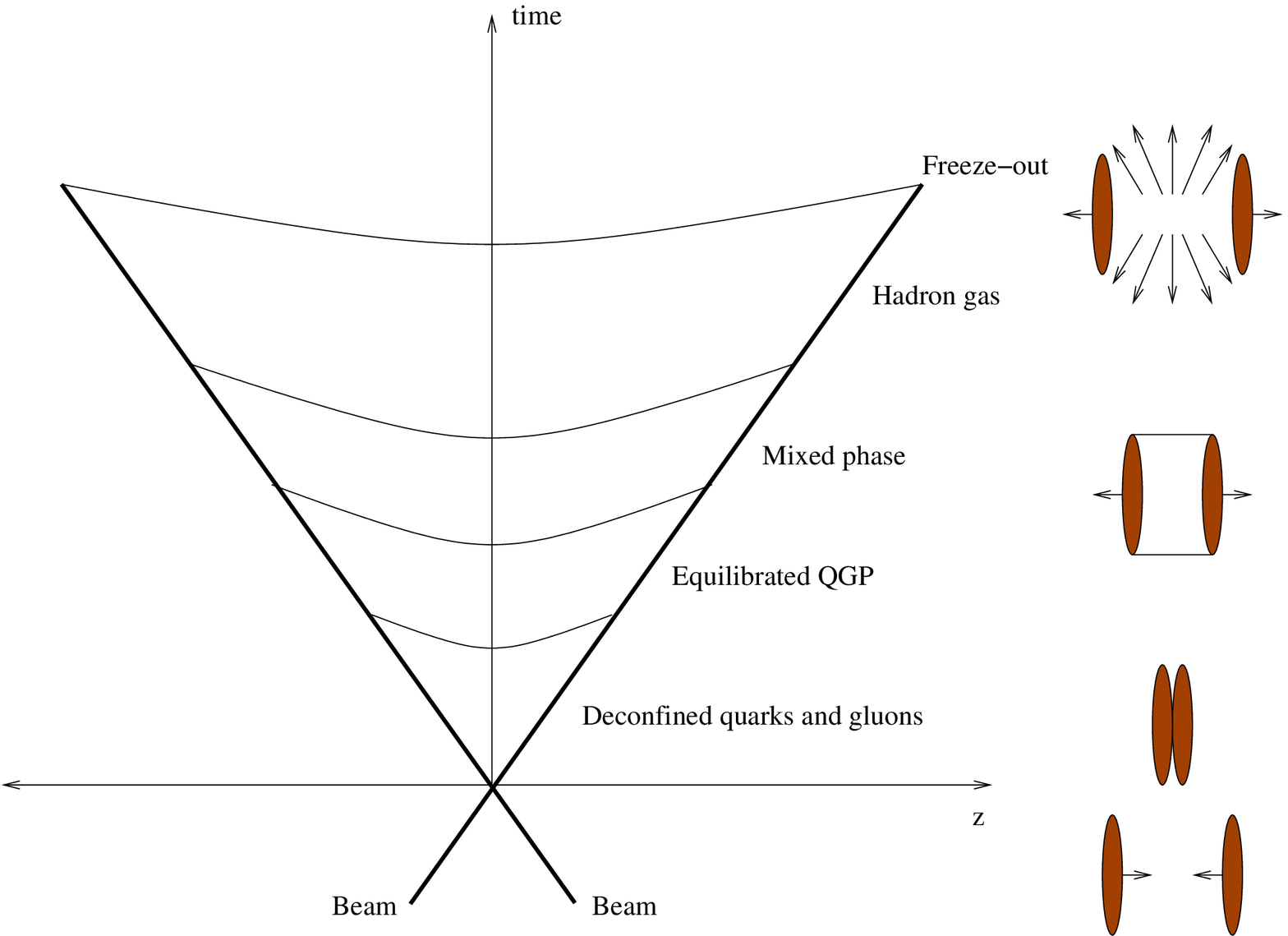}{12cm}
	{The Bjorken space-time scenario for a heavy ion collision.}
	{The Bjorken space-time scenario for a heavy ion
collision. The two colliding nuclei resembles two flat discs because of the
Lorentz contraction in the laboratory frame. The parabola indicate the
constant proper times.}
\label{HIC_bjorken}
\efig

\section{The experimental observables}
\label{HIC_observables}
In order to establish experimentally the properties of the hot and
dense partonic matter created in heavy ion collisions, a wide range of
variables of the system have to be measured. 
Due to the short existence and limited spatial extend of the generated
plasma, however, basic properties such as volume, 
temperature, density of the plasma state and the masses of the
quarks contained in it, cannot be measured directly. Instead, it must
be derived from the remnants of the collision, i.e. the final state
particles which after the freeze-out-stage has reached the
detectors. Several observables have been suggested and identified,
which needs to be evaluated individually and/or in
combination with other probes.

In general, the observables in a heavy ion collisions can be divided
into three main categories: 
\begin{itemize}
\item Hadronic observables.
\item Electromagnetic observables.
\item Hard probes.
\end{itemize}
Each of the observables are characteristic of a certain stage in the
collision, but they are not completely independent of each other.
The hadrons emerge only in the final stage of the 
collision after they freeze-out from the hadron gas, and thus carry
information about the system at the time of freeze-out. 
The
electromagnetic observables on the other hand will, because of their
long mean-free-path relative to the size of the QCD medium, manage to
escape from the system without any further interaction, and thus
emerge predominately from the earlier, hot stage of the collision. 
Lastly, the initial stage of the collision is dominated by the
collision dynamics of the produced partonic system, and the study
of hard processes enable to probe the very early parton dynamics
and evolution of the initial stage of the system.

In the following the main observables which will be relevant at LHC
energies, and therefore will be measured by the ALICE experiment, will
be introduced. These observables are based on
theoretical predictions combined with experimental results from SPS
and RHIC.



\subsection*{Hadronic observables}
The hadronic observables are often referred to as {\it soft} probes of
the heavy ion collision, as they mostly connect to the non-perturbative
aspects of QCD. They deal with the more global characteristics of the
system such as particle production, particle abundances and spectra
and correlations.

\subsubsection{Particle multiplicity}
\label{HIC_partmult}
One of the most important and fundamental observable in a heavy ion
collision is particle
multiplicity. By measuring the number of particles produced in the
collision, one can determine the energy density of the system. From a
theoretical point of view, this is important since it enters the
calculation of most other observables. On the experimental side, the
particle multiplicity fixes the detector performance, and thus the
accuracy with which many of the observables can be measured. 

The particle multiplicity in heavy ion collisions is very
difficult to predict, since it cannot be calculated from first
principles.  
\bfig[htb]
\insertplot{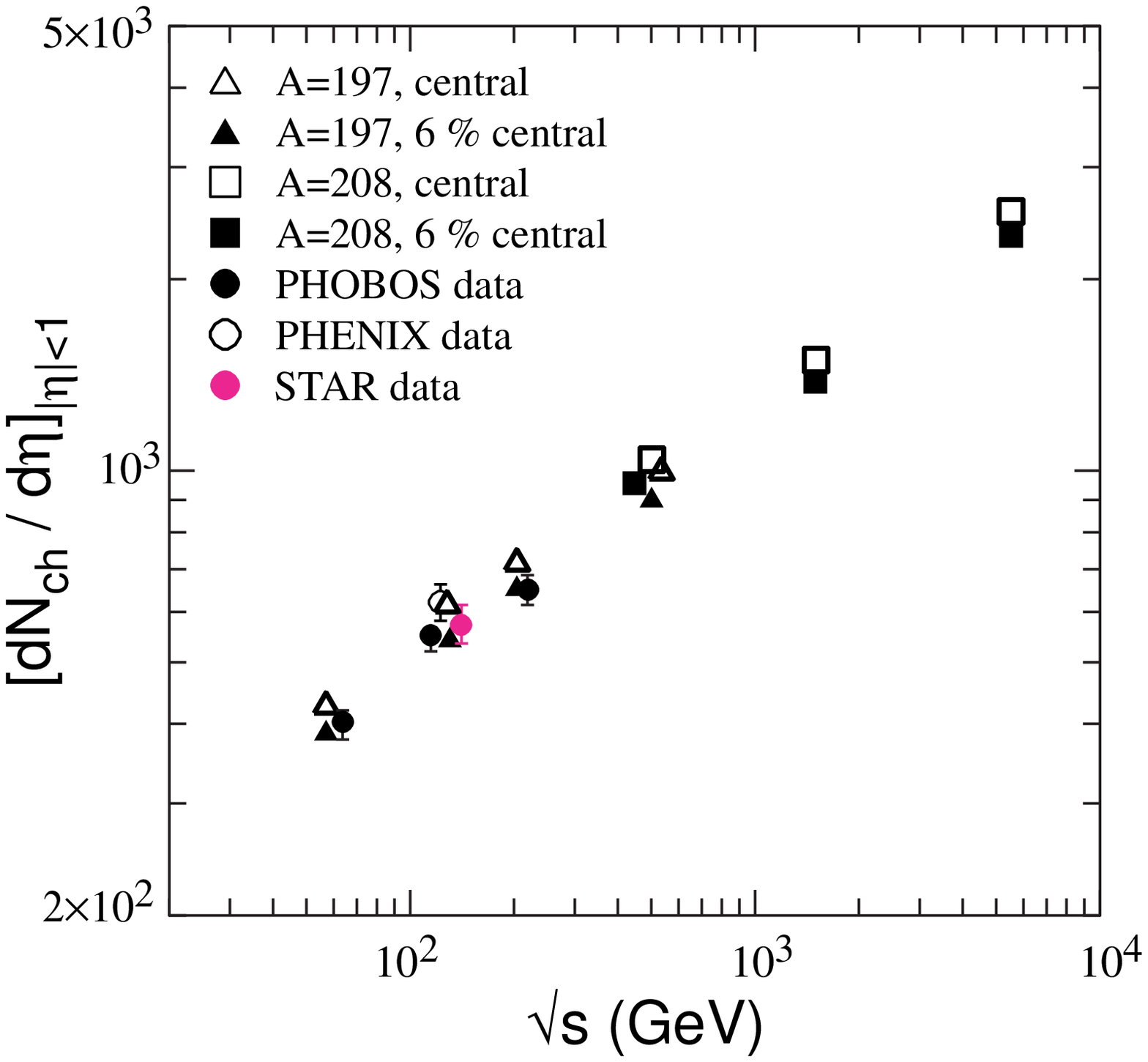}{9cm}
	{Data and predictions for charged particle multiplicity
per unit pseudo-rapidity.}
	{Data and predictions for charged particle multiplicity
per unit pseudo-rapidity~\cite{eskola}.}
\label{HIC_multpred}
\efig
The obvious approach is thus to extrapolate already measured
quantities obtained in lower energy experiments, using
different theoretical extrapolation models. In
Figure~\ref{HIC_multpred} data from RHIC and predictions
for center of mass energies up to LHC levels
($\sqrt{s}$=5.5\,TeV)~\cite{eskola} are shown. At
RHIC the charged particle multiplicity per unit pseudo-rapidity,
dN$_{\mathrm{ch}}$/d$\eta$, is measured to be 700-800 at $\eta$=0. The
predictions for LHC show that one should expect a value of
about 2200. In~\cite{amelin} the multiplicity is computed in a
two-component soft+semi-hard string model, which gives a slightly
higher density of 2600-3200, depending on the initial assumptions.



\subsubsection{Particle spectra and correlations}
Most of the particles emitted in a heavy ion collision are hadrons
which decouple from the collision region during the hadronic
freeze-out stage. Hence, by measuring the different particle spectra,
one obtains information about the chemical and kinetically freeze-out
distributions. From these
observables, one can derive quantities like the
freeze-out temperature and chemical potential, flow velocities within
the expanding system, size of the system etc.
Since these distributions
are also highly constrained by the dynamical evolution of the system,
they will also yield information about the early stages of the
collision~\cite{heinz98,stock99,stachel99}.
Furthermore, the final momentum distributions may provide detailed
information about the time evolution of the collision system~\cite{heinz01}.

Essential information about the collision system is obtained from
studying its evolution in time and space. The size and expansion results from
the work of pressure gradients within the system, and hence reflects
directly the underlying equation of state. This can be obtained
directly by particle interferometry or correlations. By these methods
one can measure the final size of the fireball, gain insight about
its expansion and phase-space density and provide information about
the timing of the hadronization.

Furthermore, the so-called {\it elliptic flow} is sensitive to the degree of
thermalization achieved in the system. In general it describes the azimuthal
asymmetry of the particle production, and builds up through
re-scattering in the evolving system which converts the spatial
anisotropy into momentum anisotropy. A rapid expansion of the hot
system will destroy the original anisotropy and reduce the following
momentum anisotropy. Thus by measuring the elliptic flow,
information about the early stage of the collision is obtained, and in particular
whether local thermalization is reached followed by a collective
hydrodynamic expansion. The observed large elliptic flow
measured at RHIC, Figure~\ref{HIC_flow}, indicates
that the hydrodynamical model is applicable for a wide range of
momenta and particle types.

\bfig[htb]
\centerline{\epsfig{file=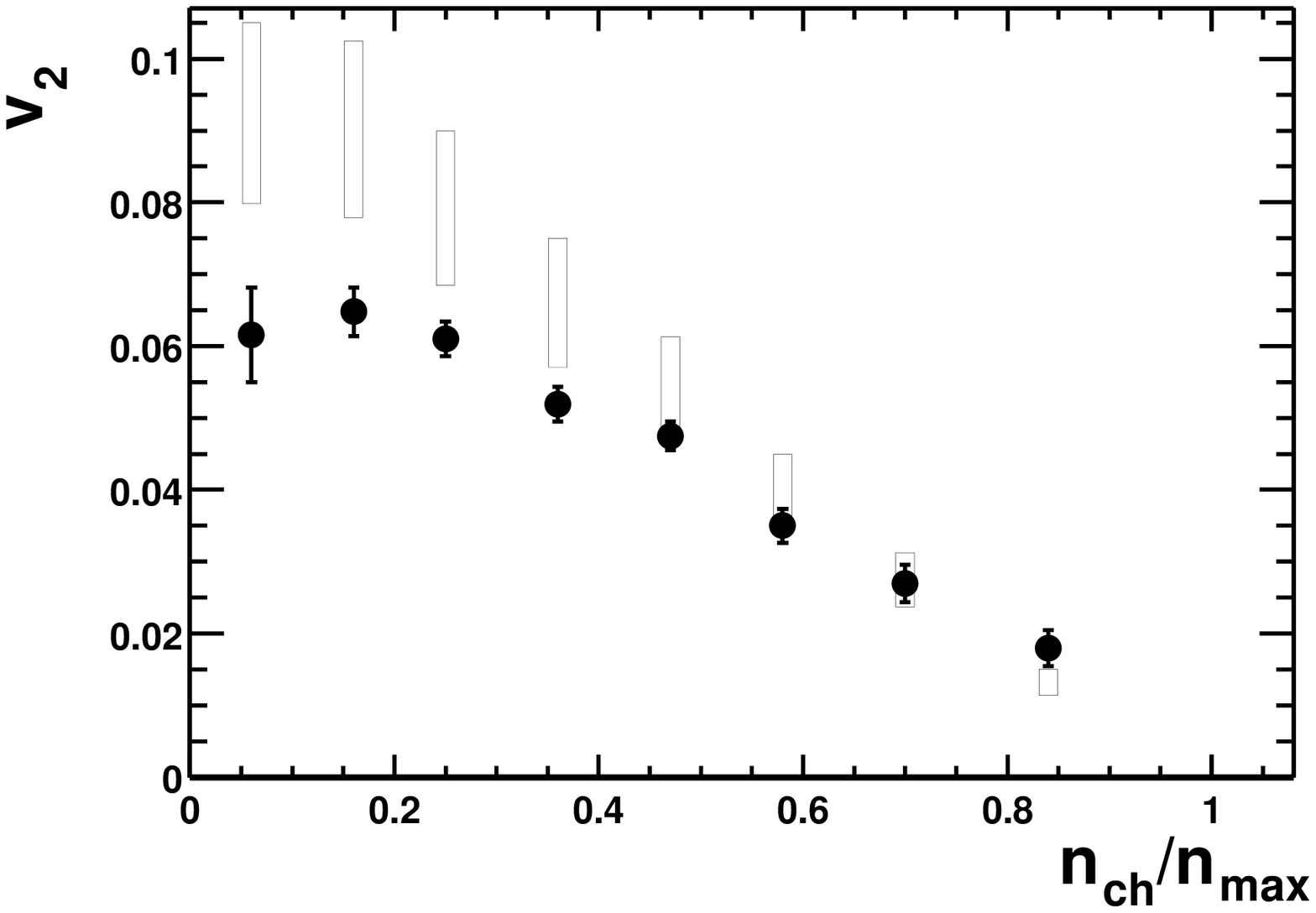,width=8cm}
\hspace{0.5cm}\epsfig{file=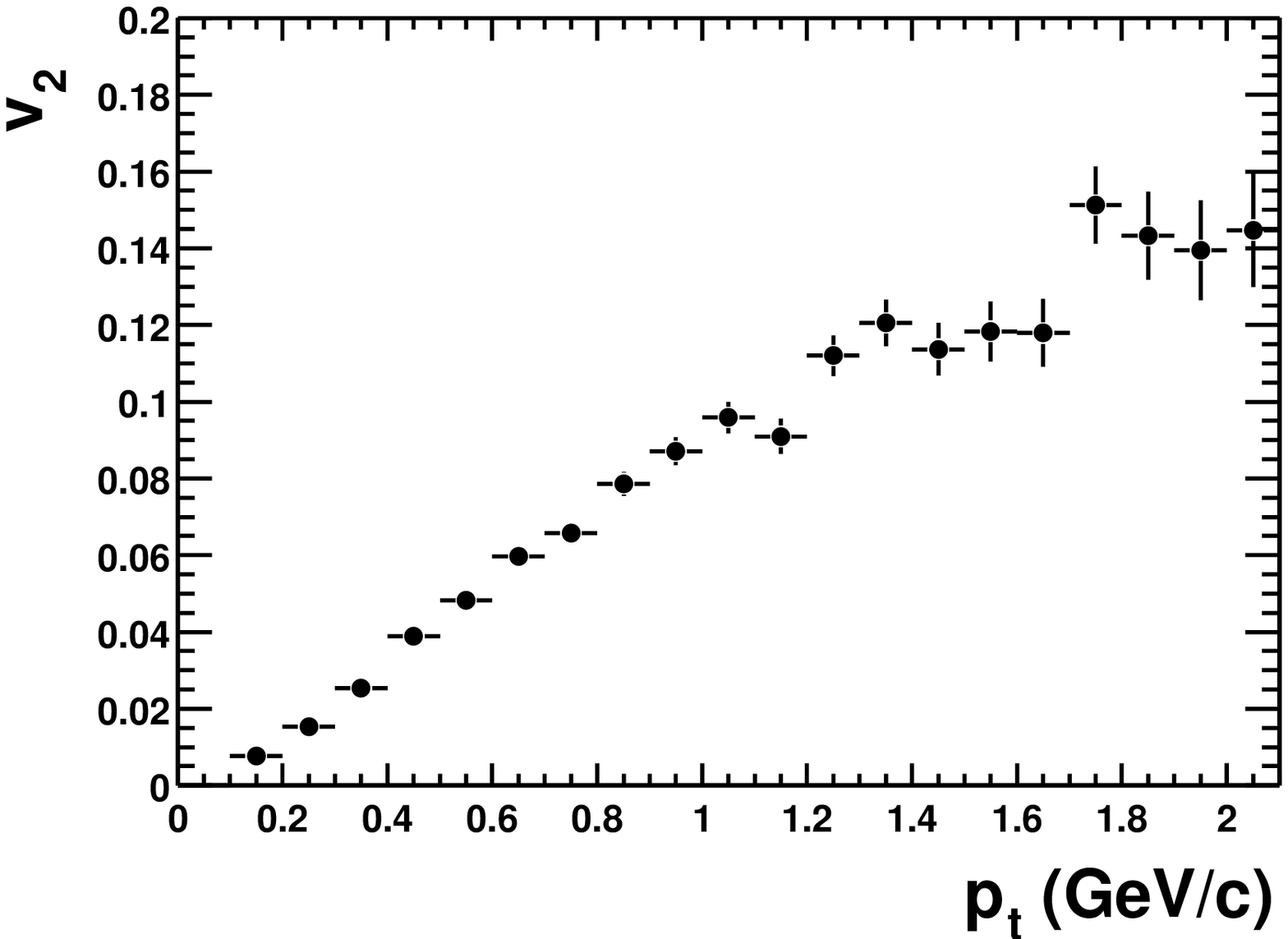,width=8cm}}
\caption[Elliptic flow as a function of centrality and transverse momenta.]
	{Measured elliptic flow at RHIC~\cite{snellings03}. Left:
Elliptic flow as a function of centrality. The open rectangles show
a range of values expected in the hydrodynamical limit. Right: Elliptic flow as a
function of transverse momentum for minimum bias events.}
\label{HIC_flow}
\efig

\subsubsection{Fluctuations}
Like any other physical measured quantities, the observables in a heavy
ion collisions are also subject to fluctuations. These fluctuations can
themselves provide useful information about the collision as they are
generally system dependent. One of these
observables is the fluctuation of certain particle ratios, as they
give access to information about the abundance of resonances at the
chemical freeze-out~\cite{becatt}. 
Furthermore, by measuring the charge fluctuations per unit degree of freedom of the
system in a heavy ion collision, one can gain knowledge whether a QGP
phase was created~\cite{jeon}. The argument is that in a QGP phase the
system would consist of quarks
and gluons which means that the unit of charge is 1/3, while in a pure
hadronic phase it will be 1. The fluctuation in the net charge depends
on the squares of the charges, and hence are strongly dependent of the
phase it originates from.


\subsection*{Electromagnetic observables}
Electromagnetic observables, like photons, carry unperturbed
information about the source in which the photons have been
produced. Since
photons are electromagnetically interacting particles, their mean free
path in the QCD medium is large enough to escape the
system without any further interaction. These so-called direct photons
provide a powerful probe of the evolution of the collision. However, the
experimental feasibility is dominated by a severe background from the
radiative decay of neutral pions
($\pi^0\rightarrow\gamma\gamma$). Results from WA98 experiment
indicates that the task of extracting the direct photons at
SPS-energies is feasible~\cite{wa98}. Recent results from the PHENIX
experiment at RHIC show a direct photon signal above the expected
background in central Au--Au events~\cite{phenixqm04}.

\subsection*{Hard probes}
\label{HIC_hardprobes}
During the initial non-equilibrated stage of a heavy ion collision at LHC,
the dynamics are dominated by hard processes within the interacting
partonic system. The study of such processes thus probes the
very early parton dynamics and the evolution of the QGP phase.
In contrast to the hadronic observables, the hard probes involve only
a limited number of energetic colliding partons, and are theoretically
treated by perturbative QCD.

\subsubsection{Jet production}
During the inter-penetration of two high energetic colliding nuclei,
the partons within the projectiles interact with each other
in hard 2 to 2 processes, and the initial parton momentum is
transferred into final state partons or
photons. Each of these final state partons will then emerge
back-to-back from the collision region and radiate energy because of
their color charges before they finally hadronize into a number of colorless
hadrons. The resulting cluster of particles is commonly referred to as jets.

High transverse energy jets produced in a heavy ion collision are 
expected to loose major
parts of their initial energy when traversing the
collisions region prior to the freeze-out phase. Studying jet
production can thus help to determine the QCD medium effects acting
on a color charge traversing a medium of color charges, in analogy to
the Bethe-Bloch physics of QED. By comparing the cross section for
jet production for that in p--p collisions at the same center of
mass energy, one can identify these medium modifications of the jet
properties which characterize the hot and dense nuclear matter in the
initial stage of the collision region. 

\bfig[t]
\insertplot{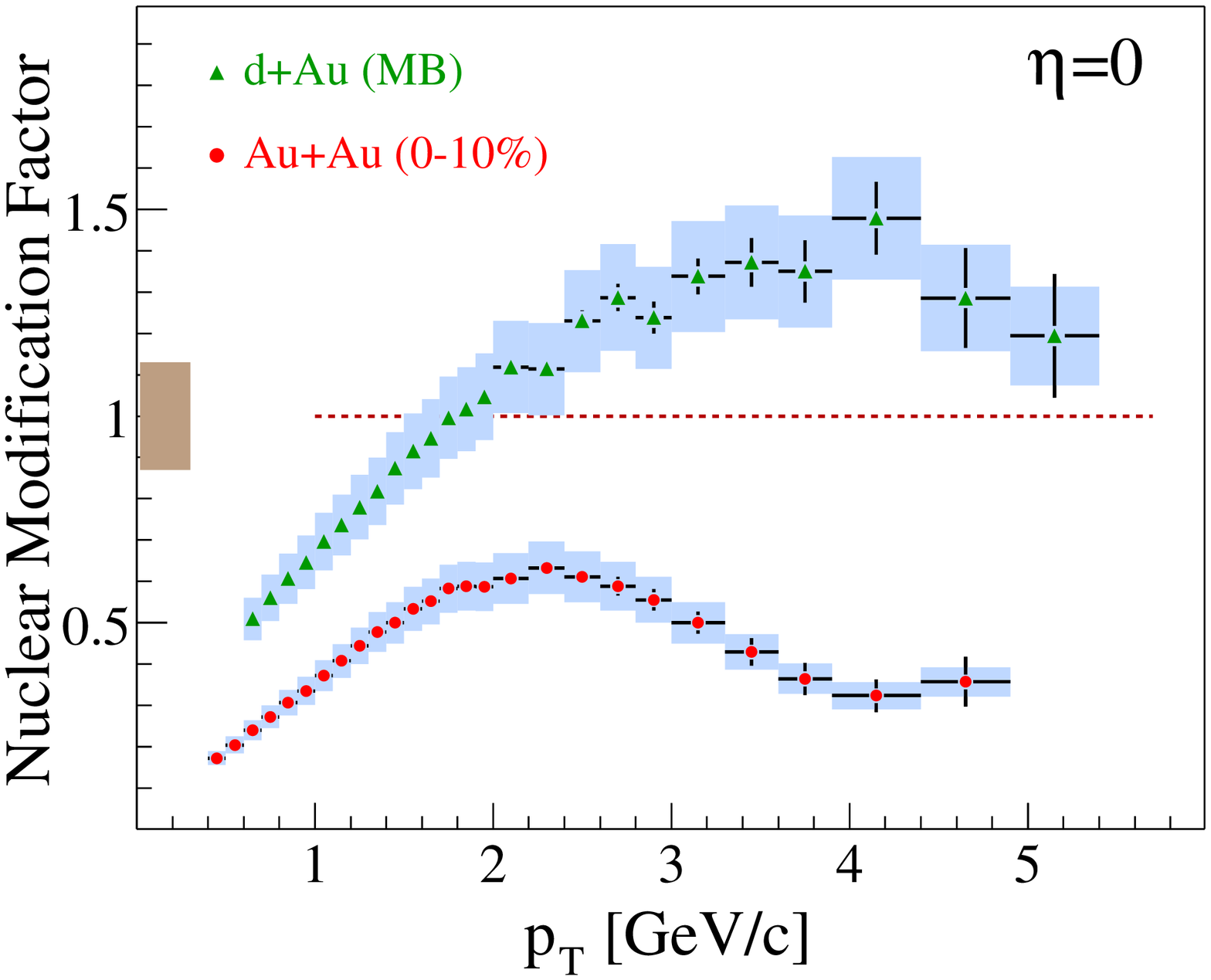}{8cm}
	{Nuclear modification factor measured for minimum biased
collisions of d--Au at $\sqrt{s_{NN}}$=200\,GeV compared to central
Au--Au collisions.}
	{Nuclear modification factor measured for minimum biased
collisions of d--Au at $\sqrt{s_{NN}}$=200\,GeV compared to central
Au--Au collisions~\cite{arsene03}. The nuclear modification factor is
defined as the ratio of the hadron yields in nucleus-nucleus
interactions and the yield in nucleon-nucleon interactions scaled by
the equivalent number of binary nucleon-nucleon collisions
to account for the collision geometry.}
\label{HIC_quenching}
\efig

Several observables has been proposed as probes for the energy loss
of the fast moving partons in the medium of deconfined color
charges~\cite{wang,wang01,doks}. In particular, this
energy loss should be visible as a 
reduced yield, or {\it quenching}, of high momentum hadrons in
central A-A collisions. This effect has indeed been observed at RHIC,
Figure~\ref{HIC_quenching}. The measurements show that central collisions
between Au--Au nuclei exhibit a very significant suppression of the
high transverse momentum component as compared to nucleon-nucleon
collisions. This observation indicates a substantial energy loss of
the final state partons or their hadronic fragments in the medium
generated by high energy nuclear collisions.

Furthermore, the production of jets has been demonstrated in angular
correlations of high transverse momentum hadrons through the
observation of enhanced correlations at $\Delta\phi\sim0$ and
$\Delta\phi\sim\pi$, Figure~\ref{HIC_dijet}. By comparing the
measurements from d--Au collisions to central Au--Au collisions one
observes a suppression of the back-to-back correlation for central
Au--Au collisions, indicating that one of the two jets is no longer present. If
this suppression would be a result of initial-state effects, it should
consequently also be observed in d--Au collisions but no such
suppression is observed. This energy imbalance thus suggests that one of
the jets which has a much longer in-medium path-length interacts with
the dense system and looses substantial amounts of its energy, which
is in agreement with the predicted jet quenching in a QGP.


\bfig[t]
\insertplot{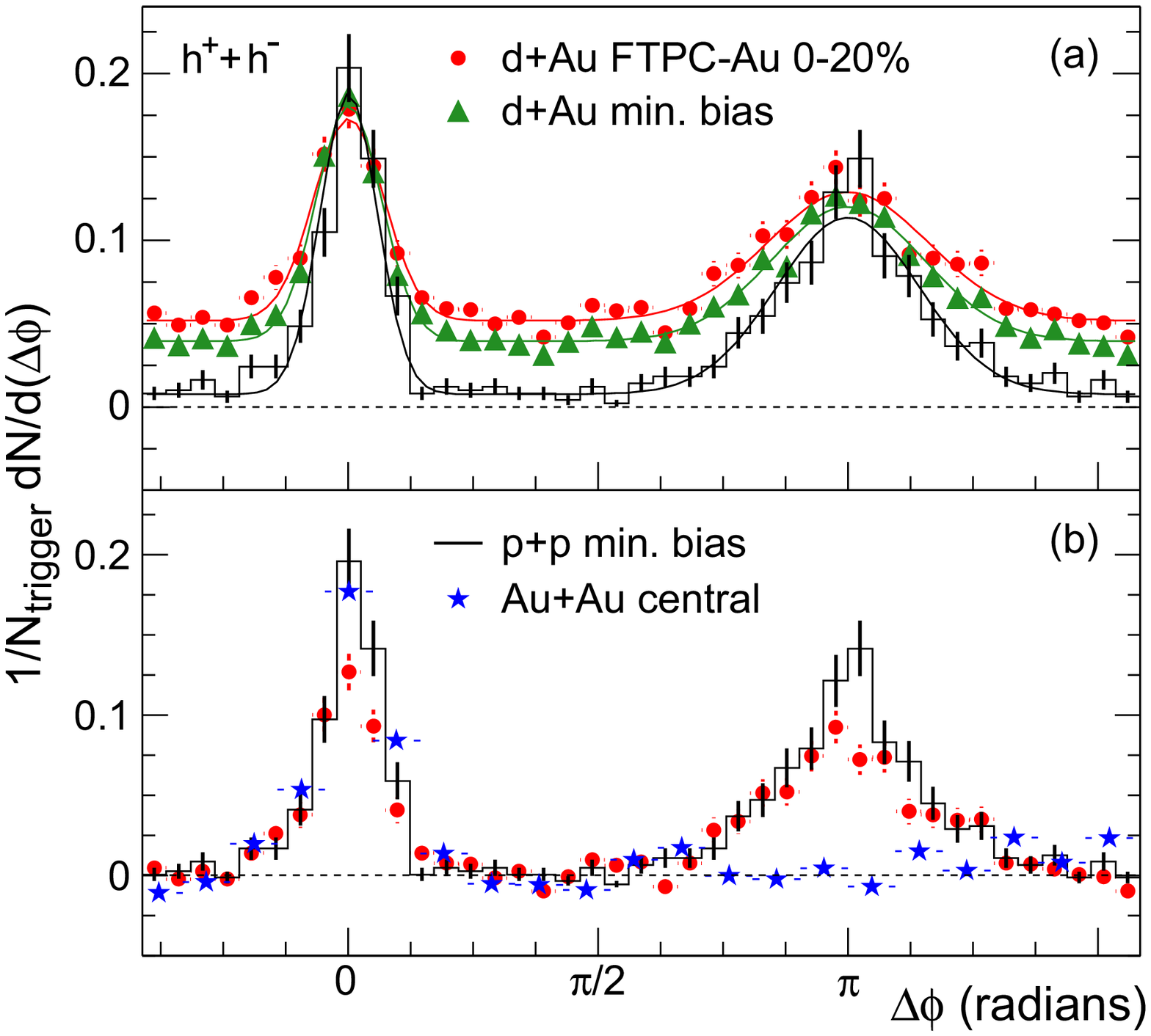}{10cm}
	{Two-particle azimuthal distributions for high transverse
momentum hadrons measured at RHIC.}
	{Two-particle azimuthal distributions for high transverse
momentum hadrons measured at RHIC~\cite{adams03}. The distributions in
d--Au collisions include a near side ($\Delta\phi\sim0$) peak similar
to that seen in p+p and Au--Au collisions and typical of jet
production, and a back-to-back ($\Delta\phi\sim\pi$) peak similar to
di-jet events seen in p+p. This second peak is suppressed in central Au--Au
collisions, indicating final state interactions with the dense system
generated in the collision.}
\label{HIC_dijet}
\efig

\subsubsection{Heavy quark production}
Heavy quarks like charm and bottom provide a probe which is highly
sensitive to the collision dynamics. Heavy quark production is an
perturbative phenomenon which takes place on a time scale of the order
of the inverse quark mass. The relative long lifetime of the
charm and bottom quarks allows them to live through the thermalization
phase of the QGP, and thereby also be affected by its presence. Also,
heavy quark-anti-quarks may form {\it quarkonium} states with binding
energies comparable to the temperature of the QGP,
implying a large quarkonium break-up and suppression. 

Typical observables including heavy quark production are the total
production rates, transverse momentum distributions and kinematic
correlations between the heavy quark and anti-quark. These observables
have to be compared to those of p--p and p--A collisions in order
extract information on the properties of the hadronic plasma. 

The observables connected to the heavy quark production will become
increasingly important at LHC energies, as the center of mass energy
will be sufficient to copiously produce the heavy quarks charm and bottom
and their bound states.

\chapter{The ALICE Experiment at LHC}
\label{ALICE}


\section{Introduction}
The LHC accelerator at CERN is scheduled for 2007. As the only
experiment build for the heavy ion program, the {\it A Large Ion
Collider Experiment} (ALICE) experiment is optimized for the study of heavy
ion collisions at the foreseen center of mass energy of
$\sim$5.5\,A\,TeV. The main
goal of this experiment is to probe in detail the non-perturbative
aspects of QCD such as deconfinement and chiral symmetry
restoration. Extrapolating from present results, all parameters
relevant to the formation of the QGP phase will be more favorable,
and in particular the energy density and the size and lifetime of the
system should all improve by an order of magnitude compared to SPS and
RHIC.

The ALICE detectors are designed to measure most of the observables
which is relevant to the formation of a QGP phase. The experimental
capabilities to measure these observables depend both on the performance of the
detectors and the number of events which can be collected. 


\section{LHC running strategy}
\label{ALICE_lhcruns}
The heavy ion program foreseen for LHC will mainly consist of two
parts~\cite{ppr}: Colliding Pb--Pb at the highest possible energy, and a more
limited systematic study of different collision systems for different
beam energies. In addition to A--A systems, both p--p and various p--A systems will
be studied in order to study the system as a function of energy
density and to provide reference data for the Pb--Pb systems.
The ALICE running program has therefore been divided
into two phases: An initial phase which is based on the current
theoretical understanding and results from RHIC, and
a second phase where a number of different running options will be
considered depending on the outcome of the initial results. 

The first data that will be taken with ALICE will be from p--p
collisions. 
The LHC will start running with several
months of proton beams, followed by the end of each year by several
weeks of heavy ion collisions. 
The effective running time per year is
expected to be 10$^7$\,s for proton and 10$^6$\,s for heavy ion
operation. During the first heavy ion run, Pb--Pb collisions at the
highest energy density is foreseen to provide global event properties
and large cross section observables. For low cross section
observables, and in particular hard processes which are the main
focus of LHC, 1-2 years of Pb--Pb runs at the highest possibly
luminosity are required to collect sufficient amount of
statistics. In the later running phase, p--Pb collision will be run in order
to provide reference data for Pb--Pb systems. Further on, energy
dependencies will be studied by using lower-mass ion systems such as
Ar--Ar.

\bfig[t]
\insertplot{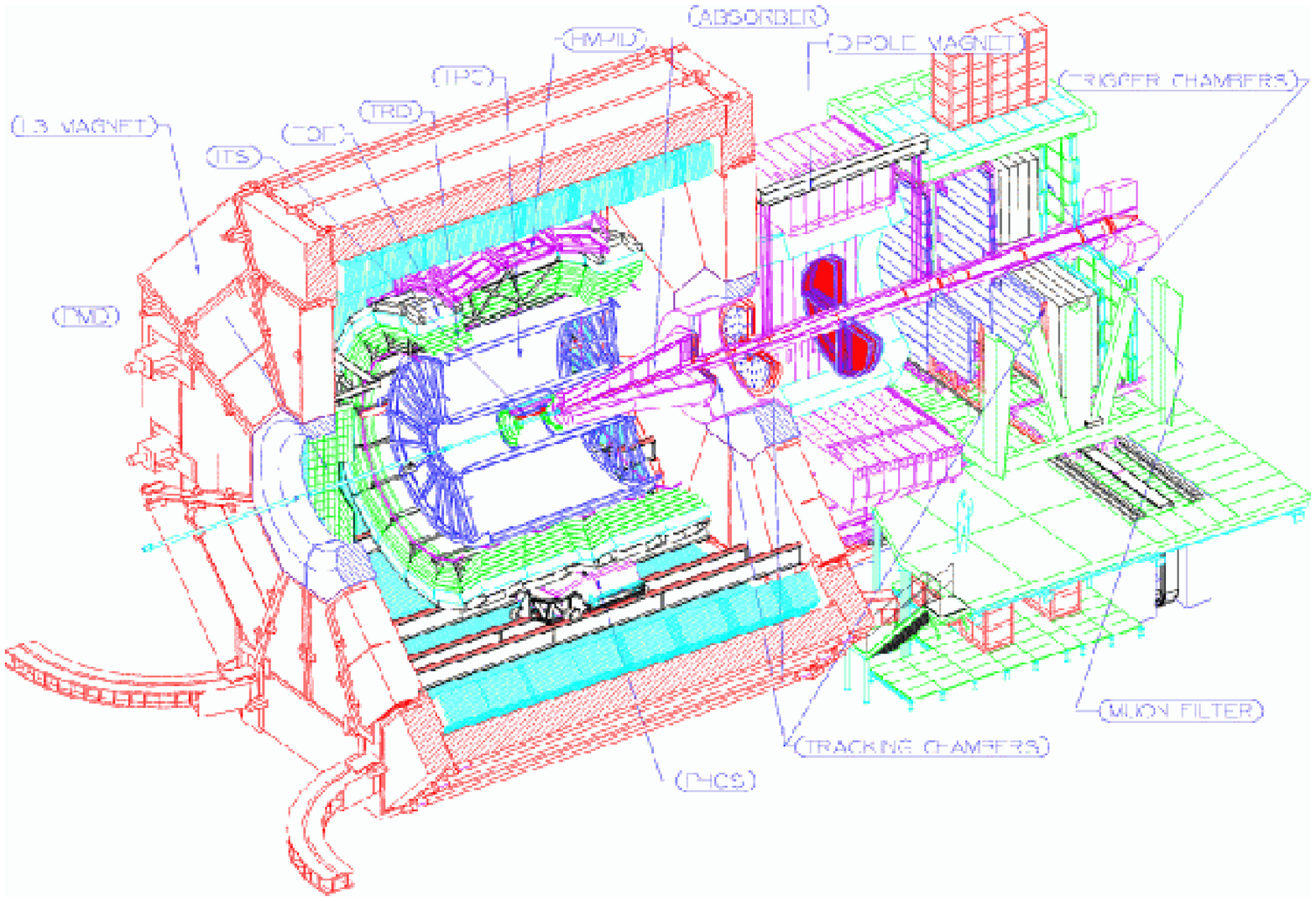}{11cm}
	{The ALICE detectors.}
	{The ALICE detectors.}
\label{ALICE_detectors}	
\efig

\section{Detector layout}
\label{ALICE_expsetup}
The complete layout of the ALICE detector as proposed initially
together with the physics objectives are
described in the ALICE Technical Proposal~\cite{alicetp,alicetp2,alicetp3}. Since then,
some of the sub-detectors have been modified to meet the new experimental
goals set by the most recent results from RHIC and latest theoretical
developments. Most of the individual sub-detectors are described in
detail in their respective technical design
reports~\cite{alicetdr,trd,tof1,tof2,pmd,muon1,muon2,its,zdc,phos,hmpid}.
In the following a brief description of the general ALICE detector layout
and a introduction to the different sub-systems, is given.

The experimental setup of the ALICE detectors,
Figure~\ref{ALICE_detectors}, is mainly composed by three parts:
\begin{itemize}
\item The central barrel which is contained in the L3 magnet. The
detectors in the central barrel region have an acceptance in
pseudo-rapidity of $|\eta|<$\,0.9 over the full
azimuth angle. These detectors will probe hadronic signals,
di-electrons and photons.
\item The forward muon spectrometer for detecting muon pairs from the
decay of heavy quarkonium in the interval 2.5\,$<\eta<$\,4.0.
\item The forward detectors, $\eta>$\,4, which will be used to determine the
multiplicity. These detectors will also be used as a fast centrality
trigger.
\end{itemize}

\subsubsection{The Inner Tracking System}
The Inner Tracking System (ITS) is designed and optimized for reconstructing
secondary vertices from hyperon and charmed meson decays, and
precision tracking and identification of low \pt particles. 
The detector consists of 6 layers of high resolution silicon
detectors, located at innermost radius 4\,cm to outermost 44\,cm. 
The different layers are designed to achieve an
impact parameter resolution of 100\,$\mu$m within the expected
particle density. Hence the innermost layers consists of pixel
detectors, silicon drift detectors for the following two, and the two
outer layers are equipped with double-sided silicon micro-strip detectors.

\subsubsection{The Time Projection Chamber}
The main tracking device in ALICE is a cylindrical Time Projection
Chamber (TPC). Its main purpose is thus to provide charge particle
momentum measurement over the central rapidity region and particle
identification via dE/dx. In addition, it will use information from
the ITS, Transition Radiation Detector (TRD) and Time Of Flight (TOF) detector
(see next section) in order to obtain a more accurate vertex
determination, particle identification and two track separation. The
TPC has an inner radius of 90\,cm which is given by the maximum
acceptable hit density, and an outer radius of 250\,cm defined by the
length required for a dE/dx resolution of $<$10\%.
The overall acceptance is 
$\vert\eta\vert<$\,0.9 and thus matches that of the ITS, TRD and TOF.

\subsubsection{Detectors for Particle Identification}
Particle identification (PID) for a large part of the phase space is
obtained by a combination of dE/dx from the ITS and TPC, and time of
flight information from the Time of Flight (TOF) detector. 

Electron identification above 1 GeV/c is provided by the Transition
Radiation Detector (TRD). The TRD will in conjunction with ITS and TPC
provide electron identification in order to measure, in the
di-electron channel, the production of light and heavy meson
resonances as well as to study the di-lepton continuum. 

For the high momentum PID a Ring Imaging Cherenkov (RICH) detector
will be used. The detector covers 5\% of the acceptance of the
central detectors, and allows PID of hadrons up to 5\,GeV.

\subsubsection{The Photon Spectrometer}
The measurement of direct photons, $\pi^0$ and $\eta$ is provided by
a high-resolution electromagnetic calorimeter, the Photon Spectrometer
(PHOS). The detector is located on the bottom of the ALICE
experimental assembly, and is built from scintillating lead-tungstate
crystals coupled with photo-detectors. The readout electronics provides
both energy and time information to reject anti-neutrons and trigger
for high \pt photons. 

\subsubsection{The Muon Arm}
The forward muon spectrometer will allow study of vector resonances
via the $\mu^+\mu^-$ decay channel. It is placed outside the L3
magnet, and consists of a composite absorber close to the interaction
point in order to reduce the $\mu$ background from $\pi$ and $K$
decays. The spectrometer magnet is a large dipole magnet with a
nominal field of 0.7\,T. Tracking is performed within 10 planes of
thin multi-wire proportional chambers with cathode readout. 

\subsubsection{The Forward Detectors}
Several smaller detectors placed in the forward region, $\eta>$4, will
be used to measure global event characteristics such as the event
reaction plane, multiplicity of charged particles and precise time of
the collision. The multiplicity information is partially used to
derive a trigger.

A set of four small and very dense calorimeters, the Zero Degree
Calorimeter (ZDC) will be used to measure and trigger on the
centrality of the collisions. 

The Photon Multiplicity Detector (PMD) is a pre-shower detector which
is mounted behind the TPC opposite to the muon arm. It will measure
the ratio of photons to charged particles, the transverse energy of
neutral particles, the elliptic flow and the event reaction plane.

The Forward Multiplicity Detector (FMD) consists of silicon pad
detectors organized in five disks which is placed on both sides of the
central detectors. It will measure the pseudo-rapidity distribution of
charged particles over a large fraction of the phase space. 

The T0 counters are 24 Cherenkov radiators and will provide the event
time with a precision of 50\,ps. The V0 counters (consisting of scintillators) will
be used as the main interaction trigger and to locate the event
vertex. 

\section{The TPC detector}
\label{ALICE_tpc}
The ALICE TPC detector is the main tracking device. It is placed
inside the homogeneous magnetic field of the L3 magnet. With almost
full three dimensional coverage, the tracking detector can provide
information of the complete particle track in addition to the
particles specific energy loss, dE/dx.
Experience from previous experiments show that TPC detectors can
handle high particle multiplicities and high track densities
(EOS~\cite{eos}, NA49~\cite{na49} and STAR~\cite{star}).

\subsection{Principle of operation}
\label{ALICE_tpcprinc}
The TPC detector is a gaseous ionization detector. This group of
detectors are sensitive to the ionization electrons and
ions that are produced when charged particles traverse the gas in the
detector.

The TPC detector consists
of a large cylindrical chamber filled with gas,
Figure~\ref{ALICE_tpclayout}. A uniform
electric field, {\bf E}, is applied and directed along the detector volume. 
When charged particles traverse the gas they will ionize the gas along
their trajectory liberating electrons. 
The liberated charge is subject to the electric field, and electrons
will drift opposite the direction of {\bf E} towards the end-caps of the
chamber where their position is detected. This then yields
the two-dimensional position of a space 
point onto the end-cap plane. The third coordinate is given by the drift time
of the ionization electrons. Since all ionization electrons created in
the sensitive volume of the TPC will drift towards the end-cap,
almost a continuous sample of space points for each track is detected
allowing a full reconstruction of the particle trajectory.
Furthermore, the charge which is collected at the end-caps is proportional to the
ionization, and thus the energy loss, of the particle. The signal
amplitudes provide information on dE/dx of the
traversing particle. In conjunction with the measured momentum
obtained from the curvature of the trajectory in the magnetic field,
this enables particle identification. 


\bfig
\insertplot{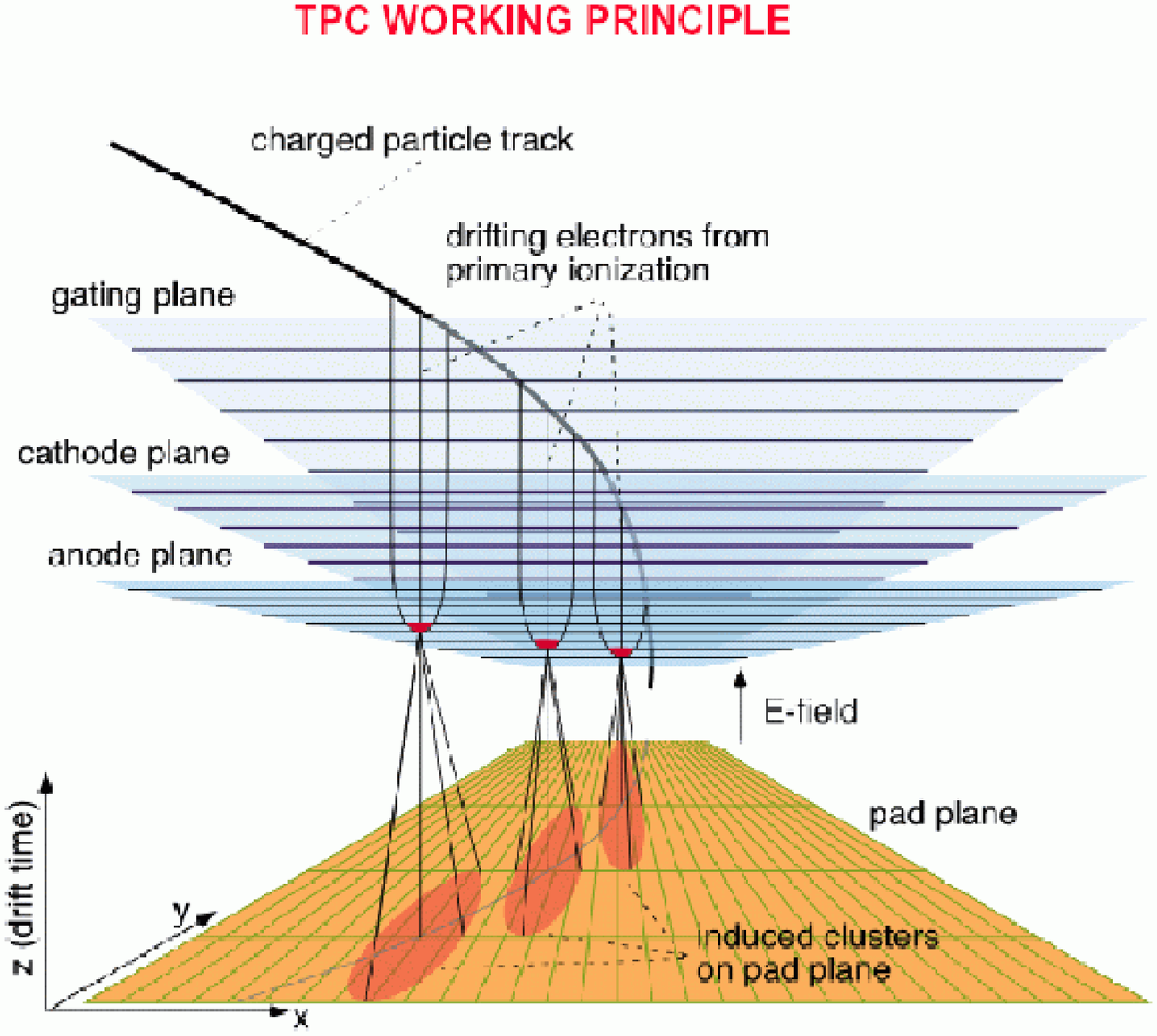}{12cm}
	{TPC principle of operation.}
	{TPC principle for detection of charged particle trajectories.}
\label{ALICE_tpcillustrate}
\efig

The detection of the drifting electrons is done by using
Multi-Wire Proportional Chambers (MWPC),
Figure~\ref{ALICE_tpcillustrate}. The primary
electrons by themselves do not induce a sufficiently large signal for
readout. The necessary signal amplification is provided by avalanche
creation in the vicinity of anode wires. The readout
chambers consist of a grid of anode wires above a
cathode pad plane, a cathode wire grid and a gating grid. A negative
voltage is applied to the cathode wires and cathode pads producing an
electric field near the anode wires
with a 1/r dependence. When the drifting electrons enters the region
behind the gating grid, they will continue to drift along the field
lines towards the nearest anode wire. Upon reaching the high field
region close to the anode wire, the electrons will be accelerated to
produce an avalanche. The positive ions liberated in the avalanche
process will induce charge on the cathode pads. This signal current is
characterized with a fast rise time and a long tail due to the motion
of the positive ions. 

The function of the gating grid is to open and close the amplification
region to the drift volume. When a trigger
signal is issued, the gating grid wires are held at the same potential, admitting the
electrons from the drift volume to enter the amplification
volume. Then, when absence of a valid trigger, the gating grid is
biased with a bipolar field which prevents the electrons from drifting
into the avalanche region. In addition, the closed gate prevents the
positive ions created in the previous event from drifting back into the
drift volume. This is important since escaping ions into the drift
volume accumulate, and can cause severe distortions of the drift field.

A precise measurement of the location of
the avalanche can be obtained if the induced signal is distributed
over several adjacent readout pads, using an appropriate
center-of-gravity algorithm. The position of the particle track in the
drift direction can be determined by sampling the time distribution of
each pad signal. The resulting two-dimensional pulse height distribution
in pad-time space is called a {\it cluster}. 




\subsubsection{Signal shape and position resolution}
\label{ALICE_signal}
The intrinsic resolution of a TPC detector is determined by the
so-called Pad Response Function (PRF). This function represent the
relative pulse height distribution of signals induced on adjacent
pads by a point-line avalanche. Its distribution is well
approximated by a Gaussian function,
\beq
P_i = C\cdot\exp\left(\frac{(x-x_i)^2}{2\sigma_{x}^2}\right),
\label{ALICE_prf}
\eeq
where $x$ is the position of the induced avalanche and $x_i$ the
respective pads.  The width of the distribution, $\sigma_x$, is
not entirely determined by the PRF.
The reason is that the drifting electrons are spread
because of diffusion when drifting towards the
end-caps\footnote{This spread is partially reduced by the parallel
magnetic field along the drift direction which confines the electrons to helical
trajectories about the drift direction.}. Thus, the distribution of the
primary electrons arriving at the anode wires cannot be considered
point-like. In addition, the finite track inclination angle with the
pad-plane spreads the ionization such that width of the initial charge
distribution represents a projection of the track segment over the
pad-length.
The resulting mean cluster width
along the pad-direction can be parameterized as~\cite{loh},
\beq
\sigma_x^2 = \sigma_{PRF}^2 + D_t^2\cdot s_{\mathrm{drift}} + \frac{l^2\cdot
\tan^2(\beta)}{12} + \frac{d^2\cdot(\tan(\alpha)-\tan(\psi))^2}{12}
\label{ALICE_clwidth}
\eeq
where $D_t$ is the diffusion constant of the gas, $s_{\mathrm{drift}}$ is the
drift distance of the drifting electrons, $l$ is the pad-length, $d$
is the distance between two anode wires and $\beta$ is the inclination
angle of the track, Figure~\ref{ALICE_pads}. 
The angle between the normal to the anode wires and the projection of
the track, $\alpha$, is in the ALICE TPC equal to
$\beta$. The Lorentz-angle, $\psi$, is defined as
the angle between the electric field and the
drift-velocity. This angle applies near the anode wires where the
electric and magnetic fields are no longer parallel, leading to a
displacement of the drifting electrons.

\bfig[htb]
\insertplot{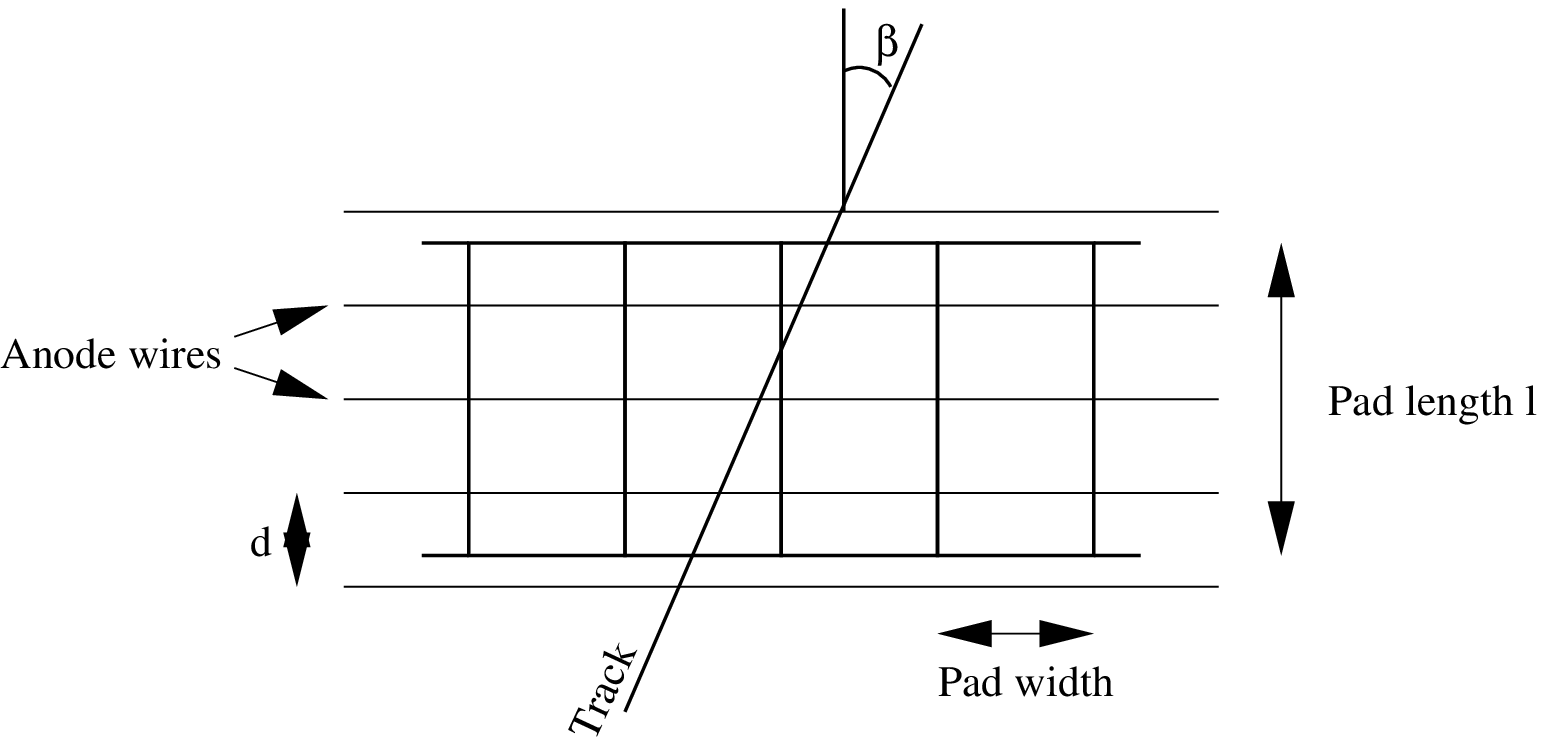}{9cm}
	{Definition of the track inclination angle in the ALICE TPC.}
	{Definition of the track inclination angle in the ALICE TPC.}
\label{ALICE_pads}
\efig



In the longitudinal direction, the width of a pad signal generated by
a single electron avalanche is given by the shaping constant of the
readout electronics. The time signal is obtained by folding
the avalanche with a Gaussian shaping function. 
Also in this direction the electron
distribution suffers from diffusion and track inclination,
and similar to Equation~\ref{ALICE_clwidth} the mean cluster width in the
longitudinal direction can be parameterized by
\beq
\sigma_L^2 = \sigma_0^2 + D_l^2\cdot s_{\mathrm{drift}} +
\frac{l^2\cdot\tan^2(\lambda)}{12},
\label{ALICE_tclwidth}
\eeq
where $D_l$\ is the longitudinal diffusion constant, and $\lambda$ is
the inclination angle of the track in the drift direction.

The cluster widths are subject to fluctuations, which depends on the
contribution of the random diffusion and the angular spread, and on
the gas gain fluctuation and secondary ionization. Furthermore,
deviations from the Gaussian shape of the clusters may occur as a
result of asymmetric distribution of the electron cluster.

The accuracy in which the centroid of the cluster can be
determined is limited by the spread of the ionization and the
subsequent diffusion
which amplifies this spread. Similar to the widths of the cluster, the
resolution therefore also depends on the track inclination angles and
the drift distance, and is theoretically given by~\cite{loh}:
\begin{eqnarray}
\delta_x^2 &=& \delta_{x,0}^2 + \frac{D_t^2\cdot s_{\mathrm{drift}}}{l\cdot n_e} +
\frac{l^2\tan^2(\beta)}{12\cdot n_{\mathrm{eff,pad}}} +
\frac{d^2\cdot(\tan(\alpha)-\tan(\psi))^2}{12\cdot
n_{\mathrm{eff,wire}}\cdot n_{\mathrm{sense}}}\nonumber\\
\delta_y^2 &=& \delta_{y,0}^2 + \frac{D_l^2\cdot
s_{\mathrm{drift}}}{l\cdot n_e} + \frac{l^2\tan^2(\lambda)}{12\cdot n_{\mathrm{eff,pad}}}
\label{ALICE_residual}
\end{eqnarray}
Here, $n_e$ is the number of ionized electron-ion pairs per cm track length,
$n_{\mathrm{eff,pad}}$ is the number of primary electrons in the gas
``column'' below the pad, $n_{\mathrm{eff,wire}}$ is the number of
primary charged units along an anode wire on which the cluster charge
is deposited and $n_{\mathrm{sense}}$ is the number of anode wires
crossing the pad. 

\subsubsection{Occupancy}
The performance of a TPC depends highly on the detector occupancy. The
occupancy is a measure 
of the track density with respect to the intrinsic detector
resolution. In general, one can define it as the probability of having
a signal above threshold,
\beq
O = \frac{N_{\mathrm{above}}}{N_{\mathrm{all}}},
\eeq
where $N_{\mathrm{above}}$ is the number of signals above threshold and
$N_{\mathrm{all}}$ is the total number of time-bins. 
For the TPC, the signals are the bins in pad-row-plane. 
The number of active signals is a function
of the particle density, $F$,$\ $ and on the effective cluster area,
$s_{\mathrm{eff}}$, and can be expressed as 
\beq
O = 1 - \exp(-F\cdot s_{\mathrm{eff}})
\eeq
Since the occupancy is a function of effective cluster area, an
optimization of the detector parameters in terms of cluster widths is
necessary. As described above the cluster widths in general depends on
diffusion, response functions and on the pad-length. For a given gas
and drift field the diffusion is no longer a variable factor, and the
cluster size is thus determined by the geometry of the pad.

\subsection{Detector layout}
\label{ALICE_tpcdesign}
The overall design of the ALICE TPC detector consists of a cylindrical
chamber with an inner radius of $\sim$90\,cm, an outer radius of $\sim$250\,cm,
and a length of $\sim$500\,cm, Figure~\ref{ALICE_tpclayout}. A thin high
voltage electrode divides the
cylinder in two, and provides a uniform electric drift towards the
end-caps. The readout chambers which cover the
end-caps of the TPC cylinder, consist of conventional MWPC
with cathode pad readout. The azimuthal segmentation of the
readout plane follows that of subsequent ALICE detectors, which leads
to 2$\times$18 (both sides of the TPC) trapezoidal sectors, each
covering 20$^{\circ}$ in azimuth. 
\bfig
\insertplot{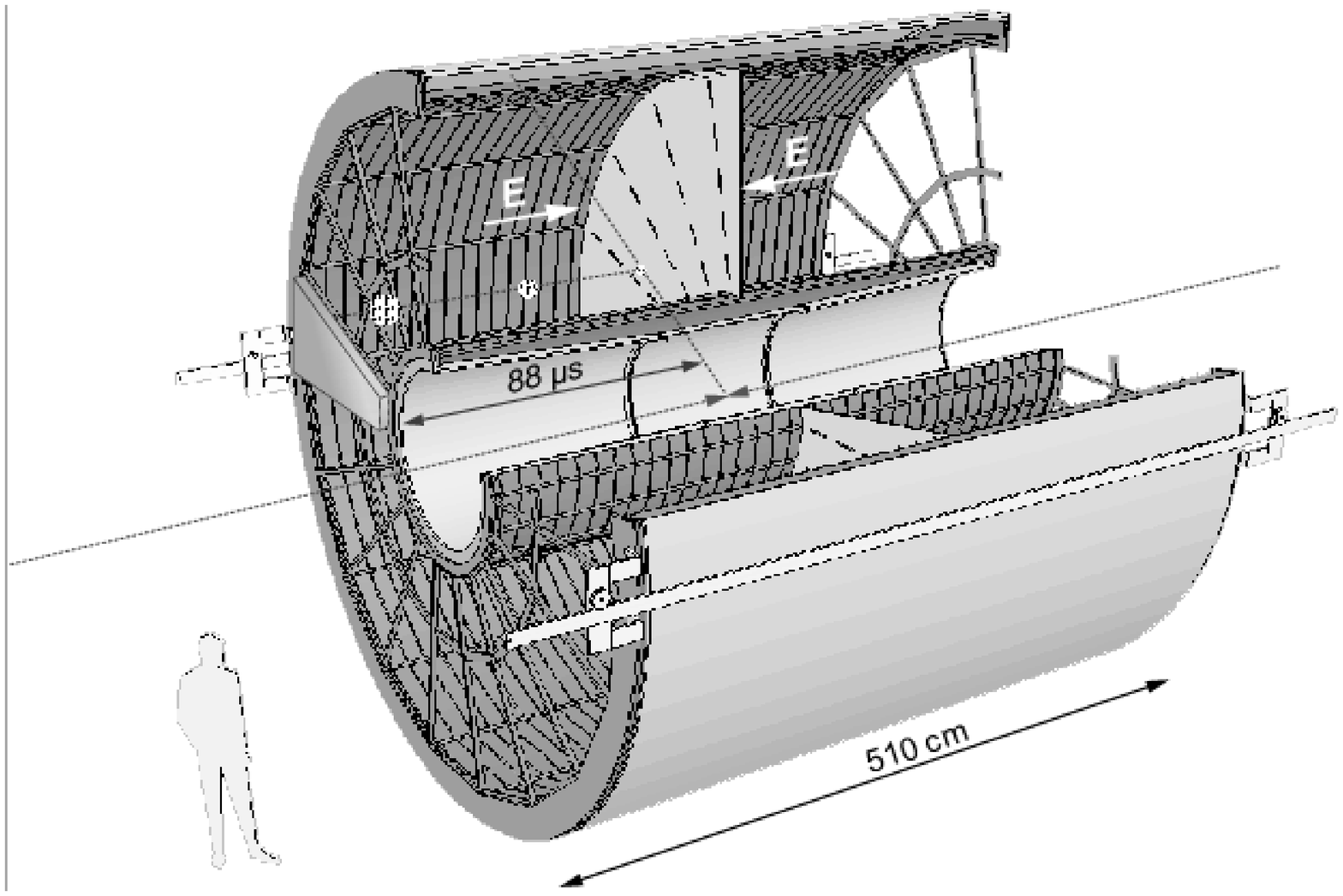}{12cm}
	{ALICE TPC schematic layout.}
	{ALICE TPC schematic layout.}
\label{ALICE_tpclayout}
\efig
\noindent The overall acceptance covered by the TPC is $|\eta|<$\,0.9 for full
radial track length, and to about $|\eta|<$\,1.5 for reduced track
lengths and poorer momentum resolutions.

The detector parameters are chosen to minimize the detector occupancy
under the expected high track density. Based on calculations of the 
PRF for different pad and wire geometry, a rectangular pad shape has
been chosen for the ALICE TPC, Figure~\ref{ALICE_pads}. Because
of the cylindrical volume of
the ALICE TPC, the track density has a radial dependency and is
proportional to $r^{-2}$. This leads to different requirements for the
pad sizes and the corresponding readout chambers as a function of
distance from the interaction point. 
Therefore, the readout is segmented radially into two separate
readout chambers with slightly different
wire geometry and pad sizes. In total there are 557\,568 readout pads of
three different sizes, Table~\ref{ALICE_tpcparams1}. 
\begin{table}
 \begin{center}
   \begin{tabular}{|l|c|c|c|}
    \hline 
     &\multicolumn{1}{|l|}{\bf Inner chambers} &
\multicolumn{2}{|c|}{\bf Outer chambers} \\ \hline
     Pad size  & 4$\times$7.5\,mm & 6$\times$10\,mm & 6$\times$15\,mm\\
     Total number of pad-rows & 63 & 64 & 32\\
     Total number of pads    & 5504 & 4864 & 5120\\ \hline
    \end{tabular}
    \caption[ALICE TPC design parameters of the readout chambers.]
            {ALICE TPC design parameters of the readout chambers~\cite{ppr}.}
 \label{ALICE_tpcparams1}
 \end{center}
\end{table}
\noindent The radial distance of the active area is
from 84.1\,cm to 132.1\,cm and from 134.6\,cm to 246.6\,cm for the inner
and outer chambers respectively, while the total area is 32.5\,m$^2$.

The drift gas is optimized for drift speed, low diffusion, low
radiation length and hence multiple scattering, small space-charge
effect and aging properties. The parameters related to the gas and the
drift field is listed in Table~\ref{ALICE_tpcparams2}.

\begin{table}
    \begin{center}
    \begin{tabular}{lc}
    \hline
    Detector gas & Ne/CO$_2$ (90/10)\\
    Gas volume   & 88\,m$^3$\\
    Drift length & 2$\times$250\,cm\\
    Drift field  & 400\,V/cm\\
    Drift velocity & 2.84\,cm/$\mu$s\\
    Drift time     & 88\,$\mu$s\\
    Total HV       & 100\,kV\\
    Diffusion      & 220\,$\mu$m/$\sqrt{\mathrm{cm}}$\\
    \hline 
    \end {tabular}
    \caption[ALICE TPC design parameters of the gas volume.]
            {ALICE TPC design parameters of the gas
volume~\cite{ppr}. The choice of gas mixture is currently under
discussion, and may also include N$_2$.}
    \label{ALICE_tpcparams2}
\end{center}
\end{table}


\subsection{Readout}
\label{ALICE_tpcreadout}
The front-end electronics of the detector is responsible for reading
out the charge induced on each of the cathode pads. Each of these
readout channels is comprised of three basic units~\cite{musa}: A charge sensitive
PreAmplifier/ShAper (PASA), a 10-bit Analogue to Digital Converter
(ADC), and a digital signal processing circuit. 

The charge induced on a pad is amplified and integrated by the
PASA. A single channel is designed to have a noise value
(r.m.s.) $\leq$1000$e$. Immediately after the PASA, the 10 bit ADC
samples the signal at a rate of 5-6\,MHz. The digitized signal is
then processed by a set of circuits contained in a single chip named
ALTRO (ALice Tpc ReadOut). Each ALTRO contains 16 channels that operate
concurrently to digitize and process the input signals. Baseline
shifts due to signal pile-ups are removed.
After the processing, the ALTRO chip performs zero-suppression. 
Zero-suppression means that a base-line ADC-value corresponding to 2-3
times the RMS-value above noise is subtracted to
correct for signal baseline instabilities. Hence, ADC-values smaller
than a preset constant threshold value are rejected.
In addition, a filter checks for a consecutive
number of samples above the threshold in order to identify the
sequences corresponding to the pulses. The zero-suppressed data are
then formatted into 32-bit words according to a back-linked data
structure. The ALTRO also contain a multiple-event buffer for storing
trigger-related data. When a L1
trigger signal is received (Section~\ref{ALICE_triggersystem}),
the data is stored in memory, and upon arrival of the second level
trigger (L2 accept or reject) the latest event in the data stream is either
frozen in the data memory until complete readout takes place, or
discarded. 


The complete readout chain is contained in the Front-End Cards (FEC)
plugged into crates and attached directly to the detector. Each FEC
contains 128 channels and is connected to the cathode readout plane by
means of 6 cables. A number of FECs are controlled by a Read Control
Unit, which interface the FECs to the DAQ, the trigger
system and the Detector Control System (DCS). The data is shipped to
the DAQ using optical fibers called Detector Data Link (DDL). 
Each of the 36 TPC sectors are read out by 6 RCUs and 6 corresponding
DDLs. 


\section{Data volumes and data-acquisition}

\subsection{Data rates}
\label{ALICE_datasizes}
The data rate produced by the detectors is a function of both event
rate and event data size. The event rate is given by the running
luminosity, while the event data size is defined by the granularity of
the detectors and the particle multiplicity. The maximum usable luminosity is
limited by both the LHC 
accelerator and the detector dead times. Given the amount of readout channels
the biggest amount of data is by far produced by the TPC detector. 

\subsubsection{Event rates}
From a detector point of view, the maximum usable luminosity is
limited by the the time it takes to read out the detectors.
In particular, the TPC detector, which is the slowest
detector, needs $<$90\,$\mu$s for the electrons to drift to
the end-caps. If the luminosity is high enough, additional events may
occur within TPC frame during readout causing several superimposed
events which are shifted in the time direction. These {\it
pile-up}\ events will contribute to the track 
density and the detector occupancy, and consequently may lead to a
loss in tracking performance.
At an average luminosity of 10$^{27}$\,cm$^{-2}$\,s$^{-1}$, the
minimum bias rate for Pb--Pb is 8\,kHz for a hadronic interaction cross
section of 8\,barn, giving a probability of having a double event
within the TPC frame of 76\%~\cite{ppr}.
The remaining ``single'' minimum biased Pb--Pb event
rate is thus limited to 2\,kHz, and the central event rate to
200\,Hz. 


In the case of p--p runs the situation is different. A single p--p
event has a very low multiplicity compared to
Pb--Pb, thus the TPC can tolerate several pile-up events without
suffering any significant loss of tracking performance. In order to
keep the pile-up at an 
acceptable level, the luminosity during p--p runs will be limited to 
$\sim$3$\times$10$^{30}$\,cm$^{-2}$\,s$^{-1}$ which corresponds to an
interaction rate of $\sim$200\,kHz. At this rate, there will on
average be $\sim$25 piled-up events in the TPC. 

Furthermore, the maximum possible event rate for both minimum biased
Pb--Pb and p--p interactions is limited by the maximum TPC gating frequency to
approximately 1\,kHz. Considering the luminosity, event pile-up conditions and the
maximum TPC gating rate, estimates of the maximum event
rates can be obtained, Table~\ref{ALICE_rates}.

\subsubsection{Event sizes}
The event sizes for Pb--Pb interactions are directly proportional to
the multiplicity produced in the collision. This makes them very difficult to
calculate as the multiplicities are hard to predict.
For Pb--Pb collisions at 5.5\,TeV predictions range
from 2000 to 8000 particles per unit rapidity for central Pb--Pb
collisions at LHC~\cite{ppr}, while extrapolation from RHIC gives
values around 2000-3500 (Section~\ref{HIC_partmult},
page~\pageref{HIC_partmult}). During the
design of ALICE the value
8000 was used as a baseline in order to provide a safety margin
on the detector performance. 

Simulations~\cite{alicetdr}
indicate that the average TPC occupancy will be about 25\% for the
highest multiplicity. 
Multiplying the number of readout channels with the number of time-bins
and taking into account the 10 bit ADC dynamic range, this 
leads to an event size directly at the detector readout of $~$350\,MB. 
The ADC conversion gain is typically chosen so that
$\sigma_{\mathrm{noise}}$ corresponds to one ADC count. This means that the
relative accuracy increases with the ADC-values, and is not needed for
the upper part of the dynamic range. The ADC-values can therefore be
compressed non-linearly from 10 to 8 bits leading to a constant
relative accuracy over the whole dynamic range. 
By compressing the ADC-values from 10 to 8 bits, the event size will be
reduced to about 290\,MB. In addition, since it is problematic to
resolve individual tracks that have a low \pt and cross the TPC volume
under small angles relative to the beam axis, a 45$^{\circ}$ cone will
be cut out of the data resulting in the rejection of all particles
which are not in the geometrical acceptance of the outer
detectors. This will reduce the data size further by a factor of
40\%. Finally, after zero-suppression the raw event size is expected to
be $\sim$75\,MB.
If running at the central Pb--Pb interaction rate of 200\,Hz, this
corresponds to a TPC data rate of $\sim$15\,GB/s. 

Regarding p--p interactions, the estimated TPC event size for a single
p--p collision is approximately 60\,kB. In this case one also
has to take into account the additional data coming from the pile-up events.
The total data volume, including the piled up events, is
estimated to be of the order of 2.5\,MB. This event size is
estimated assuming a coding scheme for the TPC data well adapted to a
low occupancy and without any data compression. If running at the
foreseen maximum TPC rate of 1\,kHz, this would produce a total data
rate $\sim$2.25\,GB/s. 
In Table~\ref{ALICE_rates} the expected event and data rates for the
different interactions are summarized.
\begin{table}
    \begin{center}
    \begin{tabular}{|c|c|c|}
    \hline
    {\bf Collision} & {\bf Event rate} & {\bf Data rate (approx.)}\\
    \hline
	p--p & 1\,kHz & 2.25\,GB/s\\
	Min. bias Pb--Pb & 1\,kHz & 22\,GB/s \\
	Central Pb--Pb & 200\,Hz & 15\,GB/s \\
    \hline
    \end {tabular}
    \caption[Expected ALICE event and data rates for the different LHC runs.]
            {Expected ALICE event and data rates for the different LHC
runs~\cite{ppr}.}
    \label{ALICE_rates}
    \end{center}
\end{table}

\subsection{The trigger system}
\label{ALICE_triggersystem}
As described in Section~\ref{ALICE_lhcruns}, the ALICE experiment will
operate under different beam conditions. The
trigger system is responsible for selecting the different types of
events and enable readout of the detector when certain criteria are
met. The ALICE trigger system is foreseen to operate in three different
levels~\cite{alicetp}: Level 0 (L0), Level 1 (L1) and Level 2
(L2). These different levels correspond to criteria imposed from 
different detectors, where the selection criteria gets stronger as
the trigger number increase. Correspondingly, the rates at which each trigger
level is operated decreases at higher levels.

The L0 and L1 trigger are both fixed-latency triggers, which means
that their rate is constant. The main
difference between the two is that the different detectors need
trigger decision to strobe the electronics at different times after
the interaction. The main task of the L0 trigger is to signal that an
interaction has taken place at the earliest possible time, which is after
about 1.2$\mu$s\footnote{The L0 latency is an estimate based on the
expected transmission time in the cables}. This trigger is based
exclusively on the information from the T0 and V0 counters and checks for
the following features:
\begin{enumerate}
\item The interaction vertex is close the the nominal collision point.
\item The forward-backward distribution of tracks is consistent with a
colliding beam interaction.
\item The measured multiplicity is above a given threshold.
\end{enumerate}
No strong centrality condition is made at L0, as non-central events
giving di-muon triggers are also required. At L1 decision, which is
made after about 6.5$\mu$s\footnote{The L1 latency is estimated from
the expected time is takes for the muon arm and the ZDC to issue a
trigger signal, including a safety margin of $\geq$20\%}, more stringent
centrality requirements are made. Its selection is based on
information from the muon system, PHOS, and on the centrality detectors, FMD
and ZDC. At this time all the remaining detectors are strobed. In
particular, the TPC gate is opened which leads to the requirement that
the L1 trigger can have a maximum frequency of 1\,kHz. 

During the drift time of the TPC ($\sim$100$\mu$s) the L2 decision
is made. Based on the data extracted from the different trigger
sub-detectors, more selective algorithms are applied (e.g. a mass cut
on the di-muon system). Also, during this time a reset can be issued as
a result of pile-up events (only when running with Pb--Pb
interactions). 
Since the selection algorithms will differ in processing time,
the latency of the L2 trigger is not fixed, but has an upper bound as
defined by the TPC drift time.
After the L2 trigger, the data are all read out from the front-end
electronics into the DAQ and High Level Trigger system. 

\subsection{The DAQ system}
\label{ALICE_daq}
The data acquisition system~\cite{hltdaqtrig} is responsible for collecting the data from
all the sub-detectors and assemble the sub-event data blocks into full
event before sending the data to mass storage. 
The architecture of the
system is based on PCs connected by a commodity network, most likely
TCP over Gigabit Ethernet. 
The data transfer from
the front-end electronics of the detectors are initiated by a L2
trigger accept. The data is then transferred in parallel from all
sub-detectors using special optical links, called Detector Data Link
(DDL), into the Local Data Concentrators (LDC) where the sub-event
building takes place. Parallel to the DAQ, the data is also shipped to
the High Level Trigger system by duplicating the data stream
(Figure~\ref{HLT_flowscheme}). The sub-events
prepared by the LDCs are transferred to one Global Data Concentrator
(GDC) where the full event can be assembled. The event building is
managed by the Event Building
and Distribution System (EDBS), which is a protocol running on all the
machines (LDCs and GDCs). A GDC destination for a particular event is
determined by the EDBS which communicates this decision to the LDCs. 
The fully assembled events are finally shipped to permanent storage
for archiving and further offline analysis.

The DAQ system is designed to be flexible in order to meet the requirements
for the different data taking scenarios. As the p--p interaction produce only
data rate of 1/5 relative to Pb--Pb interactions, the requirement on
the system is defined by the expected data rate 
from the heavy ion runs. In the heavy ion mode, two main types of
events have to be handled. The first consists of central Pb--Pb events
at a relatively low rate but with a large event size. The second one
concerns the events containing a muon pair which has been reported by
the trigger and is read out with a reduced detector subset, including
the muon arm. Much higher trigger rates are required in the latter
case, typically up to 1\,kHz. 


\subsection{The High Level Trigger}
\label{ALICE_hlt}
In the ALICE Technical Proposal~\cite{alicetp}, the collaboration
estimated that a bandwidth of 1.25\,GB/s to mass storage would
provide adequate physics statistics. As seen from
Table~\ref{ALICE_rates} the expected data rate from the detector
exceeds this number by an order of magnitude. This has lead to the
proposal and inclusion of the ALICE High Level Trigger (HLT) system.
The task of this system is to reduce the data rate to an
acceptable level in terms of DAQ bandwidth and mass storage costs, and
at the same time provide the necessary event statistics. This is
accomplished by performing online processing of the data, allowing
partial or full event reconstruction in order to select interesting
events or sub-events, and/or to compress the data efficiently using
data compression techniques. Processing the detector information at
a bandwidth of 10-20\,GB/s requires a massive parallel computing
system. 
The functionality and architecture of the HLT system are
topics of Chapter~\ref{HLT}.


\chapter{The ALICE High Level Trigger System}
\label{HLT}

\section{The necessity of a High Level Trigger}
The ultimate goal of the ALICE detector is to detect and investigate
the QGP phase of nuclear matter. This task can only be solved by a coherent
measurement of a wide range of observables from both peripheral to central
heavy ion collisions. An essential part of this measurement is the
collection of enough events in order to obtain sufficient statistics
for the physics analysis. On the other hand, hard processes such as
heavy quarkonium and jet production corresponds to relatively small
cross-sections, and consequently one needs to consider a large number
of events to provide the adequate statistics for these observables.
The systematic analysis of such hard signals therefore calls for
running the detector at the full available luminosity, which makes it
necessary to consider all interactions at the full available central
event rate of 200\,Hz. However, as described in Section~\ref{ALICE_datasizes},
the foreseen data rate in such a scenario exceeds the planned mass
storage bandwidth by an order of magnitude. 
It is therefore necessary to introduce some kind of online data
reduction into the output data stream. Such a reduction should as far
as possible reduce the data readout rate to match the DAQ and mass
storage bandwidth, and at the same time allow ALICE to acquire
sufficient statistics for the different physic observables. This has
lead to the preparation of the ALICE High Level Trigger (HLT) system,
whose prime task will be to enhance event selectivity and/or reduction
of the event size by partial or full online event reconstruction.


\section{Functionality}
Data reduction in ALICE can in general be accomplished in two ways:
\begin{itemize}
\item[--] Event rate reduction
\item[--] Event size reduction
\end{itemize}
The first method implies that only a fraction of the available events
are sent to mass storage. This option would also be used without any
HLT system being present, as the readout
rate coming from the detectors would have to be decreased in order to
meet the data rate limitation of the mass storage. However, by introducing
the HLT system, data can be processed online and event selections may be
performed on the basis of physics observables. Thus, the introduction
of the HLT system will enable an event rate reduction
and at the same time improve the event statistics
needed for the different physic programs. 
In the latter case, selections of Region of Interest (ROI) and data
compression techniques can be used to reduce
the event size itself and thus increase the possible event rate being
sent to mass storage. 

In both cases online processing of the data is required,
requiring pattern recognition in order to reconstruct the event. In the
following the two cases will be referred to as running in trigger mode
and data compression mode, respectively. In this context trigger means
selection or rejection of events or sub-events on the basis of a
specific physics analysis. In the data compression mode, the chunk of
data representing an (sub)event is compressed by applying appropriate data
compression techniques.

\subsection{Trigger mode}
The HLT trigger running mode can be divided into two subclasses: Complete event
selection/rejection, and Region Of Interest (ROI) readout. Both of
them are based on the online identification of some predefined certain
physical event. Depending on the topology of the trigger signals,
either full or partial event reconstruction is required for this mode
of operation. 

Although the hard probes of QCD are rare, and at the same time require
high statistics for a systematic study, they provide to a large degree
the most topologically distinct tracking signatures in the
TPC. Therefore, most of the online HLT trigger algorithms will be
based on online tracking of the TPC data. Further refinement may also
result from using early time information of the ITS and TRD
systems. The different feasible trigger modes envisaged to date are
described in detail in the HLT Technical Design
Report~\cite{hltdaqtrig}. A brief summary is given in the following.

\subsubsection{Jet trigger}
The study of jet production at LHC energies is one of the interesting
probes of the strongly interacting QCD matter
(Section~\ref{HIC_hardprobes}, page~\pageref{HIC_hardprobes}), but a high
number of collected events are required in order to provide the necessary
statistics. Estimations based on scaling from p--p collisions indicate
that around 10$^8$ inspected events in the TPC are required to
collect an amount of 10$^4$ jet events with $E_t>$100\,GeV. One year of
Pb--Pb acquisitions result in about 2$\times$10$^8$ events at 200\,Hz
TPC rate, but only a fraction of these events can be written to mass
storage. Employing online jet-finder tracking algorithms within HLT,
inspection of all central Pb-Pb events at 200\,Hz is however feasible,
and will thereby enhance the yield of jet events by a factor 10.

Jets with high transverse energy ($E_t>$\,100\,GeV) have a on average
a unique charged track topology.
Furthermore, they have a sufficiently charged track multiplicity
to stand over the fluctuating mini-jet
background in central Pb--Pb collisions.
The stiff nature of these tracks and their relatively close
proximity allow for the implementation of a specific and fast local
tracking in the TPC. 

\subsubsection{Open charm trigger}
The measurement of particles carrying open charm (such as $D$-mesons)
provide a probe which is sensitive to the collision dynamics at both
short and long time scales. This observable will become increasingly
important at LHC energies, and its detection and systematic analysis
is one of the main goals of the ALICE experiment.
The physics of open charm cross-section analysis requires 20\,Hz of
central Pb--Pb for 10$^6$ seconds, i.e. one month of Pb--Pb acquisition, which
amounts to 2$\cdot$10$^7$ events~\cite{paic00}. If all of these events
should be written to tape, this would require $\sim$850\,MB/s (65\%
of the available DAQ bandwidth) for this observable alone. Thus any
means for reducing the required amount of data is desired to increase
the statistics and free bandwidth for other observables.

The open charm meson D$^0$ decays via a weak decay into kaon and a
pion with a branching ratio of 3.83\%. The resulting yield in a
central Pb-Pb collisions has been estimated to $dN(D^0\rightarrow
K\pi)/dy$=0.53.
The impact parameter of the decay
products is typically about 100\,$\mu$m. To detect these
decays one has to compute the invariant mass of tracks originating
from displaced secondary vertices. In order to reduce the
combinatoric background various kinematic and secondary vertex
topology selections has to be performed.

From a HLT point of view, the foreseen event selection strategy
proceeds in two steps: A momentum filter which reduces the data
volume, and secondly, an impact parameter analysis rejecting events with
no D$^0$ candidate. Figure~\ref{HLT_pikaon} shows the distribution of
charm meson decay products into pions and kaons together with the
background (pions and kaons from the underlying event).
\bfig
\insertplot{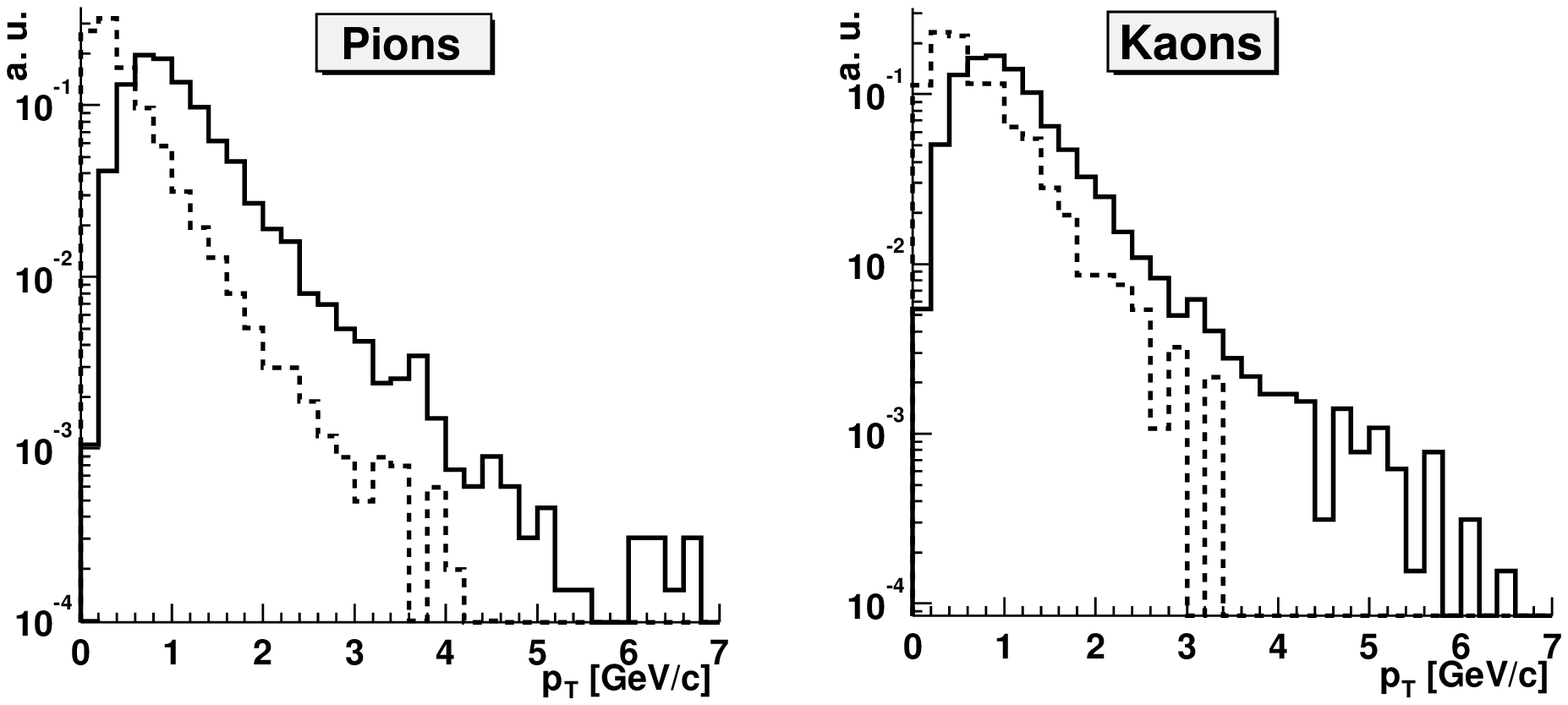}{10cm}
	{Distribution of charm meson decay products into pions and kaons.}
	{Distribution of charm meson decay products into pions and
kaons. The solid lines represent the signal while the dashed line is
the background~\cite{hltdaqtrig}.}
\label{HLT_pikaon}
\efig
\noindent For example, if the $<p_t>$ of the D$^0$ is
1\,GeV/c, the relevant decay channels, $K\pi\pi$ (for charged D), and
$K\pi^0$ (for D$^0$), have a majority of their tracks above
0.8\,GeV/c. Similarly, if the $<p_t>$ of the
underlying ``soft'' event is 0.4\,GeV/c, the fraction of tracks
with $p_t>$\,0.7\,GeV/c is about 15\% of the total\footnote{The estimate
was obtained from the \pt distribution resulting from a simulation of
central Pb-Pb event using the HIJING event generator}. Reconstructing all
tracks in the TPC online (with an emphasis on a high efficiency at
high $p_t$), and keeping only the raw data along regions revealing high
\pt trajectories, can reduce the data volume by a factor of 5-6.
Applying additional kinematical selection and secondary
vertex topology criteria would further improve the selection of
possible $D^0$ events.
Simulations show that
signal--to--event of 0.0013 and a background--to--event of
0.0116~\cite{andrea} should be obtainable in ALICE. HLT can
potentially reduce the data rate
needed for the open charm program by a factor 5-10, thus increasing
statistics and at the same time release DAQ bandwidth.

\subsubsection{Di-electron trigger}
The muon arm measures the $J/\psi$ and $\Upsilon$ spectra
via the di-muon channel. A complementary study of these particles will
be performed by reconstructing their leptonic decay into $e^+e^-$ and tracking
these di-electrons through the TPC, TRD and ITS. The TRD will trigger on
high \pt tracks by online reconstruction of particle trajectories in
the TRD chambers, and on the electron candidates by measuring of the
total energy loss and the depth profile of the deposited energy. The
true quarkonium trigger rate however is small (e.g. signal rate of $\Upsilon$
is $\approx$10$^{-2}$\,Hz) and the trigger is dominated by the
background. Depending on the set of cuts being used, a trigger rate of
di-electron pairs of 300-700\,Hz at dN$_{\mathrm{ch}}$/dy=8000 is expected~\cite{trd}.
The main contributions to the background comes from:
\begin{itemize}
\item Electron pairs from Dalitz decays of $\pi^0$, $\eta$, $\rho$,
$\omega$, $\phi$ and semi-leptonic decays of $B$ and $D$ mesons.
\item Electrons or positrons from gamma conversions, Bremsstrahlung,
and secondary interactions.
\item Pions misidentified as electrons.
\item Fake tracks from combinations of clusters from different tracks.
\end{itemize}
The HLT can be used to reject background events by two methods:
\begin{itemize}
\item Combining TRD tracklets with TPC and ITS tracking. The combined
track fit allows for a more accurate determination of the momentum
than by the TRD alone, and thus HLT will reject secondary electrons by
sharpening the momentum cut.
\item Utilizing dE/dx in the TPC. By identifying the particles
using dE/dx information from the TPC, the background from
misidentified pions can be reduced. 
\end{itemize}
Simulations indicates that event rate reduction by a
factor of ten can be achieved.

\subsubsection{Di-muon trigger}
The forward muon arm is designed to detect vector resonances via the
$\mu^+\mu^-$ decay channel, and will
run at the highest possible rate in
order to record all muons with the lowest possible dead-time.
The task of the di-muon trigger system is to select
events containing the di-muon pair from the decay of
$J/\psi$ and $\Upsilon$, where the background is mainly coming from the muons
due to $\pi K$ decays. 

The first level of the di-muon trigger consists of a transverse
momentum selection based on the information from two dedicated trigger
chambers. The trigger is optimized for two
different \pt thresholds in order to select low ($>$1\,GeV/c) and high
($>$2\,GeV/c) \pt muons
from the $J/\psi$ and $\Upsilon$ resonances, respectively. However, the
coarse-grained segmentation of these trigger
chambers does not allow a sharp $p_t$-cut, resulting in a rather large
background trigger rate. The \pt resolution can be improved
by performing an additional tracking step within
HLT using information from the muon tracking chambers, and thus achieve
higher trigger selectivity, Table~\ref{HLT_muonrates}.
\begin{table}
    \begin{center}
    \begin{tabular}{|c|c|c|}
    \hline 
		& {\bf L0} & {\bf HLT}\\
	\hline
	Low p$_t$ cut  & 2000\,Hz & 500\,Hz\\
	High p$_t$ cut & 550\,Hz &  Few\,Hz\\
   \hline
   \end {tabular}
    \caption[Expected trigger rates of the di-muon detector.]
            {Expected trigger rates of the di-muon
detector~\cite{manso02}. The two \pt-cuts correspond to the selection
of $J/\psi$ and $\Upsilon$ resonances, respectively.}
    \label{HLT_muonrates}
    \end{center}
\end{table}
The expected background rejection factor by
inclusion of HLT algorithm is 5-100.


\subsubsection{Pileup removal in pp}
In the case of p--p running, the foreseen running luminosity of
2$\times$10$^{30}$\,cm$^{-2}$\,s$^{-1}$ will result in an interaction
rate of about 200\,kHz. During the TPC drift time of
about 90\,$\mu$s, around 25 superimposed events will be captured in
the TPC frame, leading to about 95\% overhead in the data stream.
These additional piled-up events will be displaced along the
beam axis, and will not be used during offline analysis. Using HLT to
reconstruct all tracks online, the tracks corresponding to the
original triggered event can be identified while the tracks belonging
to the pile-up events can be disregarded from the readout data
stream. Simulations indicate that an overall event size reduction of
$\sim$3/25 can be achieved while retaining an efficiency of more than 95\%
for the primary tracks of the event.

\subsection{Data compression mode}
The option to compress the data online provides a method that can
improve the physics capabilities of the experiment in terms of
statistics, even without performing selective readout. If the
compression factor is high enough ($\geq$10), the full event rate can in
principle be written to mass storage. Any data compression has to be
performed with caution to assure the validity of the measured physical
observables.

The TPC detector produces by far the largest amount of data in terms
of event sizes, and any data compression scheme should therefore
be optimized to efficiently and reliably compress the TPC data.
TPC data are first compressed in the
the TPC front-end electronics by the zero suppression,
Section~\ref{ALICE_tpcreadout}. Here, pedestal subtraction (setting
threshold on the ADC
values) and identification of the sequences in time direction is performed. The
zeros between these sequences are then compressed by Run-Length
Encoding (RLE), which means that the distance between the sequences are
stored rather than storing the zeros themselves. 

The data may be further compressed by applying standard data
compression techniques such as entropy coding. 
These algorithms may be directly applied on the RLE ADC-data and allow
bit-by-bit reconstruction of the original data set. 
Since these techniques normally use some
form of coding table, they are not very computationally demanding, and can
even be performed on dedicated hardware such as Field Programmable
Gate Arrays (FPGA). Extensive studies of TPC
data in the NA49 experiment, and simulated TPC data for ALICE, show that
compression factors of $~$2 can be achieved using these
techniques~\cite{berger02}.
The most efficient data compression, however, is obtained by using
compression algorithms which are highly adapted to the underlying TPC
data. Such methods exploit the fact that the relevant information is
contained in the reconstructed cluster centroids and the track charge
depositions. These parameters can be stored as deviations from a
model, and if the model is well adapted to the data the
resulting bit-rate needed to store the data will be small. Since the
clusters in the TPC critically depends on the track parameters, the
reconstructed tracks and clusters can be used to build such
efficient data models. In contrast to the entropy coding
algorithms, such techniques do not keep the original data
unmodified as the clusters are coded rather than the ADC
data. However, from a data analysis point of view, only the effects on
the physics observables are of importance. Studies carried out in this
work indicate that compression factors of 6-10 may be achieved using
such a compression scheme. The different available TPC data
compression schemes and their performance are discussed in detail in
Chapter~\ref{COMP}.

\section{Architecture}
\label{HLT_arch}
The HLT system will have to process an expected data rate of 10-20\,GB/s.
Given this large amount of data, and the complexity of the processing
task, a massive parallel computing system is required. The HLT
system is therefore planned to
consist of a large PC cluster farm with
several hundred up to a thousand separate nodes. The architecture of such a system
is mainly driven by two constraints. Firstly, the
data has an inherent granularity and parallelism defined by the
readout segmentation of the detectors. 
Secondly, HLT is (in trigger mode) responsible for issuing a trigger decision
based on information derived from a partial or complete event reconstruction. This
means that the reconstructed data finally has to be collected at
a global layer in which the trigger algorithms are implemented. Both
of these requirements demand a hierarchical tree-like topology with a
high degree of connectivity. 

\bfig[htb]
\insertplot{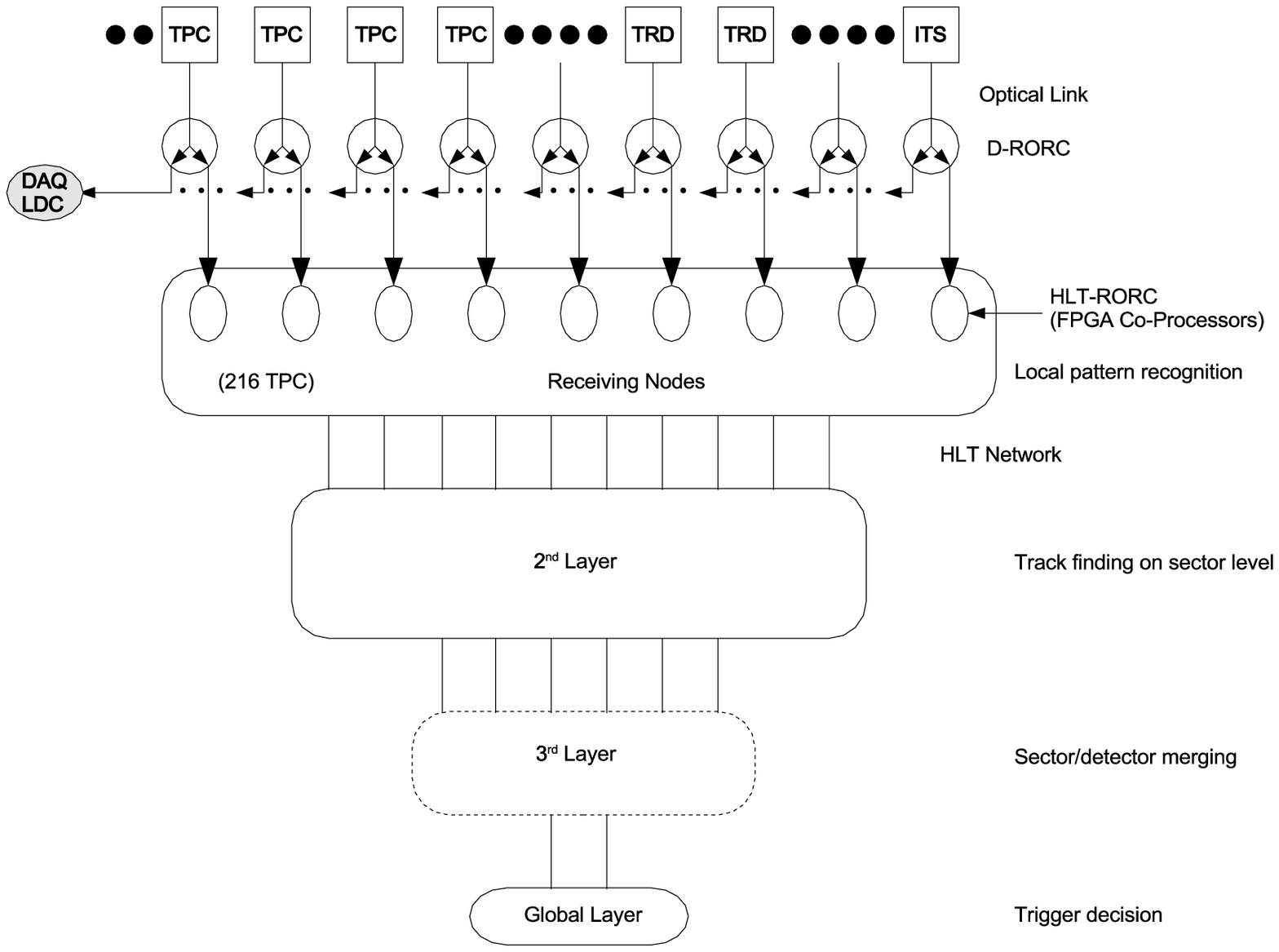}{10cm}
	{Data flow architecture of the HLT system.}
	{Data flow architecture of the HLT system~\cite{hltdaqtrig}. The detector raw
data is duplicated and simultaneously received by DAQ and HLT. The
architecture follows a hierarchical structure, adapted to the
parallelism of the data and the various tasks of the pattern recognition.}
\label{HLT_flowscheme}
\efig

In parallel to the DAQ system (Section~\ref{ALICE_daq}), the data
is received from the front-end electronics via the DDLs into the
receiving nodes of the HLT system,
Figure~\ref{HLT_flowscheme}. 
These processors constitute the first
layer of the HLT system, and are referred to as the Front-End
Processors (FEP). Each DDL is mounted on a HLT Readout Receiver
Card (HLT--RORC) which is a custom designed PCI card hosted by every
FEP. Several HLT--RORCs may be 
placed in each FEP, depending on the bandwidth and processing
requirements. Every HLT--RORC will be equipped with additional
co-processor functionality for designated pre-processing steps of the
data in order to take load off the CPUs of the FEPs.
The total number of HLT--RORCs is defined by the readout granularity of the
detectors. For the TPC detector, which is the biggest contributor, the
readout is divided into its respective 36 azimuthal sectors, where each 
sector is divided into 6 sub-sectors. Every sub-sector is read out by
one DDL, and thus there will be 36$\times$6 = 216 
DDLs for the TPC alone. Taking all detectors into account there will be a
total of about 400 DDLs.

\subsubsection{Data-flow}
All detectors will ship their data upon a receipt of a
L2--accept trigger distributed by the central trigger processor
(Section~\ref{ALICE_triggersystem}). Before
that time, the data remains within the domain of the front-end
electronics. Table~\ref{HLT_links} gives an overview of the various detector links
and their expected data payload. 
\begin{table}[htb]
    \begin{center}
    \begin{tabular}{|c|c|c|c|}
    \hline 
	        &  &\multicolumn{2}{|c|}{\bf Sub-event size per DDL}\\
			\cline{3-4}
	{\bf Detector}& {\bf Number of} & {\bf Pb--Pb central} & {\bf Pb--Pb per.}\\
		&  {\bf DDLs}  & {\bf (kB) }      & {\bf (kB)} \\
			\hline \hline
	TPC 		& 216 	& 352 & 90\\
	TRD		& 18	& 39  & 10\\
	DiMuon		& 10	& 15  & \\
	ITS		& 56	& 35  &\\
   \hline
   \end {tabular}
    \caption[Number of HLT detector links per detector and their data payload.]
            {Number of HLT detector links per detector and their data
payload~\cite{hltdaqtrig}.}
    \label{HLT_links}
    \end{center}
\end{table}
\noindent The associated L2--accept trigger
rates for the TPC are limited to 200\,Hz for Pb--Pb running and 1\,kHz
for p--p (Section~\ref{ALICE_datasizes}). The other detectors can
be triggered with 1\,kHz, when triggered without TPC
coincidence\footnote{With the exception of the Silicon Drift Detectors
(SDD).}. This
overview shows that the aggregate HLT input raw data stream amounts to a
maximum of about 20\,GB/s. The processing
rates will vary as a function of trigger type and running
scenarios. For instance in central Pb--Pb collisions the TPC trigger rate will
be limited to 200\,Hz due to pile-up protection, while for p--p
collisions the TPC trigger rate may be as high as 1\,kHz.
The maximum HLT output data rate,
however, is limited to the maximum taping rate of
1.25\,GB/s. Consequently, both the
maximum HLT input and output rates are defined almost independent of
its detailed architecture and requirements. 

The detailed data-flow within the HLT system will be defined according
to the physics requirements and operating scenarios. However, the
general structure depends to a large degree on the input and output
data flow as stated above, Figure~\ref{HLT_flowscheme}.
The data is received on the HLT--RORC where the first part of the
processing will take place using a Field Programmable Gate
Array (FPGA) co-processor. In the case of the TPC this processing
consists of local pattern recognition tasks, i.e. cluster finding
and/or the Hough Transform. The data is then
transferred into the main memory of the FEP over the PCI bus, where
further processing can be made. After this the data is shipped to a
node at the next level, which consists of as many nodes that are
necessary to perform the processing needed. Here, the processing
typically includes track finding within the TPC sectors.
The output data produced
by each level is again shipped to the next level of nodes until the
final stage has been reached. In this way, the processing hierarchy
follows a tree-like structure, where successive larger parts of an
event are processed and merged as one comes closer to the root of the
tree. At this global level all the necessary data has been collected
into the final reconstructed event, i.e. tracks from the different
sub-sectors are merged and fitted, and a final trigger decision for
the event can be taken based on selection algorithms. The decision and
the corresponding data is
then passed to the DAQ system for readout and storage. 
The interface between DAQ and and HLT will consist of a number of DDL
links between a set of HLT event merger nodes and a number of DAQ
LDCs.
The various processing tasks involved in the different levels are
described in Chapter~\ref{TRACK}.


\subsubsection{Hardware components}
The HLT system will consist of a large scale generic computation farm,
with standard PCs connected with a high bandwidth and low latency network. 
Given the huge commercial development within
this field, the project has decided to keep custom hardware development at a
minimum. This will thus enable the system to be very flexible with respect
to selecting the type of processing node and network technology, and
also delay this decision as much as possible due to the continuously
drop in prices.

The design of the data flow, and exact processing sequence, determines much of the
architecture and network topology which will be used.
It has also impact on the requirements on the communication between
each pair of the HLT nodes.
Candidates for the network technology to be used are not
yet fixed, but for the required bandwidth at least Gigabit Ethernet or
a System Area Network (SAN) dedicated to communication in cluster
solutions, is necessary. Possible choices which are considered are
among others ATOLL, SCI and Myrinet.

The HLT--RORC will implement co-processor
functionality by usage of Field Programmable Gate Array (FPGA). 
The FPGA
basically implements a large array of freely programmable logic gates
which can be connected in an arbitrary fashion. The relatively low
clock rate of less than 100\,MHz is compensated for by
a very high degree of inherent parallelism. This makes the FPGA
ideal for implementation of algorithm which can be executed in a
highly parallel fashion, such as TPC cluster finding and the Hough Transform,
Chapter~\ref{TRACK}.

\subsubsection{Communication framework}
\label{HLT_pubsub}
An essential part of the HLT system is interprocess communication 
and data transport within the system. For
this purpose a generic communication framework has been
developed~\cite{timm}. The framework has been designed with an
emphasis on three main issues:
\begin{itemize}
\item Efficiency.
\item Flexibility.
\item Fault tolerance.
\end{itemize}
Efficiency in this context is primarily the minimization of CPU cycles used for
transporting data, in order to keep as much CPU power as possible
available for processing the data. The requirement for flexibility
is a natural consequence from the fact that neither the topology nor the network
interface is defined, and will be postponed as much as possible. 
Fault tolerance means that the framework has to be able to handle and
recover from errors as autonomously as possible, and thus should not
contain any single points of failure in which a fault can disable the
whole system. 

The framework implements an interface between different readout steps,
by defining data producers -- {\it Publisher} -- and data consumers --
{\it Subscriber}. In order to be as efficient as possible the data is not
communicated between the different processes, but rather a descriptor
of the data including a reference to the actual data in shared memory is sent.
In this way data is stored in memory as long as possible, avoiding
unnecessary copying of the data.
The framework basically consists of a number of independent software
components that can be connected together in an arbitrary
fashion. The generic interface allows the processing modules to have a
common interface which is independent of the underlying network
interface.


\chapter{Fast Pattern Recognition in the ALICE TPC}
\label{TRACK}

The main processing task of the High Level Trigger system is
online event reconstruction. Both event selection and
efficient data compression needs a preceding pattern recognition
step. The algorithms which have been developed for online
reconstruction of the TPC data, together with a
performance evaluation, are presented in this chapter.


\section{Track reconstruction methods}
Every charged particle traversing the detectors leaves a number of
discrete hits that allow spatial allocation of
the particle trajectory. The task of the reconstruction algorithms is to
assign these space points to tracks and to reconstruct the particle
kinematics. Reconstructing a particle
is equivalent to finding the parameters that uniquely define its path
in space. From the reconstructed parameters one can find the physical
properties of the particles that passed through the detector. 


In general, track reconstruction is a two-folded problem:
\begin{itemize}
\item {\bf Track finding}.\\
The input to the track finder is a list of all hits in selected
regions or the complete detector.
The task is to decide which hits are made by a
specific particle, and group these hits into a number
of subsets. All the hits within a subset belong to the same particle
trajectory, and the number of subsets corresponds to the number of
particles traversing the detector.
\item {\bf Track fitting}.\\
The task of the track fitting procedure is to estimate
the parameters of the curve describing the trajectory. The input is
thus the positions of all the hits in a subset provided by the track
finder, whereas the output is a list of particles represented by an
estimate of the track parameters. 
\end{itemize}
Traditionally these two steps are performed separately, but certain
applications also combine them.

The increasing complexity in terms of track densities and data sizes
in the LHC experiments, poses high demands on
the track reconstruction algorithms. Both real-time applications and
offline analysis require fast and robust methods in order to
efficiently process the large amount of data produced. A great variety of
pattern recognition methods 
exist within the field of high-energy physics, each with their
advantages and disadvantages with respect to 
complexity and performance. In the following a
brief overview of the main track finding and fitting methods is given.


\subsubsection{Local track follower}
The principle of a track follower is to assign hits to the
track candidate under construction based on the local information of
the track. The track candidate is typically initiated by building
short track segments including only a few hits. These track
segments are subsequently extrapolated between the adjacent active detector
layers. For every step, the algorithm tries to assign a new hit to the
track, and the choice is being made by applying $\chi^2$
criteria of a temporary track fit. Once a new hit has been assigned, the track
parameters are updated and the search is continued.

The advantage with the local track follower approach is its simplicity
with respect to both data access (since it only needs to access a limited
number of hits at the same time) and the selection criteria for each
step. This makes this method efficient when the hit and track
densities are relatively low. However, when the hit density becomes high, such an
algorithm might get confused on the basis of simple $\chi^2$
criteria. This will lead to tracks being split or spliced together in
unpredictable ways, and thus effect the track finding efficiency.

\subsubsection{Track-Road methods}
The Track-Road methods are similar to the local track follower
algorithms. However, Track-Road methods build tracks
by interpolating within possible trajectory roads in the
detector. This is done by choosing hits in the outermost and innermost
layers of the detector, and interpolate between combination of these
hits to obtain the complete tracks.
Because of the combinatorial nature of these methods, they are more
computationally demanding than the local track follower algorithms.

\subsubsection{Template matching methods}
Many of the ambiguities at track crossings in detector layers that
arise from local extrapolation algorithms can be overcome with global
information of the complete trajectory. The global information can be
obtained by using different template matching methods. 
Here, a list of predefined trajectories, called {\it
templates}, are computed based on the equations of particle motion. 
The templates are then overlaid on the data from the detector, and a set
of conditions is applied to determine whether a match exists. 
Originally, the templates were static, and the
resolution of the track parameters were defined by the number and
granularity of the template set being used. In order to overcome these
shortcomings, different adaptive approaches have been developed. 

The {\it Elastic Tracking}~\cite{ohlsson,gyulassy} is an
approach that defines the templates in a dynamical way, by allowing
the predefined trajectories to deform via a set of mathematical
equations to match a pattern in the data.  
This approach has proved to be very efficient in high
track density environments~\cite{lasiuk}. It also offers the
possibility to do tracking on raw detector data without any data
pre-processing. 
However, the high efficiency comes at the expense of very high
processing time and memory requirements.

\subsubsection{Hough Transform (Histogram method)}
Histogram methods are closely related to template
matching methods. They transform particle hits
which makes up a pattern into a suitable parameter space for further
evaluation. Most common is the Hough Transform, which was initially
proposed to detect particle tracks in bubble chamber
pictures~\cite{hough}. A hit or a subset of hits on a
particle trajectory is transformed into curves in a $n$-dimensional
histogram, in which the hits give a ``vote'' to all the
possible trajectories they can possibly belong to. Once all the hits on
a trajectory have been transformed, the intersection of
these curves will correspond to the parameters of the track. In this
way the problem of recognizing global patterns is reduced to the
problem of local peak detection in parameter space. 

The advantages of the Hough Transform is its noise robustness and
its simplicity with respect to hardware implementation.
However, the transformation is also very computationally demanding,
in particular with respect to the memory requirements.

\subsubsection{Kalman filtering}
The Kalman filter method is a method for track finding and
fitting which is very commonly used in high-energy physics
experiments~\cite{bill84,fruh88,kalman}.
In the framework of the Kalman filter, the change of the parameters of
a track along its path is regarded as the dynamical 
evolution of a stochastic state vector. The algorithm basically consists of a
succession of alternating prediction and filter steps. In the
prediction step the current state vector is extrapolated to the next
detector layer, taking into account multiple scattering, energy loss
and Bremsstrahlung in the case of electrons. In
the filter step the next hit on the trajectory is 
selected based on the predictions of the state vector, and the
state vector is updated accordingly. 

The main advantage of the Kalman filter is that is a method for
simultaneous track recognition and track fitting. It also provides a
natural way of extrapolating the tracks between different
detectors. On the other hand, the Kalman filter needs an initial set
of ``seeds'' which has to be provided by a preceding pattern
recognition step. 


\subsubsection{Neural Networks}
The application of neural network methods is known to offer good
approximate solutions to different optimization
problems, among them track finding problems in high-energy
physics~\cite{hopfield}. The basic idea is to set up an {\it energy
function} whose minimum corresponds to the solution of the pattern
recognition problem. The dynamical evolution of the network should
then always decrease the energy and rapidly converge to a
solution. This method has proven to give comparable results with
other track finding approaches~\cite{abele}, and is generally less
time consuming. One major drawback, however, is
absence of a track model in the formalism, and the fact that it has to
be supplemented with a separate track fitting procedure.

\subsection*{Track reconstruction in the TPC}
\label{TRACK_tpctracking}
In case of the TPC, the signals produced by the traversing particles correspond to
two-dimensional charge distributions, called {\it clusters},
Section~\ref{ALICE_tpcprinc}. The centroid of these distributions
corresponds to the three-dimensional space points along the particle
trajectories. Pattern recognition in the TPC is often classified in
two main categories; an {\it sequential} and {\it iterative} approach.


\subsubsection{Sequential pattern recognition}
In the sequential case the pattern recognition is performed in two
sequential steps. In the first step, the cluster centroids are
calculated. A cluster finder searches the raw ADC data for local
maxima in the two dimensional charge distributions. If an isolated
cluster is found, the centroid position in pad and time is
calculated. In the case of overlapping clusters, the charge
distributions have to be separated by unfolding procedures. However,
due to the missing information about the track parameters of the
crossing tracks, neither the shape
nor the size of the clusters are known at this stage. Furthermore,
the number of tracks contributing to the distribution are unknown.
Therefore, the distributions cannot easily be unfolded,
and the resulting centroids are error-prone and may
result in a loss of tracking performance.
Together with the position of the pad-row plane, the centroids
provide the three dimensional coordinates which are interpreted as
the particle crossing point with the pad-row plane. In the second
step, the list of space
points serve as an input for the track finder which connects the space
points into track segments. The track finding problem is typically
solved by a local track follower or a Kalman filter approach.

This method is the conventional approach for track reconstruction and
has been successfully been used with TPCs in a relatively low occupancy
environment. Examples are the reconstruction programs for the
NA49~\cite{na49} and the STAR~\cite{star} experiment. In the ALICE
experiment however, the detector occupancy in the TPC may well reach
values of up to 50\% (Figure~\ref{TRACK_occupancy}).
In such a scenario the amount of overlapping
clusters will be very high, and it is generally expected that the
performance of a simple cluster finder will be low. A cluster
model for the overlapping charge distributions is likely to improve
the results.

\subsubsection{Iterative pattern recognition}
In the iterative track finding approach, the procedures of cluster
finding and track finding are not separated into two sequential steps,
but rather done in a more parallel or iterative fashion. In this case parts of
the track finding is solved prior to the cluster finding. 
In a first step ``track seeds'' are found
for instance by using a combinatorial approach based on cluster
centroids located at larger radius (where the occupancy is low)
or by applying track finding algorithms like the Hough Transform directly on the raw
ADC data. Based on the list of found track segments, one returns to the raw-data and
reconstruct the clusters along the trajectories. Since the cluster
model is a function of the crossing tracks and detector specific
parameters and the electronics, the track model provides an estimate
of shape and size of the
clusters. Hence the charge distributions can be fitted to a known
shape. Based on the additional knowledge of the tracks passing nearby
and therefore possibly contributing charge to the clusters,
overlapping clusters can be properly deconvoluted. This approach was used
in the NA49~\cite{na49}. 


\section{The ALICE tracking environment}
The experimental setup of the ALICE experiment consists of the central
barrel placed inside the L3 magnet and the forward detectors
(Section~\ref{ALICE_expsetup}). The main tracking detector is the TPC
detector, which has a coverage of $|\eta|<$\,0.9 and full
azimuth.

\subsection{Particle multiplicity and detector occupancy}
\label{TRACK_occup}
The ALICE experiment is designed to handle a charged particle
multiplicity corresponding to 8000 charged particles per unit
rapidity, dN$_{\mathrm{ch}}$/dy=8000~\cite{alicetp}. 
This estimate is regarded as an {\it extreme assumption}, and is
based on a comparison between different
event generators for Pb-Pb collisions at the LHC energy of
$\sqrt{s_{NN}}$=5.5\,TeV. The predicted charged particle multiplicity
at the mid-rapidity, however, varies strongly,
ranging from dN$_{\mathrm{ch}}$/dy of about 1500 to 7-8000, depending
on the generator used.

In order to study detector performance, it is more convenient to use
pseudo-rapidity, $\eta$, instead of rapidity since the pseudo-rapidity is a
geometrical unit while the rapidity depends on
particle composition and $p_t$-spectra. It should however be noted that
the multiplicity per unit rapidity is 10-20\% more at
mid-rapidity than the multiplicity expressed in units of
pseudo-rapidity~\cite{ppr}.
This means that a charged particle density of
dN$_{\mathrm{ch}}$/d$\eta$=8000 is 10-20\% higher than the extreme assumption of
dN$_{\mathrm{ch}}$/dy=8000. 

The ALICE Technical Proposal was written when the highest available
nucleon-nucleon center of mass energy results was
$\sqrt{s_{NN}}$=20\,GeV (CERN SPS), i.e. a factor of 300
less than the LHC energy. Since that time, new data from RHIC
at energy up to
$\sqrt{s_{NN}}$=200\,GeV has become available, and the event generators have
been correspondingly updated. In Table~\ref{TRACK_generators} the
results from the most common generators, HIJING (Heavy-Ion Jet
INteraction Generator), DPMJET (Dual Parton Model) and the SFM (String
Fusion Model) are summarized.
\begin{table}[htb]
    \begin{center}
    \begin{tabular}{|c|l|c|}
    \hline
    {\bf Generator} & {\bf Comments} & {\bf \dndy at $\eta$=0 (approx.)}\\
    \hline\hline
	HIJING 1.36 & with quenching & 6200\\
		    & without quenching & 2900\\
	DPMJET-II.5 & with baryon stopping & 2300\\
		    & without baryon stopping & 2000\\
	SFM	    & with fusion & 2700\\
		    & without fusion & 3100\\
    \hline
    \end {tabular}
    \caption[Charged particle multiplicity simulated by different
event generators.]
            {Charged particle multiplicity simulated by different
event generators~\cite{ppr}.}
    \label{TRACK_generators}
    \end{center}
\end{table}
\noindent A detailed description of the processes
included and parameters used for each of these event generators can be
found in~\cite{ppr} and references therein. Results show that
large differences for \dndy still exists, and even in the framework of a
single generator the density is strongly model dependent. Furthermore, none
of the current generators reproduce the multiplicities obtained from
the theoretical extrapolations from RHIC described in
Section~\ref{HIC_partmult} on page~\pageref{HIC_partmult}, which
indicates values of about 2000-3000. Because
of the large uncertainty in the particle
multiplicity and the fact that extrapolation from RHIC results are
still quite large ($\sim$30), the value of up to 8000 is still being
considered as the design value by the collaboration in order to
provide a absolute safety margin with respect to detector and software
design. From a pattern recognition point of view, tracking algorithms
should be developed that are capable of handling all possible scenarios. 

\bfig[htb]
\insertplot{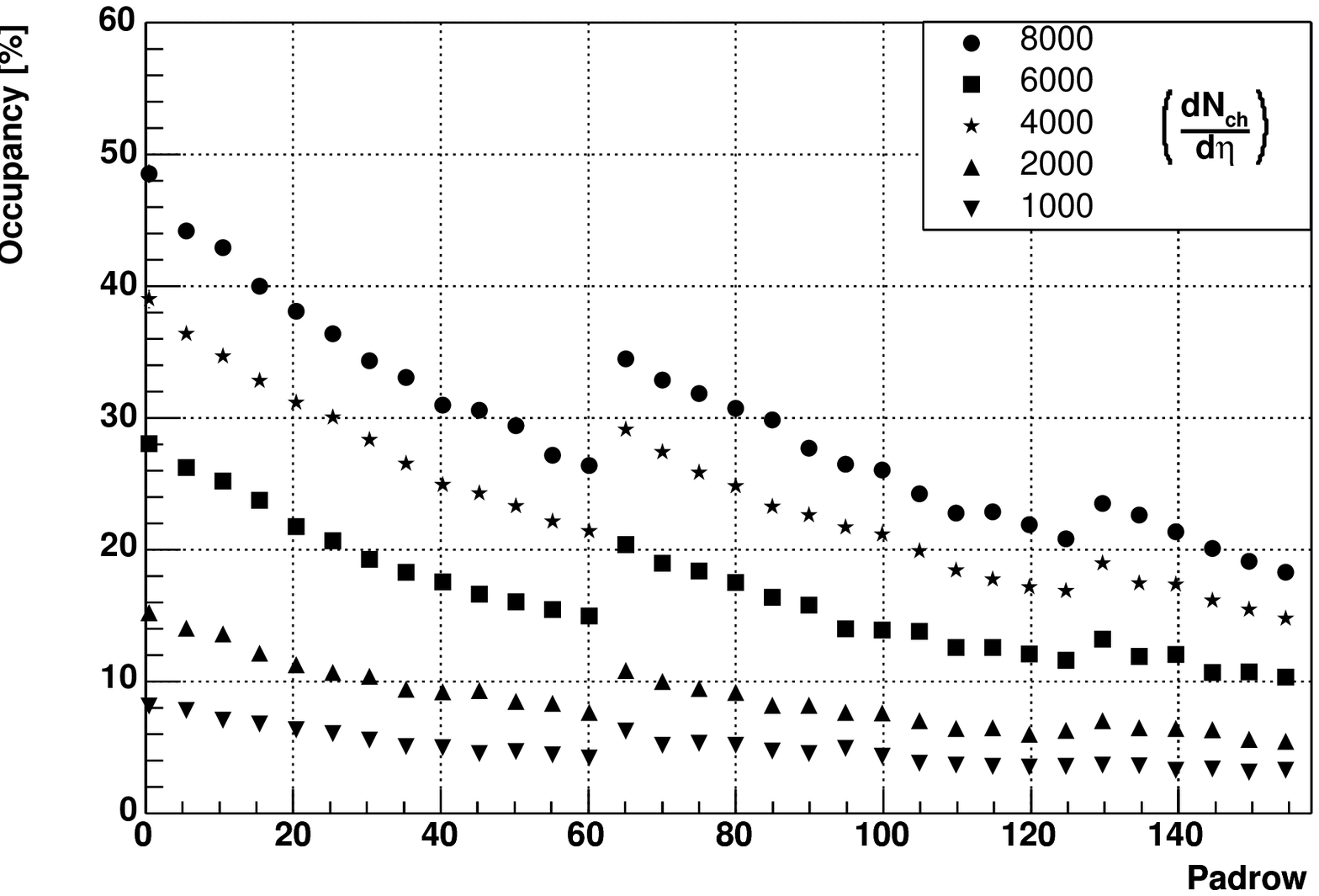}{10cm}
	{Simulated occupancy in the ALICE TPC as a function of pad-row number.}
	{Simulated occupancy in the TPC as a function of pad-row
number. The occupancy is calculated as the ratio between
signals above threshold and the total number of digits. The
steps are due to change in pad-sizes (Table~\ref{ALICE_tpcparams1}).}
\label{TRACK_occupancy}
\efig
In Figure~\ref{TRACK_occupancy} the simulated
detector occupancy for the TPC is shown as a function of radius
(represented by the pad-row number) for different multiplicities.
For dN$_{\mathrm{ch}}$/d$\eta$=8000 it reaches
$\sim$50\% for the innermost padrows, and drops to less 20\% for the outermost.

\subsection{Magnetic field settings}
The central detectors are placed inside the L3 magnet which provides
a solenoidal magnetic field. In general, the field strength selection
in a heavy ion experiment is a 
compromise between low momentum acceptance, momentum resolution and
tracking and trigger efficiency. In the ALICE experiment, the
desired low momentum cut-off is determined by the ability to probe the 
soft hadronic regime, i.e. collective effects, detection of
decay products of low \pt hyperons, and identical particle
interferometry. In the high momentum regime,
the resolution has to be sufficient to measure the leading
particles of high energy jets, and to measure the decay products of
heavy quarkonium. The ideal field strength for hadronic physics, maximizing the
reconstruction efficiency, is around 0.2\,T, while for the high \pt
observables the maximum field strength the L3 magnet is able to
produce, 0.4-0.5\,T will most likely be the best choice~\cite{ppr}. Since the high
\pt observables are the ones which are limited by statistics, ALICE
will run mostly at the higher magnetic field setting.

\subsection{Particle trajectory in a magnetic field}
A charged particle moving in a homogeneous magnetic field follows a
circular motion in the plane perpendicular to the field. This is
commonly referred to as the {\it bending plane} of the track. The
momentum component along the magnetic field is left unchanged. These
two components thus form a helix path in space. A description of
the track model and helix parameterization can be found in
Appendix~\ref{APP}. The assumption that a 
particle follows a helical trajectory is exact when the magnetic field
is constant and one neglects energy loss and multiple scattering.
For a particle with unit charge, the relation between the magnetic
field $B$\,[T], and the curvature of the track, $\kappa$\,[m$^{-1}$]
is given by
\beq
p_t = \frac{0.3B}{|\kappa|}\hspace{1cm}[\mathrm{GeV}]
\label{TRACK_kappapt}
\eeq
Once the \pt of the particle has been determined, the remaining components
of the particle momentum can be calculated from
\begin{eqnarray}
p_z = p_t\tan\lambda\nonumber\\
p = \sqrt{p_t^2+p_z^2}
\end{eqnarray}
where $\lambda$ is the dip-angle of the track.

The curvature and the dip-angle of a track segment is determined by fitting the
individual space points along the trajectory. The accuracy of which
the curvature and thus the momentum can be measured is limited by the
space point errors. The relation between the azimuthal position
resolution of a single space point and the relative transverse
momentum error can be parameterized by~\cite{gluck}
\beq
\frac{\Delta p_t}{p_t} =
\frac{\sigma_{r\phi}p_t}{0.3BL^2}\sqrt{\frac{720}{N+4}}
\label{TRACK_ptresformula}
\eeq
where $L$ is the total visible track length, $N$ is the number of
measured space points on the track and $\sigma_{r\phi}$ is the error of the
position measurement. For the ALICE TPC $L$ and $B$ are given by the
overall design parameters of the detector, while $N$ and the space point
resolution are defined by the design of the readout chambers
(Section~\ref{ALICE_tpcdesign}).

In addition to the track measurement errors, the obtained momentum
resolution will be limited by multiple scattering and energy
loss. These effects cause the particle trajectory to deviate from a
helix. Furthermore, the resolution will also be limited by detector
readout specific effects such as diffusion and angular effects etc.
(Section~\ref{ALICE_tpcprinc}, page~\pageref{ALICE_signal}), and the
noise imposed by the electronic readout chain.


\section{The AliROOT framework}
\label{TRACK_aliroot}
The complete ALICE detector setup is described and simulated within
the AliROOT framework~\cite{aliroot}, which is based on the
ROOT software package~\cite{root}. The framework consists of a set of software
tools which mimic the different steps of data processing required in
the ALICE experiment. An overview
of the functionality of the framework is shown schematically in
Figure~\ref{TRACK_alirootscheme}. 
\bfig
\centerline{\epsfig{file=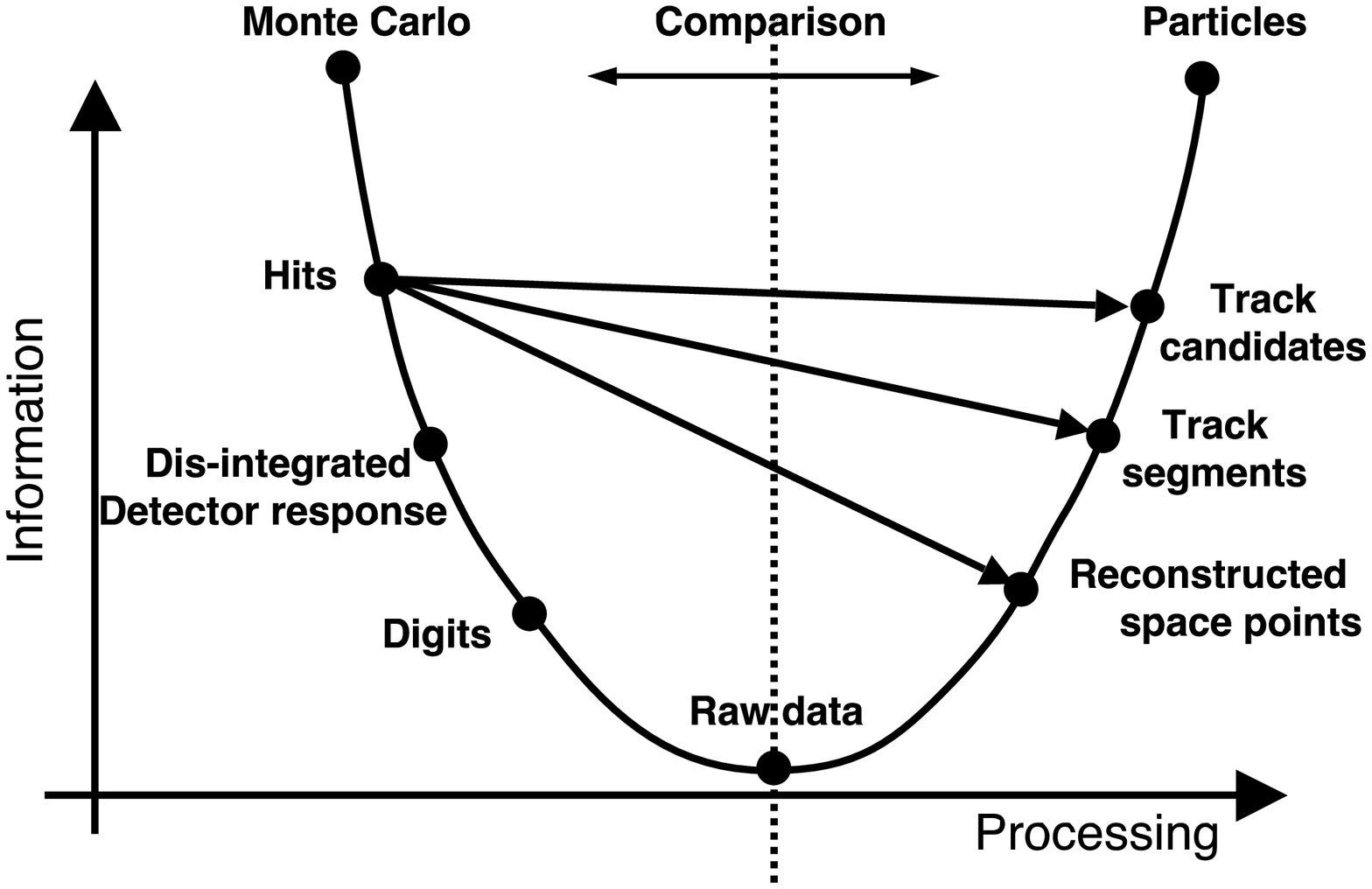,width=10cm,bb=60 70 781 568}}
\caption[Overview of the AliROOT framework.]
	{Functionality overview of the AliROOT framework~\cite{ppr}.}
\label{TRACK_alirootscheme}
\efig
\noindent Data are generated via simulation programs, i.e. event
generators and detector response simulations, and are then transformed into the
data format produced by the detector. The data produced by the event
generators contain the full information about the generated particles,
i.e. particle type and kinematics.
These data then serve as input to the detector
simulation chain, in which the information is disintegrated to that
generated by particles when traversing the detector. The outcome of
the detector simulation is representative of real data measured in a
real detector. This data can thus be used as input for the various reconstruction
and analysis chains, which can be evaluated by comparing the final
reconstructed particles with the generated ones. 
Each detector is described as an
independent module that contains the code for both simulation and
reconstruction. 


\subsection{Event simulation}
\label{TRACK_evsim}
The theoretical uncertainty with respect to the outcome of heavy ion
collisions at LHC makes it
necessary to use and compare several models for the simulation of the
final state particles in an event. Consequently, a variety of different
event generators have been implemented within the AliROOT framework. 
In order to evaluate the reconstruction and analysis
algorithms in AliROOT, a generator capable of simulating a typical ``background''
event for various multiplicities is needed. 
For this purpose a parameterization of the HIJING generator,
which is based on parameterized $\eta$-density and \pt
distributions of charged and neutral pions and kaons was encoded. In
this model, the $\eta$-distributions have been obtained from a HIJING simulation of
central Pb-Pb collisions, and
the \pt distributions of the pions are based on \pt measurements at
$\sqrt{s}$=1.8\,TeV~\cite{cdf}. The corresponding kaon \pt
distribution is obtained from the pion distribution by m$_t$ scaling. 
To simulate a multiplicity of e.g. \dndy=\,8000, the
$\eta$-distribution is scaled in such a way that 8000 charged particles
per event are produced within the range $|\eta|<$0.5.
Lower multiplicity events are scaled similarly. All the sample
events used in this work have been produced using this parameterization.

\subsection{Simulation of detector response}
\label{TRACK_detresp}
Once the collision is simulated by an event generator, the final state
particles are fed to a particle transport program. For
simulating particle transport in AliROOT the GEANT3 package is currently being
used.
The particles are transported in the material of the
detector, simulating the interaction and the energy deposition that
generates the detector response. From the energy
deposition, the ``ideal'' detector response resulting from the
traversing particles is generated. This response is subsequently
digitized and formatted according to the output of the detector front-end
electronics. Thus, the final results are expected to closely resemble
the real data produced by the ALICE detector system.


In the special case of the ALICE TPC detector, a dedicated microscopic
simulator has been implemented in AliROOT~\cite{kowalski}. All major
physical processes are incorporated in this simulator, including
parameterization of the ionization in the gas,
generation of secondary electrons, diffusion of electrons, electron
attachment, {\bf E$\times$B} effect near the anode wires, and a
complete pad and time response determined by the readout-chamber
geometry and electronics parameters.

\subsubsection{Simulation of the physical processes in the TPC}
The ionization in the gas is basically described by the generation
of primary and secondary electrons.
The electromagnetic interactions of the initial particle
with the TPC gas lead to the release of the primary electrons. If
these electrons have sufficient kinetic energy they will further ionize
the gas and produce secondary electrons, creating the
electron cluster. 
The mean distance, $D$, between two primary ionization's can be
expressed as
\beq
D = \frac{1}{N_{\mathrm{prim}}\cdot f(\beta\gamma)}.
\eeq
Here $N_{\mathrm{prim}}$ is the number of primary electrons per cm
produced by a minimum ionizing particle (MIP), and $f(\beta\lambda)$
is the Bethe-Bloch function. The energy loss function is parameterized
from a fit to energy-loss data for 90\% Ar, 10\% CH$_4$.
The energy loss released in the primary ionization to atomic electrons
has, if one neglects the atomic shell structure, a close to 1/E$^2$
dependence. For light gases the distribution has a slightly steeper
dependence, and in the simulation a 1/E$^{2.2}$ parameterization is
used. 



The total number of secondary electrons, N$_{\mathrm{tot}}$, created
in a cluster is given by,
\beq
N_{\mathrm{tot}} = \frac{E_{\mathrm{tot}}-I_{\mathrm{pot}}}{W_i} + 1,
\eeq
where $E_{\mathrm{tot}}$ is the energy loss in a given collision,
$W_i$ is the effective energy required to produce an electron-ion pair
and $I_{\mathrm{pot}}$ is the first ionization potential. The
simulated clusters are initially point-like objects, and no distinction
between primary and secondary electrons are done. 

The produced electrons drift through the gas with an constant effective
drift velocity. During the drift, the electron cloud is
subject to diffusion, which is described by a three-dimensional
Gaussian distribution,
\beq
P(x,y,z)=
\frac{1}{\sqrt{2\pi\delta_T}}\exp\left[-\frac{(x-x_0)^2}{2\delta^2_T}\right]\cdot
\frac{1}{\sqrt{2\pi\delta_T}}\exp\left[-\frac{(y-y_0)^2}{2\delta^2_T}\right]\cdot
\frac{1}{\sqrt{2\pi\delta_L}}\exp\left[-\frac{(z-z_0)^2}{2\delta^2_L}\right],
\label{TRACK_gaussdiff}
\eeq
where $(x_0,y_0,z_0)$ is the electron creation point, and the
transversal, $\delta_T$, and longitudinal, $\delta_L$, diffusion are
given by the drift length, $L_{\mathrm{drift}}$, and gas coefficients,
$D_T$ and $D_L$
\begin{eqnarray}
\delta_T &=& D_T\sqrt{L_{\mathrm{drift}}}\nonumber\\
\delta_L &=& D_L\sqrt{L_{\mathrm{drift}}}
\end{eqnarray}

During the drift through the gas the electrons can be absorbed in the
gas by the formation of negative ions. This process has been simulated
by assuming a probability of electron capture of 1\% per m drift
per ppm of O$_2$.

Near the anode wires, the magnetic field and electric fields are no
longer parallel, and because of the Lorentz force the electrons
experience a displacement along the wire direction. If an electron
enters the readout chamber at point $(x_0,y_0)$, it is displaced in
the $x$-direction (assuming that the wires are placed along the
$x$-axis). The new position of the electron is given by,
\beq
x=x_0 + \omega\tau(y-y_0),
\eeq
where $y$ is the coordinate of the wire on which an electron is
collected, and $\omega\tau$ is the tangent of the Lorentz angle.

\subsubsection{Simulation of the signal generation}
When the electron cluster enters the readout chambers, the electrons
are accelerated in an increasing electric field towards the anode
wires. Once the electric field is strong enough, an avalanche is
created. The amplitude of this avalanche is determined by the high
voltage applied to the wire, and is subject to fluctuations. The
resulting number of electrons created can be described by an
exponential probability distribution function,
\beq
P(q) = \frac{1}{\bar{q}}\cdot\exp\left(-\frac{q}{\bar{q}}\right),
\eeq
where $\bar{q}$ is the mean avalanche amplitude.

An electron which is collected on the anode wire leaves an ion behind
which induces a charge on
the pad plane. This charge is integrated over the pad area, and the
induced charge distribution in the pad plane is
determined by the Pad Response Function (PRF). In the simulation, a
two-dimensional PRF is computed for the 
pad-geometries planned for the ALICE
TPC. This calculation also incorporates possible signal measurements
from the neighboring wires (``crosstalk''). The time signal is obtained by
convolving the avalanche with the shaping function of the
preamplifier/shaper, and sampled at a given
frequency. The generated signal is then superimposed with a random
electronic noise. This noise is described by a Gaussian with RMS
equal to 1000\,$e$.
The simulated signal is finally digitized using the predefined
dynamic range of the electronics and by applying zero suppression.

\subsection{The Offline reconstruction chain}
\label{TRACK_offlinechain}
The ALICE Offline reconstruction algorithm is based on the Kalman
filtering approach~\cite{kalman}. 
The reconstruction chain starts with a cluster finder in the TPC,
which provides the space points that are used for tracking with the
Kalman filter. The overall tracking then starts with track seeding in
the outermost pad-rows of the TPC. It begins with a search for all
pairs of points which are projecting to the primary vertex. Different
combinations of the pad-rows are used with and without a primary vertex
constraint, in order to obtain both primary and secondary tracks.
When a reasonable pair of points is found, parameters of
a helix going through these points and the vertex are calculated.
These parameters and the corresponding covariance matrix are then
taken as the initial track candidate which are propagated from the
outermost point to the inner point using the Kalman filter. If at
least half of the possible points between the initial ones were
successfully associated with the current track candidate, it is stored
as a valid seed and the search continues. Each seed is then
propagated through the entire TPC.

One of the main shortcomings using the Kalman filter in ALICE is that it
depends on good seeds to start a stable filtering procedure. In the
present approach this is very computationally demanding, as the
search is done in a straightforward combinatorial way. Another
drawback is that the clusters have to be reconstructed prior to the
track finding, which is a difficult task in regions with high
occupancy and cluster overlapping. However, recent developments
deal with this problem utilizing an iterative tracking
approach, i.e. by performing cluster finding in
parallel to track finding~\cite{marian}. In this way, information about the
tracks which produce the clusters is available, and deconvolution is
more easily performed.

\section{Premises}
The pattern recognition algorithms implemented in this work are
primarily developed for the ALICE High Level Trigger system. In this
context there are two main considerations:
\begin{itemize}
\item Computing requirements.
\item Parallelization.
\end{itemize}
The computing requirements define the processing power in terms of
processing time and memory requirements needed to
reconstruct a complete event. From the High Level Trigger point of view, the
time available for this task is constrained by the event
rate. The amount of processing power needed for the full system will
critically depend on the processing requirements for the individual
processing modules. The algorithms should therefore be optimized with
respect to both efficient data organization and the minimization of
computing intensive tasks.
The latter consideration is related to the fact that the readout of
the detector is done in parallel, Section~\ref{HLT_arch}. This defines the
first level of the HLT architecture, in which the first part of the
pattern recognition is performed. Furthermore, merging the data
from different parts of the detector implies copying data over
network, and should therefore be avoided as far as possible. Effort
has therefore been made to implement the processing schemes in a
highly modular fashion, allowing for a high degree of
parallelization.

As part of this work, two different TPC pattern recognition approaches were
implemented and evaluated on simulated ALICE TPC data. They are in the
following referred to as {\it sequential tracking} and {\it iterative
tracking}, reflecting the two main TPC track reconstruction methods
introduced on page~\pageref{TRACK_tpctracking}.

\section{Sequential tracking}
\label{TRACK_seqtracking}
The implemented sequential tracking scheme consists of 4 main
successive processing modules, referred to as {\it
Cluster Finder}, {\it Track Finder}, {\it Track Fitter} and {\it Track Merger}.
The algorithms used for the Cluster Finder and the
Track Finder are based on the reconstruction scheme
implemented and used in the STAR L3 trigger~\cite{starl3}. 
All algorithms, and their implementation issues are described in the following.

\subsection{The Cluster Finder}
\label{TRACK_clfinder}
The cluster finder algorithm is an implementation of a straightforward
sequence matching technique. The main emphasis of the algorithm is
to optimize the program with respect to CPU time and 
memory access. A FPGA implementation of the algorithm has been
implemented~\cite{gaute} for the purpose of utilizing the co-processor
functionality planned for the Front-End Processors.

The basic functionality of the algorithm is to group sequences in
the pad-row-plane which belong to the the same cluster and calculate
the two-dimensional centroid,
($\lambda_{\mathrm{pad}},\lambda_{\mathrm{time}}$), by a weighted mean,
\begin{eqnarray}
\lambda_{\mathrm{pad}} = \frac{\sum_i q_i p_i}{\sum_i q_i}\hspace{1cm}
\lambda_{\mathrm{time}} = \frac{\sum_i q_i t_i}{\sum_i q_i}.
\end{eqnarray}
Here $q_i$ is the ADC-value of a given $(p_i,t_i)$ (pad,time) bin.
A flow diagram of the algorithm is shown in
Figure~\ref{TRACK_cfflow}. The input to the cluster finder is a list
of sequences for each pad. For every new sequence the centroid position in the time
direction is calculated. This temporary mean is then compared to the
sequences in the previous processed pads. During this step there are two
possible outcomes:
\begin{itemize}
\item[--] A match is found. This means that there is another
sequence on the neighboring pad that
overlaps with the current sequence. The two 
sequences are then merged and the mean in both pad and time are
calculated. 
\item[--] No match was found. In this case, the sequence is regarded as
the start of a new cluster.
\end{itemize}

The essential feature of the algorithm is that it enables the program to
handle all sequences in two distinct lists: The current pad list and
the previous pad(s) list. 
This also allows the loop over the data only
once, while performing all the calculations {\it
on-the-fly}. Consequently, it is assumed that the input data stream is
ordered in pad and time, i.e. all sequences on a single pad are
received before the successive pad. Thus, the
current pad list contains information about the sequences on the pad
which are currently being processed, while the previous pad(s) list
contains the clusters on the previous pad(s). A cluster is
considered complete once there are no matching sequences on the current
pad. 

\bfig
\insertplot{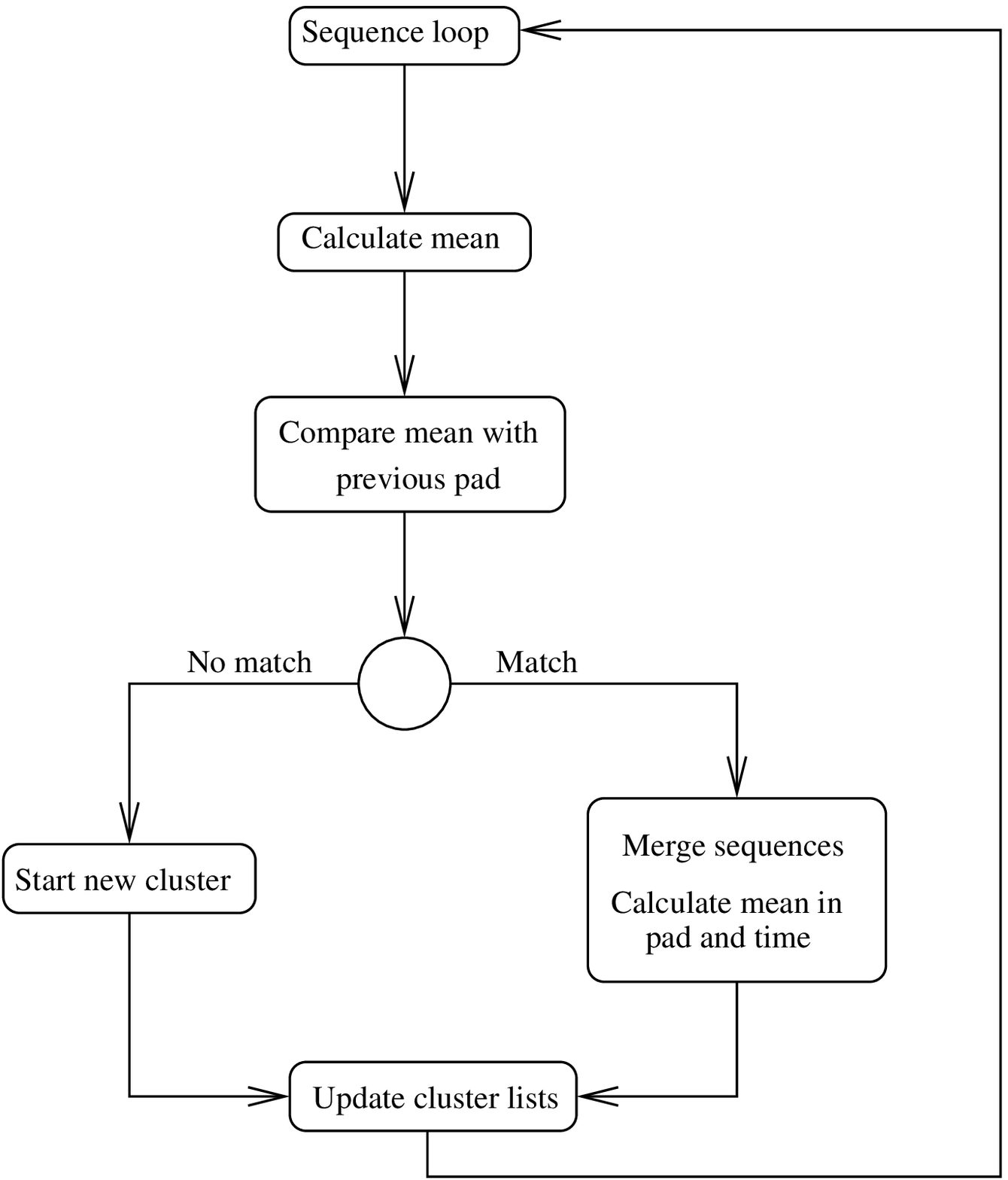}{8cm}
	{Flow diagram of the HLT Cluster Finder algorithm.}
	{Flow diagram of the HLT Cluster Finder algorithm.}
\label{TRACK_cfflow}
\efig

In the case of overlapping clusters,
a simple cluster deconvolution scheme is applied. A check is performed on each
sequence whether there exists a local minimum in the charge values. If
such a minimum exists, the sequence is separated into two sequences by
splitting at the time-bin containing the minimum charge. Following the
same procedure in the pad direction, a check is done whether there is
a local minimum in the pad charge for the cluster under
construction. If this is the
case, the matching cluster on the previous pad is not merged, but
considered complete, while the sequence on the current pad is
considered as a new cluster.

In addition to the cluster centroids, also the shape of the cluster is
needed as additional information for the estimation of the space point
errors, see next section. The shape can be represented by the widths
of the cluster in the two dimensions, and is thus calculated as the
RMS of the distributions.

In Table~\ref{TRACK_cfres} the space point resolution obtained with
the algorithm is listed for isolated clusters. The estimates are
obtained by
comparing the reconstructed centroids with pad-row plane crossings of
the the simulated particle trajectories. The results are averaged over
all primary tracks with \pt$\geq$\,0.1\,GeV which cross the entire TPC
volume ($|\eta|<$\,0.9). The values correspond to the standard
deviation obtained by performing a Gaussian fit of the
distributions. For reason of comparison, also the
results from the Offline cluster finder algorithm are shown. The
resolutions are in the order of 0.8-1\,mm and 1-1.4\,mm for the pad and
time direction, respectively. The results for the Offline cluster
finder are slightly worse than the HLT cluster finder, which may be due
to a wrong splitting of a few but large pathological clusters.
\begin{table}[htb]
    \begin{center}
    \begin{tabular}{|l|c|c|c|c|}
    \hline 
			&\multicolumn{2}{|c|}{\bf Pad direction [mm]} 
			&\multicolumn{2}{|c|}{\bf Time direction [mm]}\\
			\cline{2-5}
			& HLT & Offline & HLT & Offline\\
			\hline \hline
   Inner chambers	&0.99 & 1.10 & 1.33 & 1.50 \\
   Outer chambers	&0.88 & 0.96 & 1.14 & 1.23 \\
   \hline
   \end {tabular}
    \caption[Space point resolution obtained using the HLT Cluster Finder.]
            {Space point resolution obtained using the HLT Cluster
Finder on isolated clusters. For comparison, the results from the
Offline cluster finder algorithm are also listed.}
    \label{TRACK_cfres}
    \end{center}
\end{table}

\subsubsection{Space point errors}
\label{TRACK_clerrors}
In addition to the three dimensional space point coordinates of the
clusters, an estimate of the errors of the space points is needed for the track
finding procedure. Similar to the cluster widths, the errors are
determined by the diffusion and angular spread of the drifting
electron cloud, and are thus dependent of
detector specific parameters and the track parameters,
Section~\ref{ALICE_tpcprinc}. However, since
the cluster finder does not have any information about the tracks, the
errors are assumed to be proportional to the calculated RMS-values
of the clusters. The coefficients have been found by comparing the reconstructed space
points with the pad-row crossing points of the simulated tracks,
and parameterizing the errors as a function of the
cluster widths for both pad and time direction. 





\subsection{The Track Finder}
The track finding algorithm is a local track follower algorithm. Its
main feature is that it incorporates a transformation on the space points
commonly referred to as {\it conformal mapping}. The purpose of this
transform is to describe the circular motion of the particle trajectory in the
bending plane by a linear parameterization. Since fitting straight
lines is significantly less computationally demanding than fitting
circles, the effect of the transform is to minimize the 
amount of calculations needed for the fitting procedures of the track
finding algorithm.

\subsubsection{Conformal mapping}
The purpose of the conformal mapping is to transform the points on a
circular pattern into a
space where it can be described using a linear relation. Denoting
the conformal space by ($x',y'$), a space point ($x,y$) transforms in
according to,
\begin{eqnarray}
x' &=& \frac{x-x_t}{r^{2}}\nonumber\\
y' &=& -\frac{y-y_t}{r^{2}}\nonumber\\
r^2 &=& (x-x_t)^2 + (y-y_t)^2
\label{TRACK_confmap}
\end{eqnarray}
where the point (x$_t$,y$_t$) is a fixed point on the circle. In order
to derive the linear relationship between points in conformal space,
consider two points ($x,y$) and ($x_t,y_t$) on a
circle with center $(x_c,y_c)$ and radius $r_c$. For both points the
following relations are then valid,
\begin{eqnarray*}
r_c^2 &=& (x-x_c)^2+(y-y_c)^2\\
r_c^2 &=& (x_t-x_c)^2+(y_t-y_c)^2,
\end{eqnarray*}
and therefore,
\begin{eqnarray*}
x^2+y^2 = 2x_c(x-x_t) + 2y_c(y-y_t) + x_t^2+y_t^2.
\end{eqnarray*}
Inserting the relations,
\begin{eqnarray*}
x^2 &=& (x-x_t)^2 + 2xx_t - x_t^2\\
y^2 &=& (y-y_t)^2 + 2yy_t - y_t^2
\end{eqnarray*}
and the reverse transformation from Equation~\ref{TRACK_confmap},
\begin{eqnarray*}
x &=& x'r^2+x_t\\
y &=& -y'r^2+y_t
\end{eqnarray*}
results in,
\begin{eqnarray*}
(x-x_t)^2+(y-y_t)^2 = 2r^2[x'(x_c-x_t)-y'(y_0-y_t)],
\end{eqnarray*}
and together with the definition of $r^2$ in Equation~\ref{TRACK_confmap} this gives
\beq
y'(x') = \frac{x_c-x_t}{y_c-y_t}x' - \frac{1}{2(y_c-y_t)} =
a_{y'}x'+b_{y'}.
\label{TRACK_lconfmap}
\eeq
Equation~\ref{TRACK_lconfmap} thus provides a linear parameterization of the
circle in conformal space, Figure~\ref{TRACK_exaconfspace}.

\bfig[htb]
\centerline{\epsfig{file=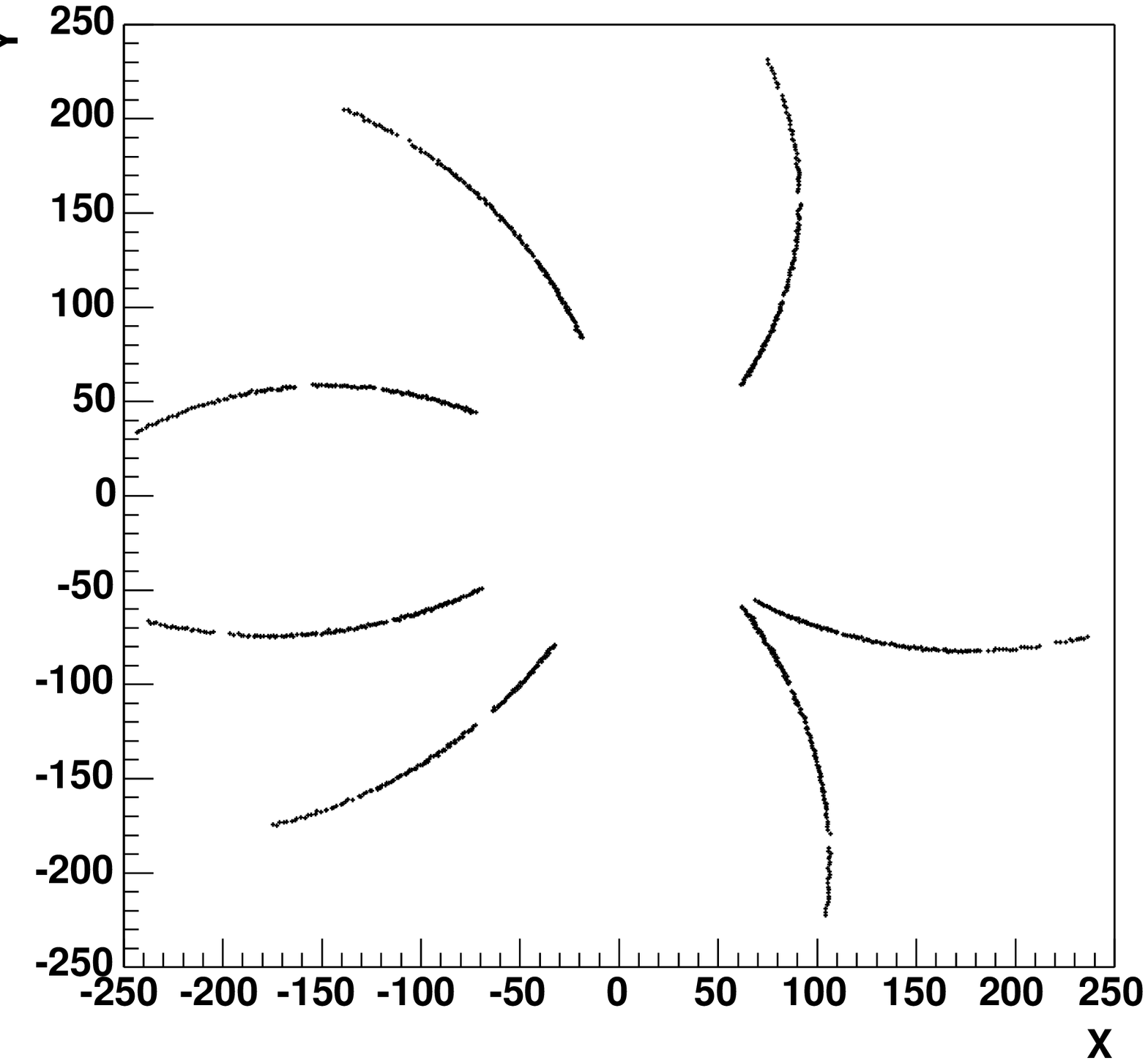,width=8cm}
\epsfig{file=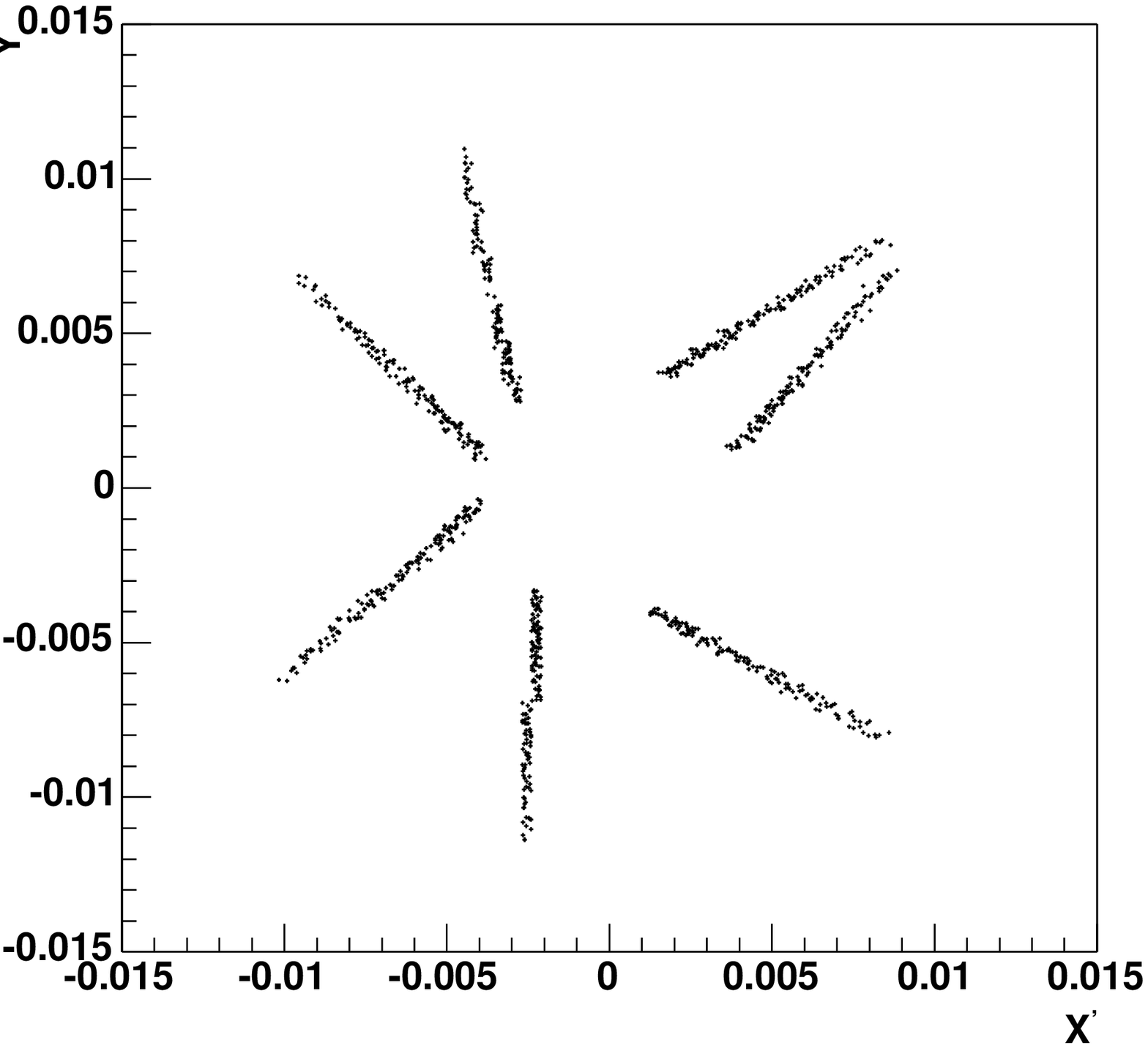,width=8cm}}
\caption[Illustration of conformal mapping of space points along circular track segments.]
	{Conformal mapping of space points lying on the circular track
segments. Each point in $xy$-space (left) is transformed into
conformal space (right) using the vertex as a fixed point on the circle.}
\label{TRACK_exaconfspace}
\efig

The transformation needs a fixed point, ($x_t,y_t$), on the circle. In
the case of track finding, this point can be replaced by the collision
vertex coordinates if the track is assumed to originate from a primary particle.
Alternatively, one can use the first point associated on the
track. In that case, one assumes that the track originate from a
secondary vertex.

\subsubsection{The track follower}
The track finding procedure is based on a {\it follow-your-nose}
method initially proposed in~\cite{pablo}. The algorithm consists of
building track segments by combining
space points which are co-linear in the conformal space. 

The first step of the algorithm consists of data organization. An
essential part of the algorithm is to organize the space points in a
efficient manner, so that the track finder has fast access to all the
space points when building the track segments. This is done by
assigning the space points to special sub-volumes in ($r,\phi,\eta$),
where $r$, $\phi$ and $\eta$\ correspond to pad-row number,
polar angle and pseudo-rapidity, respectively. A sub-volume is
denoted using the corresponding indexes
($i_r,i_{\phi},i_{\eta}$).
Every space point is
uniquely associated with one sub-volume, and one sub-volume can of
course contain many space points. 
The sub-volumes merely acts as
{\it containers} for the space points, which in the implemented program
correspond to a list of pointers to the respective space points
structures in memory.

After transforming the space points into conformal space and
organizing them into sub-volumes, the actual track finding starts. The
tracks are being initiated by building track segments, and the search
is starting at the outermost pad-rows. A space point is chosen as the
starting point, $S$, in which the corresponding sub-volume is denoted
($i_r^S,i_{\phi}^S,i_{\eta}^S$).
During the search, only space
points which are in the nearby sub-volumes are considered. More
specific, this corresponds to the sub-volumes satisfying the conditions:
\begin{eqnarray}
i^S_r - N \leq j_r \leq i^S_r - 1\nonumber\\
i^S_\phi - 1 \leq j_\phi \leq i^S_\phi + 1\nonumber\\
i^S_\eta - 1 \leq j_\eta \leq i^S_\eta + 1
\label{TRACK_volsearch}
\end{eqnarray}
where $N$ is a tunable parameter of the program. In this context,
the distance, $d$, between two points is defined as
\beq
d = |i^S_r-i_r^j|\times(|\phi^S - \phi^j| + |\eta^S
- \eta^j|).
\eeq
The space points which are closest together are linked, and
the search is continued until a certain number of space points is
reached.

Once the track segments are obtained, the corresponding points are
fitted to straight lines in conformal space and in the $(s,z)$
plane. The tracks are then build by associating new space points to
the track which are close to the fit. The search continues from the
outer pad-rows towards the inner
pad-rows. When searching for a new space point, the nearby sub-volumes
according to Equation~\ref{TRACK_volsearch} are being searched. For every step
there are three possibilities:
\begin{itemize}
\item[--] There are no points in the sub-volumes.\\
If the track already has enough space points assigned, the track is
considered a finished track, and the assigned points are removed from
the available space point list. If the
track does not have enough points assigned, the track is rejected
and the space points are kept in the event.
\item[--] There is one space point in the sub-volumes.\\
If the space point gives a updated track fit with acceptable $\chi^2$,
the point is included and the track parameters are correspondingly
updated. Otherwise the track
is considered a final track if the number of already assigned space points is
sufficient. Otherwise the track candidate is deleted and the space
points made available for further tracking.
\item[--] There are several space points in the sub-volumes.\\
In this case the program can either search through all the space point
candidates and choose the one which gives the best $\chi^2$ of the track
fit, or it can search through the list until a space point which gives
good enough fit is found.
\end{itemize}
A flow diagram of the algorithm is illustrated in
Figure~\ref{TRACK_flowtracker}. 
\bfig[t]
\insertplot{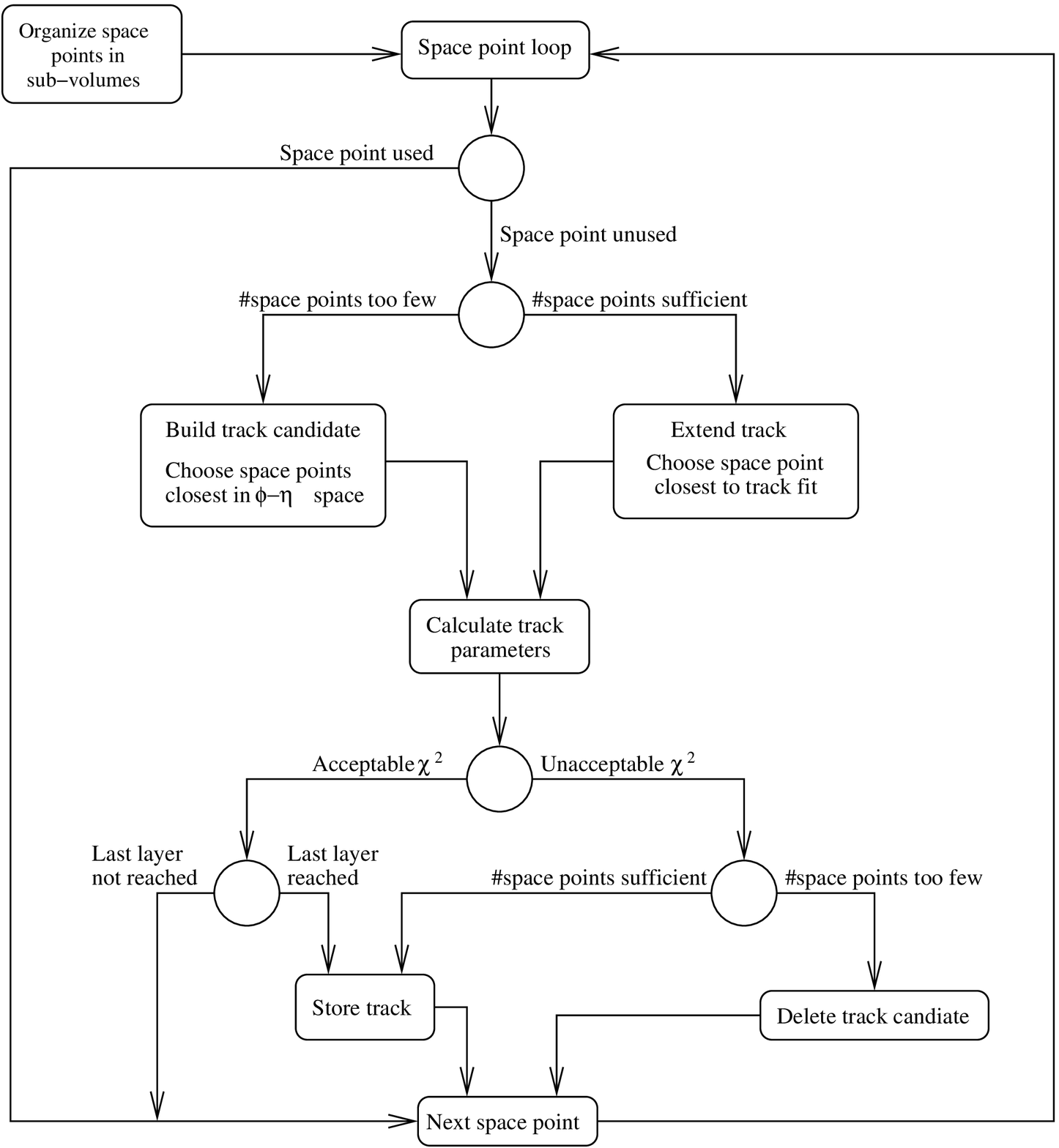}{10cm}
	{Flow diagram of the HLT Track Follower algorithm.}
	{Flow diagram of the Track Follower algorithm.}
\label{TRACK_flowtracker}
\efig

The performance of the algorithm on a given event sample is governed
by the values set for the various tracking parameters. These typically
include the granularity of the sub-volumes, maximum allowed $\chi^2$
during the track building and minimum number of hits on a valid
track.

\subsubsection{Non-vertex tracking}
The Track Finder can perform two tracking passes, where the second
pass takes the unused space points from the first pass as an input. In
this case, the Track Finder does not apply any vertex constraint when
building the tracks, and the conformal mapping is done relative to the
first point associated with the track. 

Due to the fact that most of the interesting tracks are in fact
originating from the area of the main vertex, this option is in
general not applied. Also, secondary tracks,
which have a small impact parameter, can still be detected by the Track
Finder if a sufficiently high uncertainty is assigned to the location
of the primary vertex.

\subsection{The Track Fitter}
\label{TRACK_helixfitter}
Once the Track Finder has recognized the tracks by grouping the
space points into subsets, a subsequent fit of the
subsets in space is done in order to obtain the best estimate of
the respective track parameters. 
The track fitting procedure assumes that the particles follows helical
trajectories. This assumption is valid if the magnetic field is
sufficiently uniform and energy losses are small. The helical
motion can be decomposed into a circle projected on the bending
plane and a straight line projected on the non-bending
plane (Appendix~\ref{APP_helix}). In order to optimize the fitting procedure
the track helix parameters are
evaluated in two independent procedures in which the space points are
first fitted to a circle in the bending plane and then to straight
lines in the non-bending plane. 

The circles in the transverse plane are fitted using an algorithm
presented in~\cite{oskov}. This fitting procedure is less
computationally expensive than normal $\chi^2$ minimization fitting
routines, while at the same time preserving its stability and
accuracy. 
The longitudinal part of the tracks are fitted to a straight line in
the $(s,z)$-space, which denotes the path along the helix and the
coordinate along the beam-direction, respectively.
The linear fit is a simple least squares fit to a line of the form
$y=ax+b$ using linear regression. 

\subsection{The Track Merger}
The purpose of the Track Merger is to merge multiple tracks segments belonging
to the same particle trajectory. This is necessary when a track is
being reconstructed in multiple parts corresponding to the different
sub-detectors. For the TPC this is the case if track finding is
performed independently in different sectors, that later are merged with
neighboring sectors.

The Track Merger is based on a simple matching method of track segments which
might belong to the same trajectory. The input to the Track Merger is
thus all
the reconstructed track segments within each individual sector. The
tracks are organized into individual lists corresponding to which sector they are
in. In a first step, all track segments within a list are checked
to determine if they cross the corresponding sector boundaries. This
is done by taking the intersection
of the projected helix in the transverse plane and the straight lines
representing the sector boundaries. Only the tracks for which this is
true, are further processed by the Track Merger. The
parameters of each track are calculated at
a plane perpendicular to the sector boundary, and in the
middle of the TPC cylinder. Pair of tracks within two
neighboring sectors are compared by taking the difference between
their respective track parameters at this plane. If the difference between
the parameters are less than a predefined threshold value,
the two tracks are merged into one track, and the two track
segments are subsequently removed from the lists. 
The track parameters which are evaluated are the two-dimensional crossing
point with the plane, the azimuthal angle of the track momentum,
the curvature and the dip angle.

\subsection{Data flow}
The complete track reconstruction chain described above, has been
implemented in a generic way. This allows for a high degree of
parallelization of the different processing components. The final
processing scheme will be adapted to both the inherit data flow coming from the
detectors into the HLT system, Section~\ref{HLT_arch}, and the
required data flow from the individual processing components. The
potential decomposition of each processing module depends on the
inherit locality of the algorithm with respect to required input
data. 

{\bf Cluster Finder} The Cluster Finder solves a local two-dimensional
problem, and does not 
need information from more than one pad-row at the same time. Thus the
algorithm could be broken into a number of parallel components
corresponding to the total number of pad-rows in the TPC. This allows
the algorithm to be implemented at a very early stage in the
readout scheme.

{\bf Track Finder} Since the tracks have to 
be build using space points from different layers, this processing
step needs a certain number of successive pad-rows. In principal only five
points are needed to fit a track to a helix, however,
the resolution of the obtained track parameters depends on the
track length and the number of assigned space points. Also,
when dividing the track
finding problem into smaller regions of the detector, the track
segments have to be merged across the boundaries of these regions which
again will introduce an extra processing step. Such an additional
track merging step will also prevent some split tracks from merging,
and consequently lead to a certain loss of tracking efficiency. One
therefore has to weight the trade-off between the gain of reduced
processing time with the complexity of the hardware topology and the
possible loss of tracking performance. 

{\bf Final track fitting} The final tracks can only be obtained when
the information from the tracking detectors has been
collected. In the case of TPC tracking, this corresponds to all
the pad-rows in which the tracks have produced a signal. Hence, the final
track fitting of the TPC tracks can only be done when the track and
space point information has been collected from all the TPC sectors,
and the track segments have been merged across the detector boundaries.

\bfig[htb]
\insertplot{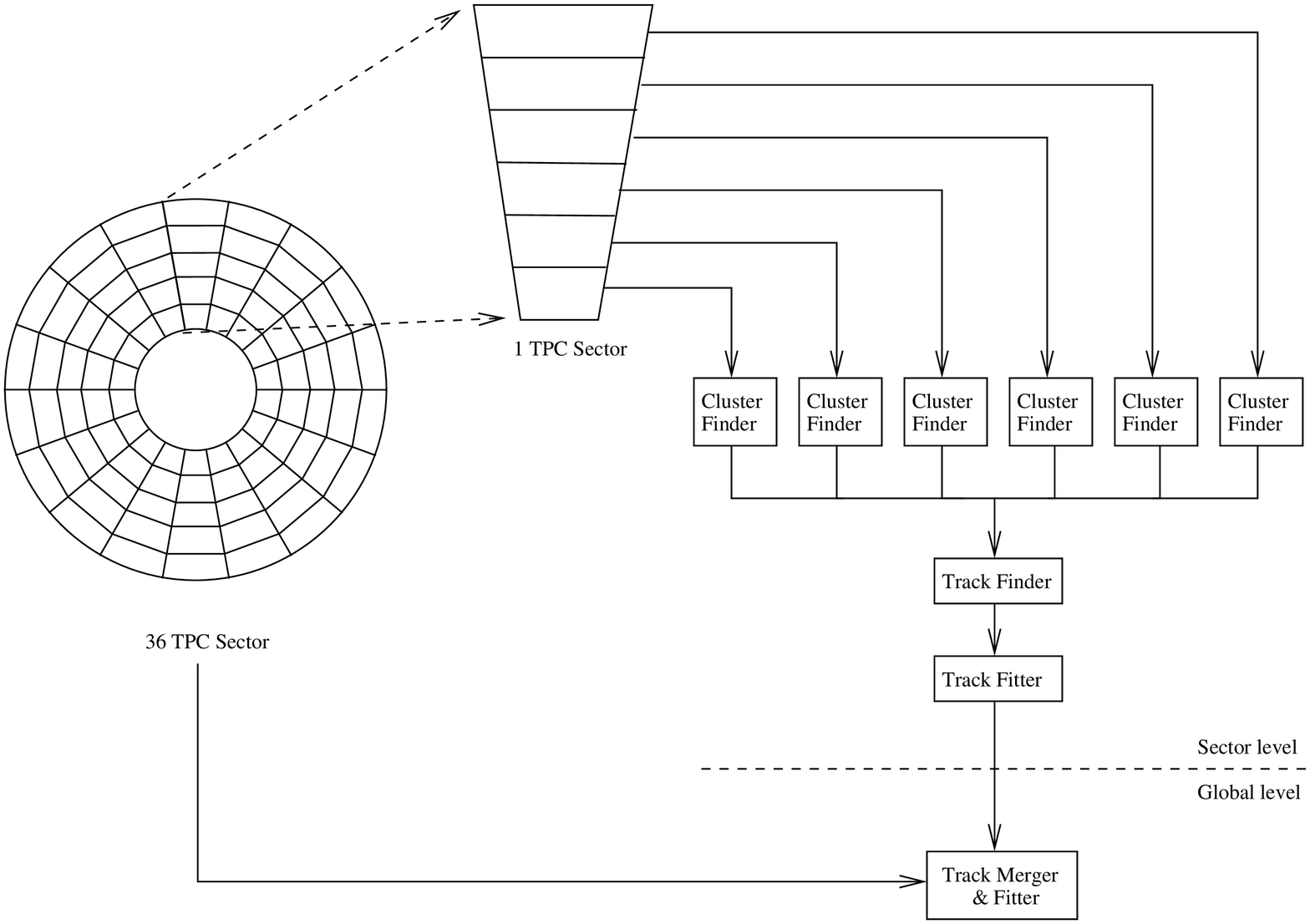}{12cm}
	{Possible data flow for the sequential track reconstruction chain
within the HLT system.}
	{Data flow for the sequential track reconstruction chain.}
\label{TRACK_topology}	
\efig

Figure~\ref{TRACK_topology} shows a possible data flow scenario of the
sequential tracking approach.
Given the inherit parallelism of the Cluster Finder a natural choice will be
to run it on the Front-End Processors (FEP) of the HLT
system, where a copy of the raw-data is received in a massive parallel
fashion. Hence there are six Cluster Finder processes running in parallel for each TPC
sector, each processing the data from one sub-sector. The output are then
collected at the TPC sector level, where the Track Finder takes as
input the list of space points from a complete TPC sector. Once the
Track Finder has found all track segments, they are passed to
the Track Fitter which performs a fit of the tracks to obtain the
helix parameters. Finally, the tracks from all 36 TPC sectors are
collected at a global layer, where they are merged across the sector
boundaries. In addition, a final track fit is performed on the merged tracks
in order to get the best estimate of the track parameters.

In this way the processing topology follows a tree-like structure,
where the output from one processing level is being merged at a higher
level until all tracks are globally reconstructed. Such a scheme
is consistent with a parallel solution implementing a generic PC cluster
of individual nodes, whose number can be adjusted according to the
computing requirements of the individual processing components.

\subsection{Performance}
\label{TRACK_seqperformance}
The implemented sequential reconstruction chain has been evaluated for
various particle multiplicities.
The event samples were generated using the HIJING
parameterization, Section~\ref{TRACK_evsim}, and four different
multiplicities corresponding to \dndy of 1000, 2000,
4000 and 6000. The magnetic field settings correspond to standard solenoidal
field map of 0.2\,T and 0.4\,T. All results have been obtained using a
data flow scheme as illustrated in Figure~\ref{TRACK_topology}.
For each event sample the results are compared to the
results obtained by the Offline TPC reconstruction chain
(Section~\ref{TRACK_offlinechain}). In the following the results are
referred to as HLT and Offline respectively. Unless otherwise stated, the
standard Offline algorithm has been used.

\subsubsection{Definitions}
\label{TRACK_trackdef}
In order to determine the performance of the track reconstruction
chain, certain definitions are required. In principle
there are no ``global'' detector independent definitions to be used
for this purpose. Instead, definitions which are agreed upon within
the collaboration, are applied. 
In particular, one has to define the quantities which enter
the equations:
\begin{eqnarray}
\mathrm{Efficiency} &=& \frac{Number\ of\ found\ good\ tracks}{Number\
of\ generated\ good\ tracks}\nonumber\\
\mathrm{Contamination} &=& \frac{Number\ of\ found\ contaminated\
tracks}{Number\ of\ generated\ good\ tracks}
\label{TRACK_effdef}
\end{eqnarray}
A {\it generated good track} refers to a generated particle whose
trajectory should be found by the reconstruction program. In general
this means
that the particle trajectory is contained within the detector acceptance, and
simultaneously produces sufficient amount of signals when traversing the
detector. For instance, a
track which traverse through a dead-zone of the TPC detector might
not produce enough clusters on the pad-rows to be
reconstructed, and thus should not be counted as a generated good
track. Furthermore, a {\it found good track} is a track which 
has been correctly reconstructed from the simulated raw-data. This
usually means that a certain number of space points were assigned to
the track, and that most of these space points were reconstructed from
clusters produced by the generated good track. Similarly, a {\it found
contaminated track} refers to a track which had enough assigned space points,
but with too many of them wrongly assigned. 

In this work the definitions from the
Offline reconstruction framework~\cite{alicetdr} are used:
\begin{itemize}
\item {\it Generated good track} -- A track which crosses at least
40\% of all pad-rows. 
\item {\it Found good track} -- A track for which the number of assigned
space points is at least 40\% of the total number of pad-rows. In addition
the tracks should have no more than 10\% wrongly assigned space
points, and half of the innermost 10\% of the space points must be
assigned correctly.
\item {\it Found contaminated track} -- A track which has sufficient number
of space points assigned ($\geq$\,40\%), but more than 10\% wrongly
assigned clusters. 
\end{itemize}
If not otherwise stated only primary tracks were considered in the
evaluations.

\subsubsection{Tracking efficiency}
The obtained tracking efficiencies as a function of \pt are shown in
Figure~\ref{TRACK_seqeffvspt} for the different event samples.
Figure~\ref{TRACK_seqfakevspt} shows the corresponding contamination.
\bfig
\centerline{\epsfig{file=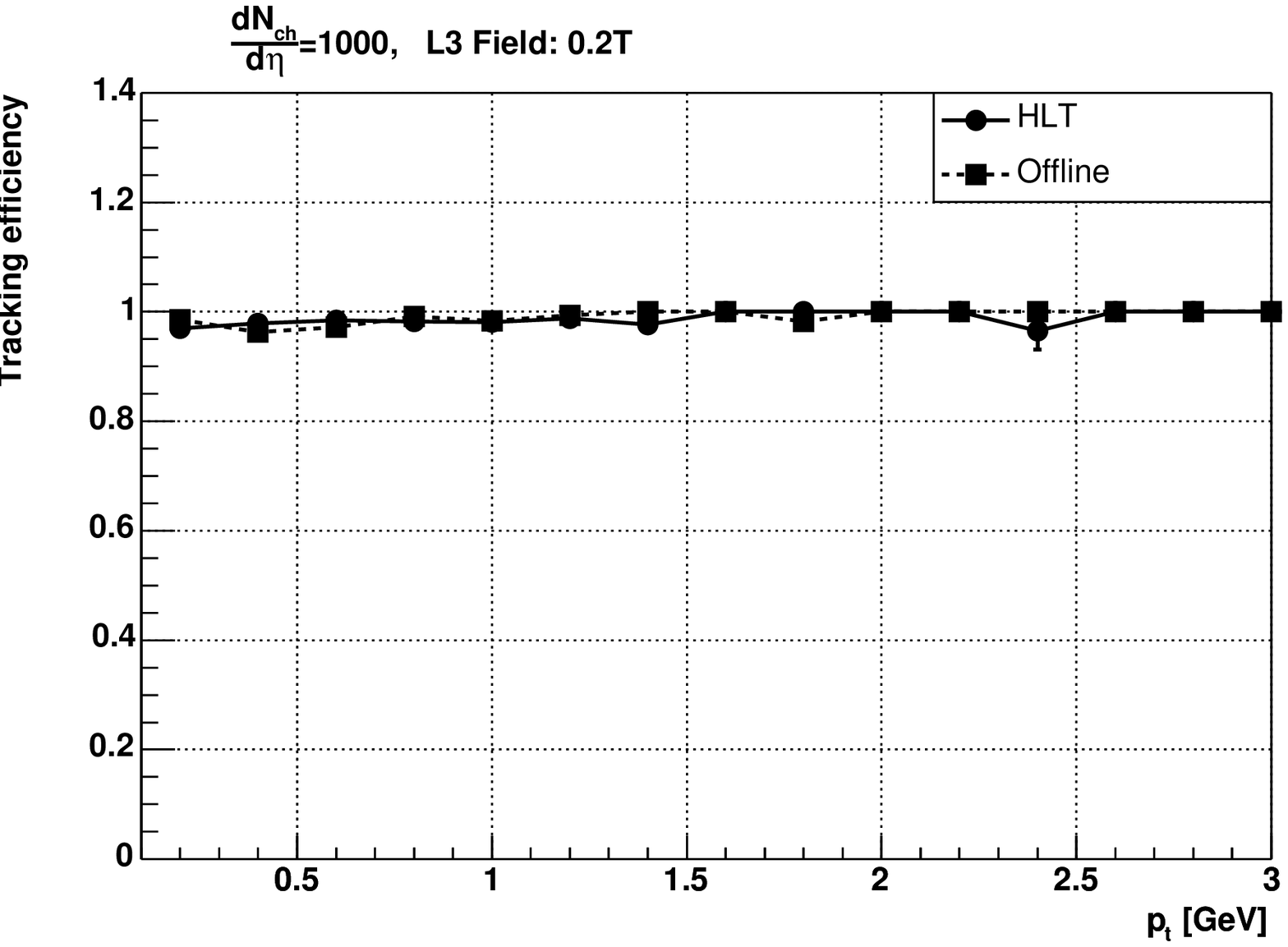,width=8cm}
\epsfig{file=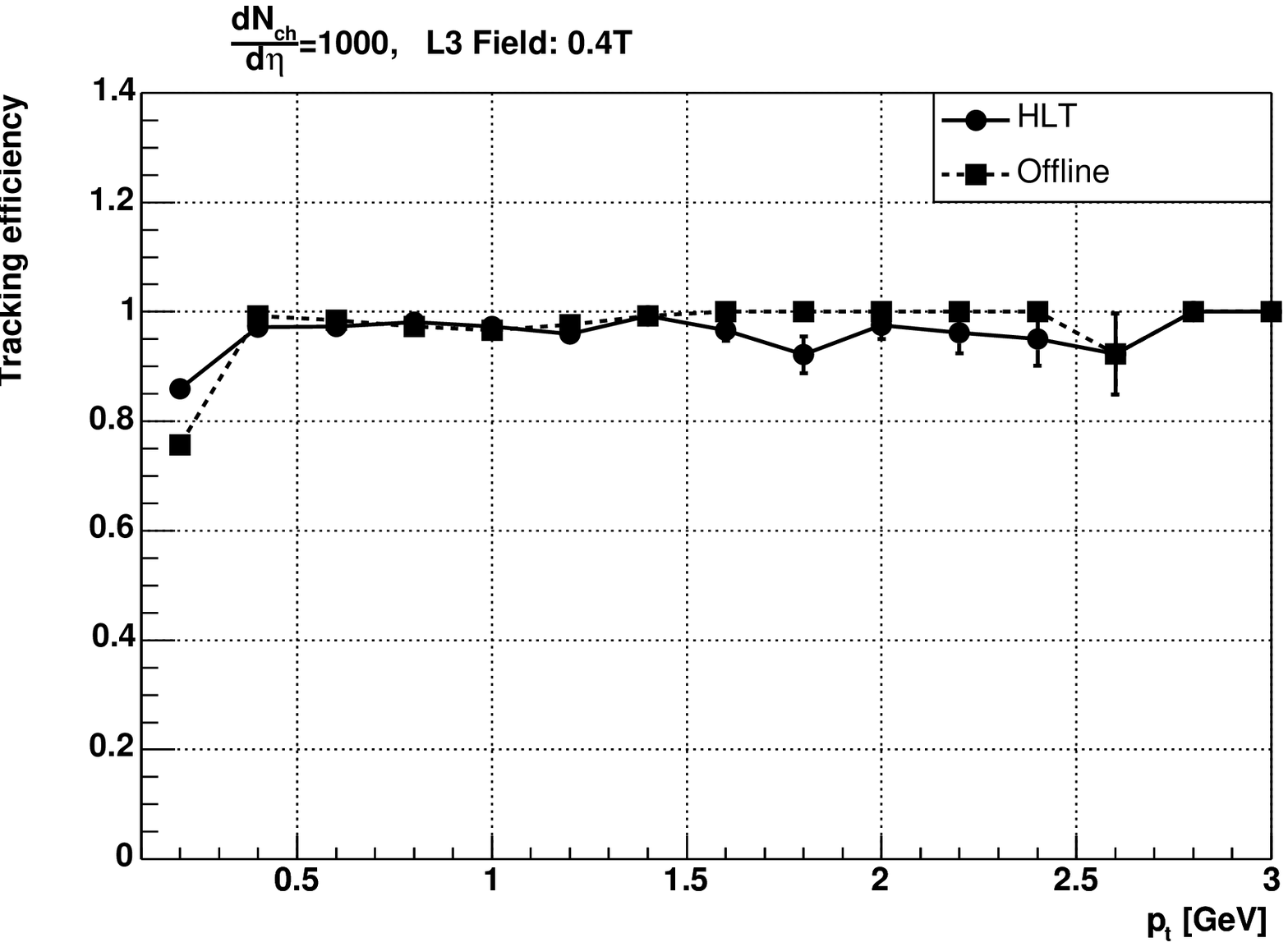,width=8cm}}

\centerline{\epsfig{file=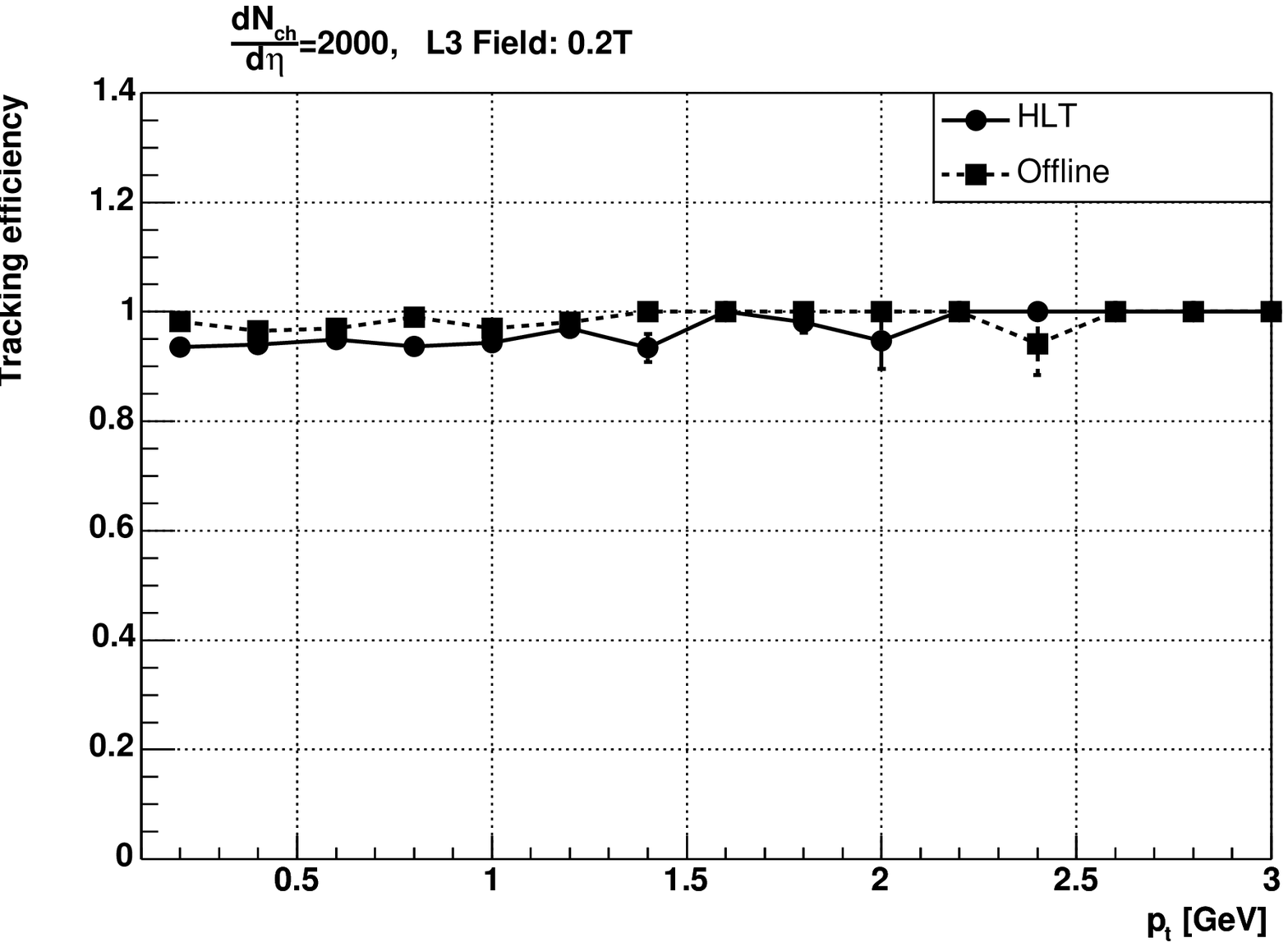,width=8cm}
\epsfig{file=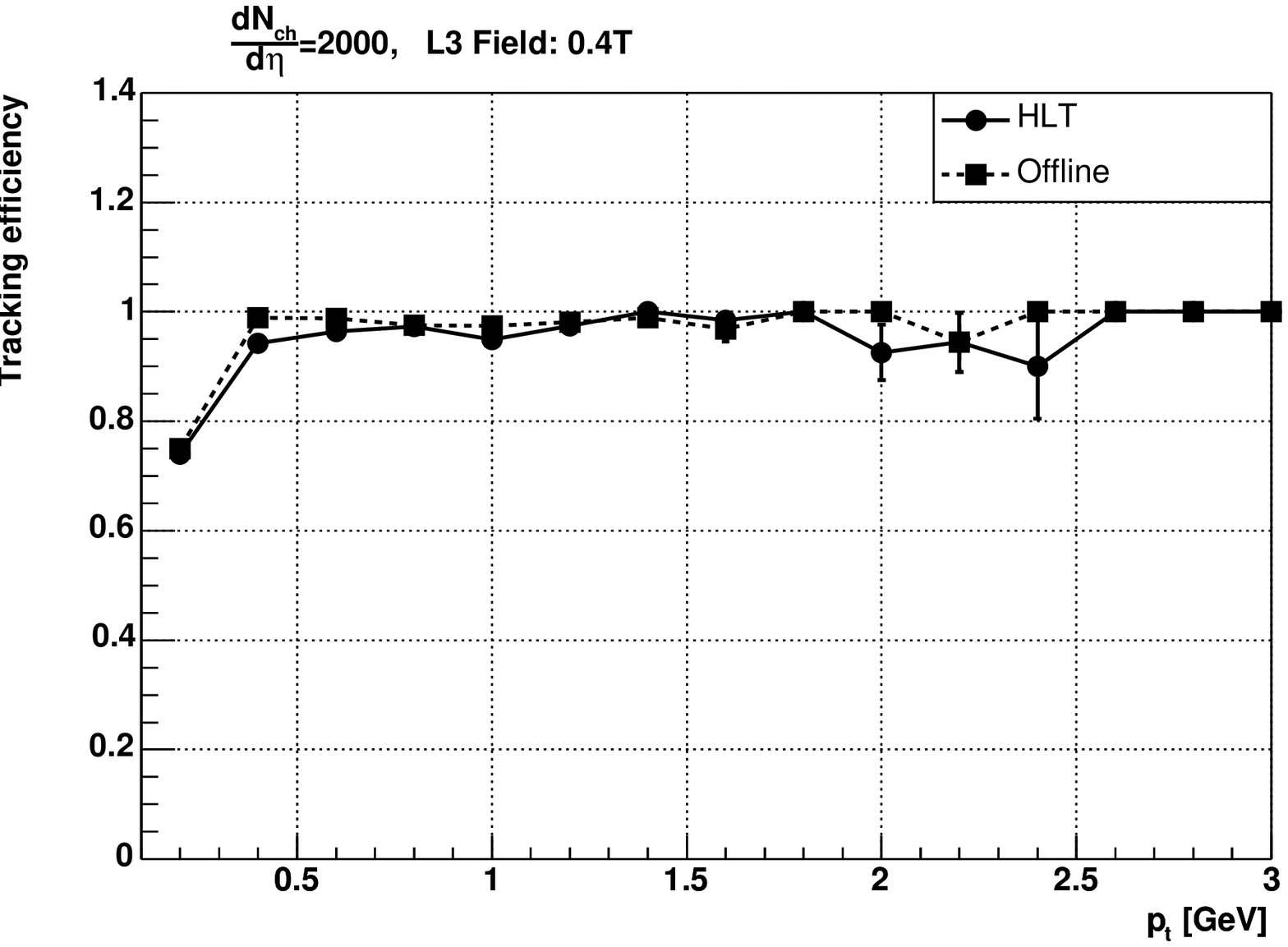,width=8cm}}

\centerline{\epsfig{file=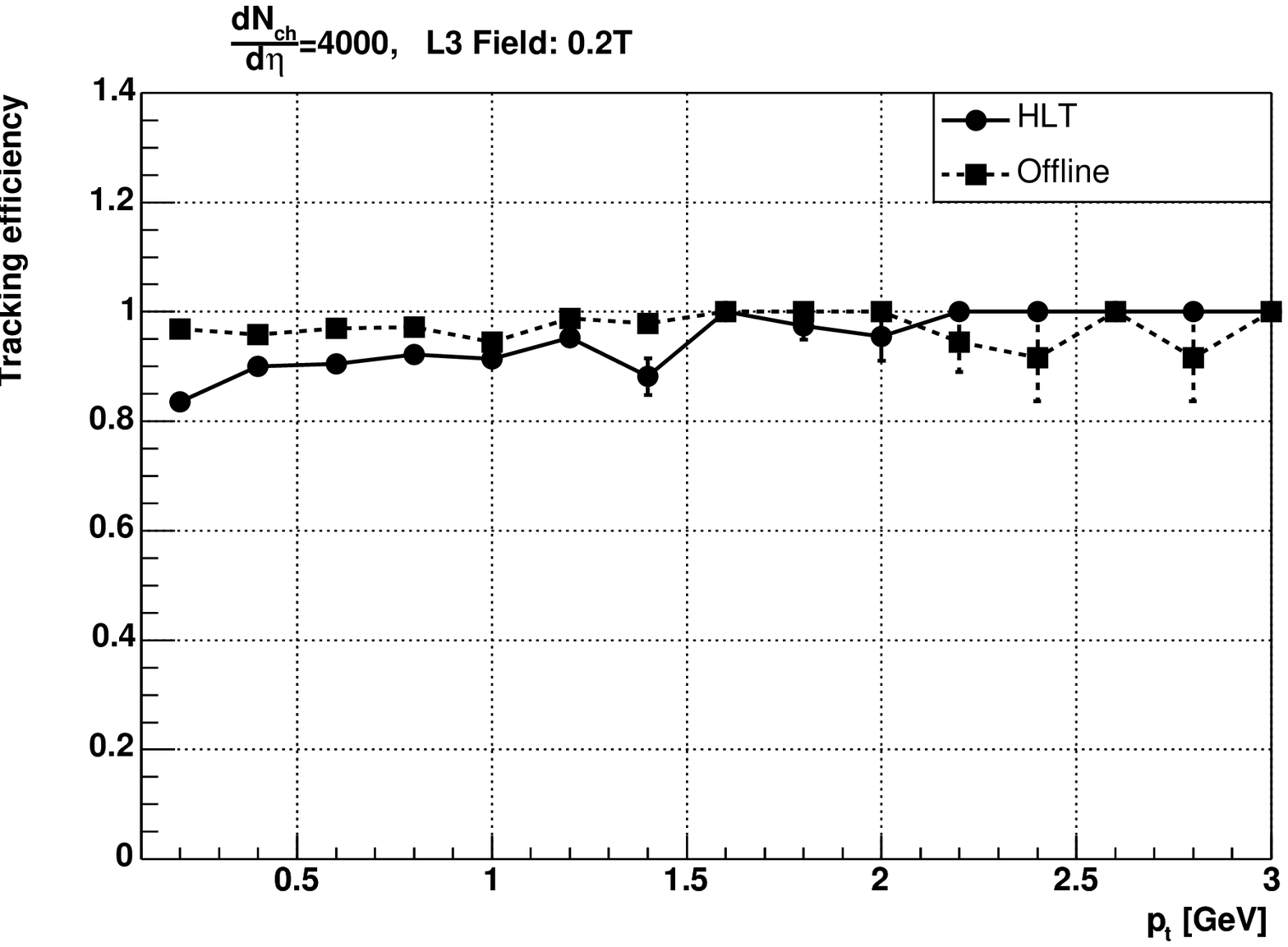,width=8cm}
\epsfig{file=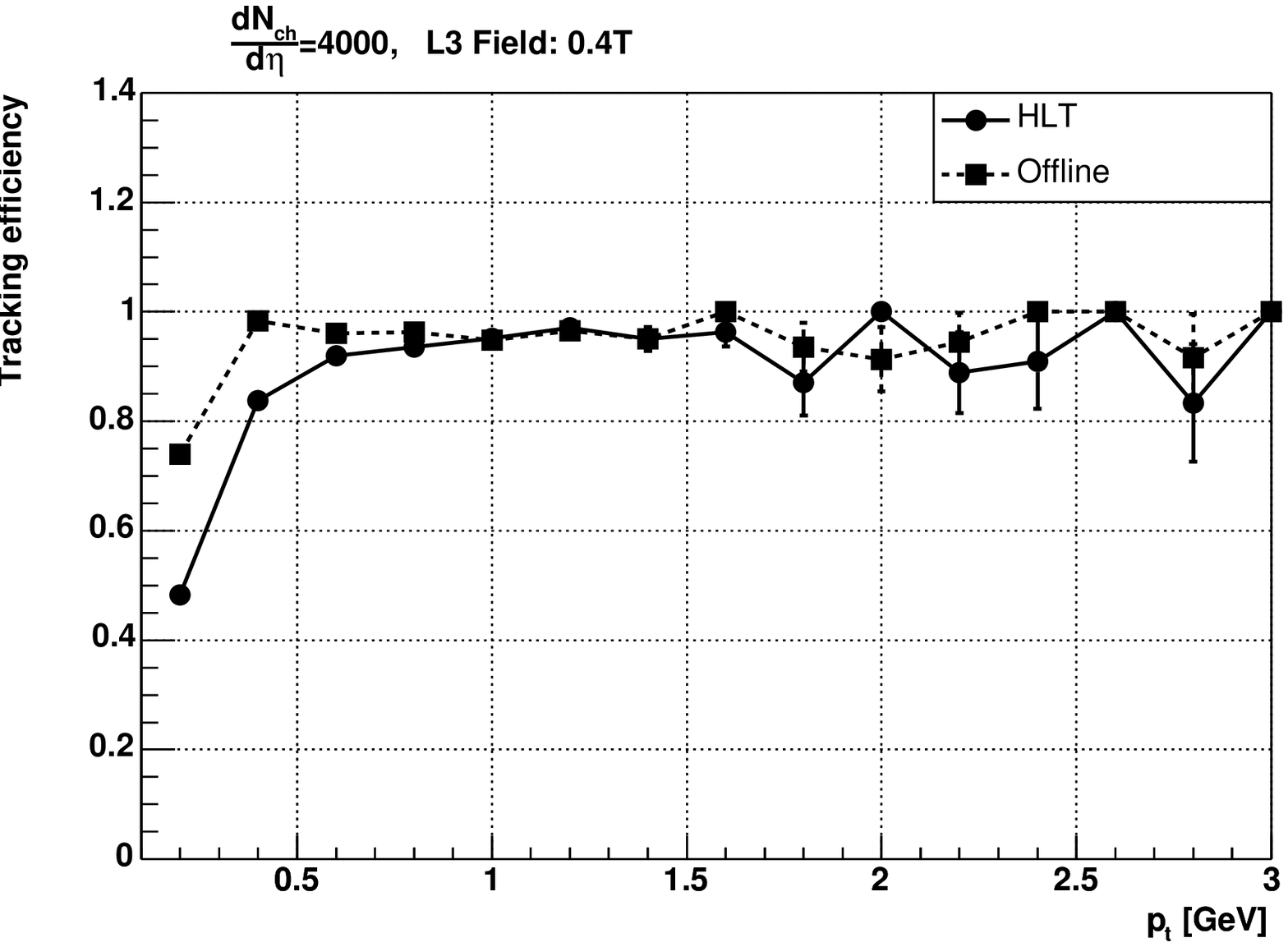,width=8cm}}

\centerline{\epsfig{file=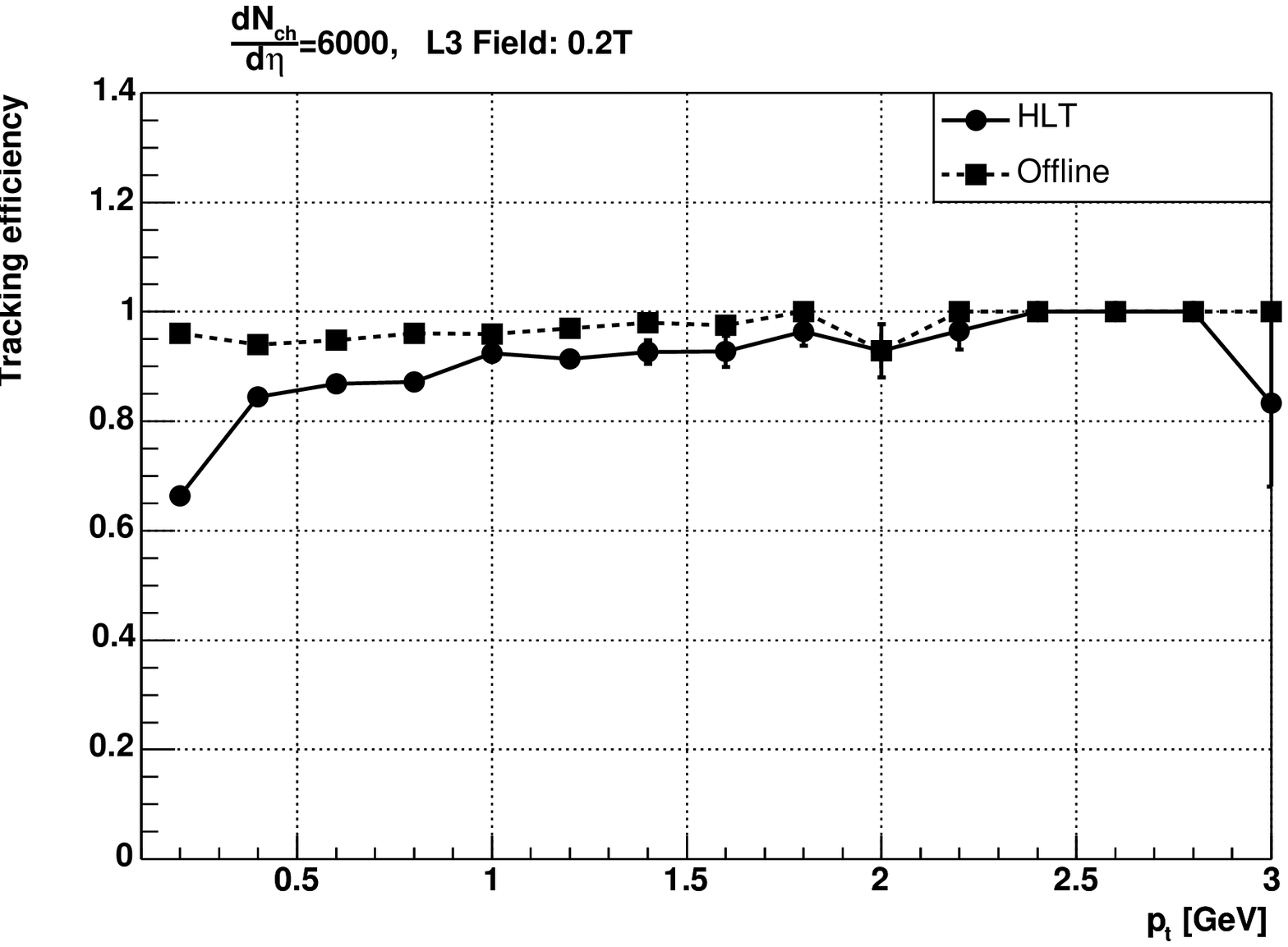,width=8cm}
\epsfig{file=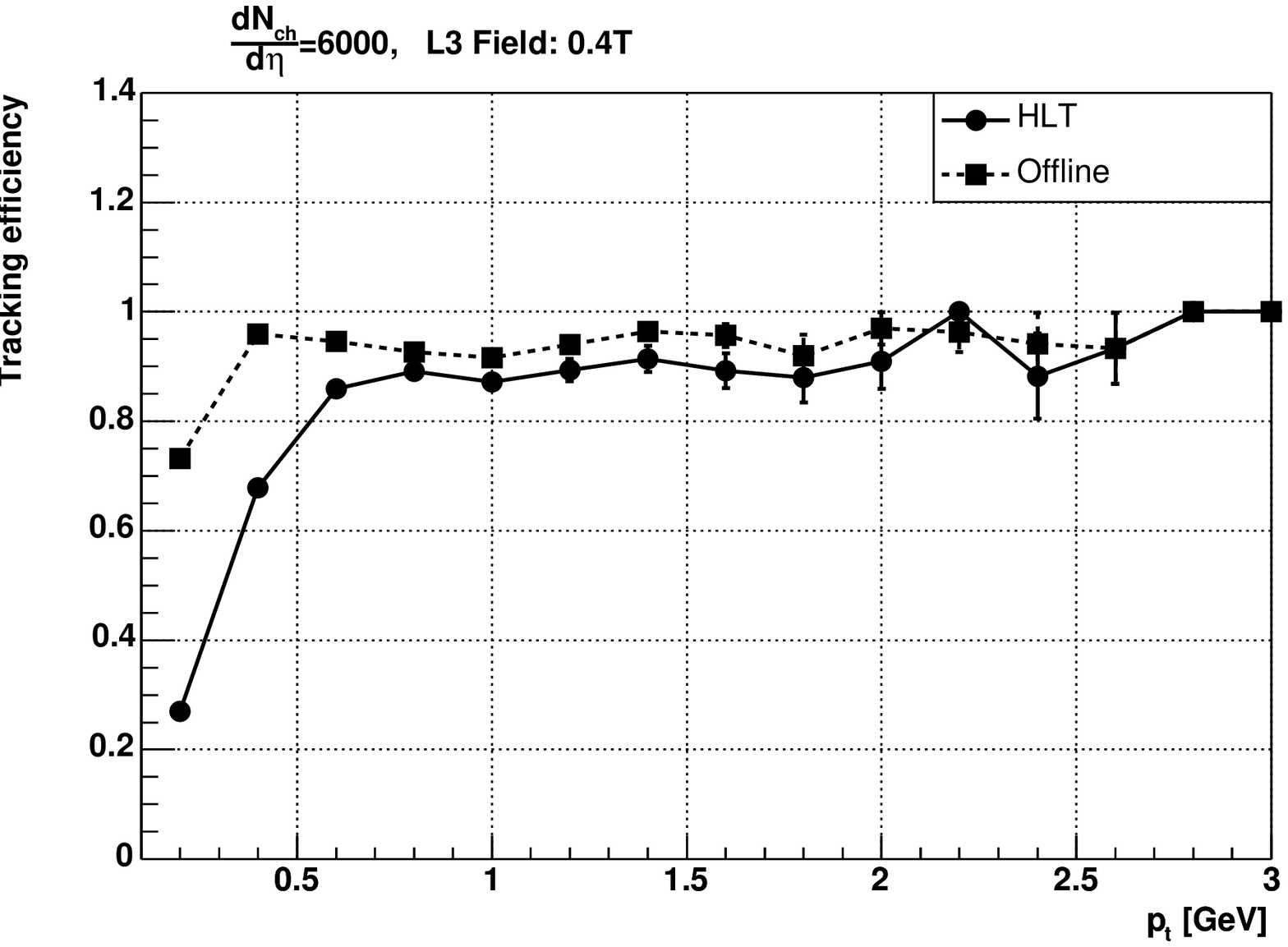,width=8cm}}

\caption[Tracking efficiencies as a function of $p_t$ for the HLT
sequential track reconstruction chain.]
	{Tracking efficiencies as a function of $p_t$. Results using
magnetic field strength of 0.2\,T (left) and 0.4\,T (right).}
\label{TRACK_seqeffvspt}
\efig
\bfig
\centerline{\epsfig{file=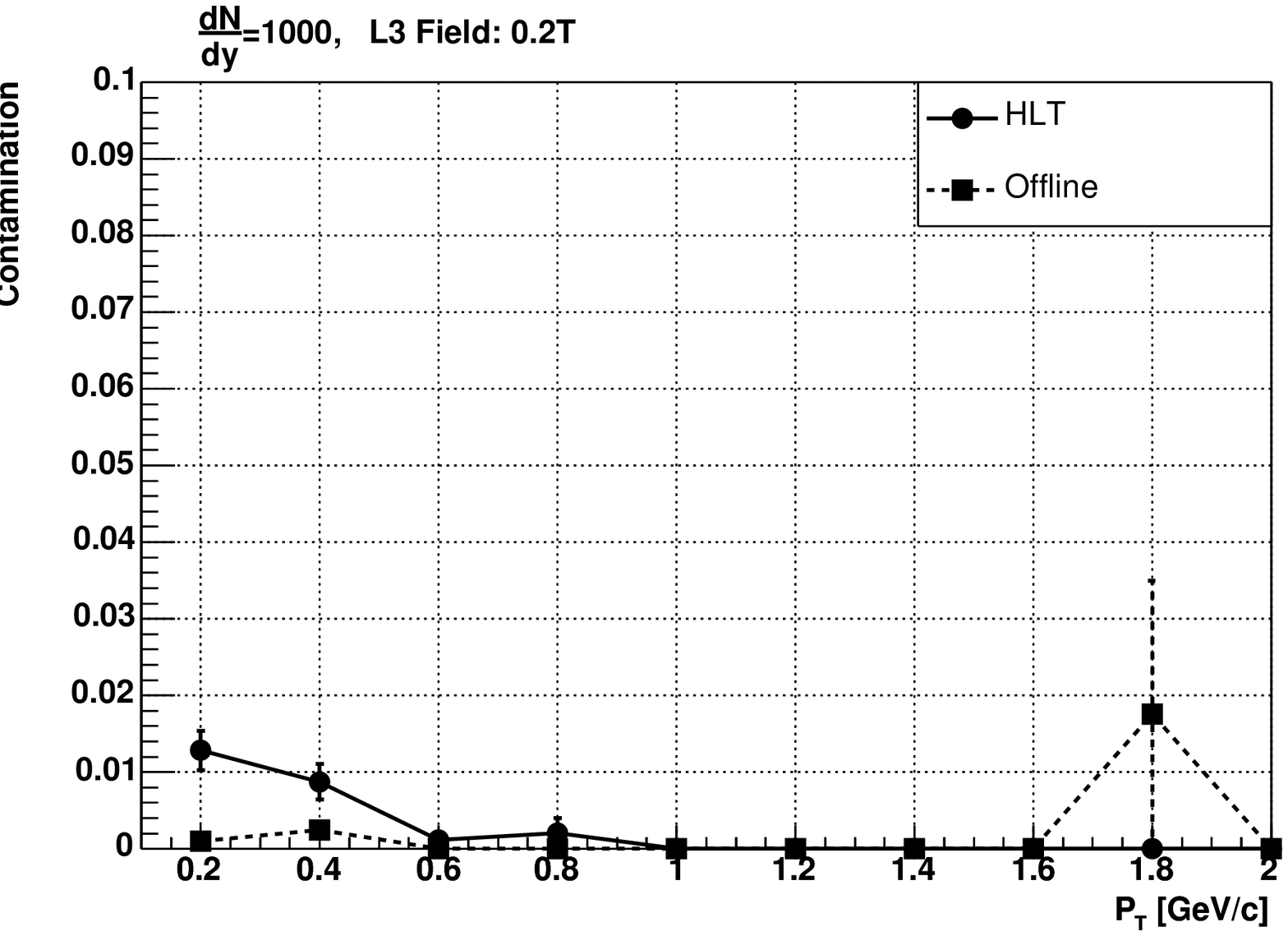,width=8cm}
\epsfig{file=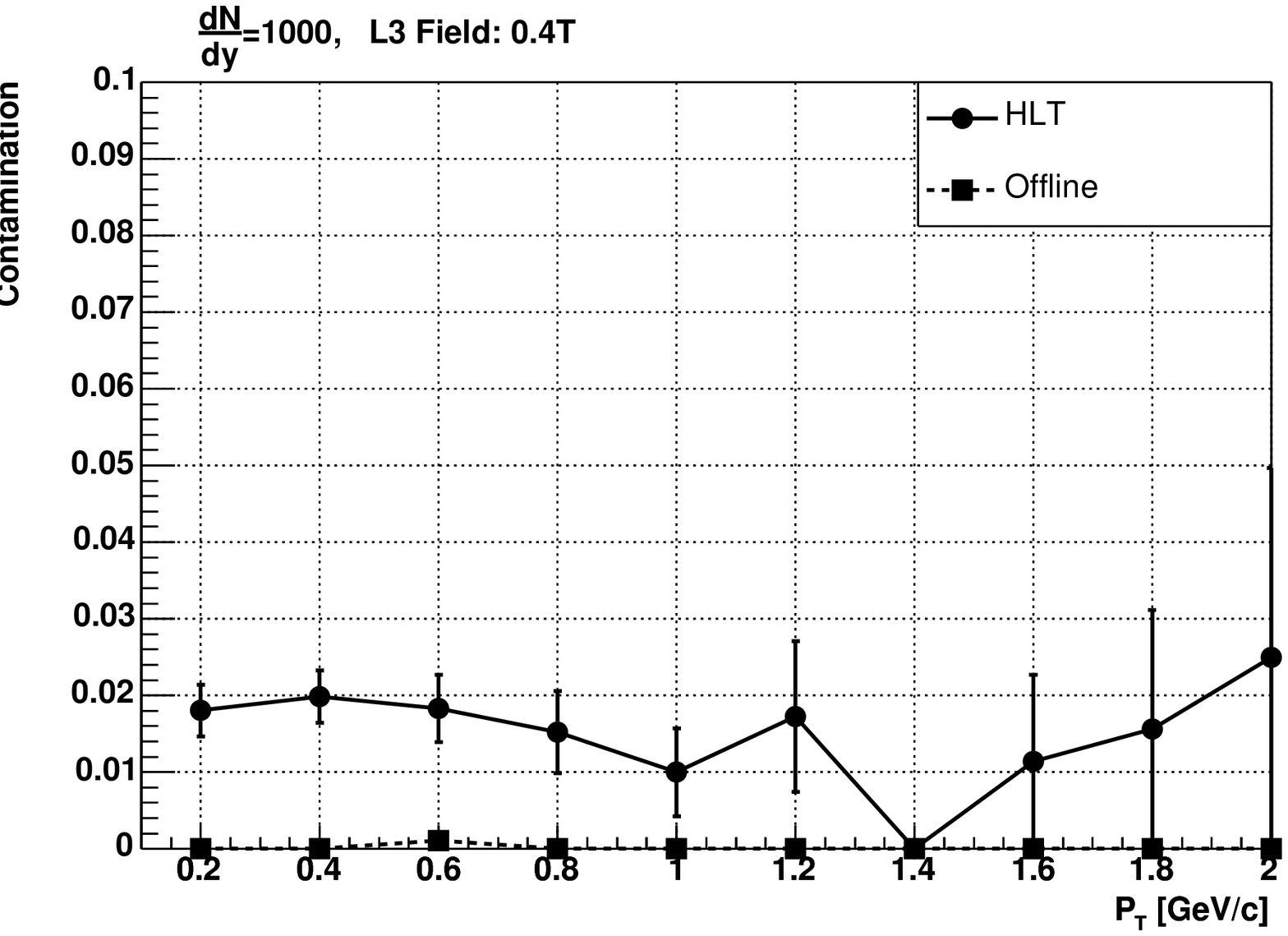,width=8cm}}

\centerline{\epsfig{file=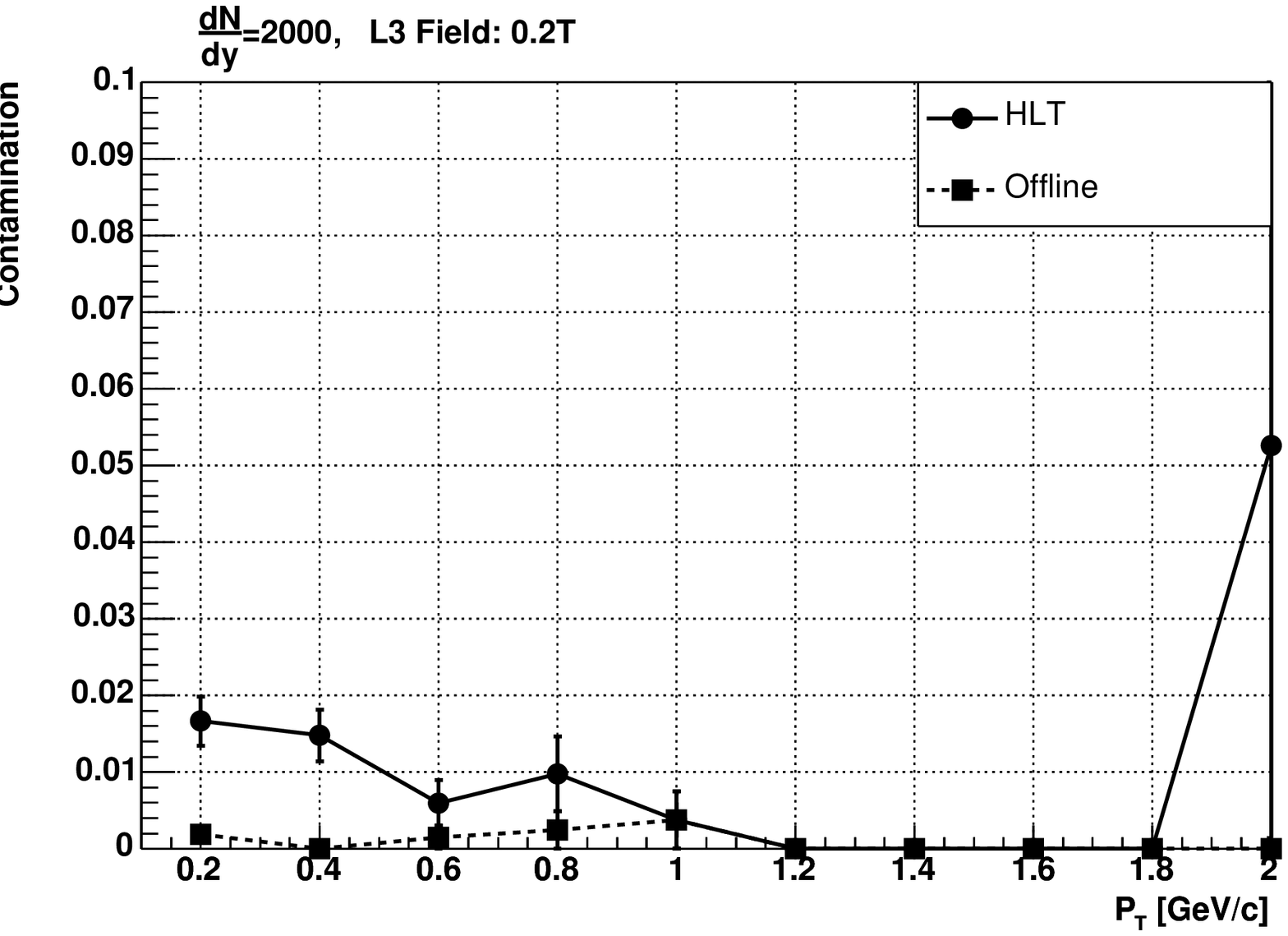,width=8cm}
\epsfig{file=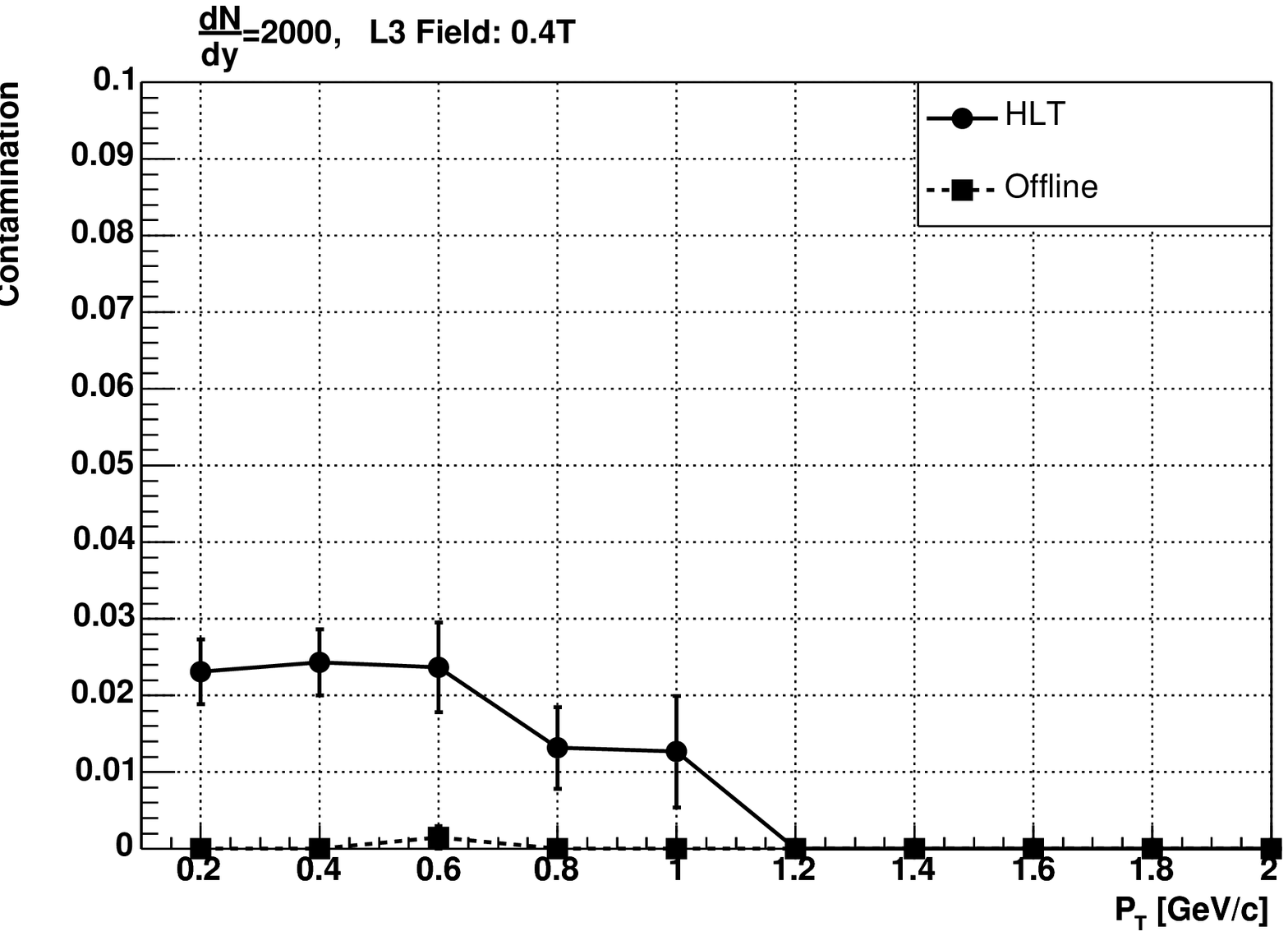,width=8cm}}

\centerline{\epsfig{file=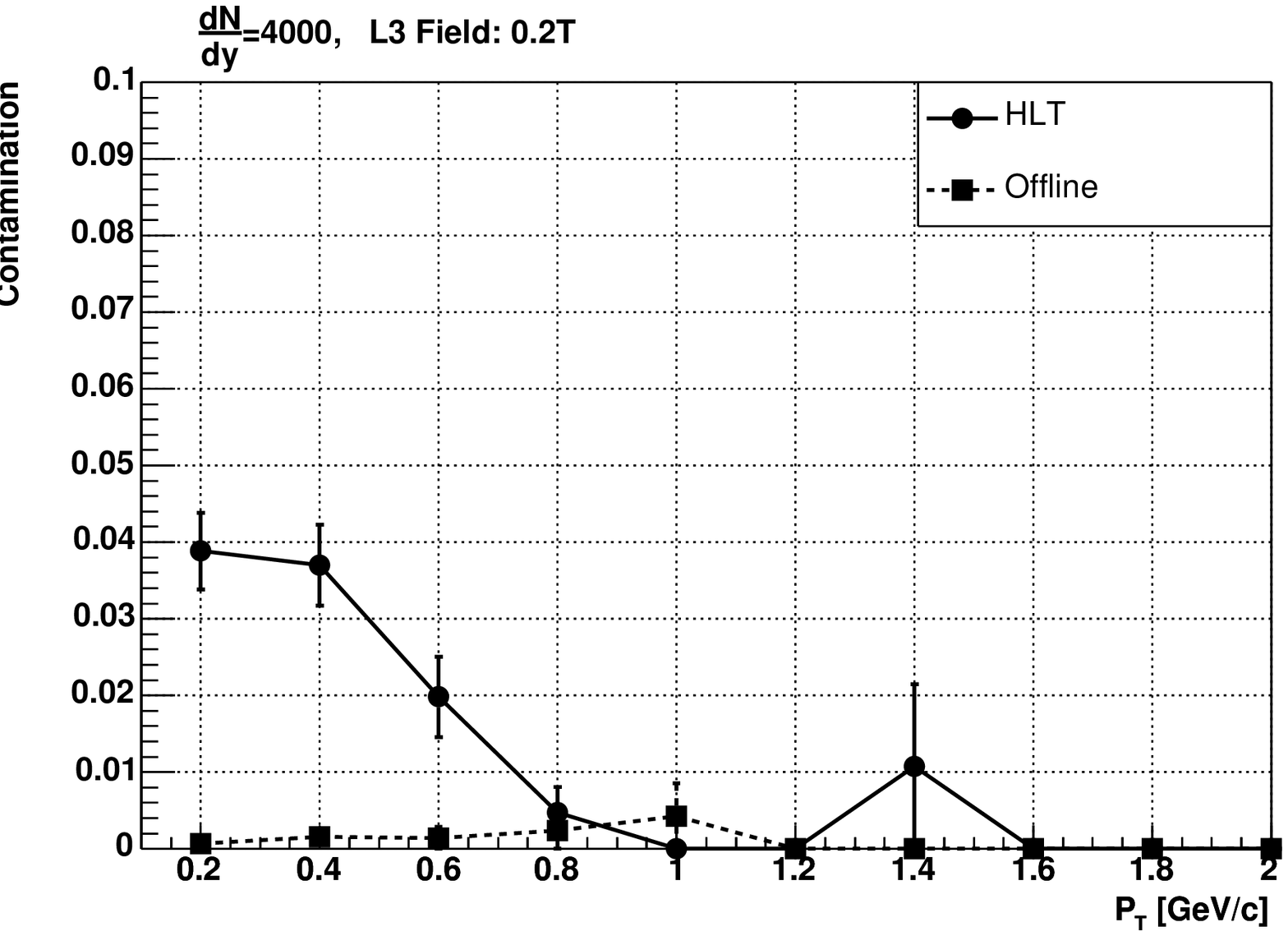,width=8cm}
\epsfig{file=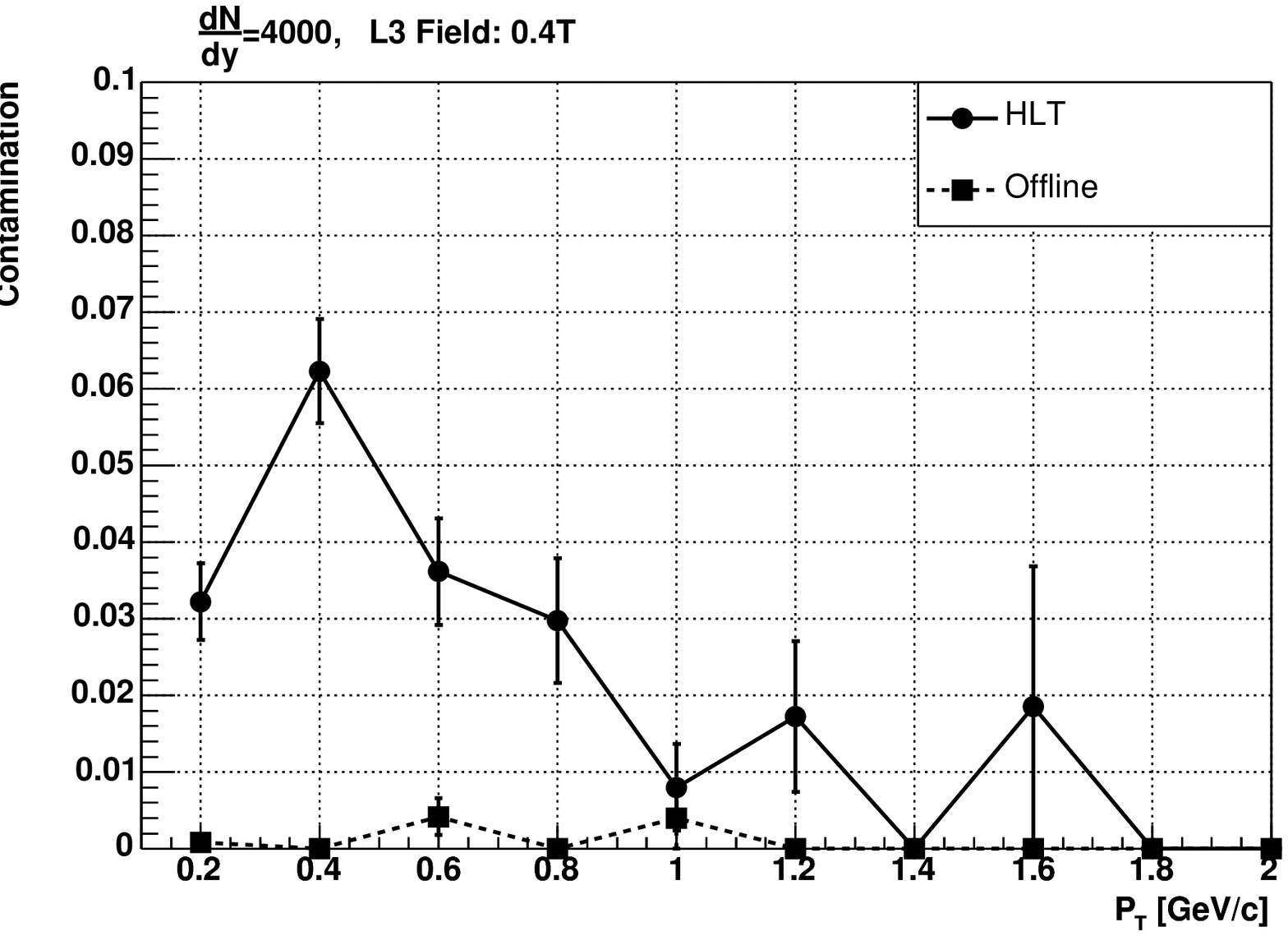,width=8cm}}

\centerline{\epsfig{file=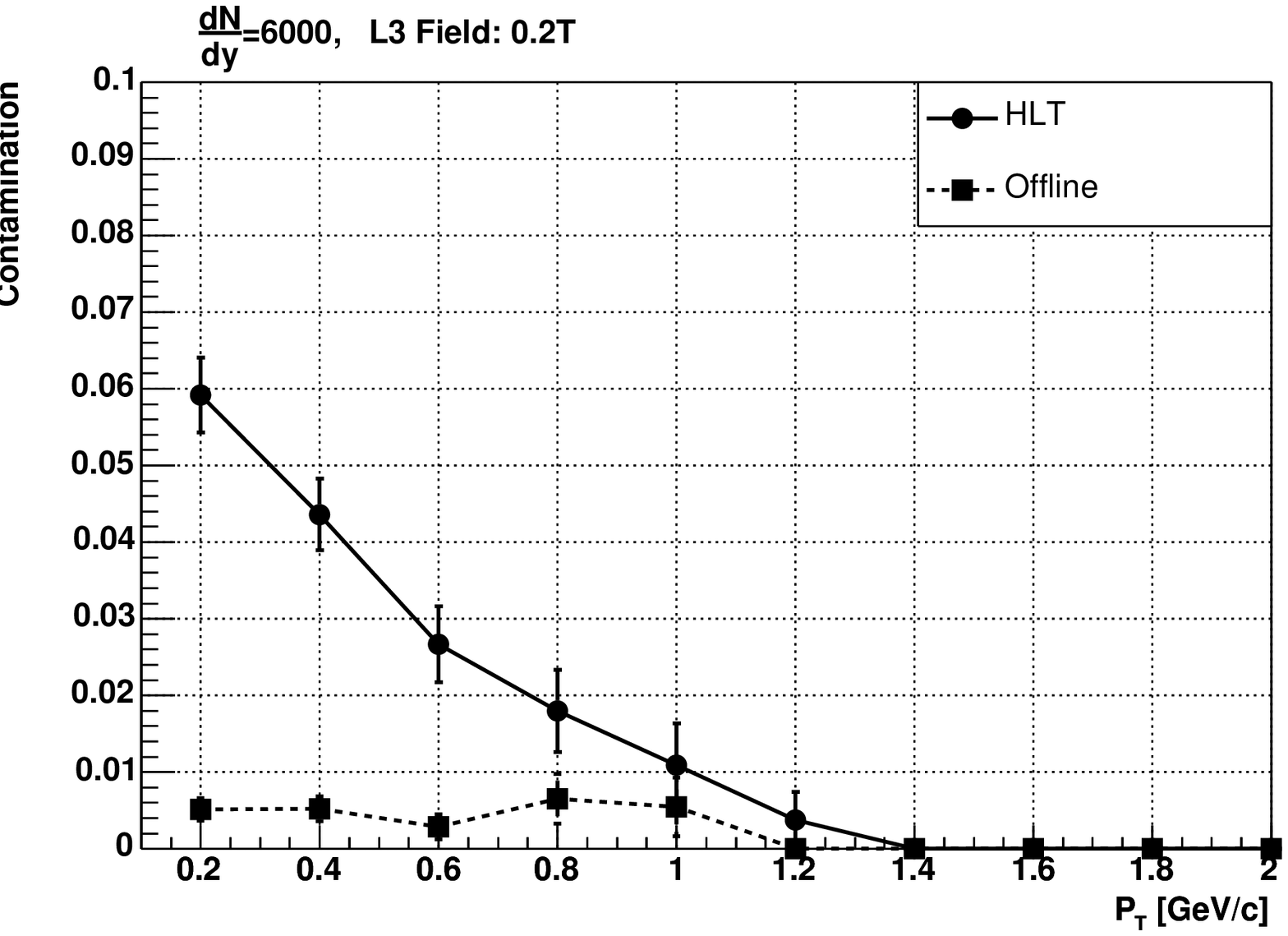,width=8cm}
\epsfig{file=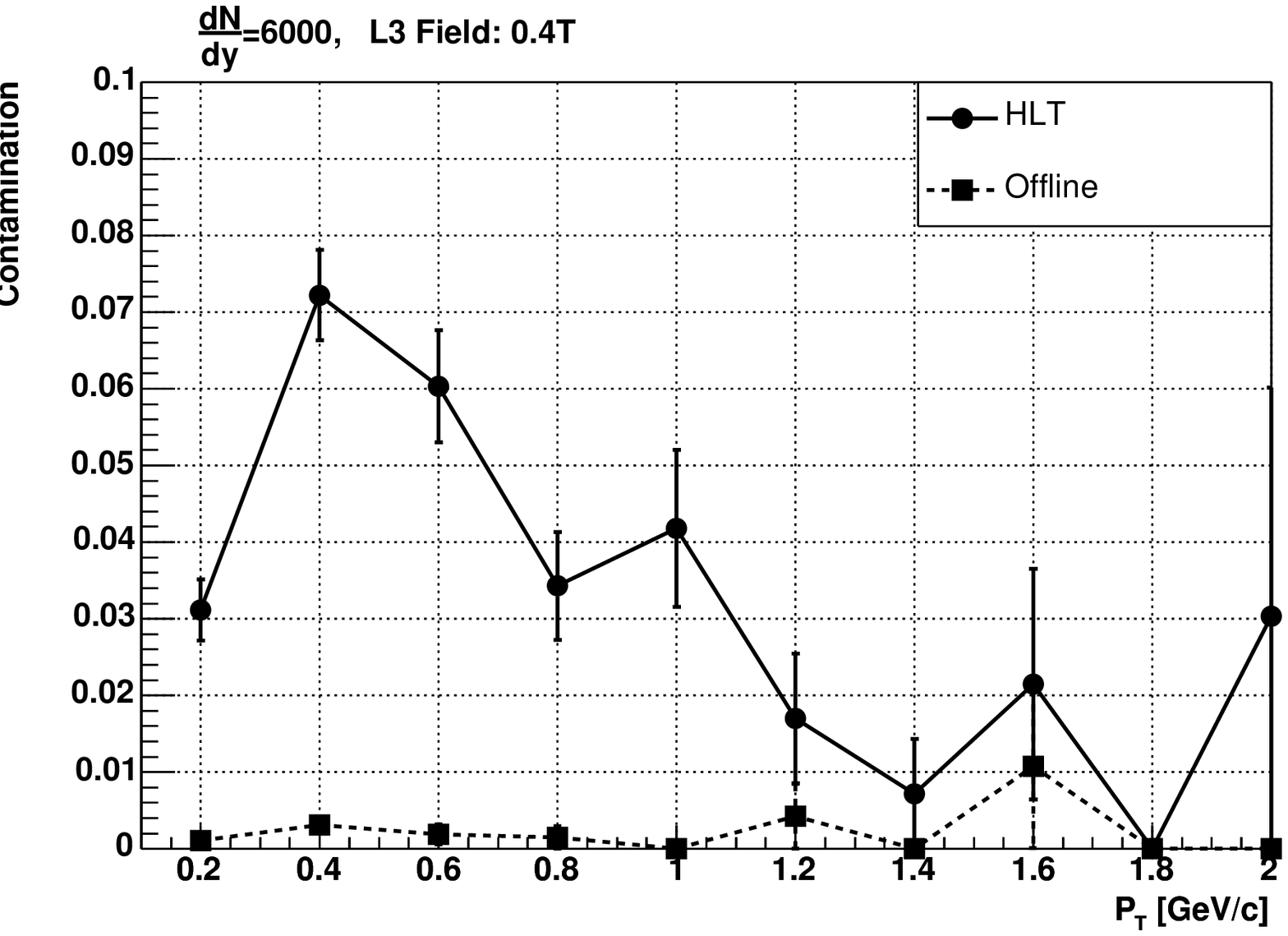,width=8cm}}

\caption[Contamination as a function $p_t$ for the HLT sequential
track reconstruction chain.] 
	{Contamination as a function of $p_t$. Results using magnetic
field strength of 0.2\,T (left) and 0.4\,T (right).}
\label{TRACK_seqfakevspt}
\efig
For a magnetic field strength of 0.2\,T the efficiency shows no
significant dependence on the transverse momentum at lower
multiplicities. At \dndy=\,6000, however, a decrease of the low
momentum efficiency is observed. This decrease is not seen in the
Offline efficiency. For a field strength of 0.4\,T, a decrease is
seen for the low \pt component for both HLT and Offline for all
multiplicities. However, the tendency is more significant for the HLT
results when going towards higher multiplicities. Up to \dndy=\,2000, HLT
and Offline results are similar, while for \dndy=\,4000 and 6000 the low \pt
efficiency for HLT is significantly lower than corresponding Offline results.


Figure~\ref{TRACK_inteff} shows the resulting
integral efficiency and contamination as a
function of multiplicity for both magnetic field strengths. An obvious
similarity is seen in the results obtained with the
two tracking approaches for multiplicities $\leq$\,2000 for both field
settings. For higher multiplicity, the efficiency of HLT algorithms
is partially lost to contaminated tracks whose relative amount
reaches 4-5\% for \dndy=\,4000-6000.

\bfig[htb]
\centerline{\epsfig{file=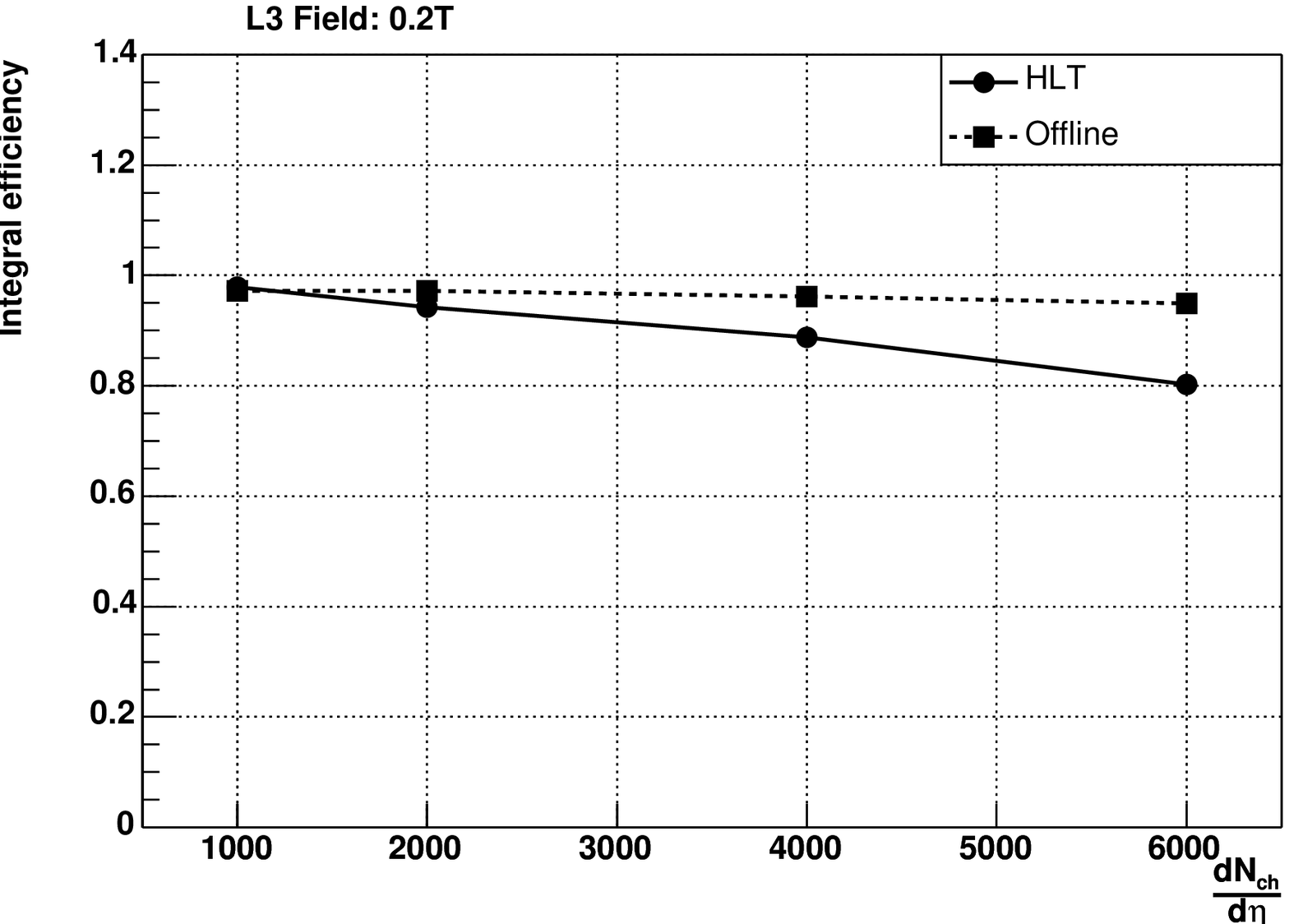,width=8cm}
\epsfig{file=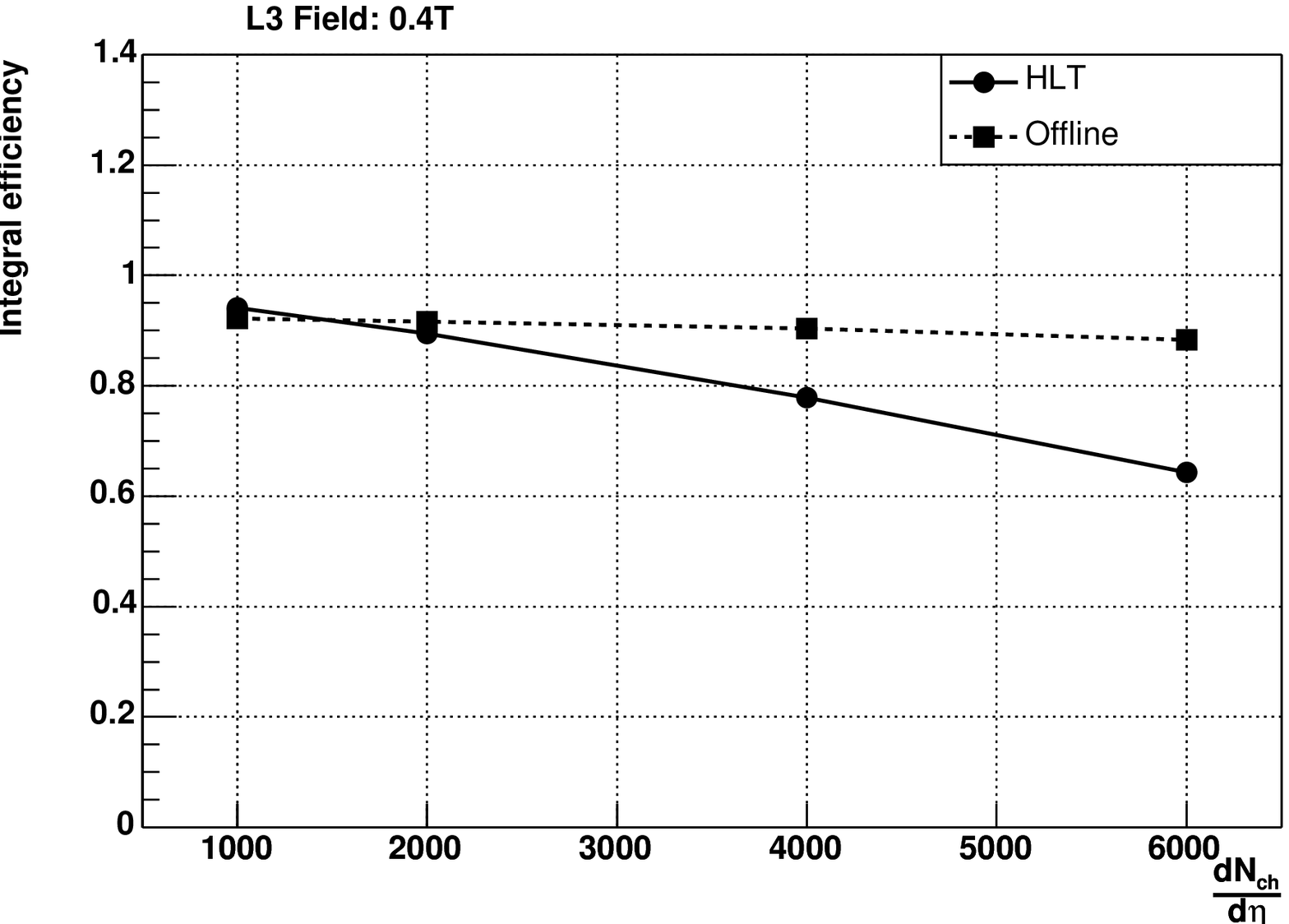,width=8cm}}
\centerline{\epsfig{file=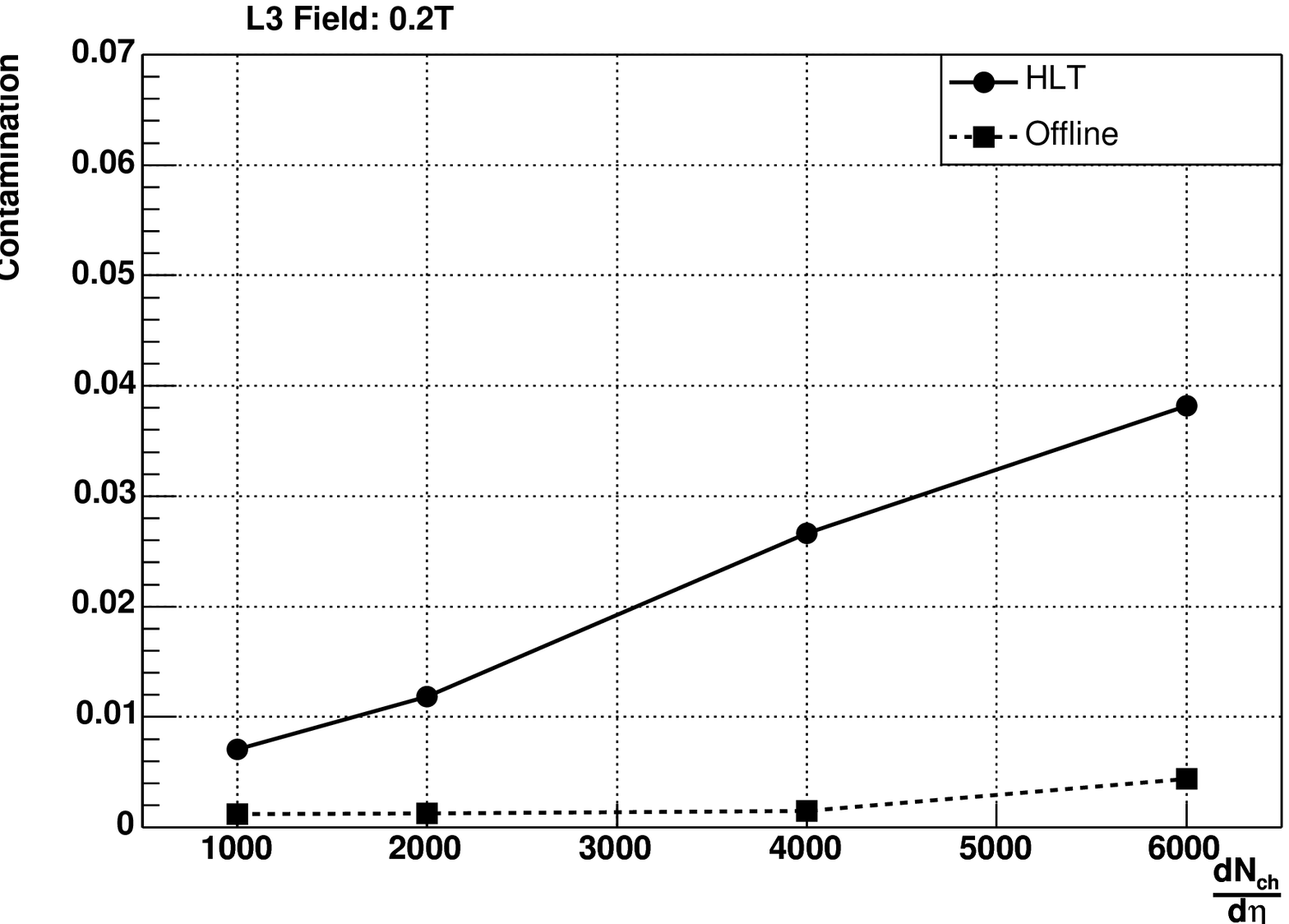,width=8cm}
\epsfig{file=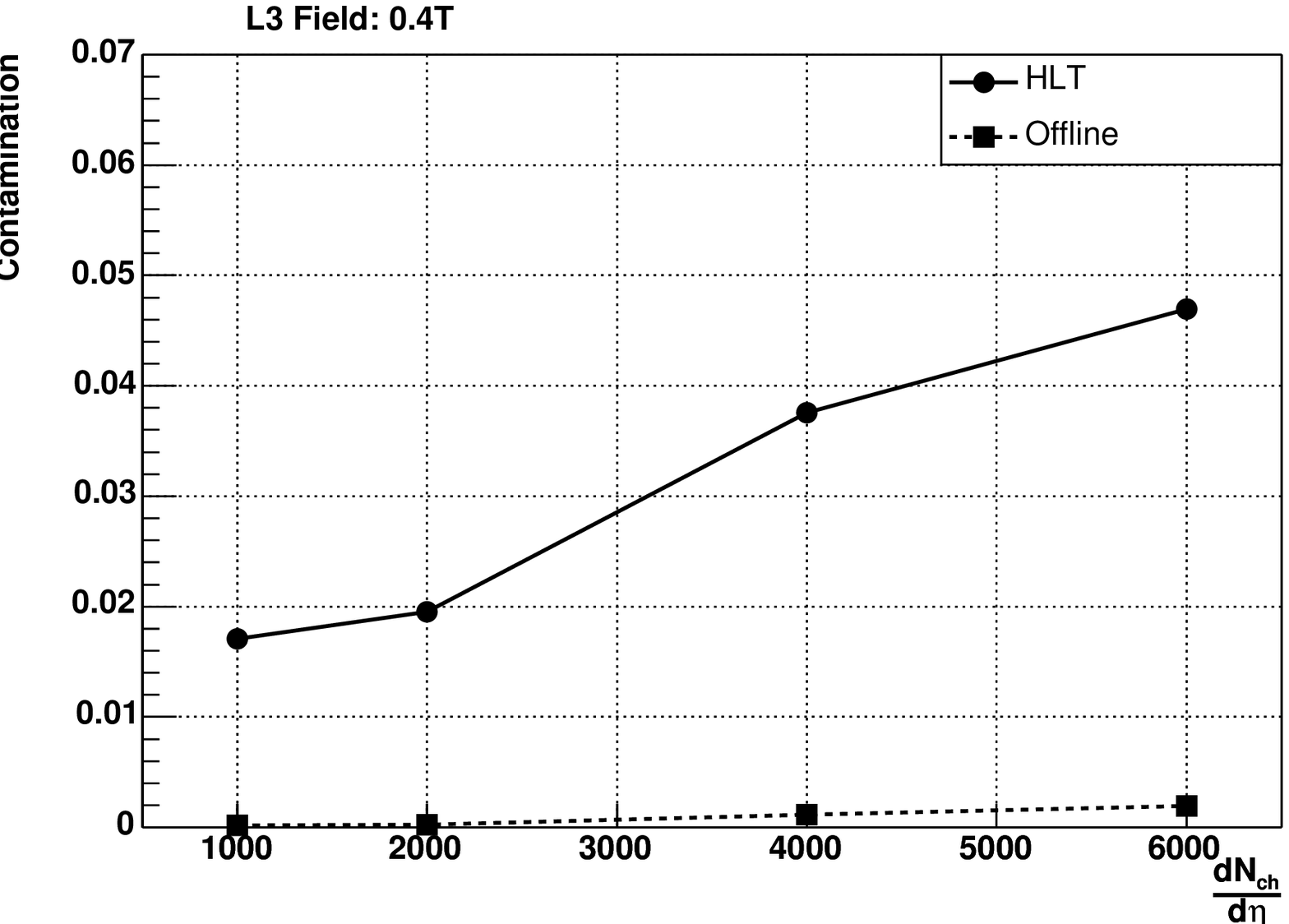,width=8cm}}
\caption[Integrated efficiency and contamination as a function of multiplicity.]
	{Integral efficiency and contamination as a function of multiplicity.}
\label{TRACK_inteff}
\efig


\subsubsection{Position resolution}
The position resolution can be estimated by calculating the
distance, $\delta$, from each space point on a track to the respective trajectory
defined by the track fit. The width of the resulting distribution of
these residuals can be taken as a measure of the position resolution
of the tracks.
In Figure~\ref{TRACK_resdist} such a distribution is shown for the
transverse and longitudinal direction for a multiplicity of
\dndy=\,1000. 
Only tracks with small dip-angle ($|\lambda|\leq 5^{\circ}$) are taken
into account for the longitudinal residuals, while all tracks with
$p_t\geq$\,0.1\,GeV are included for the transverse residuals.
The distributions include both inner and outer sectors
of the TPC, and is thus averaged over the three different
pad-geometries. 
The resolution represented by the RMS-values of the distributions
is about 1.37\,mm and 1.47\,mm for the pad and time direction, respectively. 
In Figure~\ref{TRACK_resvsmult} the obtained position resolutions are plotted as a
function of multiplicity. Here also the corresponding results from
Offline is shown. The results shows a clear deterioration of the
resolution in both directions for increasing multiplicities. Similar
effect is seen for Offline, although in this case the dependency
on multiplicity is less significant.

\bfig[htb]
\centerline{\epsfig{file=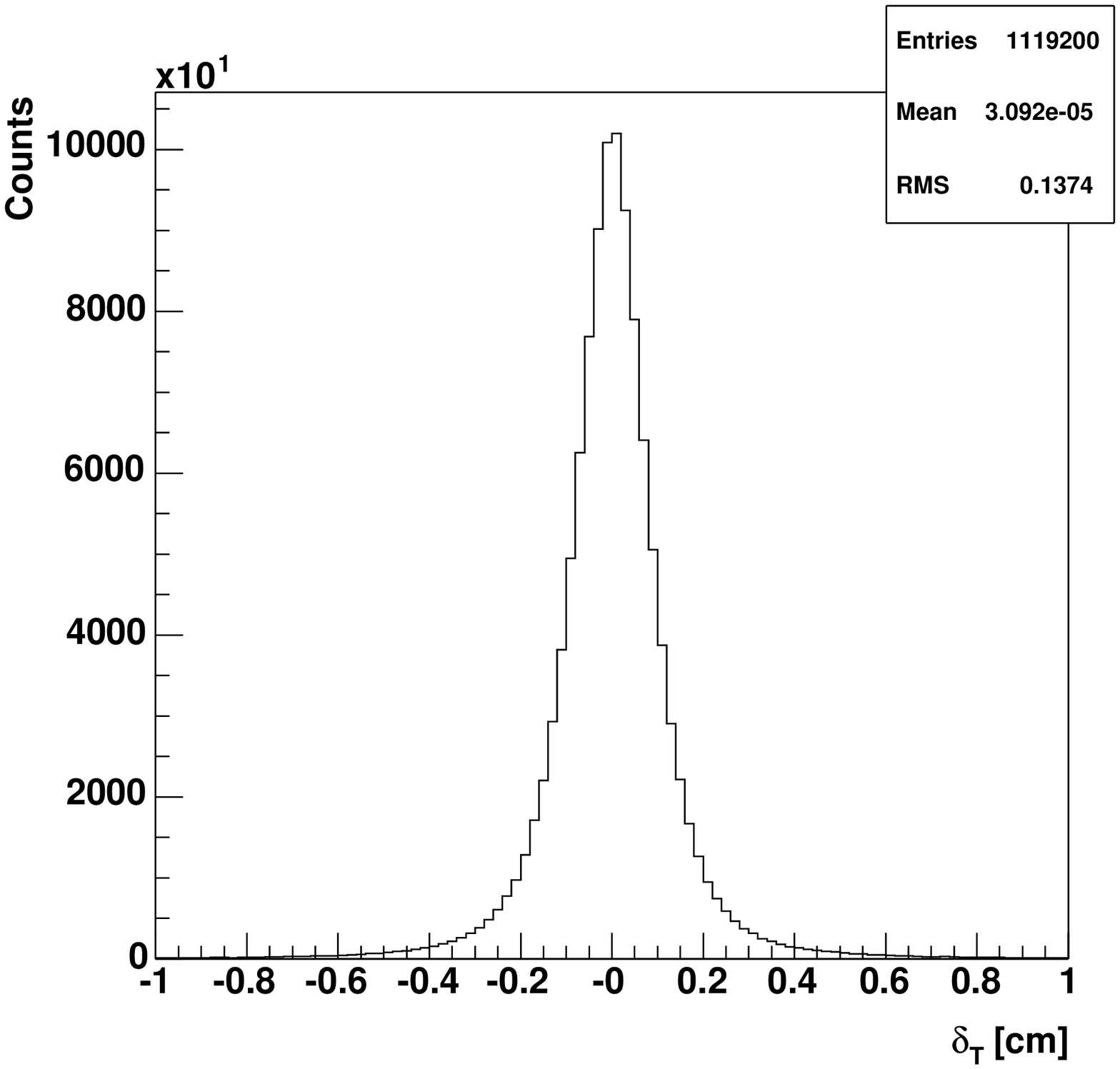,width=7cm}
\epsfig{file=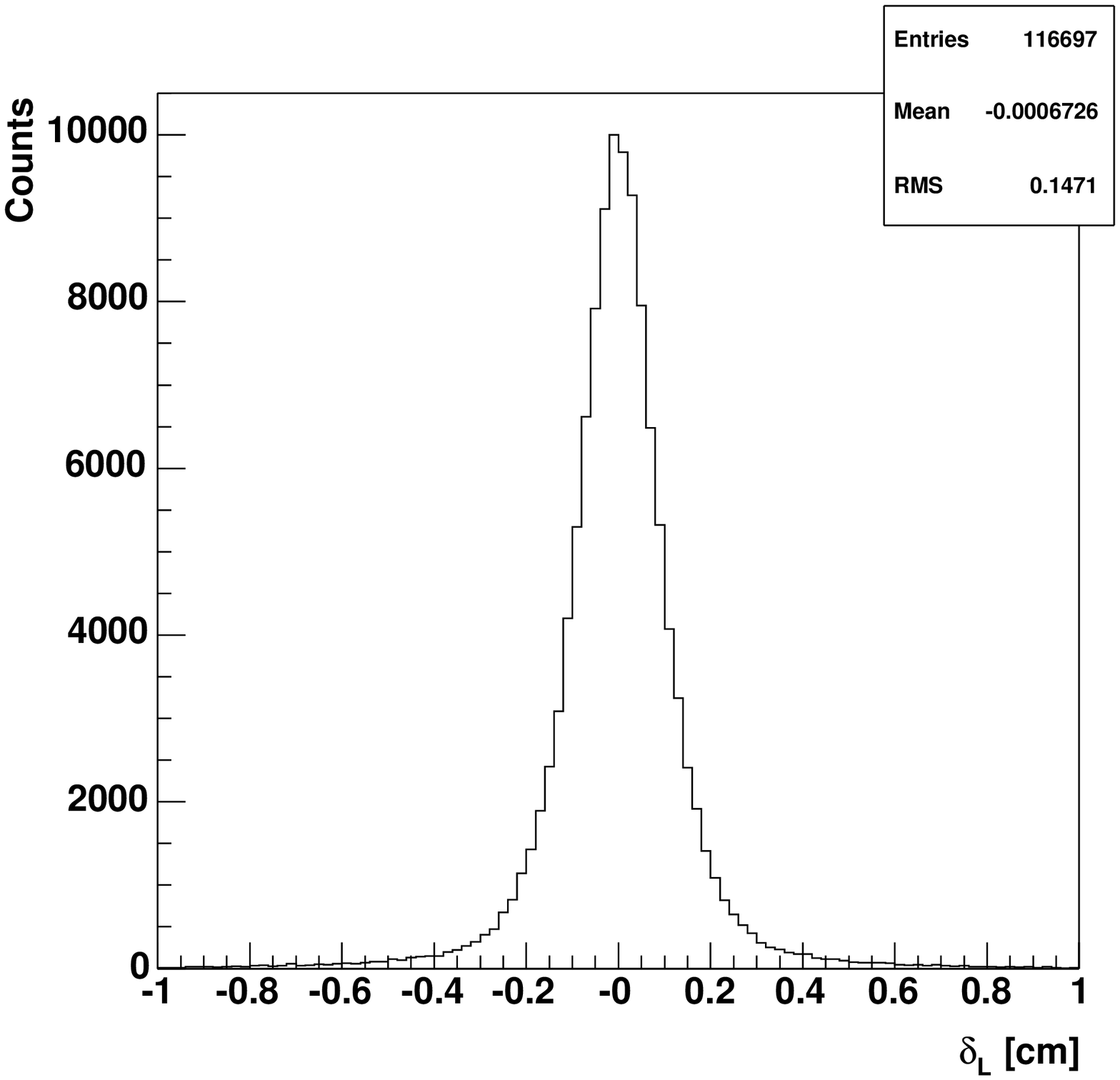,width=7cm}}
\caption[Residual distributions for the HLT sequential track
reconstruction chain for \dndy=\,1000.]
	{Distribution of transverse (left) and longitudinal (right)
residuals for dN$_{\mathrm{ch}}$/d$\eta$=1000. The distributions
includes clusters from both the inner and outer TPC readout chambers.}
\label{TRACK_resdist}
\efig 

\bfig[htb]
\insertplot{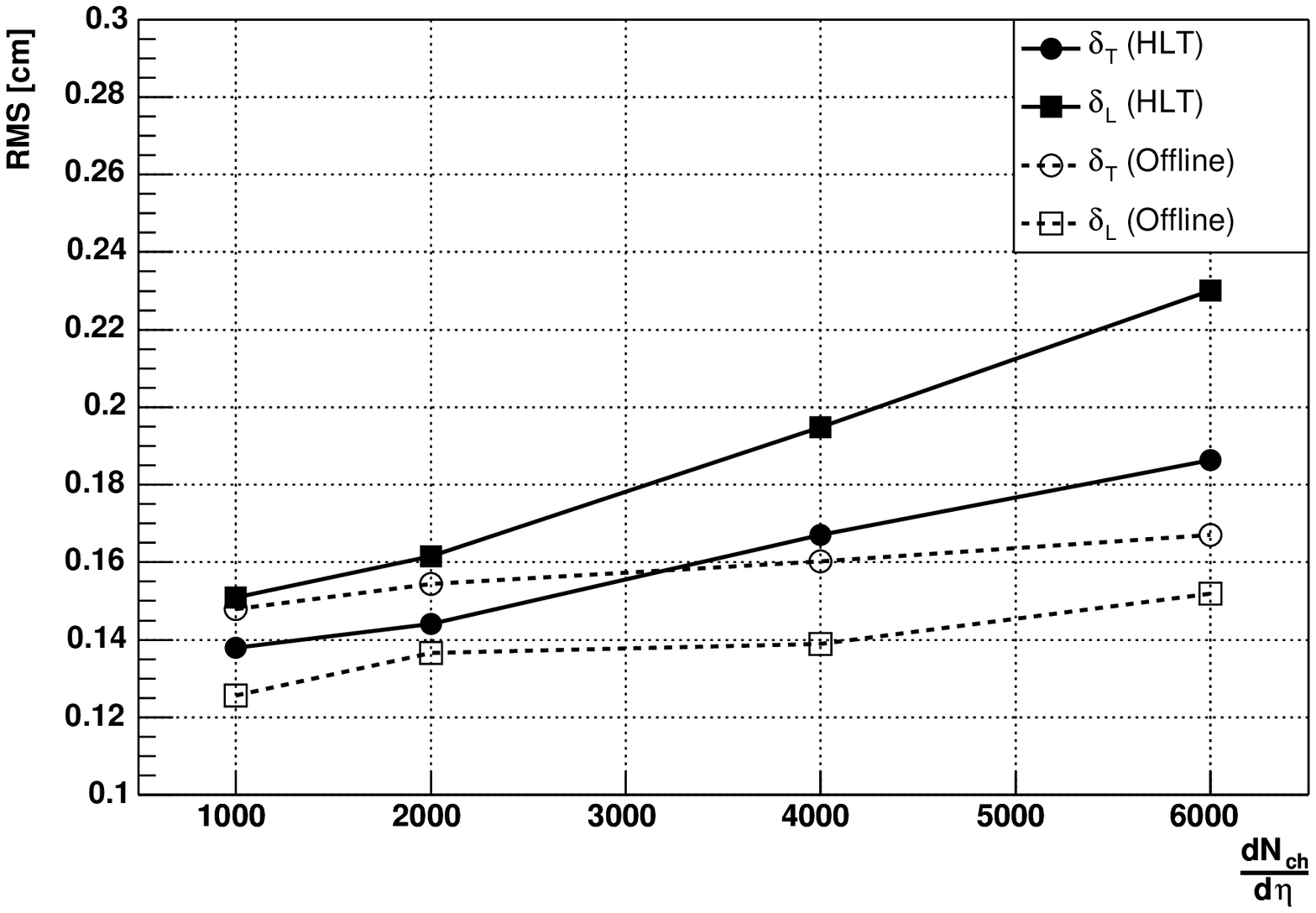}{9cm}
	{Residuals as a function of multiplicity for the HLT
sequential track reconstruction chain.}
	{Residuals as a function of multiplicity. The
results are obtained from the RMS-value of the respective distributions.}
\label{TRACK_resvsmult}
\efig

\subsubsection{Momentum resolution}
The momentum resolution of the reconstructed tracks can be estimated
by comparing the reconstructed momentum with the momentum of the
corresponding simulated particle. The resolution is a function of
$p_t$, Equation~\ref{TRACK_ptresformula}, and the relative \pt
resolution is defined as,
\beq
\Delta_r p_t = \frac{\Delta p_t}{p_t} =
\frac{p_{t,\mathrm{measured}}-p_{t,\mathrm{particle}}}{p_{t,\mathrm{particle}}},
\eeq
where $p_{t,\mathrm{measured}}\ $ is the reconstructed transverse momentum of the
track, and $p_{t,\mathrm{particle}}\ $ is the transverse momentum of the
corresponding simulated particle. In order to measure this quantity
for a given sample of tracks, a  Gauss-fit of the
distribution is commonly performed and the resulting width is taken as a
measure of $\Delta_r p_t$, see example in Figure~\ref{TRACK_intptres}.
The results for the different event samples are shown in
Figure~\ref{TRACK_ptres} as a function of $p_t$. 
\bfig
\insertplot{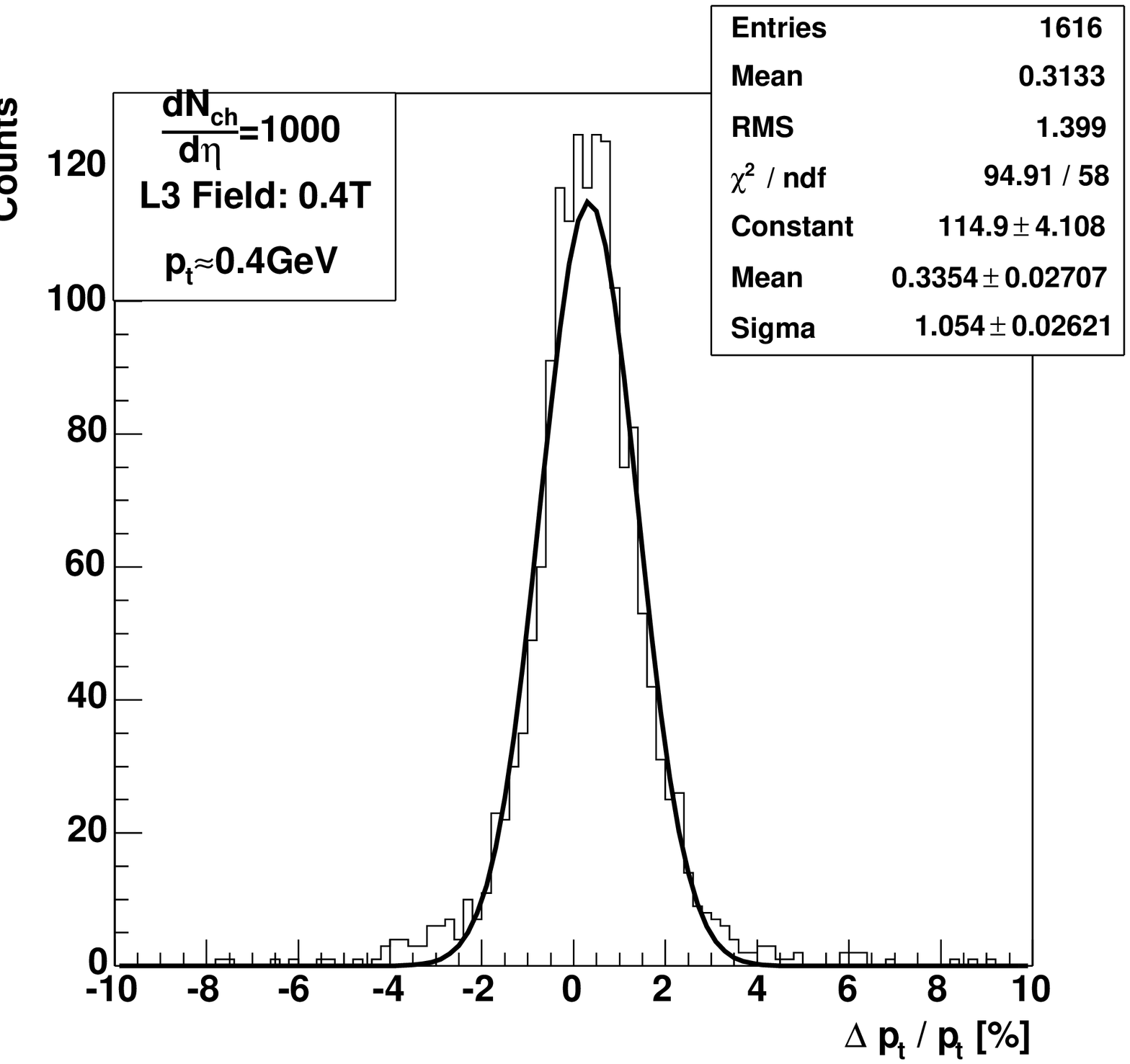}{8cm}
	{Relative transverse momentum resolution for the HLT
sequential track reconstruction chain.}
	{Relative transverse momentum resolution for tracks with
\pt$\approx$0.4\,GeV. The distribution is fitted with a Gauss function,
that gives a relative width of 1.05\%. The distribution is
not completely symmetrical, because the measured
momentum is inversely proportional to the curvature of the track.}
\label{TRACK_intptres}
\efig

For the low multiplicity regime the relative resolution is contained
within the interval 1.5-2.5\%, with an average of 1.8\% for magnetic field of
0.2\,T over the entire \pt range. For a magnetic field of 0.4\,T the
resolution is centered around 1.2\% for the same multiplicity. The
difference between HLT and Offline results in this regime is
small. For higher multiplicities the difference between HLT and
Offline becomes more significant, in particular in the higher \pt range.

\bfig
\centerline{\epsfig{file=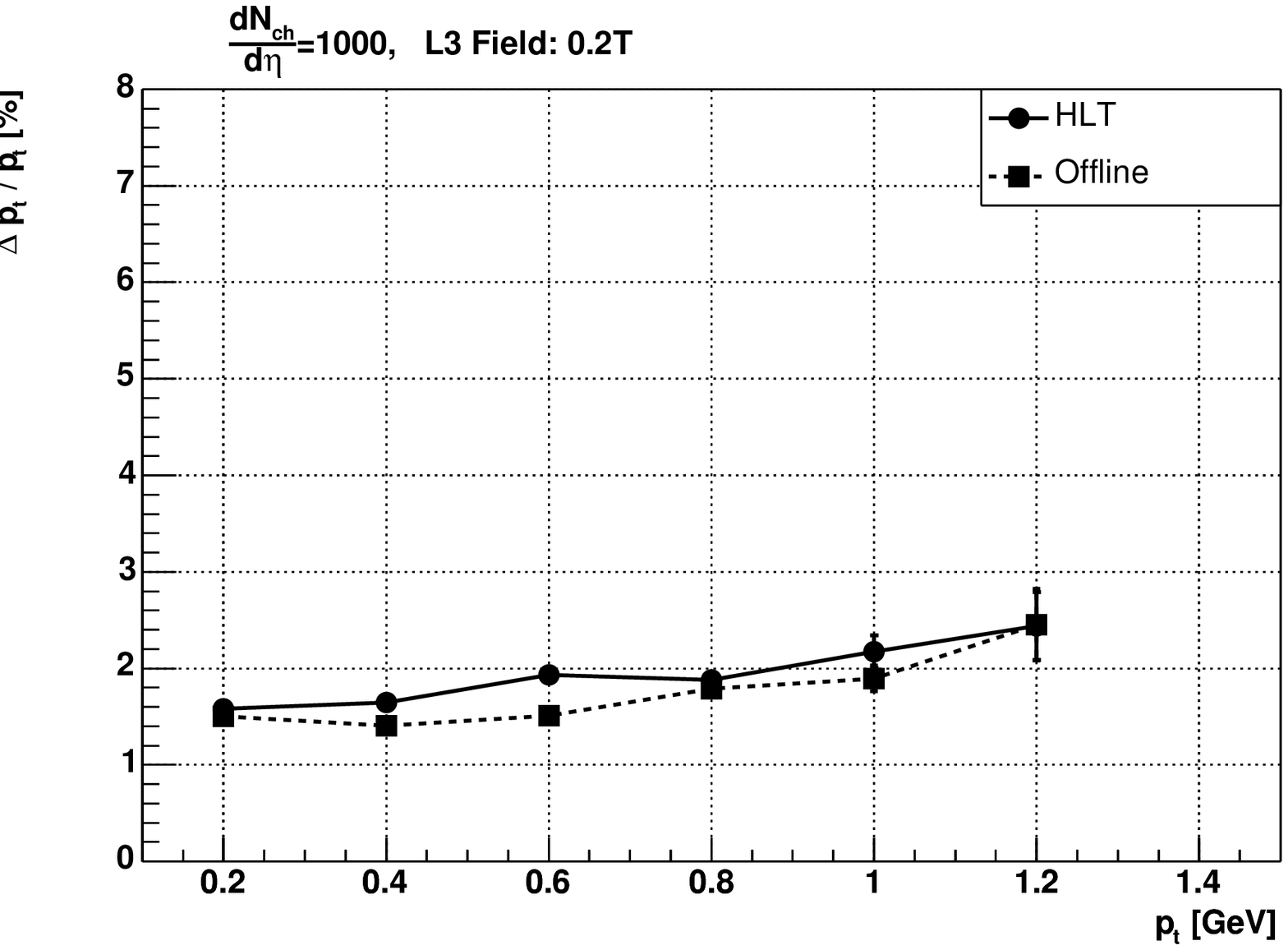,width=8cm}
\epsfig{file=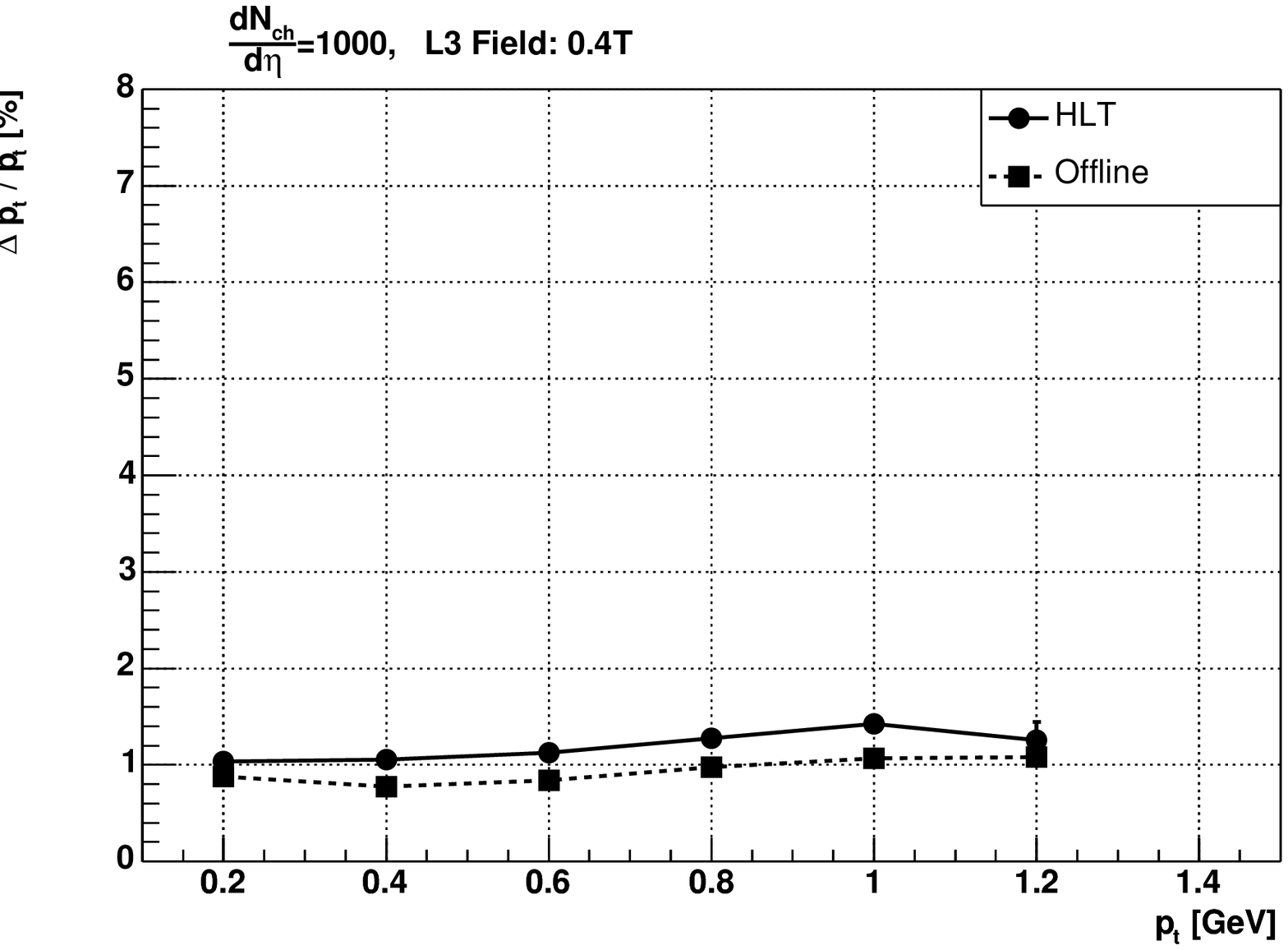,width=8cm}}

\centerline{\epsfig{file=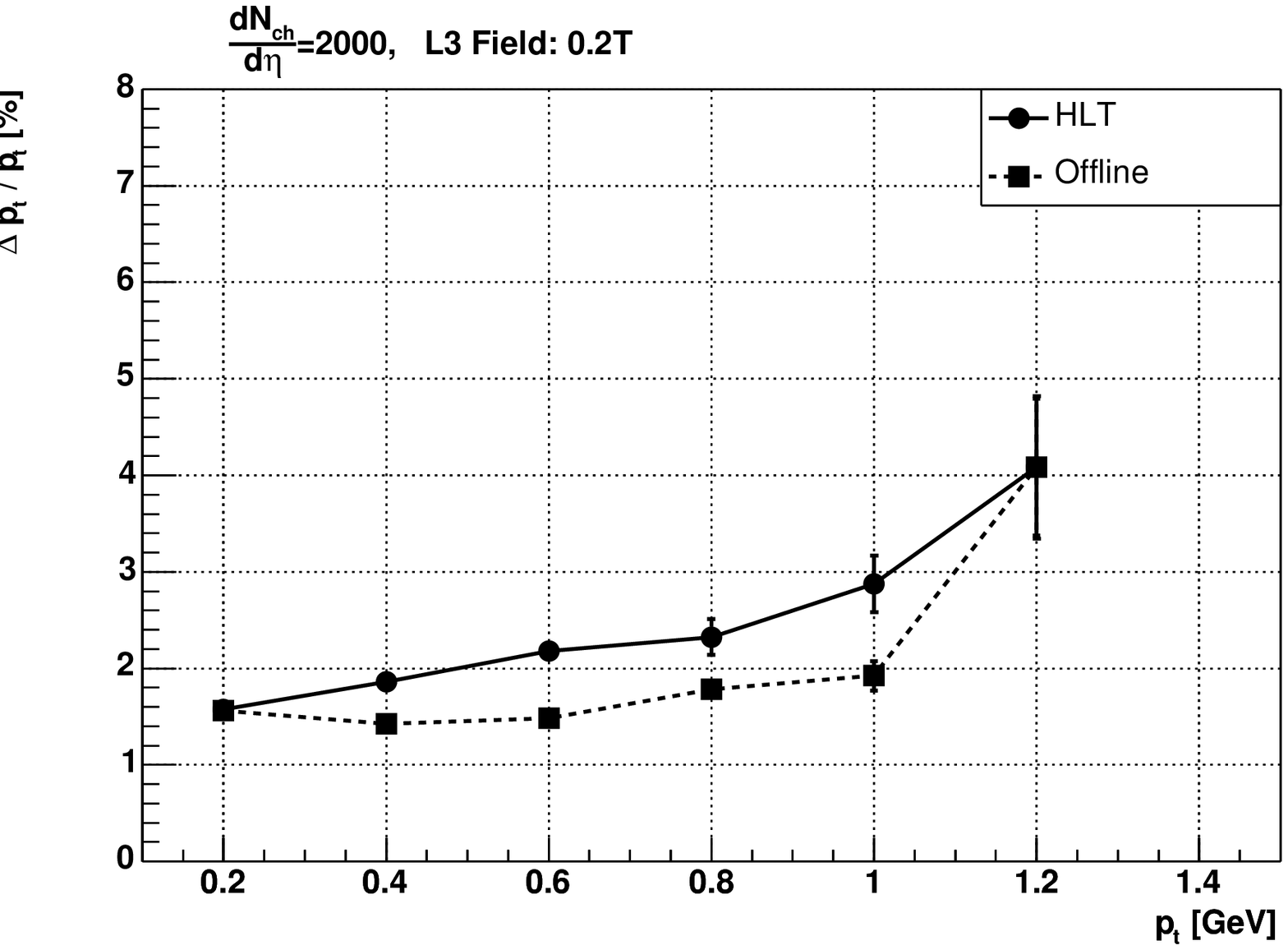,width=8cm}
\epsfig{file=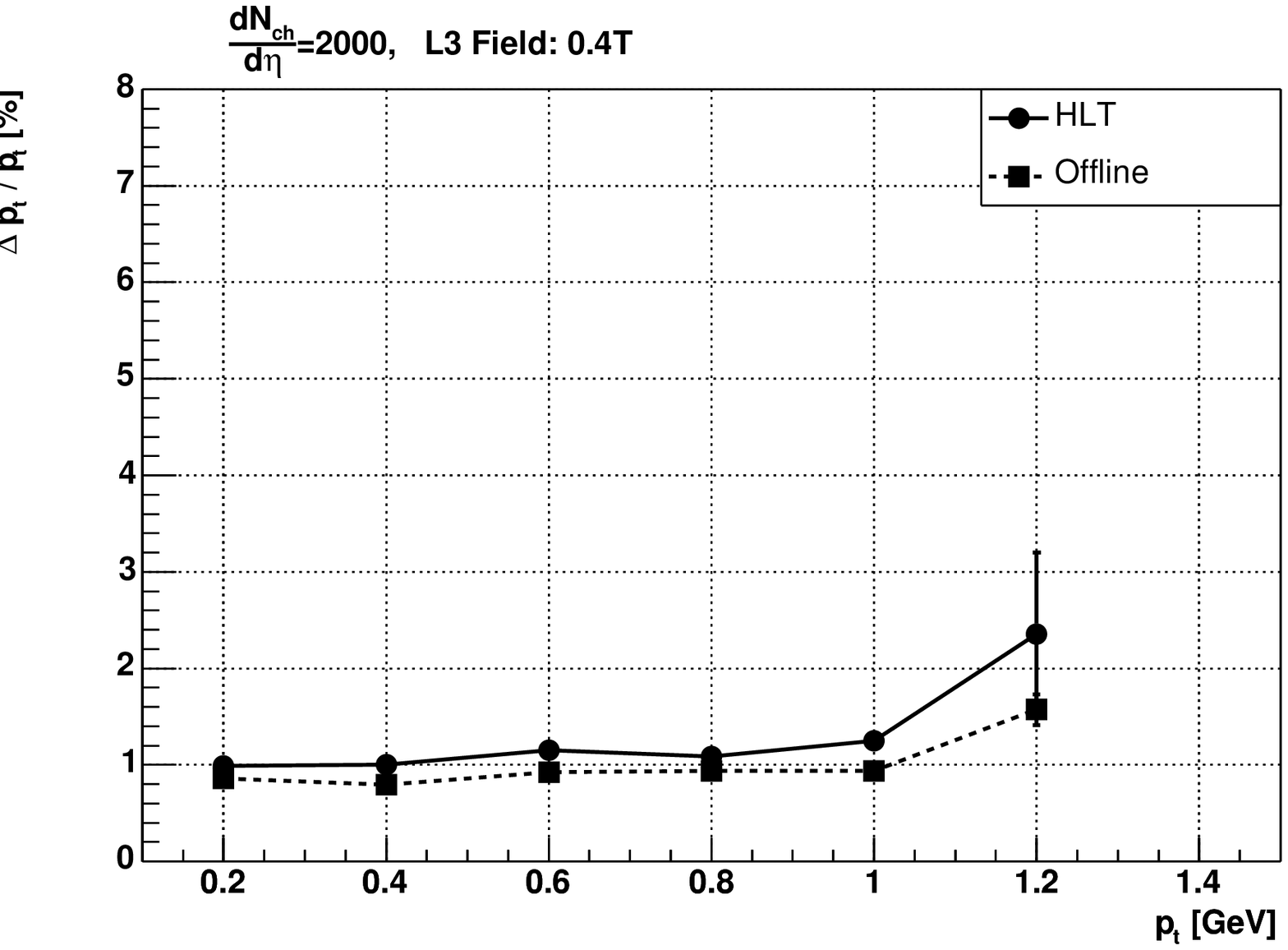,width=8cm}}

\centerline{\epsfig{file=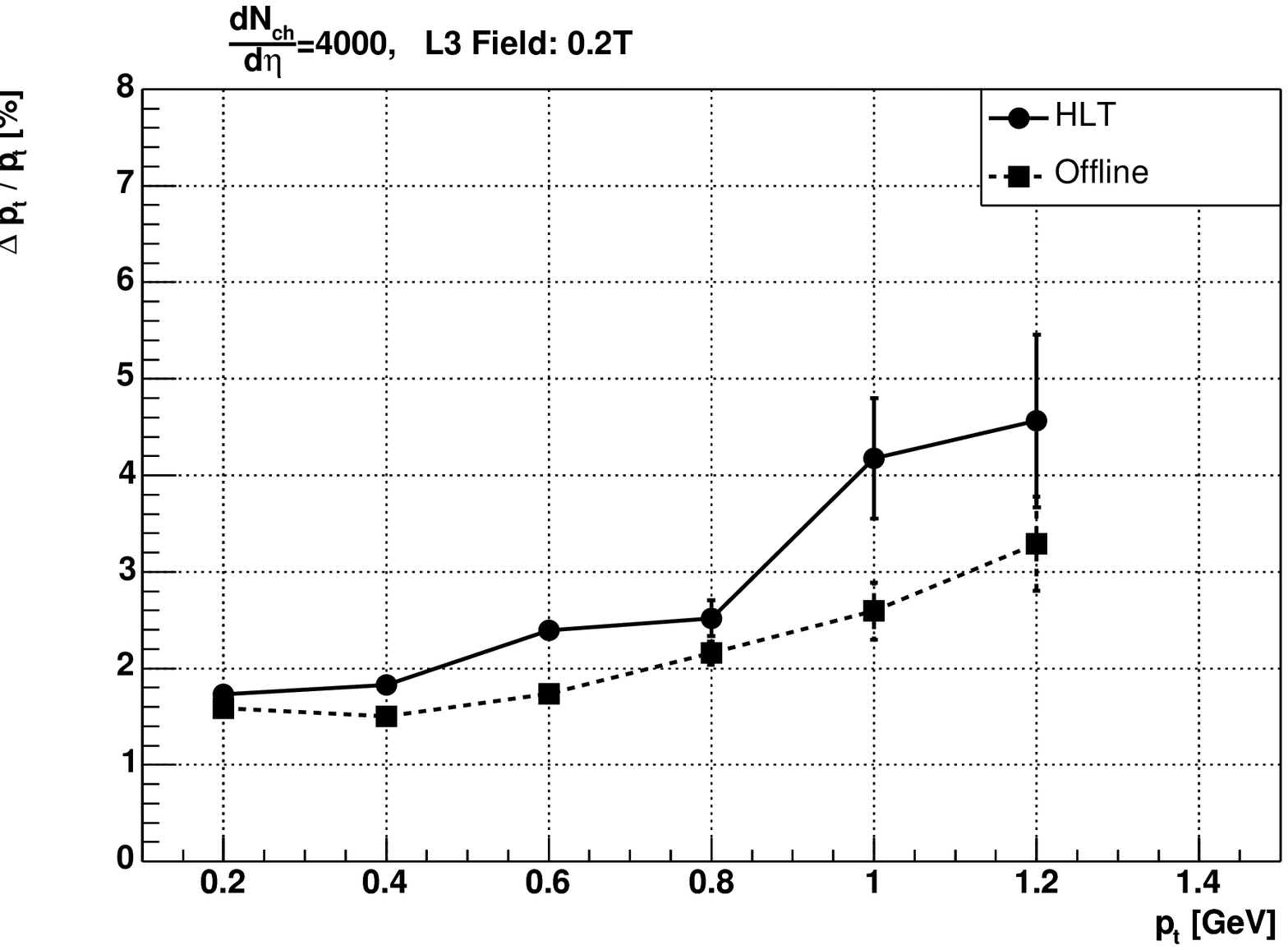,width=8cm}
\epsfig{file=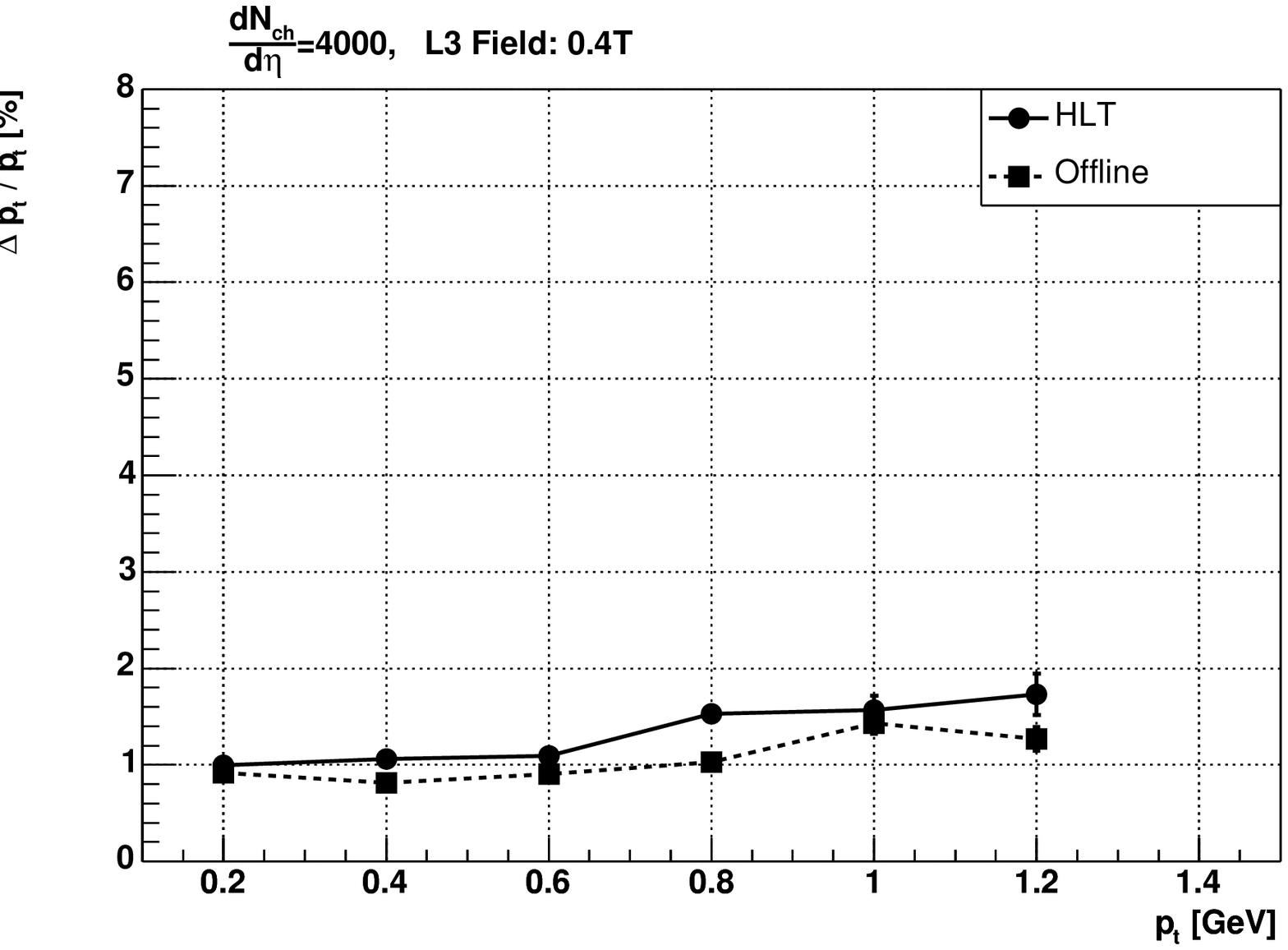,width=8cm}}

\centerline{\epsfig{file=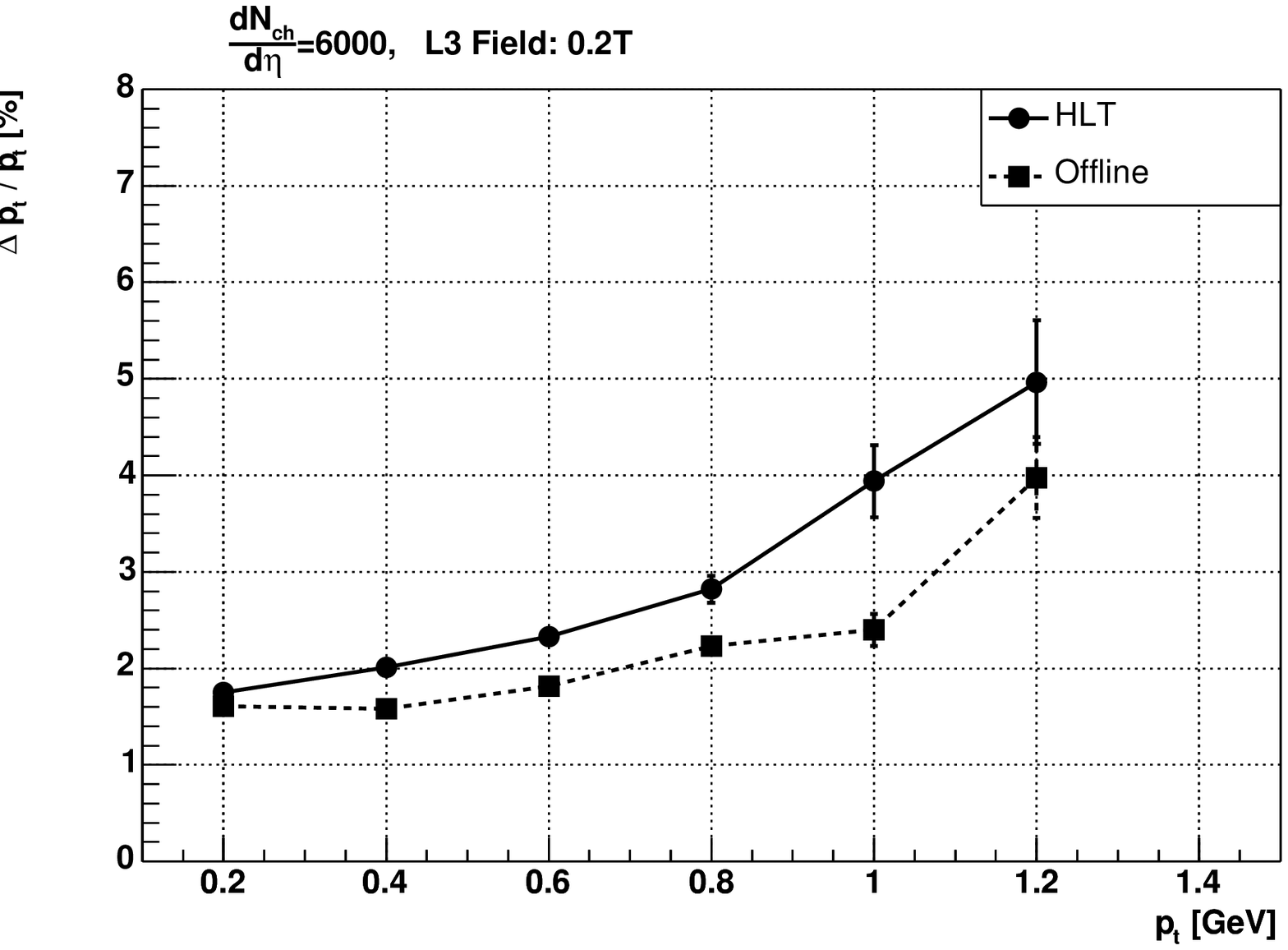,width=8cm}
\epsfig{file=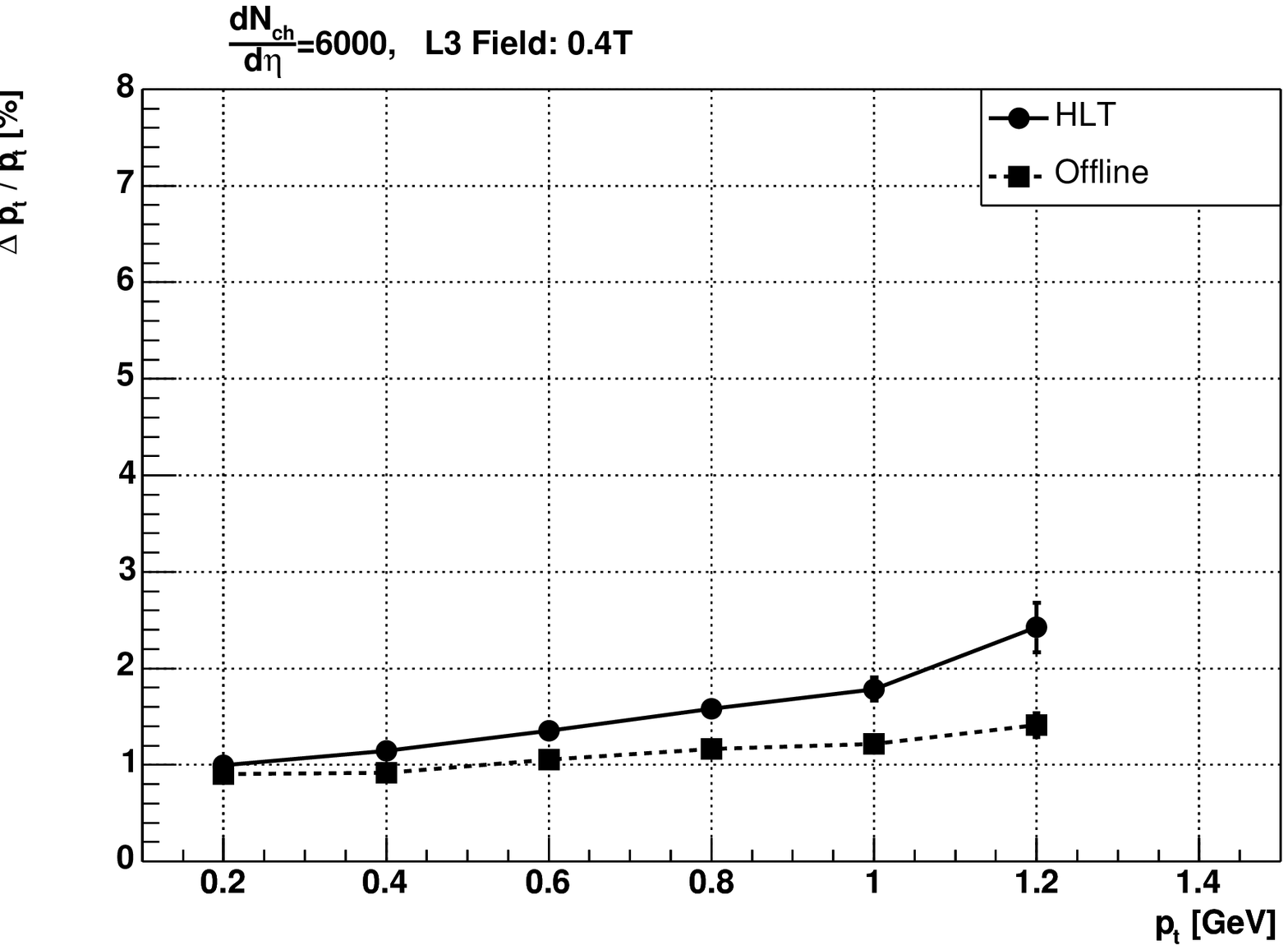,width=8cm}}

\caption[Relative transverse momentum resolution for the HLT track
reconstruction chain as a function of \pt.]
	{Relative transverse momentum resolution as a function of
$p_t$. Results obtained using magnetic field strength of
0.2\,T (left) and 0.4\,T (right), respectively.}
\label{TRACK_ptres}
\efig

\subsubsection{Remarks on the observed tracking performance}
In summary, the results from these simulations indicates a significant
loss of tracking performance in
the low momentum regime for higher multiplicities. 
These efficiencies may be explained by several factors. 
The main reason for the efficiency loss seen for the HLT
reconstruction chain compared to the Offline results
is the quality of the reconstructed clusters. The Cluster Finder
algorithm follows a rather simple sequence matching scheme, and
implements a straight-forward center-of-gravity approach to obtain the
cluster centroids. Due to the fact that the Cluster Finder has no
prior information about the individual charge distributions,
overlapping clusters are not handled very well. As a consequence, overlapping clusters are
split wrongly and the centroid calculations are biased. This is
also reflected in the observed deterioration of the space point resolution
as a function of multiplicity (Figure~\ref{TRACK_resvsmult}).
In addition, a number of {\it noise} clusters may also be wrongly
identified from very low \pt 
tracks or $\delta$-electrons which produce rather large areas of
continuous signal in the pad-row planes. These areas are interpreted by the Cluster Finder
as many overlapping clusters. Both these effects
contribute to the local track follower getting dis-oriented during
cluster assignment. As a result tracks are being split and contaminated
by wrongly assigned clusters.

Furthermore, the track finding procedure reconstructs tracks based on
a simple fit selection criteria in conformal space, and does not
include more advanced correction for energy loss and multiple scattering.
Since the impact of
multiple scattering and energy loss is higher for low energy
particles, this effect is more significant in the low momentum regime.


Parts of the efficiency loss at low momentum and high
field setting may also be explained by the fact that the Track Finder is
implemented in a parallel fashion, i.e. the
tracks are reconstructed at the TPC sector level. Consequently, the tracks
which cross the sector boundaries have to be merged by the subsequent
Track Merger. If the quality of the individual track segments are poor and the
track density is high, the Track Merger may not be able to correctly
merge the corresponding track segments. 

\subsubsection{Secondary tracks}
The tracking performance discussed above were determined exclusively for primary
tracks. In order to get an estimate of the reconstruction capabilities
of secondary particles, a complementary was
done including a selection of secondary particle tracks.
In this case the {\it generated good tracks} in
Equation~\ref{TRACK_effdef} was defined as particles which are decay
products of $K$ and $\Lambda$ particles. In addition, the respective
tracks are required to be contained within the TPC acceptance and
furthermore cross at least 40\% of all padrows (same requirement as
was made for the primary tracks above). The selected
particles thus consists of
\begin{eqnarray*}
K&:&\ \mu,\ e,\ \pi\\
\Lambda&:&\ p,\ \pi
\end{eqnarray*}
for the two cases respectively.

In this case, a second pass was done during the track finding step, where
the second tracking pass took the unused clusters from the first pass
as input. As explained previously, no vertex constraint was imposed on these
secondary tracks, and thus the conformal mapping of the space points was
calculated relative to the
first assigned cluster to a track. In Figure~\ref{TRACK_seceff} the
resulting efficiency as a function of \pt is shown for events with
multiplicity of \dndy=\,1000. For reason of comparison, the efficiency obtained
with an recently developed version of the Offline chain is also
shown. This new Offline approach consists of an improved Kalman filter~\cite{marian}
which includes optimization for detection of secondaries. The
tracking efficiency is $\sim$60\% for the low momentum bin, while at
higher \pt it increases to $>$80\%. The difference between Offline and
HLT is significant at the low momentum, whereas the difference
decreases at higher momentum. It should be noted that the current HLT algorithm was
not tuned for the detection of secondary tracks with the exception of
the second tracking pass adaption.

\bfig[t]
\insertplot{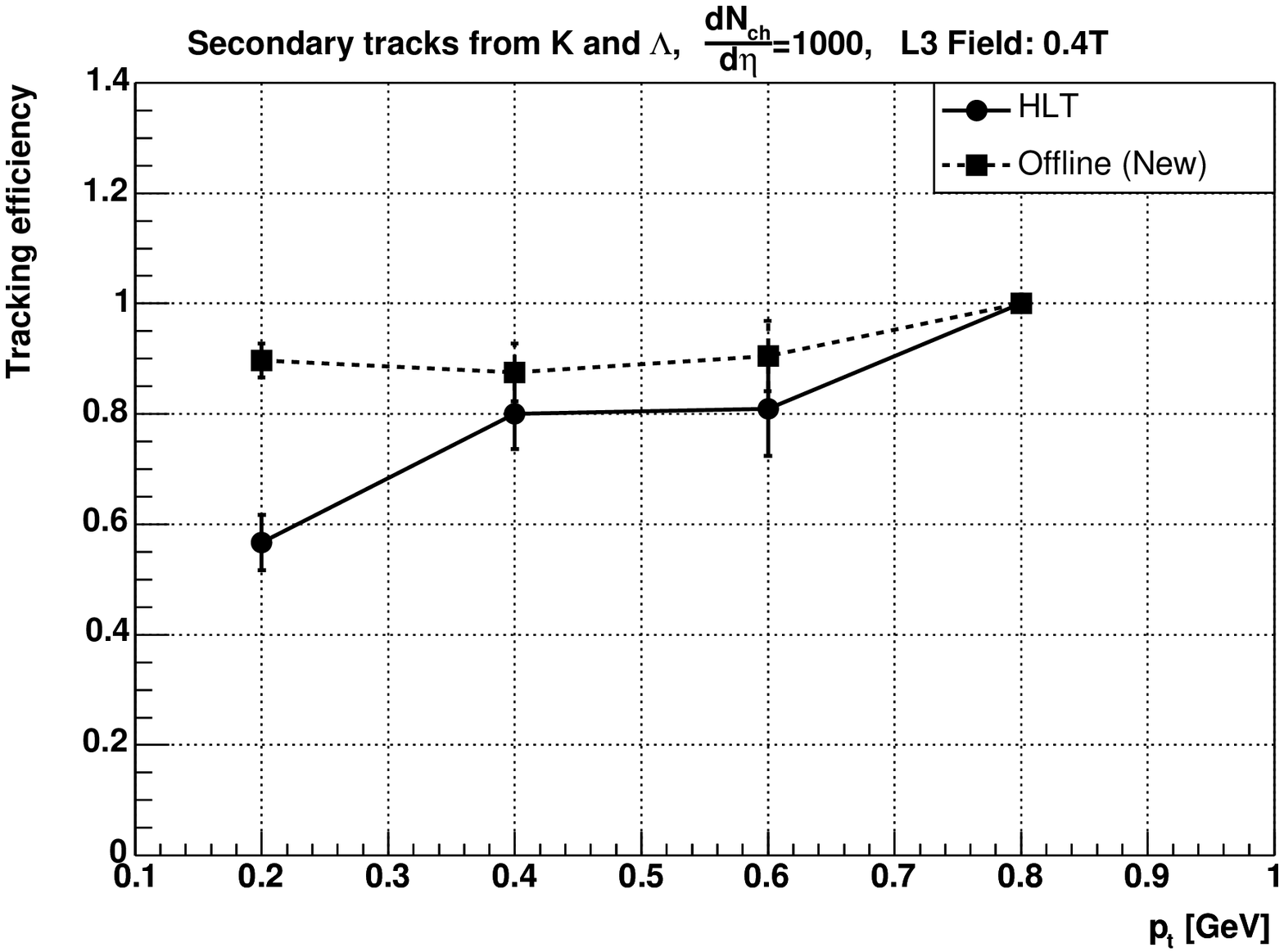}{9cm}
	{Tracking efficiency of secondary tracks for the HLT
sequential track reconstruction chain.}
	{Tracking efficiency of secondary tracks from decay of $K$
and $\Lambda$ particles using HLT (solid line) and an improved Offline
chain (dotted line).}
\label{TRACK_seceff}
\efig

\subsubsection{Computing requirements}
\label{TRACK_seqtiming}
The computing requirements of the various processing tasks can be estimated by
measuring the CPU time needed for the different processing
modules. Such measurements may indicate the amount of computing power
needed to implement the algorithm to run on the High Level Trigger
system, and in particular which processing steps represent the
bottlenecks. In the same way as the tracking performance depends on
the particle multiplicity, the required processing time is
correlated with the number of space points and tracks which have to be
reconstructed.  

The processing time of all the individual process components was
measured for various event multiplicities.
For the test, a standard PC consisting of a 800\,MHz Twin Pentium
III with a Serverworks Chipset, 256\,kB L2 Cache running a
Linux kernel v2.4 was used. All measurements were performed while
the data was located within the memory, i.e. no overhead from any
external device access (disk, network interface etc.) was included.
The left plot in
Figure~\ref{TRACK_timing1} shows the measured processing times for
three different multiplicities. The measurements are resolved with
respect to the different computing steps, where the values corresponds
to the time needed by a single processing module as implemented in the
scheme shown in Figure~\ref{TRACK_topology}. For instance, for a multiplicity
of \dndy=\,1000 the Cluster Finder needs on average about 8\,ms to process the
pad-rows which is contained within a single sub-sector, while the
Track Finder needs about 130\,ms to reconstruct the track segments
within a complete TPC sector. The Track Finder step includes both the
transformation of the space points into conformal space and the track
following procedure.
\bfig[htb]
\centerline{\epsfig{file=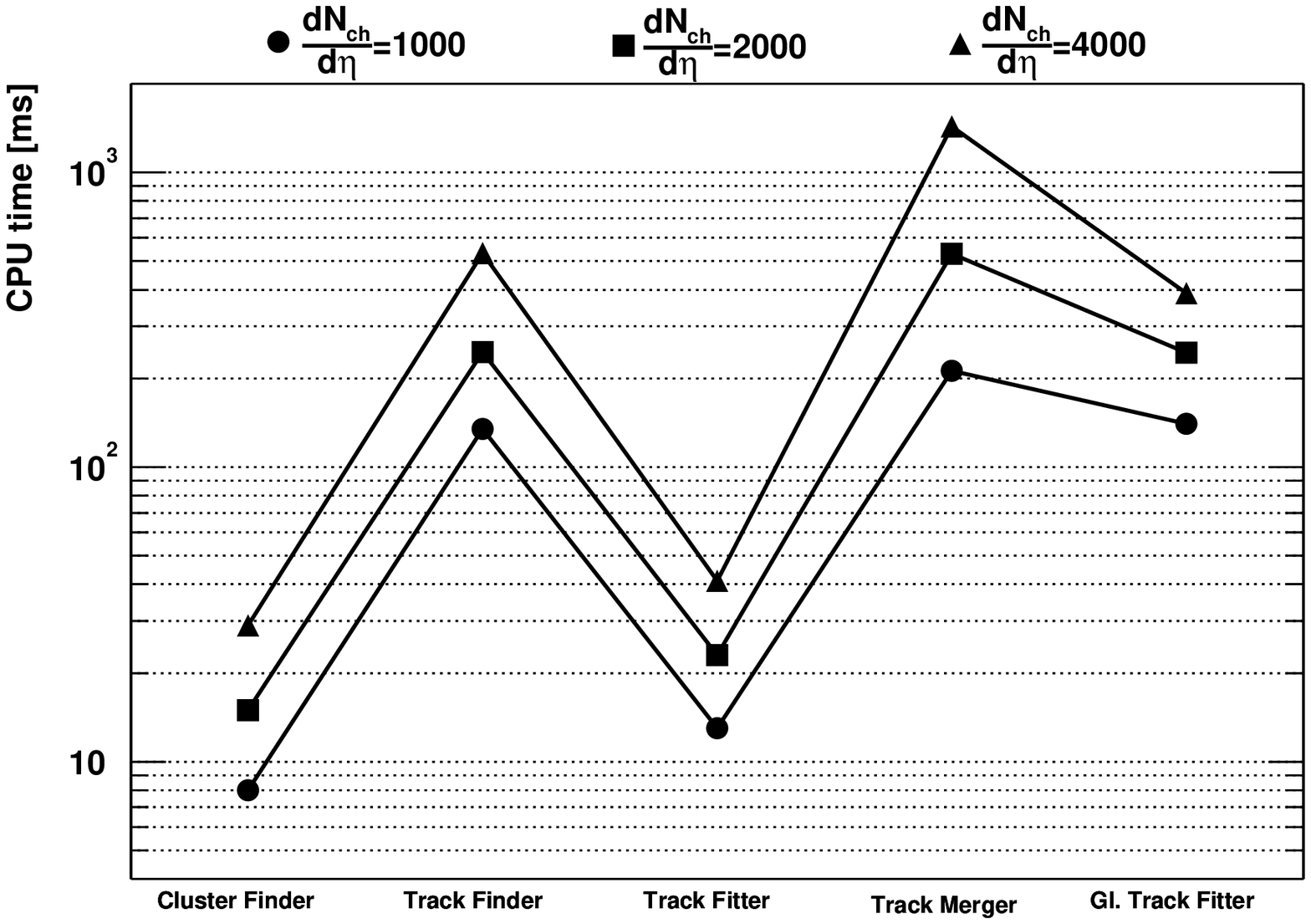,width=8cm}
\epsfig{file=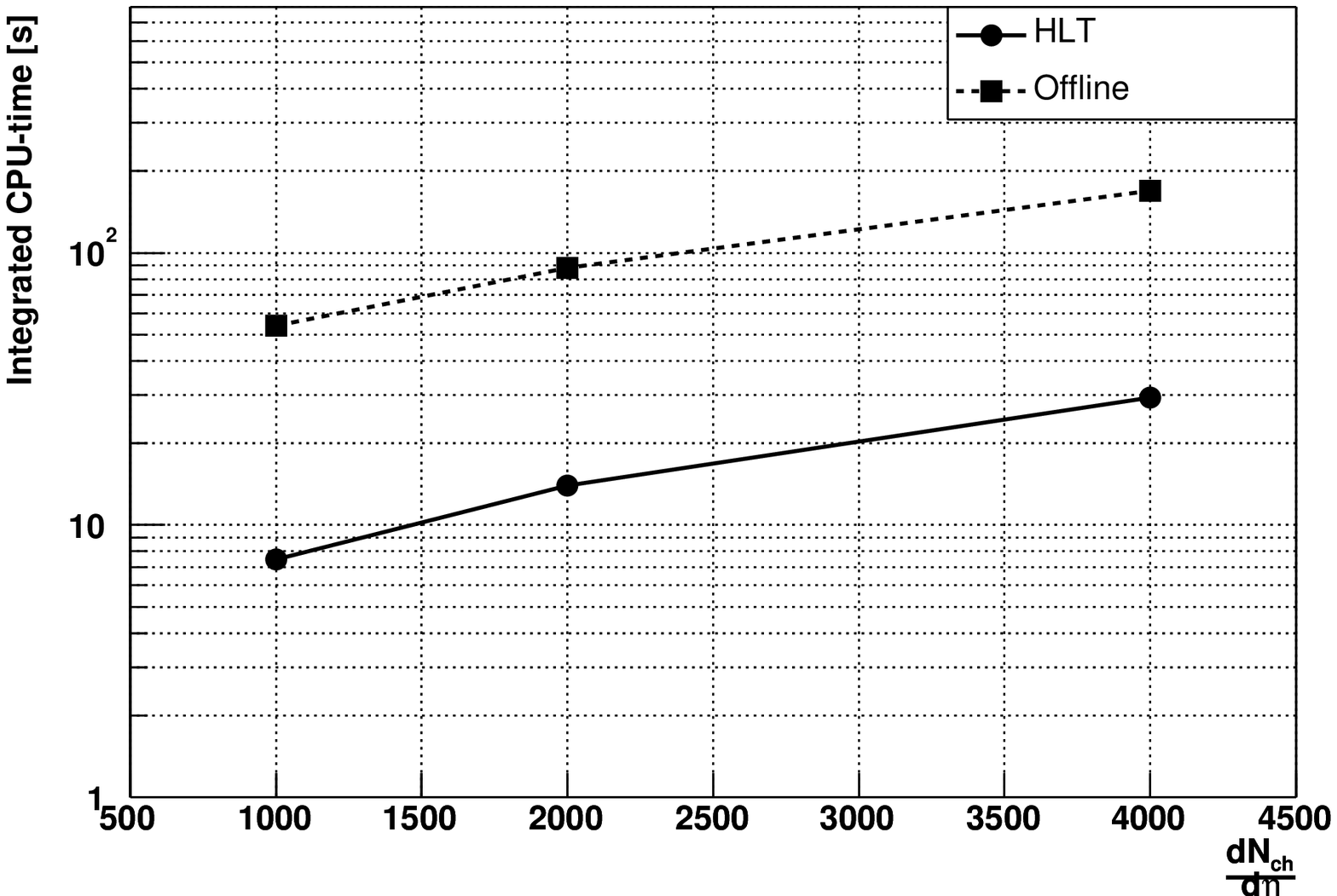,width=8cm}}
\caption[Measured CPU-time for the HLT sequential track reconstruction chain.]
	{Compute time requirements measured on an 800\,MHz processor
for three different particle multiplicities. Measured CPU-time
for the sequential tracking chain resolved with respect to the
different processing steps (left). Measured CPU-time integrated over
all the processing steps performed sequentially on a single CPU
(right). For reason of comparison the corresponding CPU-time measured for the Offline
reconstruction chain is shown (dotted line).}
\label{TRACK_timing1}
\efig

In the right plot in Figure~\ref{TRACK_timing1} measured CPU-time
integrated over all the processing steps performed sequentially on a
single CPU is shown as a function of multiplicity. The measured
CPU-time needed by the Offline
reconstruction chain is also shown. For the Offline reconstruction
chain, data loading into
memory is included in the measurement, as no clear separation between
loading and processing of data is available in its current implementation.
At \dndy=\,4000 the HLT chain needs a total of about 30\,s to for a full
TPC event reconstruction, while Offline needs about 170\,s.


\section{Iterative Tracking}
In the sequential tracking approach described in the previous section,
the cluster centroids are obtained using a straight-forward
center-of-gravity calculation. Such an algorithm has obvious
limitations when applied to a high occupancy environment, since the lack
of information about the tracks bias the centroid calculation in the
case of overlapping charge distributions. The main objective of
iterative tracking is to provide the track information prior to the
cluster finding in order to better fit and unfold the overlapping
clusters. From this the correct cluster centroids should be obtained.

In this approach, the pattern recognition scheme consists of two main
parts: Track candidate finding and cluster fitting. In a first step an
implementation of the Hough Transform is applied to the
raw ADC-data in order to obtain a list of track candidates. These track
segments serve as input for the {\it Cluster Fitter}, which
reconstructs the cluster centroids along the particle trajectories by
fitting the respective charge distributions to a parameterized
shape. Finally, the assigned clusters are fitted to a helix in order to obtain
the best estimate of the track parameters.

\subsection{The Cluster Fitter}
\label{TRACK_clusterfitter}
The implemented fitting routines were initially developed for TPC
data in the NA49 experiment~\cite{na49}. It assumes that the clusters
can described using a two-dimensional Gauss function.

\subsubsection{The cluster model}
The shape of the clusters is given by the convolution of the response
functions of the readout pads, Section~\ref{ALICE_tpcprinc}. Both
the pad and time response functions have a close to Gaussian shape, as
well as the spread of the electron clouds due to diffusion.
The cluster model can thus to a good approximation be
described with a two-dimensional Gauss-distribution whose widths depend on geometry of
the pads and the track parameters.
A total of five independent parameters are needed to fully describe
the model:
\begin{itemize}
\item The two-dimensional position $(p,t)$ in pad and time direction.
\item The widths $(\sigma_{\mathrm{pad}},\sigma_{\mathrm{time}})$, in pad and time.
\item The amplitude of the distribution.
\end{itemize}
These parameters vary for each cluster, so without any prior
knowledge they would all have to be fitted for. In the case of overlapping
clusters such a fitting procedure would be a very demanding task
as no information about the number of contributing tracks,
nor the shape for each individual charge distributions, is known.
The shape of the clusters in the TPC depends on the drift length
(diffusion of the drifting electrons in the gas), and track crossing
angles with the pad-row-plane (spread of primary ionization
electrons), and can be parameterized according to~\cite{loh}:
\begin{eqnarray}
\sigma_{\mathrm{pad}}^2 &=& \sigma_{PRF}^2 + D_T^2\cdot s_{\mathrm{drift}} + \frac{l^2\cdot
\tan^2(\beta)}{12} + \frac{d^2\cdot(\tan(\alpha)-\tan(\psi))^2}{12}\nonumber\\
\sigma_{\mathrm{time}}^2 &=& \sigma_0^2 + D_L^2\cdot s_{\mathrm{drift}} +
\frac{l^2\cdot\tan^2(\lambda)}{12}
\label{TRACK_clwidths}
\end{eqnarray}
where the various terms are described in Section~\ref{ALICE_tpcprinc},
page~\pageref{ALICE_signal}.
This parameterization indicates that any prior knowledge of track parameters
could provide the fitting procedure with an improved initial guess of
the parameters, or even reduce the number of parameters
to vary in a fit.

Although the average cluster widths can be described with the
parameterization in Equation~\ref{TRACK_clwidths}, their properties are
determined by stochastic processes and are thus also
subject to fluctuations. 
The fluctuation of the shape depends on the
contribution of the random diffusion and angular spread of the
ionization, and also on the gas gain fluctuations and secondary
ionization. 

\subsubsection{Fitting procedure}
The implemented procedure fits the clusters to the model
\beq
q(p,t) = \sum_{k=1}^K A_k\exp\left[-\frac{1}{2}\left(\frac{p-p_k}{\sigma_{\mathrm{pad},k}}\right)^2
-\frac{1}{2}\left(\frac{t-t_k}{\sigma_{\mathrm{time},k}}\right)^2\right]
\label{TRACK_fitmodel}
\eeq
which is a sum of $K$ two-dimensional Gauss-distributions, with the variables $A_k$,
$(p_k,t_k)$ and $(\sigma_{\mathrm{pad},k},\sigma_{\mathrm{time},k})$
denoting the amplitude, position
and widths of the individual distributions, respectively. A $\chi^2$
fit of the model to the cluster
data is done by applying the {\it Levenberg-Marquardt
method}~\cite{levmarq} to minimize
the least square error in the fit\footnote{The Levenberg-Marquardt method (also
called Marquardt method) is a standard algorithm for the minimization
of the least square error in the fit of nonlinear models.}.
All five parameters may be free to vary in
the fit, or only a subset of them may be fitted while the rest are
held fixed at their input values. In either case, the 
fitting routine returns the best fit values for the free parameters in
the fit. Once each cluster has been fitted to a Gaussian distribution, the total
charge of the cluster can be obtained from the relation
\beq
Q_{\mathrm{total},k} = 2\pi\sigma_{\mathrm{pad},k}\sigma_{\mathrm{time},k} A_k.
\eeq

The procedure can also be extended to other models for the cluster
shape, e.g. to account for asymmetric Gaussian distributions. In that
case the two-dimensional Gaussian given above may be replaced with
an alternative model, and the fitting procedure is done accordingly.

\subsubsection{Deconvolution of overlapping clusters}
In the case of overlapping clusters, $K\geq2$, the corresponding
charge distribution is fitted to a sum of $K$ individual
Gauss-distributions. If the track parameters are known from a
preceding pattern recognition step, all 5$\times K$ parameters may be
provided with initial values. In addition, deconvolution can now be done
by fixing the widths of the individual distributions and only letting
the positions and the amplitude parameters vary freely.

An example of deconvolution of two overlapping clusters is shown in
Figure~\ref{TRACK_fitexample}. 
\bfig[htb]
\centerline{\epsfig{file=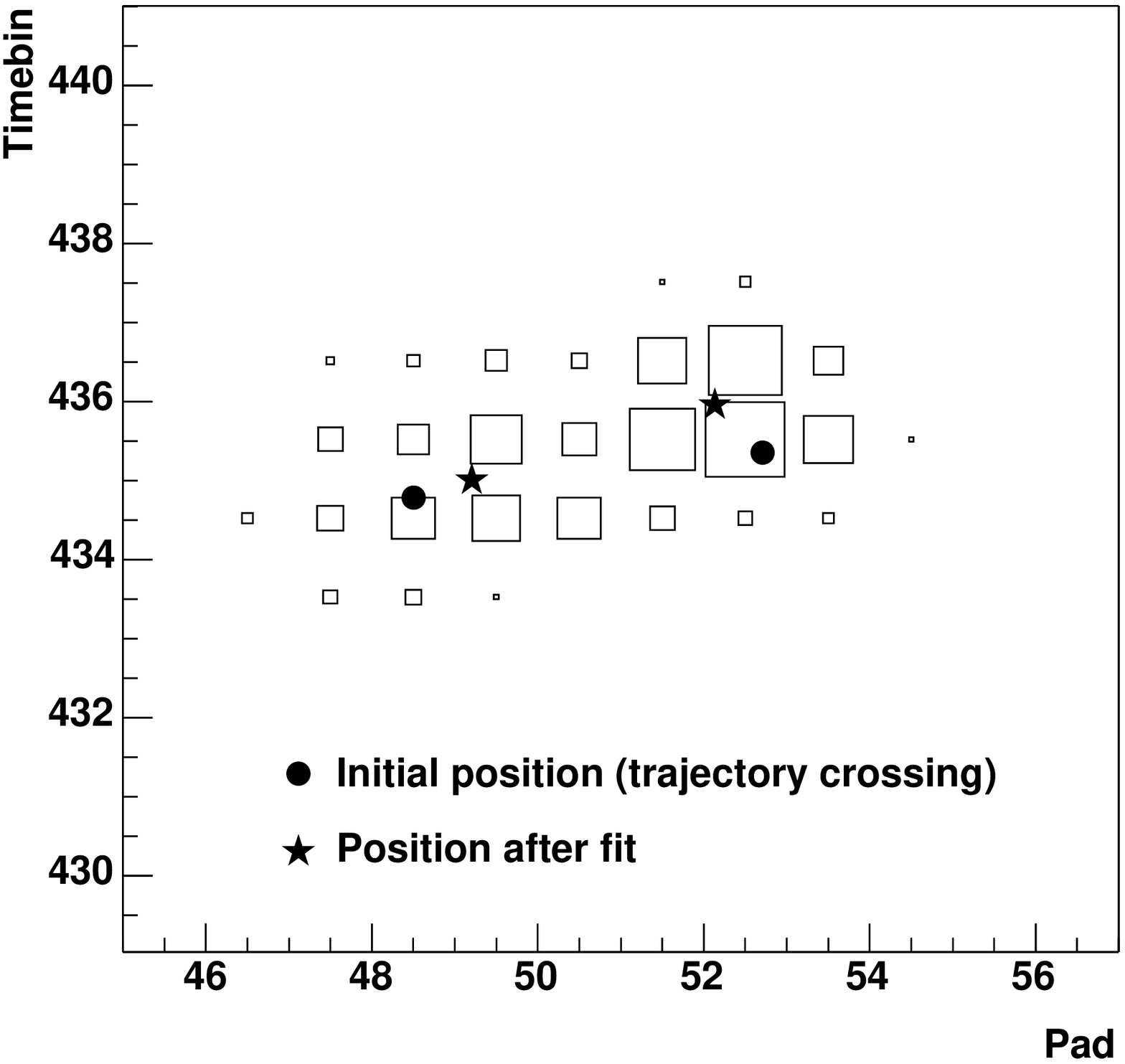,width=7cm}
\epsfig{file=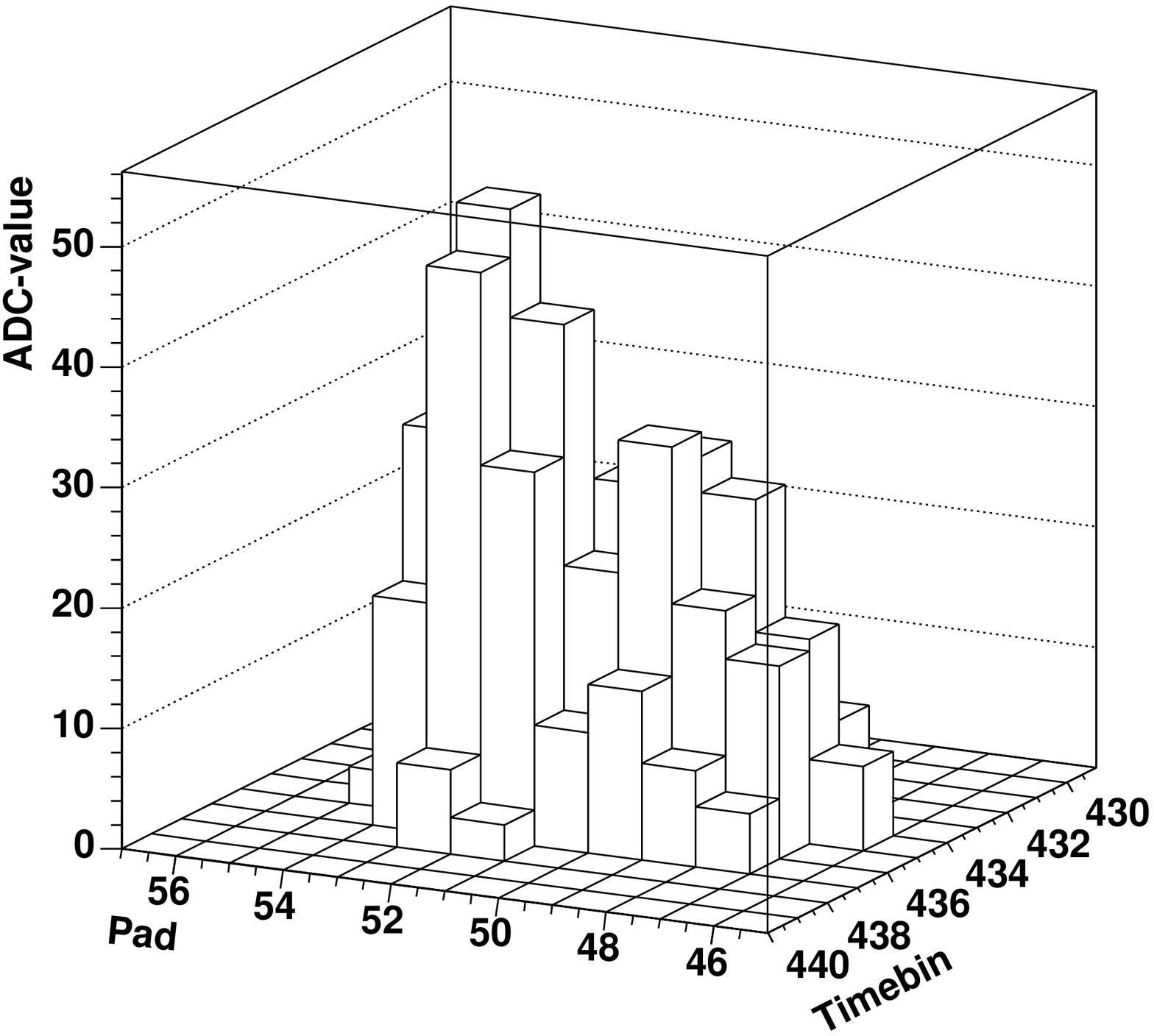,width=7cm}}
\caption[Example of fitting and deconvolution of overlapping clusters.]
	{Illustrating fitting and deconvolution of overlapping
clusters. Left: The circles mark the initial position given as input
to the fitting routine, while the star marks the best-fit values
returned from the fit. Right: Same clusters shown using a lego-plot.}
\label{TRACK_fitexample}
\efig
\noindent In this case the input to the fitting
routine are the two sets of 5 initial parameters, each consisting of the
position in the pad-row-plane (illustrated by the circle
markers), the widths of the distribution
in two dimensions and the amplitude. The initial values of these
parameters are determined from the results obtained by the
preceding tracking finding algorithm. The initial positions are taken as
the crossing point of the computed tracks with the pad-row-plane, while the
widths are calculated from Equation~\ref{TRACK_clwidths} with the
track inclination angles and initial positions as input. During the fitting
procedure, the widths are kept fixed at their input values, while the
remaining parameters are free to vary. This means that in this case
there are 3+3 parameters
which are being fitted for. The fitting routine finally returns the best-values
obtained for each position (illustrated by the star markers) and the
respective amplitudes of the individual distributions. In this way the
initial positions serve as the prediction of the cluster centroids,
while the returned values from the fit are the corrected,
final reconstructed cluster centroids. The improved parameter
estimates are subsequently returned to the track model. 

\bfig[t]
\insertplot{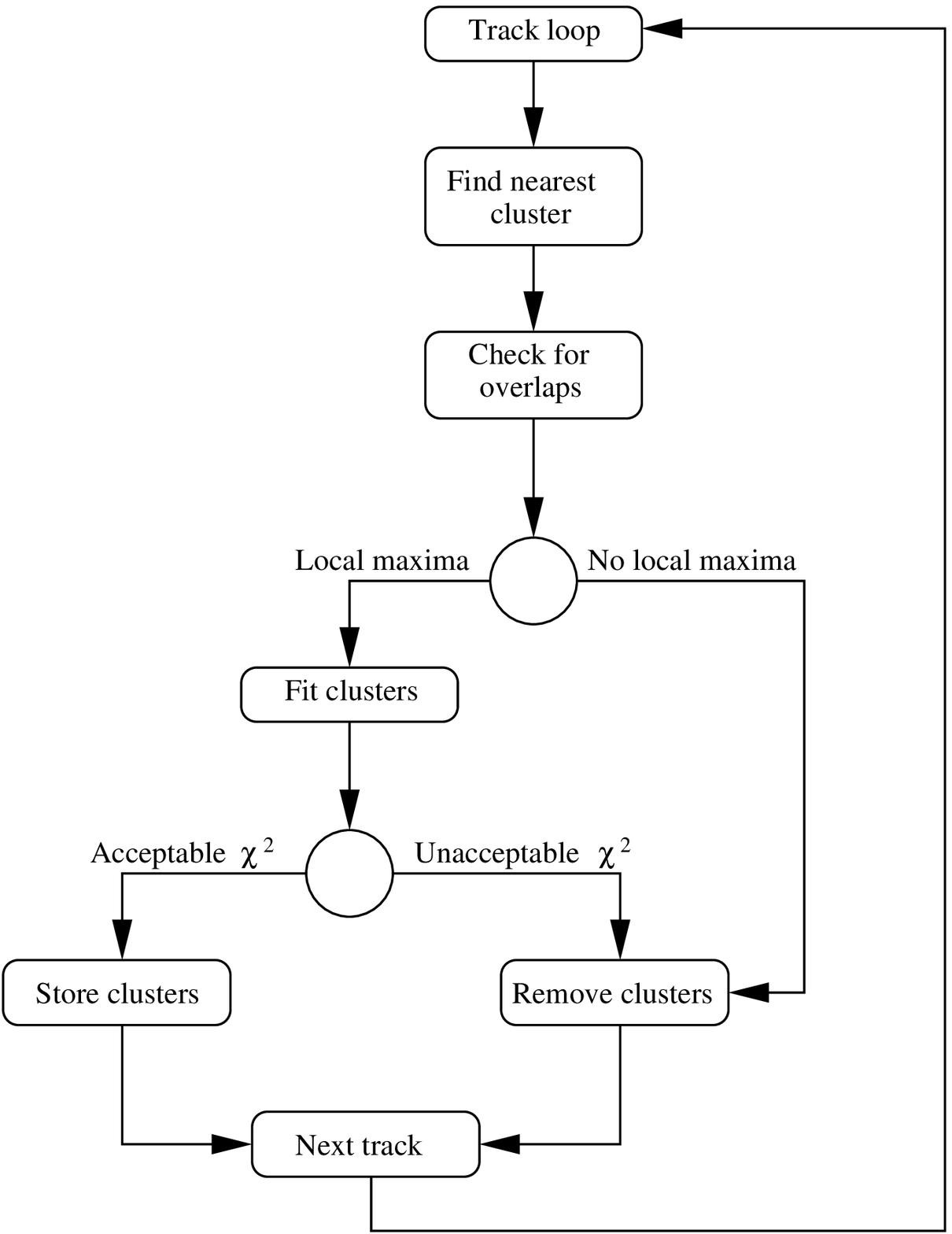}{7cm}
	{Flow diagram of the HLT Cluster Fitter algorithm.}
	{Flow diagram of the Cluster Fitter algorithm.}
\label{TRACK_clfitter_flow}
\efig

\subsubsection{The algorithm}
The implemented Cluster Fitter algorithm assumes that the track parameters
have been obtained
by a preceding track finding algorithm, and thus takes a list of {\it track
candidates} as input. In a first step each input track is propagated
through the TPC detector and the intersection between the
helical trajectories and the pad-row-planes is calculated and stored
internally for each track candidate. 

The algorithm then processes each pad-row separately as illustrated
by the flow diagram in Figure~\ref{TRACK_clfitter_flow}. A loop is performed over
all track intersection points with the current pad-row. For each track,
a region around its intersection point in the pad-row-plane is searched for a
nearest cluster. In addition, any potential additional tracks
contributing to the given cluster is identified. If a local maxima is
found within the cluster distribution, the clusters are fitted. A
track candidate is only taken into account in the fit if it can be associated
with a local maxima in the distribution. The reason for this criteria
is to avoid taking into account {\it fake} tracks, i.e. tracks which
may have been falsely identified as a track candidate during the
preceding track procedure. If the resulting fit returns
an acceptable $\chi^2$ the clusters are stored with the values
obtained in the fit, otherwise the clusters are removed. During the
fit the widths are always held fixed at their input values, while the
position and amplitude are fitted for.

In Table~\ref{TRACK_spres} the space point resolutions achieved with
the Cluster Fitter are listed for both isolated ($K$=1) and
overlapping clusters ($K\geq$\,2).
The input parameters to the fit were in this case provided by the respective
simulated particle trajectory
producing the cluster. 
\begin{table}
    \begin{center}
    \begin{tabular}{|l|c|c|c|c|}
    \hline 
			&\multicolumn{2}{|c|}{\bf Pad direction [mm]} 
			&\multicolumn{2}{|c|}{\bf Time direction [mm]}\\
			\cline{2-5}
	{\bf $K$=1}	&  Gauss-fit & CoG & Gauss-fit & CoG\\
			\hline \hline
   Inner chambers	&0.97 & 0.99 & 1.48 & 1.33 \\
   Outer chambers	&0.90 & 0.88 & 1.27 & 1.14 \\
   \hline\hline
	{\bf $K\geq$2}	& & &  & \\
			\hline \hline
   Inner chambers	& 1.90 & 2.90 & 2.33 & 3.54\\
   Outer chambers	& 1.78 & 4.68 & 1.92 & 4.36\\
	\hline
   \end {tabular}
    \caption[Comparison of the space point resolution obtained by
the HLT Cluster Fitter and the HLT Cluster Finder on both
isolated and overlapping clusters.]
            {Comparison of the space point resolution obtained by
applying the fitting procedure and the center-of-gravity (CoG),
corresponding to the Cluster Finder, on both
isolated ($K$=1) and overlapping ($K\geq$\,2) clusters.}
    \label{TRACK_spres}
    \end{center}
\end{table}
\noindent For comparison, the corresponding resolutions achieved using 
the Cluster Finder in the sequential tracking scheme
(Section~\ref{TRACK_clfinder}) are also listed.
The results show
that for isolated clusters the transverse resolution obtained in the
two approaches are similar, while the longitudinal resolution is
slightly worse in the fitting approach. In the case of overlapping
clusters, the Cluster Fitter achieves 30-50\% better resolution
compared to the Cluster Finder. 

\subsection{The Hough Transform}
The Hough Transform (HT) is a widely adapted algorithm in image analysis
related fields. Originally it was used to
recognize linear patterns in binary images~\cite{hough}, but the principle can
easily be implemented to any kind of pattern. A complete review of HT 
methods and main developments can be found in e.g.~\cite{illkitt}
and~\cite{leav}.

The basic idea of the HT is easily understood by
considering a straight line, $y=ax+b$, where the parameters $a$ and $b$ define the
slope and intercept respectively. The HT uses the idea that this
equation can be reformulated into $b=-ax+y$. Thus a point ($x,y$) is seen
to define a straight line in a two-dimensional {\it parameter
space}. In the space of parameters $a$ and $b$, this line corresponds
to all possible straight lines going through the point $(x,y)$ in the
$xy$-plane. If several points $(x_i,y_i)$ are along a
straight line in the $xy$-plane, the transformed lines in parameter
space will intersect at the point $(a,b)$ which defines the actual parameters
of the line. By identifying this point of intersection, one has
identified the line itself. Thus the effect of the HT is to reduce the
recognition of global patterns in a {\it image space} to a local peak
detection in parameter space. This is illustrated in
Figure~\ref{TRACK_illustrateline}.

\bfig
\insertplot{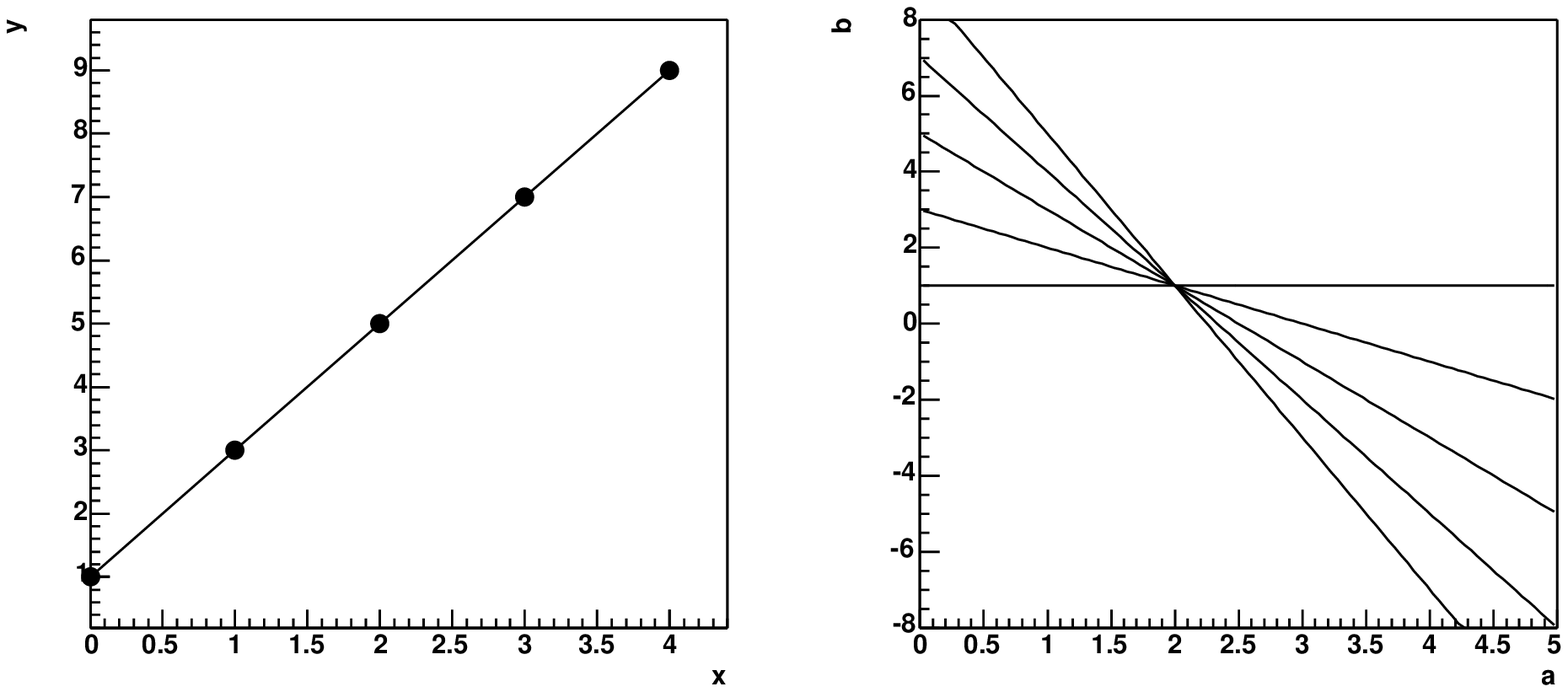}{14cm}
	{Illustration of the HT applied to a straight line.}
	{Illustration of a simple HT applied to points
lying on the straight line $y=2x+1$ (left). Each
point transforms to a line in the parameter space (right), and the
intersection of these lines defines the parameters of the line.}
\label{TRACK_illustrateline}
\efig


Formally one can describe the transform as follows.
Let $X\ $ be a point in image space and $\Omega$ a point in parameter
space. A set of image points which lie on a curve can then be defined by
the function $f$ such that
\beq
f(X,\Omega) = 0.
\label{TRACK_ht}
\eeq
In the case of a straight line one can define $X$=$(x,y)$ and $\Omega=(a,b)$.
The equation for a straight line then becomes
$f(X,\Omega) = y - ax - b = 0$ and maps each value of the
parameter combination $(a,b)$ to a set of image points. 
The mapping is
one to many from the space of possible parameters values to the space
of image points. Equation~\ref{TRACK_ht} can be viewed as a mutual
constraint between image points and parameter points and therefore it
can be interpreted as defining a mapping from a set of image points to
a set of possible parameter values. 

The HT can be defined for arbitrary parametrically defined image
curves. When considering curves characterized by $N$ parameters, the HT will
map out of hyper-surface in a $N$-dimensional parameter
space. Correspondingly the intersection of these surfaces marks a
possible candidate for the curves in image space. The HT for a set of
image points, $\{X_1,X_2,...,X_n\}$, is defined as
\beq
H(\Omega) = \sum_{j=1}^{n}p(X_{j},\Omega)
\label{TRACK_sht}
\eeq
where
\beq
p(X,\Omega)=\left\{\begin{array}{ll}
			1 &\mbox{if $f(X,\Omega)=0$}\\
			0 &\mbox{otherwise} \end{array} \right.
\eeq


In practical implementations of the HT, the continuous parameter space
is usually considered to be composed of the union of finite-sized
regions. One then makes parameter space discrete and defines a
counter for each cell in the parameter space. Each cell is associated
with a element in a multi-dimensional array, and the respective
counters are incremented when a hyper-surface from the transform of an
image point passes through the region of parameter space associated
with the element. This array is in literature referred to as the
Hough {\it accumulator array}. 
If C$_{\Omega}$ denotes a element in the accumulator array
which corresponds to a cell in the parameter space centered at $\Omega$,
then consequently
$p(X,\Omega)$ is 1 if any curve corresponding to the
parameter space 
cell C$_{\Omega}$ passes through the point, $X$, in image space.
This can be considered as every image point giving
``votes'' to all the possible parameter combinations it can possibly
belong to. The detection of the intersection of the surfaces is
done by detection of local peaks in the number of accumulator
counts. The size and shape of these peaks depend on several factors,
such as the number of transformed image points, the errors of the
image points and the size of the parameter cells.

The standard Hough transform as described above is a {\it divergent}
transform, i.e. a point is mapped onto a curve in parameter space. It
is however possible to reformulate the transform so that it becomes a
{\it convergent} transform. In this case, 
instead of transforming single image points into curves in parameter
space, subset of points is mapped. The advantage of
this approach is that each mapping is into a smaller subset of the
parameter space. However, since all combinations of possible subsets of image points
has to be mapped in the transform, the combinatorics of such algorithm
makes it very computing intensive.

\subsection*{Implementation issues}
\label{TRACK_houghimplement}
As described in the previous section, the underlying idea of the HT is
to search for some kind of predefined parameterized pattern in an
certain image space. In the case of the ALICE TPC, the image space
consists of the digitized charge distributions sampled from the
ionization along the particle trajectory, while the patterns to be
parameterized are defined by the motion of the charged particles in the
homogeneous magnetic field. The main implementation issues of such an algorithm
consists of defining a proper parameterization and its corresponding
HT, together with the adaption of a optimal parameter space. These
issues are discussed in the following.

\subsubsection{Parameterization and derivation of the HT}
A charged particle which moves in a homogeneous magnetic field follows
a helical trajectory. This corresponds to a circular motion
in the plane perpendicular to the magnetic field, and a non-curved
motion in the plane along the magnetic field.
In total five independent parameters are needed to fully describe the helix in
three dimensions (Appendix~\ref{APP_helix}). Under the assumption that
the helix originates from the vertex, the following parameterization
can be used,
\begin{eqnarray}
x(t) &=& \frac{1}{\kappa}\{\sin(\psi_0+ht)-\sin\psi_0\}\nonumber\\
y(t) &=& \frac{1}{\kappa}\{-\cos(\psi_0+ht)+\cos\psi_0\}\nonumber\\
z(t) &=& \gamma t,
\label{TRACK_helixparam}
\end{eqnarray}
where $\kappa$ is the curvature, $\psi_0$ is the emission angle with the
x-axis, $h$ is the sense of rotation, $\gamma$ governs the
longitudinal translation and $t\in[0,\pi]$.
In this scenario the number of
required parameters are reduced to the 3 independent parameters,
$(\kappa,\psi_0,\gamma)$, where the two former describe the circle
in the transverse plane and the latter the straight line in the
longitudinal plane.

Now let $X_i=(x_i,y_i,z_i)$
denote a point lying on the helix, and $\Omega$ =
$(\kappa,\psi_0,\gamma)$ the parameters of a helix. Following the
terminology introduced above, $X_i$ and $\Omega$
represent points in image space and parameter space respectively. 
Let $M_i(X_i,\Omega)$ denote the minimal squared distance
between a helix and a point $i$,
\beq
M_i(X_i,\Omega) = M_i^{xy}+ M_i^{z} = (x_i-x)^2+(y-y_i)^2+(z-z_i)^2,
\eeq
where $M_i^{xy}$ and $M_i^z$ denote the transverse projection and the
longitudinal distance respectively.
In order to derive the standard HT
for this helix parameterization one has to solve the equation
\beq
M_i(X_i,\Omega) = 0.
\label{TRACK_helixdistance}
\eeq
This formula allows the determination of the set of all helix
parameters which correspond to the given point $X_i$. The
transformation defines a mapping of one point in our image space into
a three dimensional curve in the parameter space span by
$(\kappa,\psi_0,\gamma)$. 
The problem can be simplified by considering the transverse and the
longitudinal part of the helix separately:
\begin{eqnarray*}
(x,y)&\leftrightarrow&(\kappa,\psi_0)\nonumber\\
(z)&\leftrightarrow&(\gamma)
\end{eqnarray*}
The transform in the transverse plane is then defined by the equation
$M_i^{xy}$=0, and
and maps the transverse projection of a point into a curve in the
two-dimensional parameter space span by $(\kappa,\psi_0)$. In the latter
case the transform is defined by $M_i^z$=0 and maps a point into a
one-dimensional parameter space span by $\gamma$.

Substituting $x$ and $y$ from Equation~\ref{TRACK_helixparam}
into $M_i^{xy}$ in Equation~\ref{TRACK_helixdistance} gives
\begin{eqnarray*}
\frac{1}{\kappa^2}\left\{1-\sqrt{(\kappa x_i + \sin\psi_0)^2 +
(\kappa y_i - \cos\psi_0)^2}\right\}^2 = 0.
\end{eqnarray*}
Taking the square root and changing to polar coordinates
($x_i,y_i)\rightarrow(r_i,\phi_i)$ gives
\begin{eqnarray*}
\pm\frac{1}{\kappa}\left\{1-\sqrt{\kappa^2 r_i^2 + 2\kappa
r_i(\cos\phi_i\sin\psi_0 - \sin\phi_i\cos\psi_0)+1}\right\} = 0\nonumber\\
\frac{1}{\kappa}\left\{-\kappa^2 r_i^2+2\kappa
r_i\sin(\phi_i-\psi_0)\right\} =0
\end{eqnarray*}
which finally gives the expression
\beq
\kappa = \frac{2}{r_i}\sin(\phi_i-\psi_0)
\label{TRACK_circlehough}
\eeq
Equation~\ref{TRACK_circlehough} is thus the standard HT for the
parameterization of a helix going through (0,0) projected into the transverse
plane. According to this transform a
single image point will by mapped into a sinusoidal
curve in parameter space.

Similarly, substituting $z$ from Equation~\ref{TRACK_helixparam} gives
\beq
\gamma = \frac{z_i}{t}.
\label{TRACK_lambdahough}
\eeq
This transform thus corresponds to a mapping into a one-dimensional
parameter space span by $\gamma$.

Equations~\ref{TRACK_circlehough} and~\ref{TRACK_lambdahough} 
defines the HT for the transverse and longitudinal pattern of the
helix parameterization in Equation~\ref{TRACK_helixparam}.
Using this definition two separate accumulator arrays are required
which can be denoted as
$H(\kappa,\psi_0)$ and $H(\gamma)$ respectively. The two sets are
correlated, i.e. a valid peak in $H(\kappa,\psi_0)$ will have a
corresponding peak in $H(\gamma)$. This can be utilized by defining
subsets in $\gamma$, and thereby transforming all points located
within a given subset into $H(\kappa,\psi_0)$. A helix whose
points are confined within the subset, will consequently create a peak
in $H(\kappa,\psi_0)$ corresponding to the parameters of the helix in
the transverse plane. The longitudinal parameter of the helix
can thus be determined from the values of the subset
being transformed.
For the purpose of defining such sub-volumes, the
pseudo-rapidity, $\eta$, has been used. This is a geometrical quantity
defined by the angle between the trajectory and the beam axis, $\theta$, as
\beq
\eta = \ln(\tan\frac{\theta}{2})^{-1}.
\label{TRACK_defeta}
\eeq
$\theta$ is related to the dip-angle of the track, $\lambda$, as
$\theta = \frac{\pi}{2} - \lambda$. 
The pseudo-rapidity of a image point $(x,y,z)$ is defined as
\beq
\eta = \frac{1}{2}\ln\frac{r_3+z}{r_3-z}
\label{TRACK_pointeta}
\eeq
where $r_3=\sqrt{x^2+y^2+z^2}$.
The standard HT (Equation~\ref{TRACK_sht}) for a set of image points,
$\epsilon_k=\{X_1,X_2,...,X_n\}$, is then defined as 
\beq
H_k(\kappa,\psi_0) = \sum_{j=1}^n p(x_j,y_j,\kappa,\psi_0) 
\eeq
where
\beq
p(x,y,\kappa,\psi_0)=\left\{\begin{array}{ll}
			1 &\mbox{if $\kappa=\frac{2}{r}\sin(\phi-\psi_0)$}\\
			0 &\mbox{otherwise} \end{array} \right.
\eeq
and
\beq
\epsilon_k\in[\eta_k,\eta_{k+1}] 
\label{TRACK_longsegm}
\eeq
$\epsilon_k$ corresponds to the set of points located within a
sub-volume in the longitudinal plane,
Figure~\ref{TRACK_helixexample}. Each sub-volume thus defines an
image space which contains circular patterns corresponding to the
projected tracks in the transverse plane.



\bfig[htb]
\insertplot{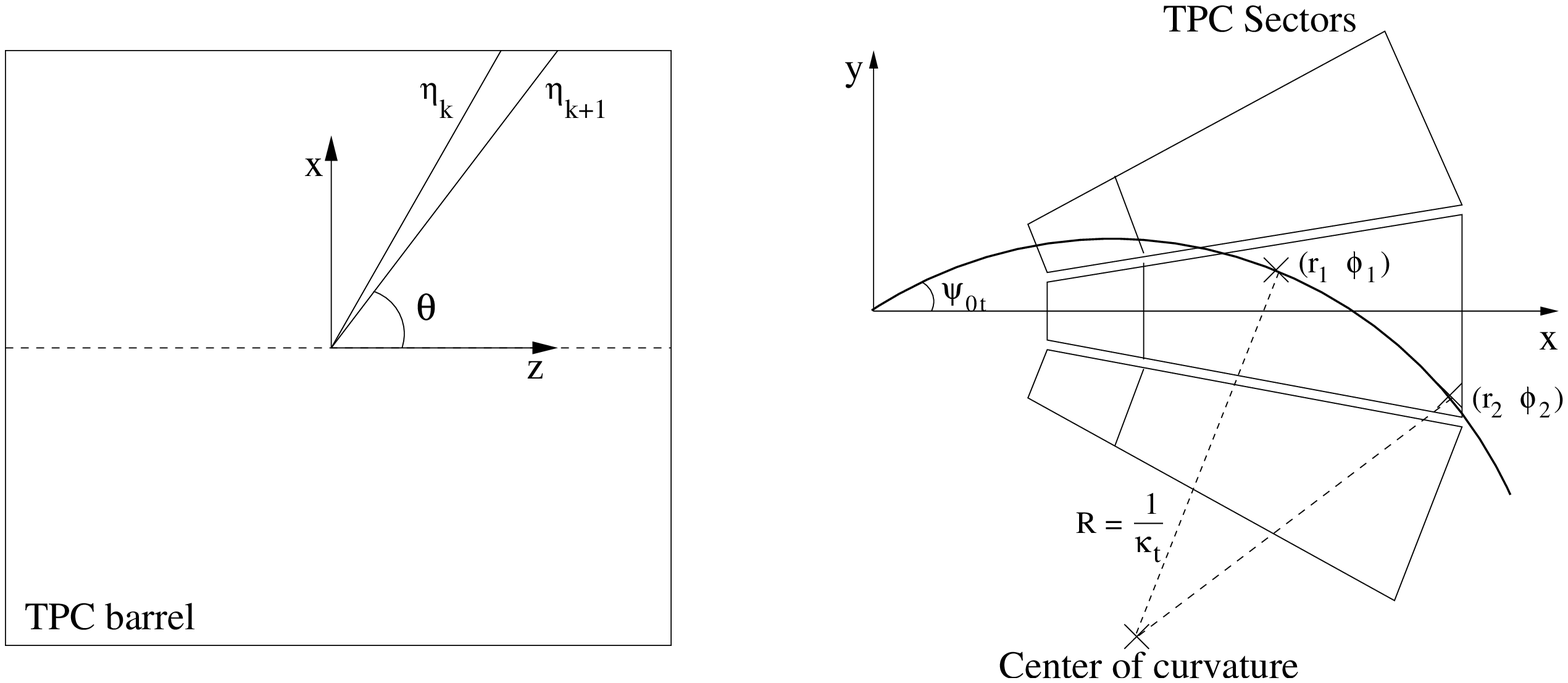}{12cm}
	{Definition of the image space in the HT.}
	{Definition of the image space in the HT. The image space
(right) is defined as a sub-volume characterized by the
pseudo-rapidity, $\eta_k$ (left).}
\label{TRACK_helixexample}
\efig

\subsubsection{On choosing the quantization scheme for $H(\kappa,\psi_0)$}
To compute the HT both the ranges and the quantization steps must be
selected for $H(\kappa,\psi_0)$. The proper choice of these parameters is
an essential part of the algorithm as they directly affect the memory
size, processing times and the final accuracy of the obtained track
parameters. The optimal
number and size of bins in the parameter space
depend generally on aspects like the number and density of the data
points, the required resolution, the noise level and computing
requirements. For example, choosing a bin size which is too big will
restrict the achievable resolution on the reconstructed tracks.
On the other hand, if the bin size is too small the image points on a single
track will be spread throughout too many bins and may not create a
distinct peak at all.

In order to choose the set of properties for the parameter space, an
investigation of the characteristics of the formed peak is necessary.
The shape and size of the peak is determined by two main factors: The
sampling of the parameter space, and the ``noise'' of the
image points. In this context noise includes all factors which contribute to the
image points not being perfectly aligned on the circle
parameterization, and consequently would lead to the intersection point
in parameter space being smeared over a finite area.


The intrinsic accuracy of the transform is limited by the sampling intervals
chosen for the accumulator array, $H$. 
Moreover, the formation of peaks in the parameter space is directly
influenced by how the space is quantized.
\bfig
\centerline{\epsfig{file=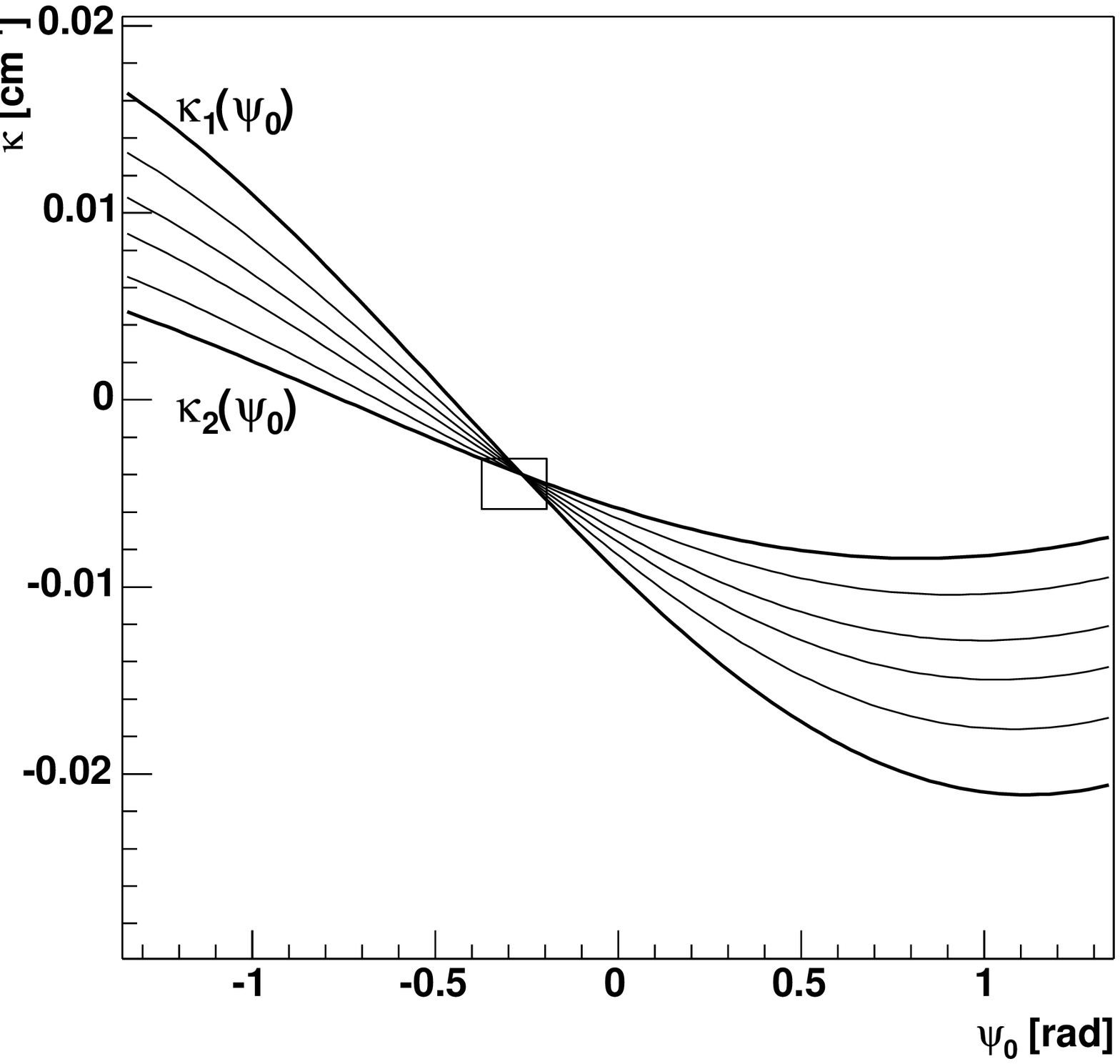,width=8cm}
\epsfig{file=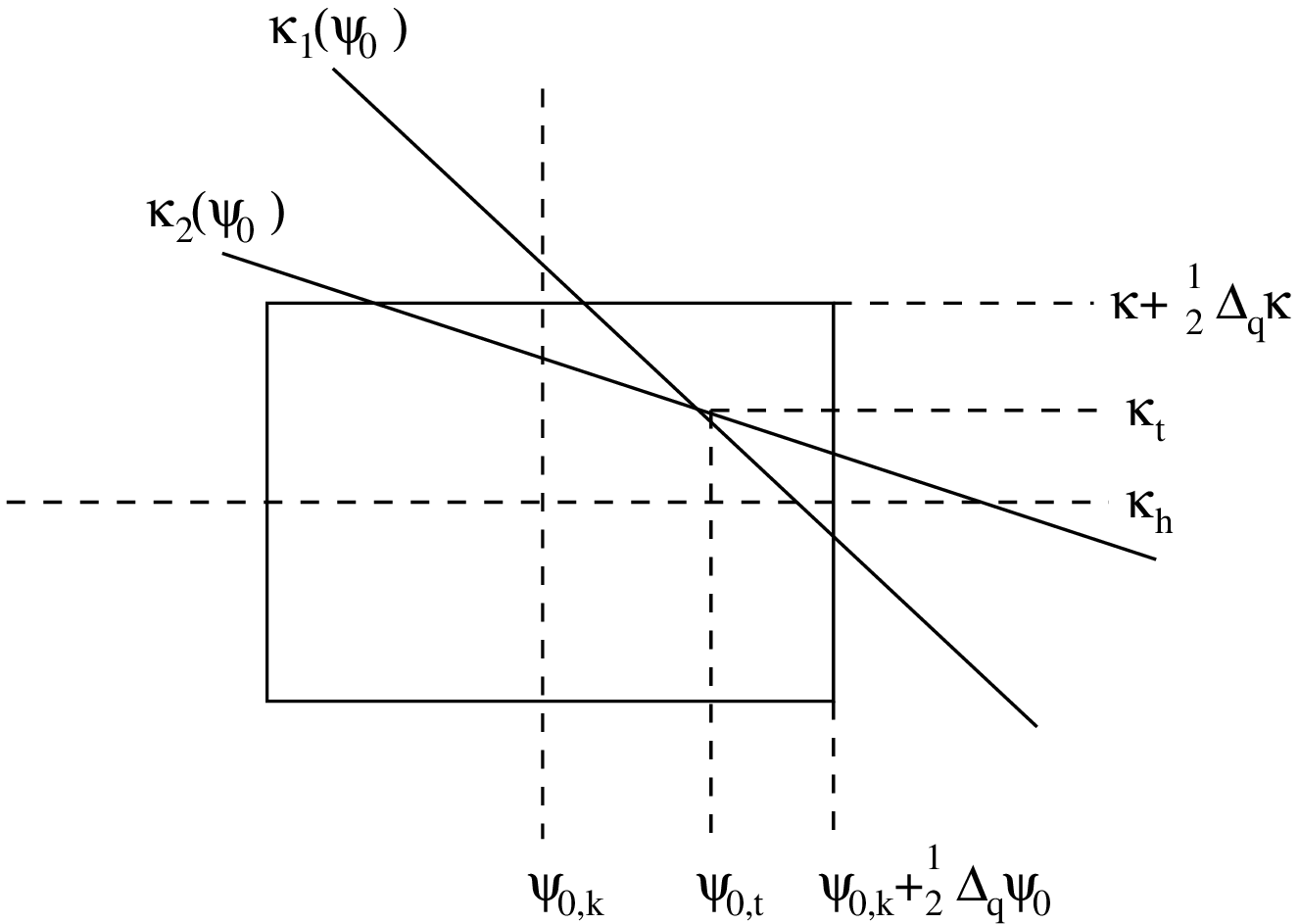,width=8cm}}
\caption[Peak formation in the HT parameter space.]
	{Peak formation in the HT parameter space. The transformation
of the points lying in a circular segment define sinusoids in the
parameter space (left). The curves corresponding to two end-points on
the segment are denoted $\kappa_1$ and $\kappa_2$. The peak is spread
due to the sampling of the parameter space (right).}
\label{TRACK_houghcell}
\efig
Following the procedure introduced in~\cite{lam}
one can
investigate the effect on formation of the peaks in the parameter space.
Consider an ideal circular trajectory defined by the parameters
$(\kappa_t,\psi_{0,t})$ with two points 
$(r_1,\phi_1)$ and $(r_2,\phi_2)$ as illustrated in
Figure~\ref{TRACK_helixexample}. These two
points will under the HT define two sinusoids in the parameter space,
$H(\kappa,\psi_0)$ (Figure~\ref{TRACK_houghcell}, left),
\begin{eqnarray}
\kappa_1(\psi_0) &=& \frac{2}{r_1}\sin(\phi_1-\psi_0)\nonumber\\
\kappa_2(\psi_0) &=& \frac{2}{r_2}\sin(\phi_2-\psi_0),
\end{eqnarray}
with the derivatives
\begin{eqnarray}
\frac{\partial\kappa_1}{\partial\psi_0} &=& -\frac{2}{r_1}\cos(\phi_1-\psi_0)\nonumber\\
\frac{\partial\kappa_2}{\partial\psi_0} &=& -\frac{2}{r_2}\cos(\phi_2-\psi_0),
\end{eqnarray}
In addition, let $(r_0,\phi_0)$ represent a point on the circle
between the two points.
Assuming that $|\phi-\psi_0|<\pi/2$ and $r_1<r_0<r_2$
gives
\begin{eqnarray}
\kappa_1(\psi_0)>\kappa_0(\psi_0)>\kappa_2(\psi_0)\hspace{0.5cm}\mbox{for $\psi_0<\psi_{0,t}$}\nonumber\\
\kappa_1(\psi_0)<\kappa_0(\psi_0)<\kappa_2(\psi_0)\hspace{0.5cm}\mbox{for $\psi_{0,t}<\psi_0$}
\end{eqnarray}
The peak formed by the transformation exhibits a typical {\it cross
shape} (Figure~\ref{TRACK_houghcell}, right), where the spread will be
confined by the the curves of $\kappa_1$ and $\kappa_2$.
Let now $\psi_0$ be sampled with a uniform interval $\Delta_q\psi_0$ and
assume that $\psi_{0,t}$ lies between
$\psi_{0,k}-\frac{1}{2}\Delta\psi_0$ and
$\psi_{0,k}+\frac{1}{2}\Delta\psi_0$ where $k\in[0,m]$. The
uncertainty between $\psi_{0,k}$ and $\psi_{0,t}$ can be given by
\beq
-\frac{1}{2}\Delta_q\psi_0\leq\delta\psi_{0,k}=\psi_{0,t}-\psi_{0,k}
\leq\frac{1}{2}\Delta_q\psi_0.
\eeq
Each of
the accumulators can be regarded as small rectangular window overlaid
the continuous parameter space. Assume that the true peak is
located within the window of size $\Delta_q\psi_0\times\Delta_q\kappa$
with center $(\psi_{0,k},\kappa_h)$ where $h\in[0,n]$. The
uncertainty between $\kappa_h$ and $\kappa_t$ is then
\beq
-\frac{1}{2}\Delta_q\kappa\leq\delta\kappa_h=\kappa_t-\kappa_h
\leq\frac{1}{2}\Delta_q\kappa.
\label{TRACK_errorkappa}
\eeq
Depending on the relative size of $\Delta_q\psi_0$ and
$\Delta_q\kappa$, the peak will be distributed or spread out over
several bins in the accumulator array. The total spread in the
$\kappa$-direction, $\Delta_{ps}\kappa$ at the sampling interval
$\psi_{0,k}$ is given by 
\beq
\Delta_{ps}\kappa(\psi_{0,k}) = \kappa_1(\psi_{0,k}) - \kappa_2(\psi_{0,k})
\eeq
Substituting $\delta\psi_{0,k}=\psi_{0,t}-\psi_{0,k}$ gives
\begin{eqnarray}
\Delta_{ps}\kappa(\psi_{0,k}) &=&
\frac{2}{r_1}\sin(\phi_1-\psi_{0,t}+\delta\psi_{0,k}) -
\frac{2}{r_2}\sin(\phi_2-\psi_{0,t}+\delta\psi_{0,k})\nonumber\\
&=& \cos(\delta\psi_{0,k})\left\{\frac{2}{r_1}\sin(\phi_1-\psi_{0,t}) -
\frac{2}{r_2}\sin(\phi_2-\psi_{0,t})\right\} \nonumber\\
&& + \sin(\delta\psi_{0,k})\left\{\frac{2}{r_1}\cos(\phi_1-\psi_{0,t}) -
\frac{2}{r_2}\cos(\phi_2-\psi_{0,t})\right\}\nonumber\\
&=& \cos(\delta\psi_{0,k})\left\{\kappa_1(\psi_{0,t})-\kappa_2(\psi_{0,t})\right\}\nonumber\\
&&+\sin(\delta\psi_{0,k})\left\{
\left(\frac{\partial\kappa_2}{\partial\psi_0}\right)_{\psi_0=\psi_{0,t}}
-\left(\frac{\partial\kappa_1}{\partial\psi_0}\right)_{\psi_0=\psi_{0,t}}\right\}
\end{eqnarray}
In the last equation the first term equals zero since
$\kappa_1(\psi_{0,t})=\kappa_2(\psi_{0,t})=\kappa(\psi_{0,t})$ which gives
\beq
\Delta_{ps}\kappa(\psi_{0,k}) =\sin(\delta\psi_{0,k})
\left\{\left(\frac{\partial\kappa_2}{\partial\psi_0}\right)_{\psi_0=\psi_{0,t}}
-\left(\frac{\partial\kappa_1}{\partial\psi_0}\right)_{\psi_0=\psi_{0,t}}\right\}
\eeq
Maximum spreading of the peak occurs when (see Equation~\ref{TRACK_errorkappa})
\beq
\mbox{max}(|\delta\psi_{0,k}|) = \frac{1}{2}\Delta_q\psi_0
\eeq
which yields 
\beq
(\Delta_{ps}\kappa(\psi_{0,k}))_{\mbox{max}} = \sin(\frac{1}{2}\Delta_q\psi_0)
\left\{\left(\frac{\partial\kappa_2}{\partial\psi_0}\right)_{\psi_0=\psi_{0,t}}
-\left(\frac{\partial\kappa_1}{\partial\psi_0}\right)_{\psi_0=\psi_{0,t}}\right\}
\label{TRACK_maxspread}
\eeq
Consequently, the following condition on the bin size will allow a
maximum collection votes without peak spreading:
\beq
\Delta_q\kappa = (\Delta_{ps}\kappa(\psi_{0,k}))_{\mbox{max}}.
\label{TRACK_quantrel}
\eeq

Equation~\ref{TRACK_quantrel} shows that values of quantization steps combine to
influence the spreading of the peak, which indicates that they cannot
be selected fully independently. In particular, one can differ between
the two cases,
\begin{eqnarray}
\Delta_q\kappa &<&
(\Delta_{ps}\kappa(\psi_{0,k}))_{\mbox{max}}\nonumber\\
\Delta_q\kappa &>& (\Delta_{ps}\kappa(\psi_{0,k}))_{\mbox{max}},
\label{TRACK_psisampling}
\end{eqnarray}
which can be interpreted as {\it under-sampling} and {\it oversampling}
of $\psi_0$ with respect to $\kappa$ respectively. The effect of the
former case is to split the votes to neighboring accumulators in the
$\kappa$-direction, while the latter corresponds to the case
where the peak is distributed over several neighboring bins in the
$\psi_0$-direction. 

This derivation shows that for an ideal circle segment, the
resulting peak will be spread with a characteristic form due to
the quantization of the parameter space. The implicit assumption is that all the
image points on the circle are transformed into a single bin if the
proper quantization is chosen. 
However, the image points, represented by the digitized charge distribution in the
TPC detector, do not lie on an ideal circular trajectory due to the
energy loss and multiple scattering of the particles and the intrinsic
detector noise.
As a consequence of the uncertainty of the image points,
$(r_i,\phi_i)$, on a circular track,
the transformation of these points gives a region in the
$(\kappa,\psi_0)$-space rather than a point. The uncertainty of the
calculated $\kappa$ can be obtained by taking the Taylor expansion of
the transformation Equation~\ref{TRACK_circlehough} around $r$ and $\phi$.
The first order approximation yields
\begin{eqnarray}
\delta\kappa &=&
\left(\frac{\partial\kappa}{\partial\phi}\right)_{r,\psi_0}\delta\phi +
\left(\frac{\partial\kappa}{\partial r}\right)_{\phi,\psi_0}\delta r\nonumber\\
&=& \frac{2}{r}\cos(\phi-\psi_0)\delta\phi
-\frac{2}{r^2}\sin(\phi-\psi_0)\delta r.
\label{TRACK_noisespread}
\end{eqnarray}
This relation shows that spread of the peak in $\kappa$ is influenced
by the noise of image points, but also on the location of the points in the image.
Due to the $r^{-2}$ dependence in the second
term, the error will be dominated by the error in $\phi$.

In general, the characteristics of the formed peaks
in the parameter space depend on both the location of the image points and the
track segment in space, and the errors of the individual image points.
For a given input image containing
several track segments it is therefore generally impossible to quantize
$H(\kappa,\psi_0)$ optimally such that all peaks formed by the transform have
the same extension and shape. It is therefore desirable to construct a
quantization scheme that as far as possible allows an average peak
which keeps a sufficient peak significance with respect to the background.

\subsubsection{Accumulation}
The HT itself is a pure accumulation procedure. The transformation of
each image point consists of incrementing the corresponding counters
in a multidimensional array representing the cells in the parameter
space. Depending on the nature and topology of the image, different
methods can be applied. Many implementations of the HT have suggested using
weighting factors so that the most prominent or most certain image
features contribute more to the accumulator cells than less certain
data. The effect of such schemes is evident: By assigning
higher weights to image points lying closer to the parameterization, the
corresponding bins will receive a higher vote and will consequently
lead to a sharpening of the peaks in parameter space. Formally this is
introduced by modifying Equation~\ref{TRACK_sht} to
\beq
H(\Omega) = \sum_{j=1}^{n}w(j)p(X_{j},\Omega)
\label{TRACK_htw}
\eeq
where $w(j)$ represents the weighting factor for each image point. 

In the case of TPC data, each image point has an
associated ADC-value. These values represent the charge distributions
of the ionized charge in each
pad-row-plane, where the centroid of the distribution denotes the space
point of the crossing particle track. In this context, each image
point can be regarded as having a certain {\it weight}, corresponding
to its ADC-value. Thus $w$ in Equation~\ref{TRACK_htw} is
replaced by the ADC-value of the time-bin to be transformed. The
implementation of the resulting accumulation algorithm can be
illustrated in pseudo-code as follows:
\begin{tabbing}
{\tt Initialize $H(\kappa,\psi_0)$ to zero;}\\
 {\tt for} \= {\tt i:=1 to npoints do}\\
   \> {\tt for} \= {\tt k:=1 to m do} \+ \\
      \> {\tt $\kappa_t=\frac{2}{r_i}\sin(\phi_i-\psi_k)$;}\=\+\\
         {\tt $\kappa_h = Qntz(\kappa_t,\Delta_q\kappa)$;}\\
	 {\tt $H[\psi_k][\kappa_h] = H[\psi_k][\kappa_h]$ + ADC-value;}\-\\
      {\tt end for}\-\\
{\tt end for}\\
\end{tabbing}
$Qntz()$ denotes the quantizing function which locates the
corresponding bin in the $\kappa$-direction. It is thus defined by
\begin{eqnarray*}
Qntz(\kappa,\Delta_q\kappa) = \left\{\begin{array}{ll}
				\kappa_{p-1} &\mbox{if
$\kappa-\kappa_{p-1}\leq\Delta_q\kappa/2$}\\
				\kappa_p &\mbox{otherwise}\end{array}\right.
\end{eqnarray*}
where $p\in[0,m]$.

Each of the accumulator arrays, $H(\kappa,\psi_0)$, can be interpreted
as a two-dimensional histogram, Figure~\ref{TRACK_paramspace}, where
each bin in the histogram corresponds to an element in the accumulator
array.



\bfig[htb]
\insertplot{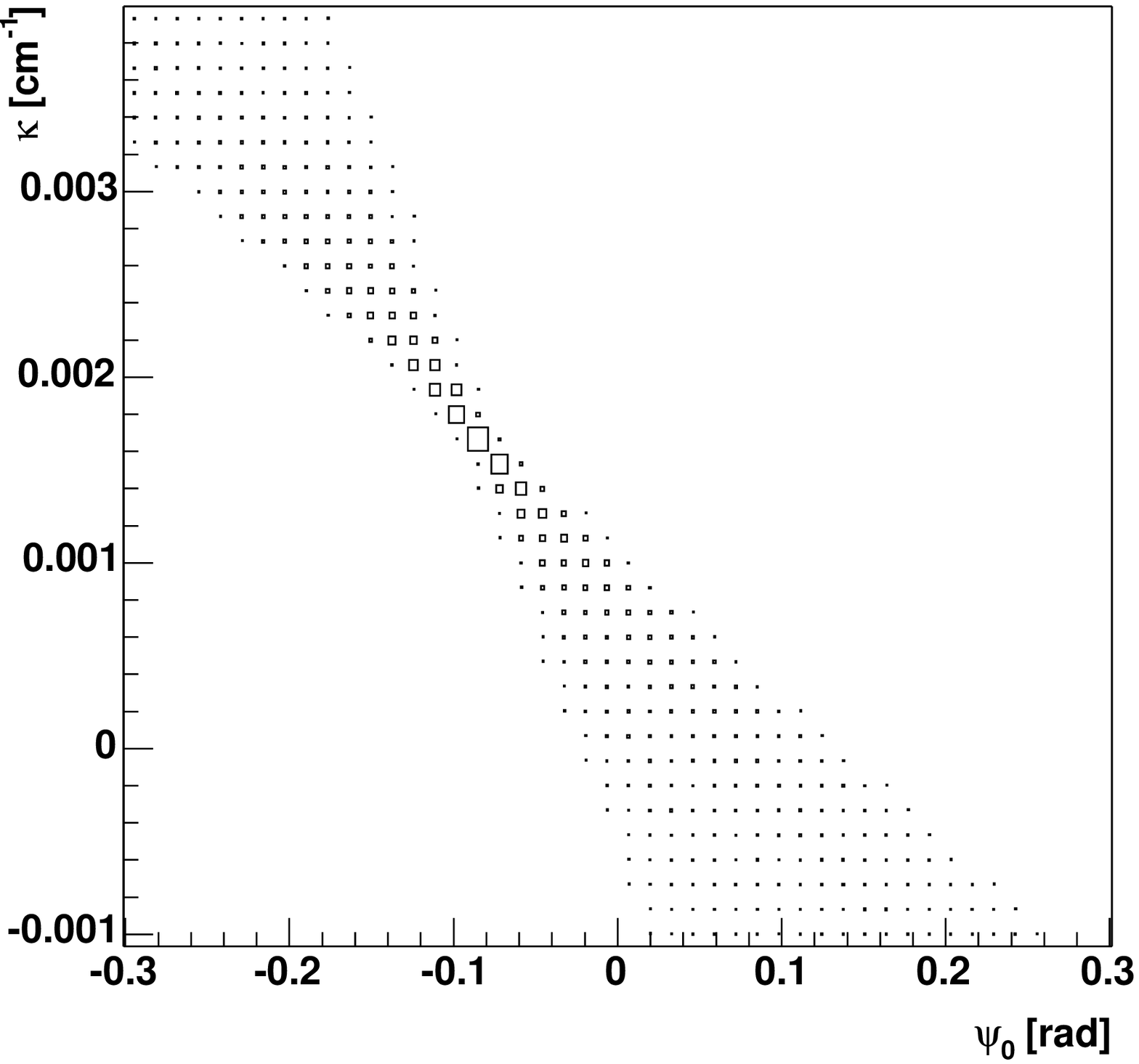}{8cm}
	{Example of the HT parameter space.}
	{Example of the HT parameter space after transforming all
points lying along a track with \pt=0.5\,GeV.}
\label{TRACK_paramspace}
\efig

\subsubsection{Peak finding}
Once the two-dimensional histograms have been filled according to the
transformation outlined above, the histograms have to be searched for local peaks
which corresponds to potential track candidates. Among these peaks
there will in general also be a certain number of {\it false}
peaks which are originating from artificial features in the
image. Such features may e.g. be noisy data points and tracks
which are outside the parameter range which is searched. In
both cases structural backgrounds are created in the parameter
space.

Prior to the peak search, a {\it threshold} operation is applied to the
histogram. The reason for this is twofold. Firstly, the number of bins which have to
be searched by the peak finder, are reduced. Secondly, such a threshold
eliminates the contribution from peaks that are too small to
correspond to a real track
segment. In the simplest case a local peak finder identifies a single
histogram bin
which corresponds to a local maxima. The requirement for such a local
maxima is that it has a value that is larger than its immediate
neighbors in the histogram. However, since the shape of the peaks is
to the first order known, the peak finder can take advantage of this
knowledge in order to better be able to exclude the true peaks from
the background.

The goal of the peak finder is to estimate the coordinates of
the crossing point, $(\kappa_t,\psi_{0,t})$, of the peak shape.
The implemented procedure starts by locating the initial peaks within a
sliding window in the $\kappa$-direction,
Figure~\ref{TRACK_pfexa}. The length of the window, $n_{\kappa}$, is
determined by the expected spread of the peak. The idea is thus to
choose $n_{\kappa}$ large enough so that it captures the full peak
spread in $\kappa$ near $\psi_{0,t}$, and at the same time small
enough so that it captures the spread only near $\psi_{0,t}$. 
If the value of $n_{\kappa}$ is too small the window will not take into
account all the transformed points along the track segment, while a
window that is too large will include more of the background and may
create ambiguities. The complete procedure can be outlined as follows:
\begin{enumerate}
\item Location of the initial peaks.\\
For each bin in the $\psi_0$-direction the sum of values within the
sliding window in $\kappa$ is calculated. If a local maxima is found,
this window is stored in a temporary list.
\item Location of the two-dimensional peaks.\\
Each of the local maxima is compared with the local maxima's in the
neighboring $\psi_0$-bins in order to validate the peak in the two
dimensions. If a sufficient number of matching windows are found, the
peak is passed to the next step, otherwise the window is deleted from
the list.
\item Evaluation of the peaks.\\
The two-dimensional peaks are evaluated by performing a simple check
on their shape. This is done by comparing the sum of values within
regions on both sides of the maximum bin in $\psi_0$. If the shape
exhibits a clear asymmetry, the peak is stored in a final peak list.
\end{enumerate}
The bin with the maximum value within the peak-region defines the
final location of the peak.

\bfig[htb]
\insertplot{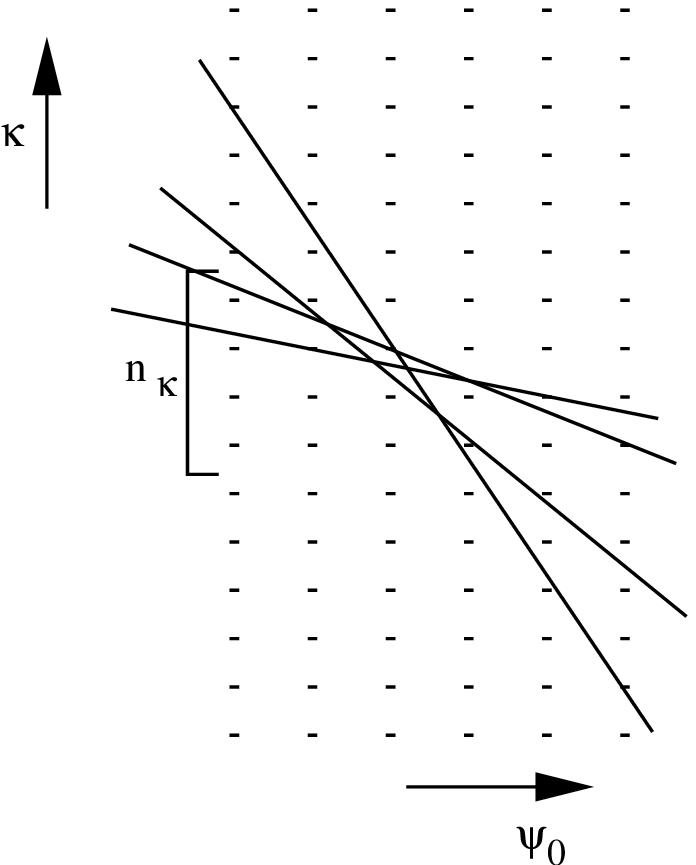}{4cm}
	{Illustration of the peak finding method in the HT parameter
space.}
	{Illustration of the peak finding method.}
\label{TRACK_pfexa}
\efig

\subsection*{Choice of parameters}


\subsubsection{Image boundaries}
The critical aspect in choosing the boundaries of the image space is
the requirement for the image to contain enough signals from each of
the track segments in order to produce a clear peak in the parameter
space. In particular, the image must contain a sufficient number of
pad-rows in which the track produces a signal in order for the
accumulator array to pick up enough votes for the tracks. This is
necessary for the formed peaks to be significant compared to the
background structures. From a implementation point of view, the choice
of image space defines the required data volume access.
Thus the segmentation in the transverse direction is a choice based on a
trade-off between two main factors: Required data flow and performance
of the algorithm. 

As described in Section~\ref{HLT_arch}, the input data-flow in the HLT system is
defined by the granularity of the detector. The TPC is read out in
sub-sectors, where each sub-sector corresponds to 1/6 of a
complete TPC sector. The data from each sub-sector is transferred
directly into the first level nodes of the HLT system (FEPs). In order
to minimize the required connectivity in terms of bandwidth between
the HLT nodes, an attempt has been made to do as much as possible of
the processing directly on the FEPs. In this approach, the image
in the transverse direction will be defined by the boundaries of the
sector readout chambers, Figure~\ref{TRACK_houghparam}.

The performance of the algorithm depends on the topology of the
parameter space, and the ability to distinguish the local maxima from
the background. In this context, the prime task is the selection of an
image space which allows for a sufficient accumulation of hits along
the particle trajectory. The
inherit spread of the peak in parameter space is a function both of
the sampling intervals and the location of the image points in
space. In particular, the {\it opening angle\ } between the arms of the
peak cross shape is dependent on the distance between the two
end-points of the track segment, Figure~\ref{TRACK_houghcell}. If the
points are close in space
this angle is small, and $\Delta_q\kappa$ needs to be chosen
correspondingly small in order to have a sharp peak. Furthermore,
$\Delta_q\psi_0$ has to be chosen according to the conditions in
Equation~\ref{TRACK_psisampling} in order to avoid under- or
over-sampling. On the other hand, if $\Delta_q\kappa$ is significantly smaller
than the spread caused by the noise of the image points, the peak will
be smeared out over a large number of bins and may loose its significance with
respect to the background.

Based on the above analysis, it has proven necessary to include more
than one sub-sector in the HT in order to gain sufficient peak
significance versus background. Since the HT is a pure accumulator
procedure, this can be implemented by adding the accumulator arrays
from the different sub-sectors. Hence the HT is 
performed locally on a sub-sector, and in subsequent step the
accumulator arrays from all successive sub-sectors within a complete
sector are added,
\beq
H^{\mathrm{sector}}(\kappa,\psi_0) = \sum_{i=1}^6 H_i^{\mathrm{subsector}}(\kappa,\psi_0)
\label{TRACK_histadder}
\eeq
Here, $H^{\mathrm{sector}}(\kappa,\psi_0)$ is the resulting accumulator
array obtained by adding all the non-zero entries from the individual sub-sectors,
$H_i^{\mathrm{subsector}}(\kappa,\psi_0)$.

\subsubsection{Longitudinal segmentation}
In the longitudinal direction, the data volume is divided into
sub-volumes in pseudo-rapidity,
Figure~\ref{TRACK_helixexample}. This segmentation of the data
serves two purposes: Firstly, it allows the determination of the
parameter governing the longitudinal motion. Secondly, it reduce the
density of tracks within one image.
The size of the $\eta$-volumes is critical to the performance of the
algorithm. If the
size is too small, the tracks may be split into several
sub-volumes. This splitting causes the signal of a single track to
appear in several neighboring sub-volumes, and may prevent a
track with too few hits in a single image, to produce a clear peak
in the corresponding parameter space. On the other hand, choosing a
sub-volume which
is too wide leads to a higher track density within an image and produces
a correspondingly higher occupancy in the parameter space. This makes
it difficult to correctly identify the peaks.
In addition, a very course segmentation
will lead to a low resolution in determining the longitudinal
parameter of the trajectory.

Figure~\ref{TRACK_deltaeta} shows the average spread in pseudo-rapidity,
$\Delta\eta$, for a given particle as a function of $p_t$.
\bfig
\centerline{\epsfig{file=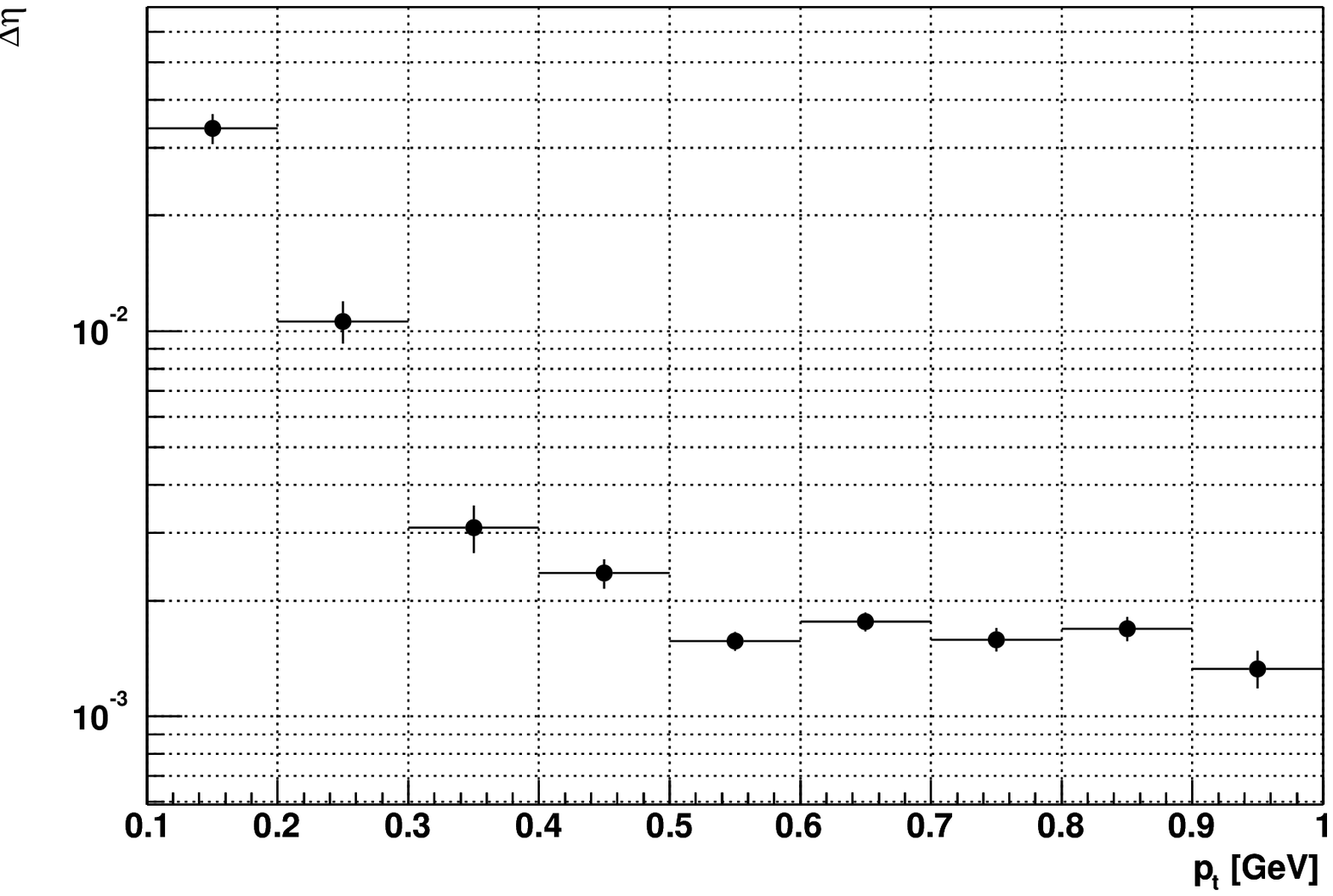,width=8cm}}
\caption[Average spread in pseudo-rapidity,$\Delta\eta$, as a function of $p_t$.]
	{Average spread in pseudo-rapidity, $\Delta\eta$, of signals
produced by a given particle as a function of $p_t$.}
\label{TRACK_deltaeta}
\efig
\noindent The spread is obtained from the $\eta$-distribution of the
signals produced by a given particle. 
Such a plot thus gives an estimate of
how much of the signal from a particle will be contained
within a given $\eta$-sub-volume.
$\Delta\eta$ increases below
\pt of 0.3-0.4\,GeV, while for higher momentum the spread saturates at
$\Delta\eta\sim$0.001-0.003.
This observed increase at lower momentum is mainly due to the fact
that particle energy loss and multiple scattering becomes more
significant for lower momenta. 

Based on the simulated spread, a uniform
segmentation of $\eta$ with intervals of size 0.01 has
been done. This leads to about 100 sub-volumes in the central cone of
$|\eta|<$\,0.9.




\subsubsection{Parameter space}
\label{TRACK_htparamspace}
The properties of the parameter space are determined by the choice of
image space. In particular, the range of the histogram axis is defined
by the possible values of the track parameters. The parameters consist of the
curvature of the track, $\kappa$, and the emission angle with the
$x$-axis, $\psi_0$. The curvature of the track is inversely proportional
to the transverse momentum of the particle,
Equation~\ref{TRACK_kappapt}, and the maximum value of $\kappa$ is thus
defined by the minimum value of $p_t$:
\beq
|\kappa|_{\mathrm{max}} = \frac{0.3B}{p_{t,\mathrm{min}}}.
\label{TRACK_kappamax}
\eeq
In this context, $p_{t,\mathrm{min}}\ $ corresponds to the lower limit of the
desired \pt range to be measured.
Once the $\kappa$-range is defined, the range in $\psi_0$ can be given by
taking the inverse of Equation~\ref{TRACK_circlehough},
\beq
|\psi_0|_{\mathrm{max}} = \phi_m -
\sin^{-1}\left(\frac{r_m}{2}\kappa_{\mathrm{max}}\right),
\eeq
where $(r_m,\phi_m)$ is the polar coordinates of a point on the line
which confines the image space, Figure~\ref{TRACK_houghparam}. 
\bfig
\insertplot{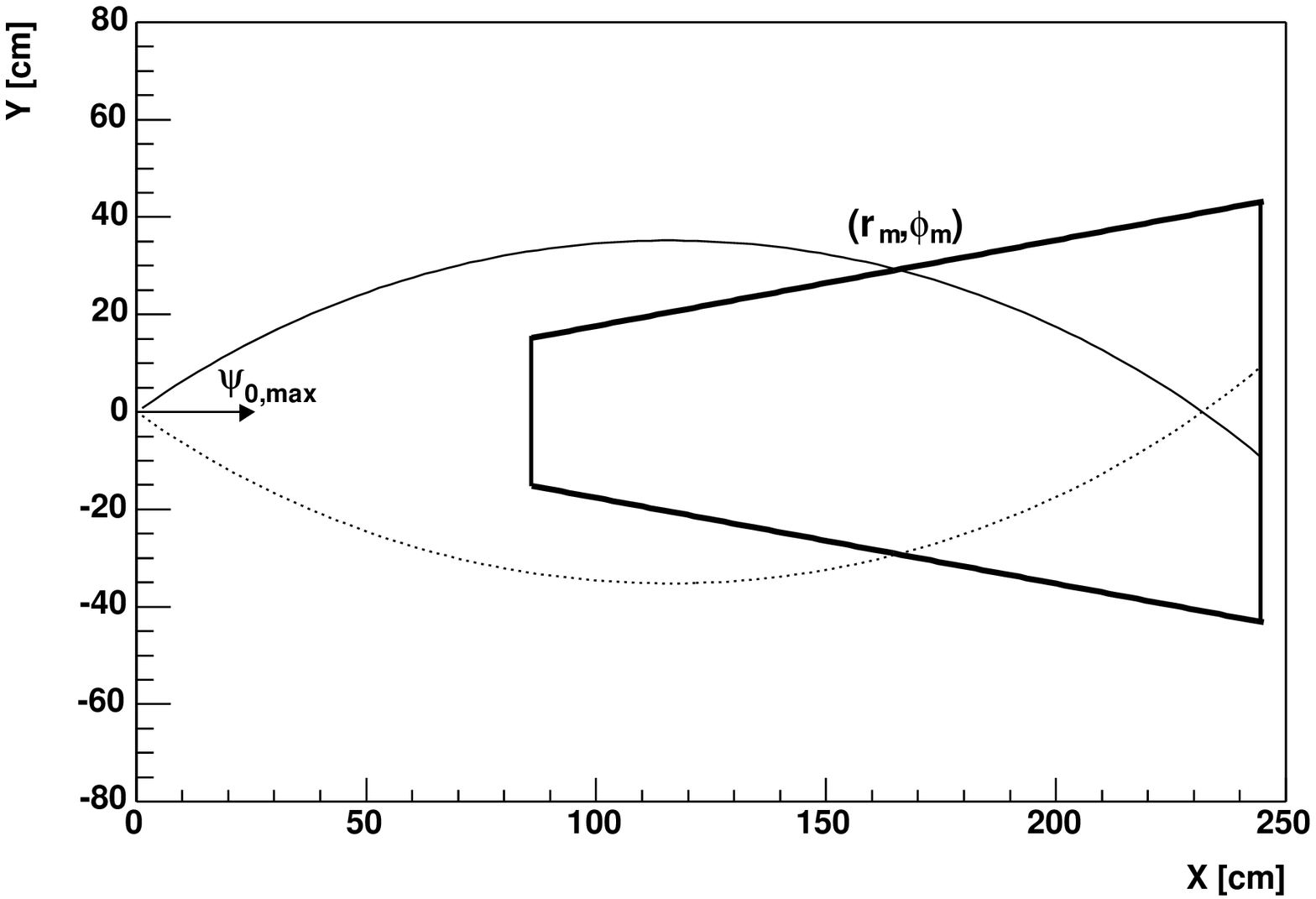}{8cm}
	{Parameterization of the HT image space.}
	{Parameterization of the HT image space. The parameter range of
the HT parameter space is defined by the minimum \pt (maximum
curvature, $\kappa$) and a point $(r_m,\phi_m)$ on
the line which confines the image space in the transversal direction
(this line corresponds to the TPC sector boundary).}
\label{TRACK_houghparam}
\efig
\noindent This line corresponds to the azimuthal angle of the boundary
of the TPC sector. The point is chosen to be in the middle of the TPC
barrel, in order to avoid an unequal distribution of the track
segments in neighboring sectors. 

The fact that the maximum values of $\kappa$ are defined by the minimum
value of $p_t$, makes it possible to apply a lower $p_t$-cut in a
straight-forward way. Particle tracks with a higher curvature than the
range of the histogram axis will not create a valid peak in the parameter
space, and will thus be excluded from the pattern recognition. This can be
exploited by the algorithm, which can be optimized to search for
tracks within a certain $p_t$-range.

The sampling intervals of the parameter space were chosen according to
the average spread of the formed peaks in the
simulations. The relative size of the intervals were selected in order to avoid
possible under- or over-sampling, Equation~\ref{TRACK_psisampling}.


\subsection{Data flow}
Similar to the sequential tracking code, the iterative reconstruction
chain has been implemented in modules in order to be highly
configurable with respect to different process topologies. Each
processing step has however certain requirements with respect to
locality and data flow.

{\bf HT} The HT is a pure accumulation procedure, and
consists of incrementing an array of counters for every input
point. The algorithm is parallel by nature, since the transform
of each image point is treated independently. However, the
requirement on the image space with respect to the peak significance
versus background makes it necessary to include a minimum number of data
volume as input data.

{\bf Peak Finder} Peak finding is done independently on each
accumulator array. Each peak array may thus be processed in
parallel.

{\bf Cluster Fitter} The Cluster Fitter closely resembles the
Cluster Finder as far as parallelization is concerned. They are both
restricted to the two-dimensional pad-row-plane
without any need of information from the
other pad-rows. However, the Cluster Fitter requires the track parameters as
input for the fitting procedure, and thus needs information on the tracks that
crosses the pad-row being processed.

\bfig
\insertplot{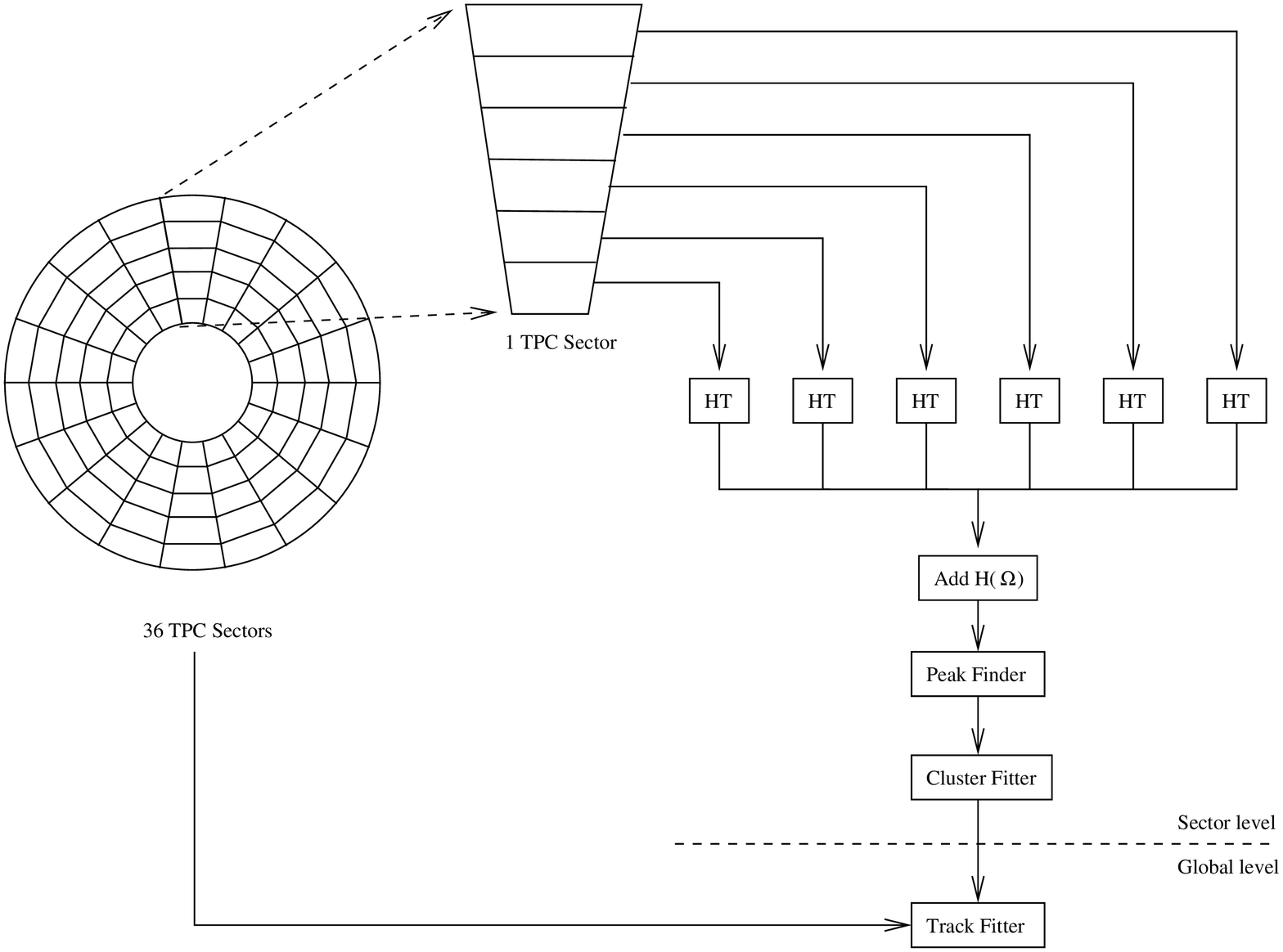}{12cm}
	{Possible data flow for the iterative reconstruction chain
within the HLT system.}
	{Data flow for the iterative reconstruction chain.}
\label{TRACK_houghtopology}
\efig

A possible data flow scheme is shown in
Figure~\ref{TRACK_houghtopology}. In this scenario the HT is performed in parallel on
each sub-sector, whereas the resulting accumulator arrays from each
sub-sector within a sector are added according to
Equation~\ref{TRACK_histadder}.
The arrays are then processed by the
peak finder, whose output is a list of track candidates. The list of
tracks are passed to the Cluster Fitter
(Section~\ref{TRACK_clusterfitter}), which
reconstructs the clusters along the helix trajectories.
In order to obtain the optimal track
parameters, the collected space points belonging to each track are fitted to
a helix using the Track Fitter algorithm, Section~\ref{TRACK_helixfitter}.


\subsection{Performance}
The complete reconstruction chain, as outlined above, has been evaluated
for different particle multiplicities.
In order to compare the results with the results from the other
tracking algorithms, the same
definitions concerning tracking efficiency as described
on page~\pageref{TRACK_trackdef} have been used. The results are
compared to both the sequential tracking scheme and the standard
Offline chain, here referred to as HLT sequential and Offline, respectively.


\subsubsection{Tracking efficiency}
Figure~\ref{TRACK_hteffallpt} shows the tracking
efficiency as a function of \pt for two event samples of
multiplicity \dndy=\,1000 and \dndy=\,4000, respectively.
All primary particles with $p_t\geq$\,0.15\,GeV were included in
this evaluation, and the boundaries of the parameter space were adapted
accordingly. At \dndy=\,1000 the efficiency is slightly lower than for both
HLT sequential and Offline. In particular, the efficiency drops in the low
momentum regime for \pt$\leq$\,0.5\,GeV. For higher \pt the
efficiencies shows similar behavior for all approaches.
At higher multiplicity, \dndy=\,4000, the track
efficiency in the iterative approach is significantly lower than both
the HLT sequential and Offline reconstruction chain.




\bfig[htb]
\centerline{\epsfig{file=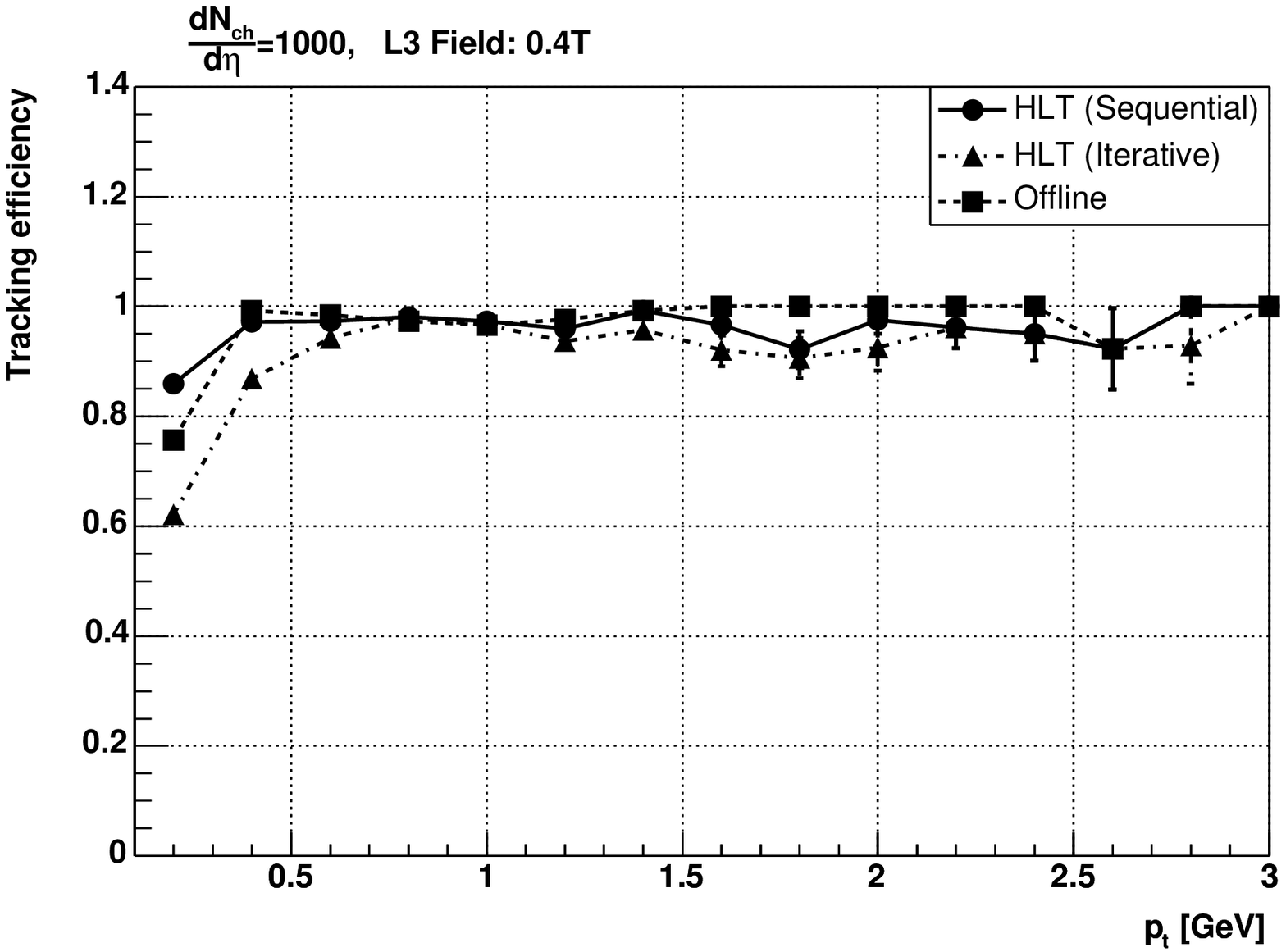,width=8cm}
\epsfig{file=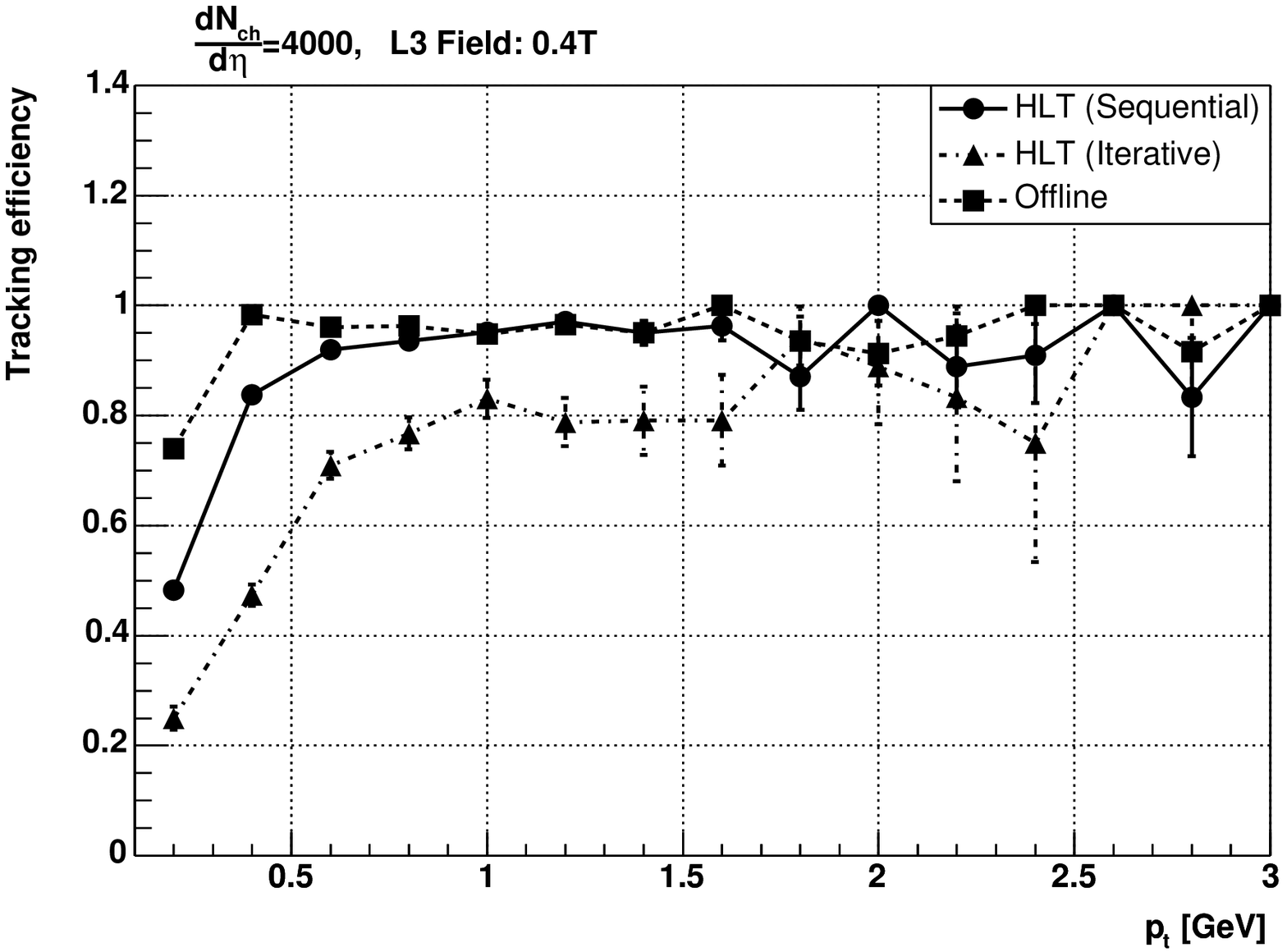,width=8cm}}
\caption[Tracking efficiencies as a function of \pt for the HLT iterative reconstruction chain.]
	{Tracking efficiencies as a function of \pt for two
multiplicity samples. All tracks
with \pt$\geq$\,0.15\,GeV have been included in the evaluation.}
\label{TRACK_hteffallpt}
\efig

In Table~\ref{TRACK_htfakeratio} the ratio between the number of track
candidates found by the HT, and the number of tracks
reconstructed by the Cluster Fitter are listed for the two event
samples. This ratio provides an estimate of the relative number of
false track candidates which originate from falsely identified peaks
in the HT parameter space. The ratios indicate that more than 70\% and
130\% of all the track candidates from the HT failed to be
reconstructed by the Cluster Fitter for the two multiplicities, respectively.
\begin{table}
    \begin{center}
    \begin{tabular}{|c|c|c|}
	\hline
	        & \multicolumn{2}{|c|}{\bf Ratio}\\\cline{2-3}
	\dndy 	& $p_{t,\mathrm{min}}\geq$\,0.15\,GeV & $p_{t,\mathrm{min}}\geq$\,0.5\,GeV \\
    \hline
    \hline
	1000 & 1.75 & 1.39\\
	4000 & 2.35 & 1.32\\
	\hline	
    \end {tabular}
    \caption[Ratio between track candidates found by the HT and tracks
reconstructed by the Cluster Fitter.]
            {Ratio between track candidates found by the HT and tracks
reconstructed by the Cluster Fitter.}
    \label{TRACK_htfakeratio}
    \end{center}
\end{table}
\noindent The right column in Table~\ref{TRACK_htfakeratio} shows the corresponding
ratios for the case where the parameter space in the HT has been
restricted to only include tracks for \pt$\geq$\,0.5\,GeV. In this case
the ratio decreased to $\sim$1.3 for both multiplicities, indicating
that the major part of the fake track candidates originates
from the lower $p_t$-range.

Figure~\ref{TRACK_hthighpteff} shows the tracking efficiencies for the
same event samples as evaluated for the sequential track
reconstruction chain, Figure~\ref{TRACK_seqeffvspt}. Due to the
problem with fake tracks at lower momentum, the parameter space in the
HT was optimized for tracks with \pt$\geq$\,0.5\,GeV, and consequently
only particle tracks with \pt$\geq$\,0.5\,GeV were included in the evaluation.
At lower multiplicities (\dndy$\leq$\,2000) the tracking efficiency of all
approaches are similar for this $p_t$-region. At \dndy=\,4000, however,
the efficiency of the iterative approach is significantly lower than
for the HLT sequential and Offline reconstruction chains.

\bfig
\centerline{\epsfig{file=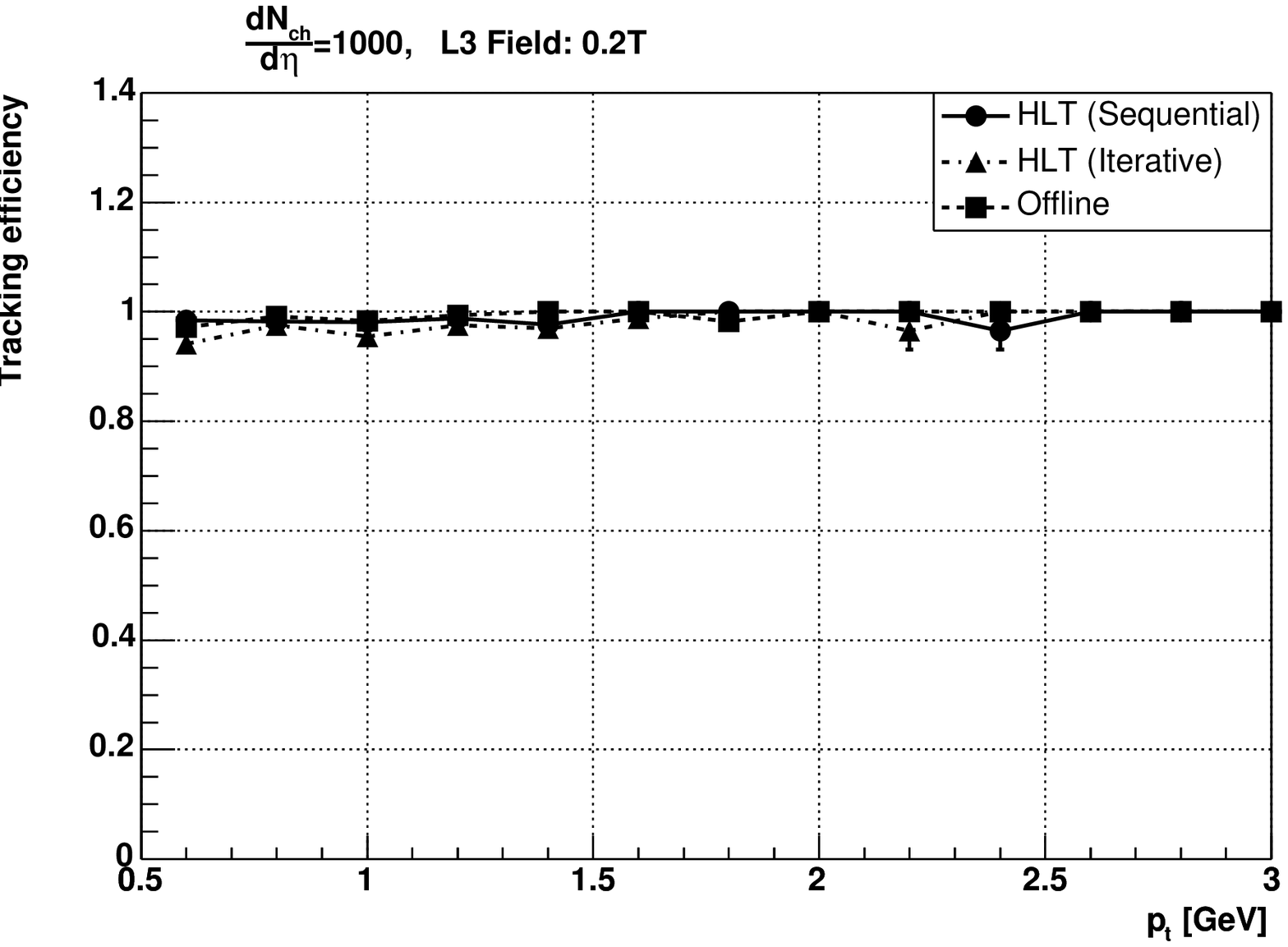,width=8cm}
\epsfig{file=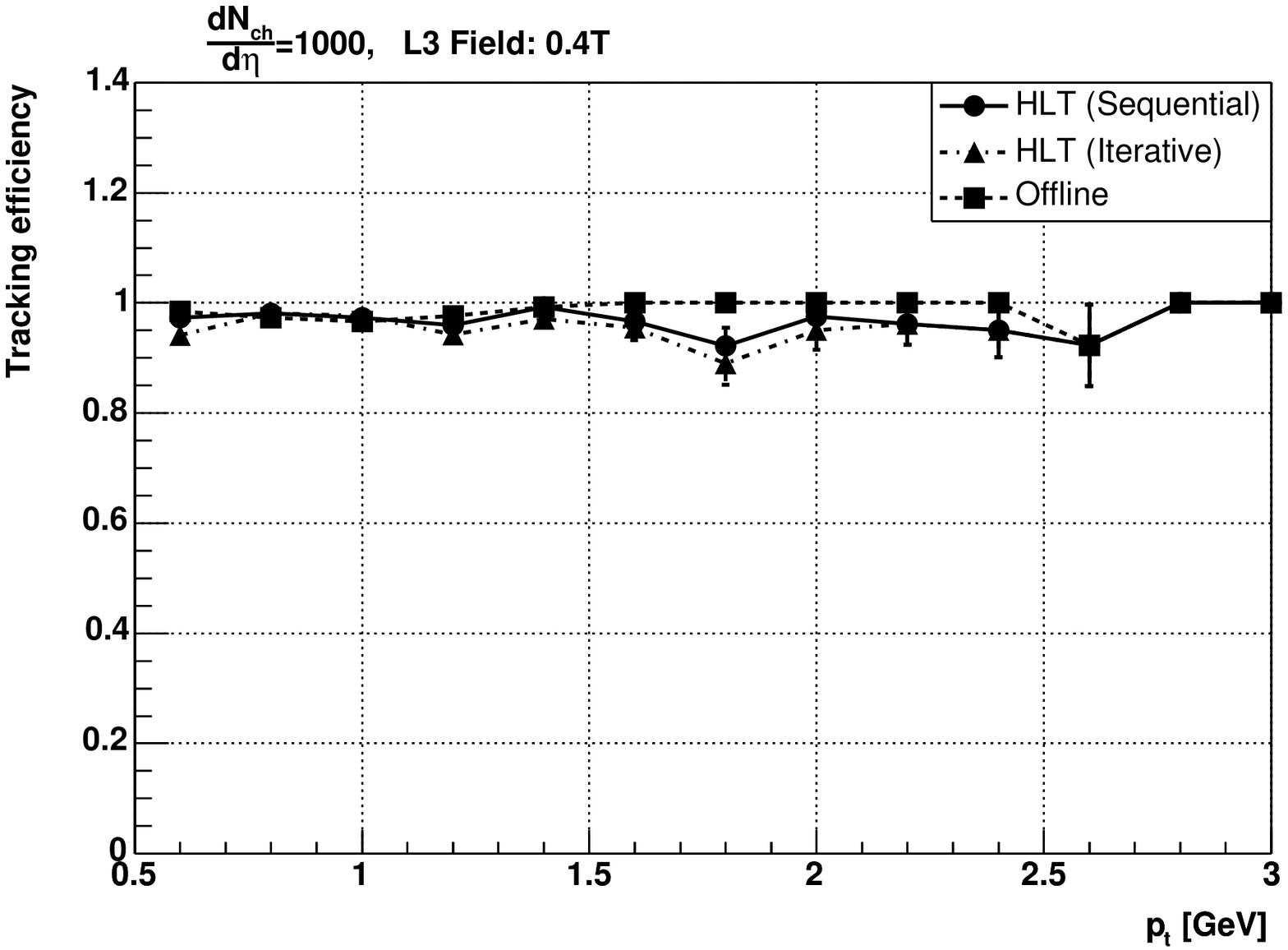,width=8cm}}

\centerline{\epsfig{file=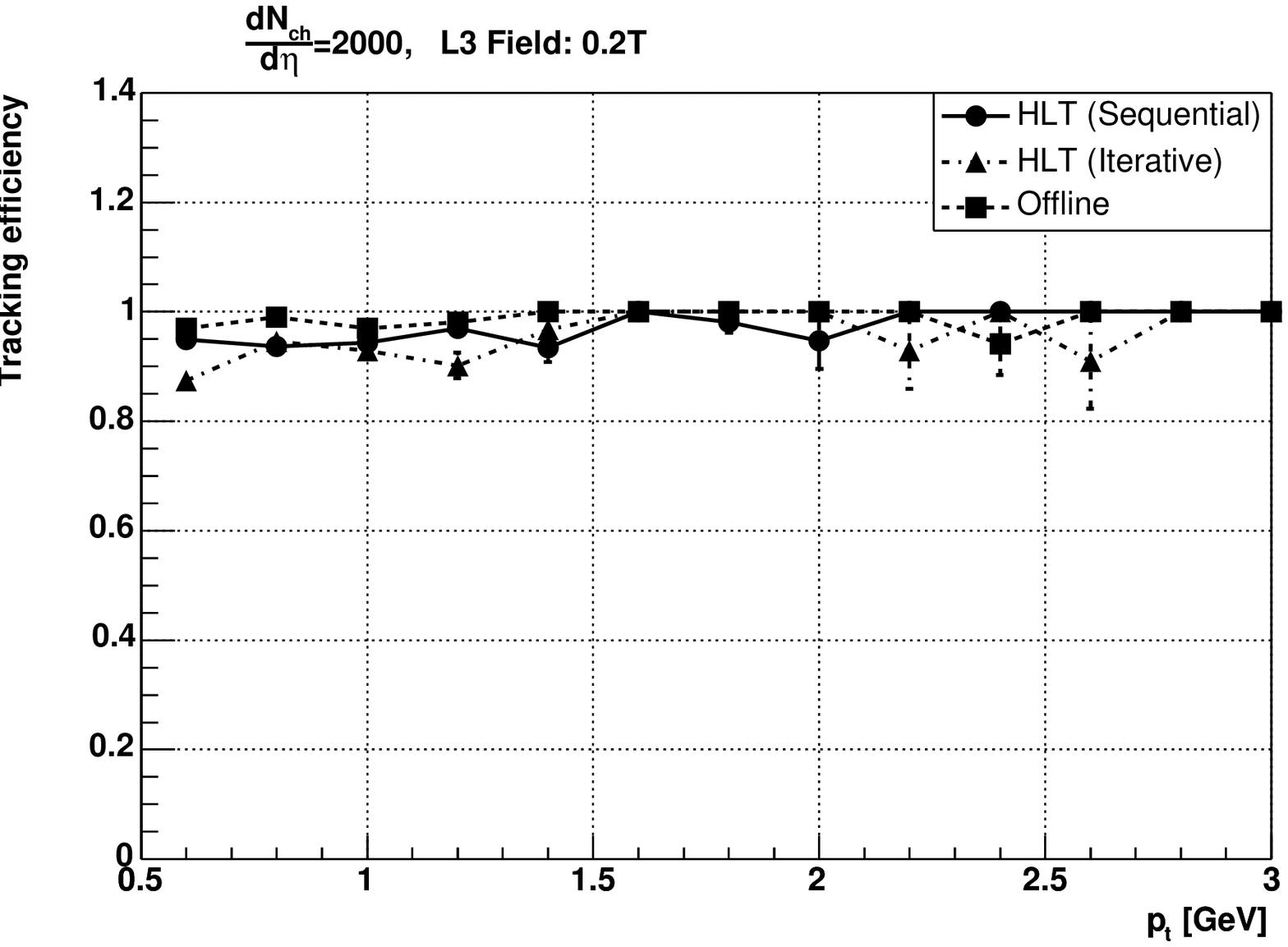,width=8cm}
\epsfig{file=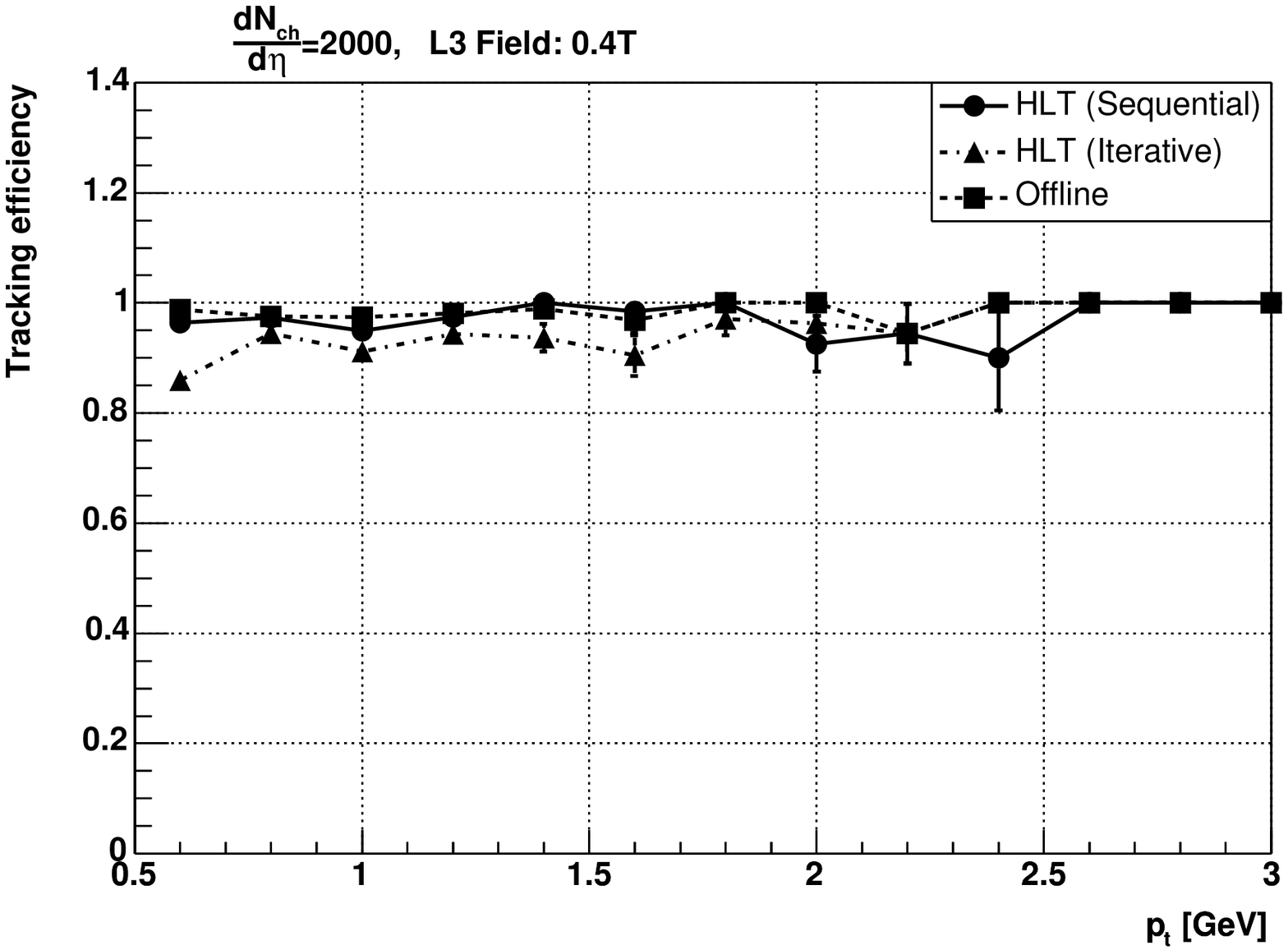,width=8cm}}

\centerline{\epsfig{file=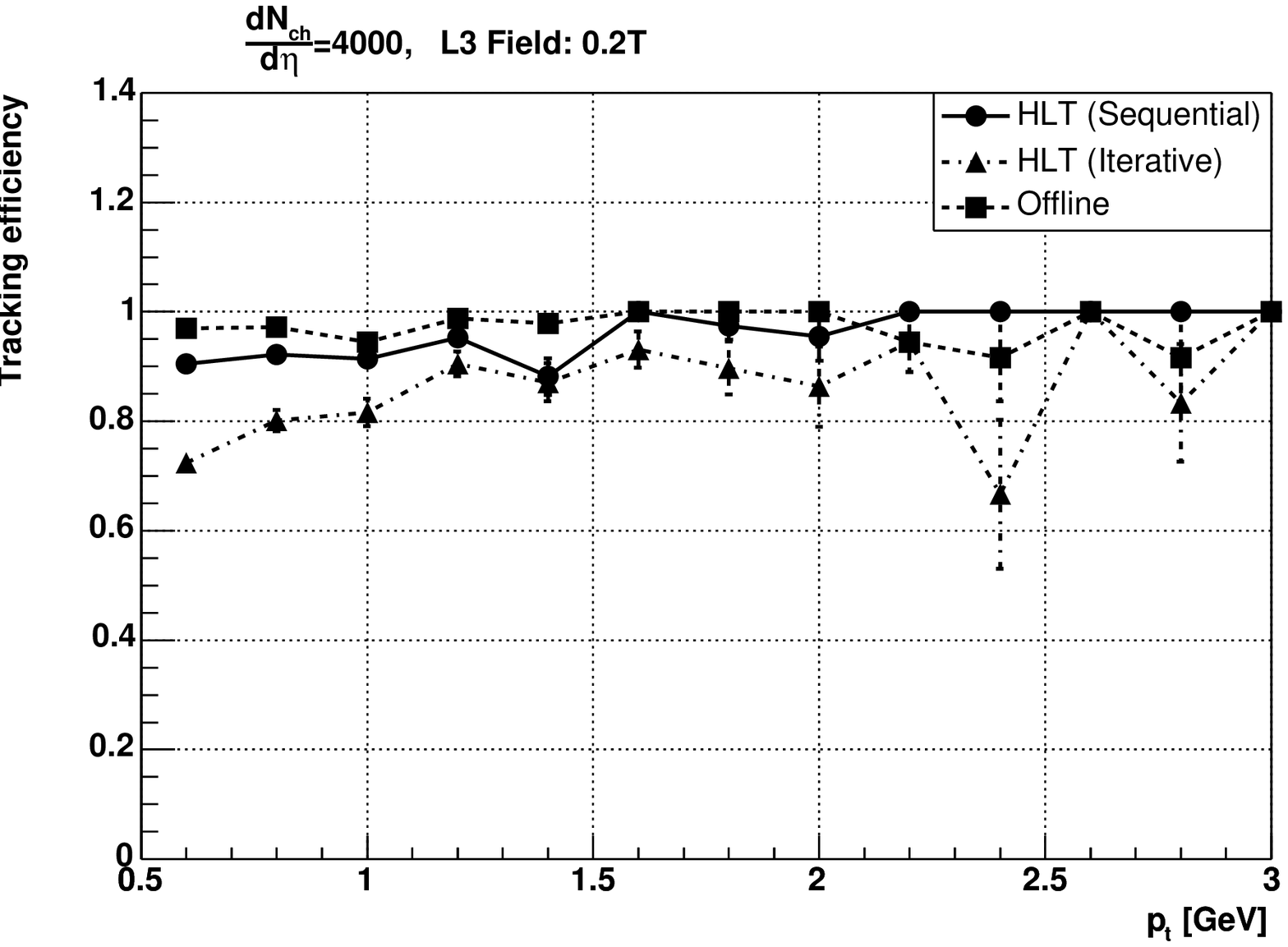,width=8cm}
\epsfig{file=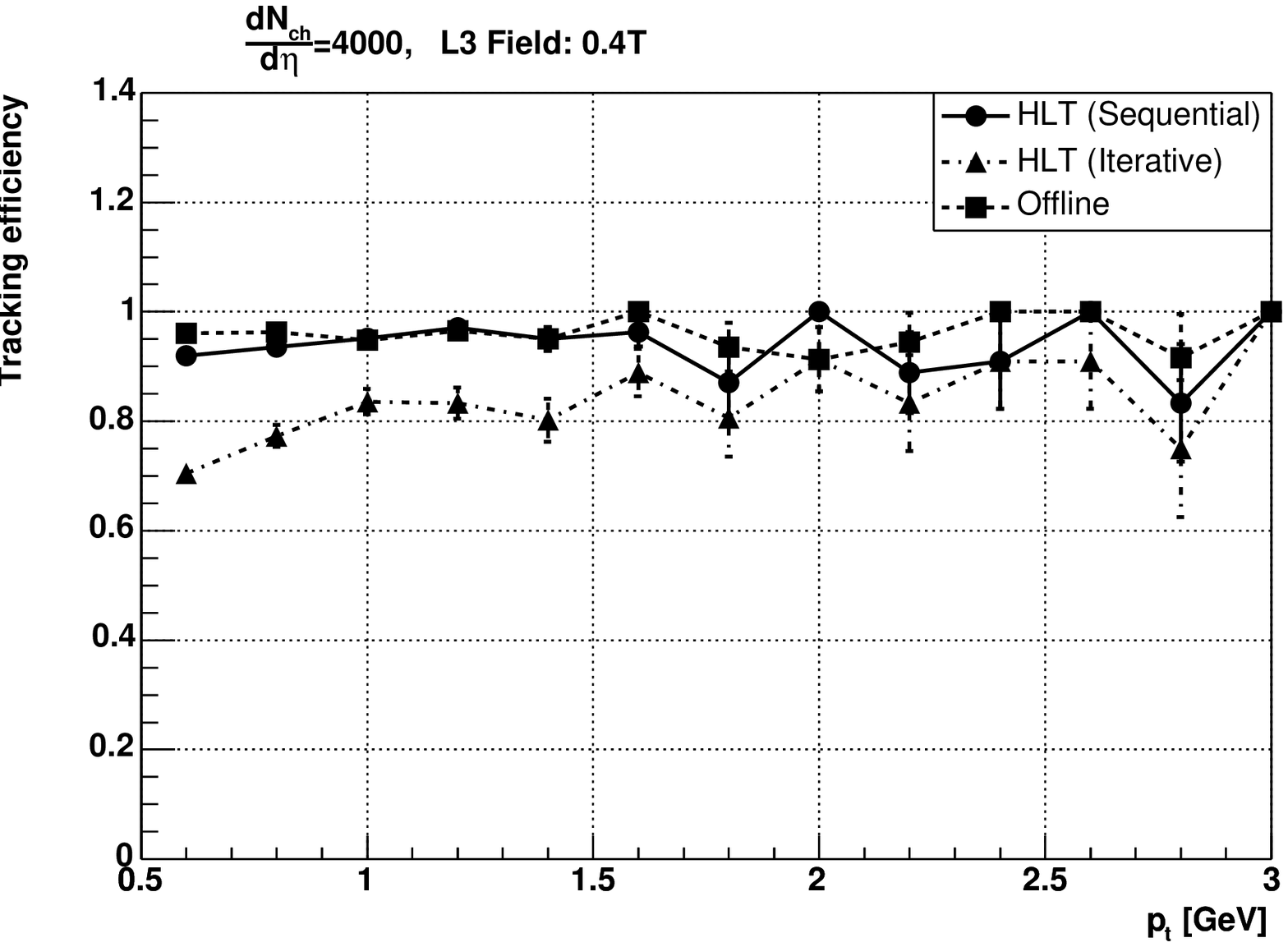,width=8cm}}

\centerline{\epsfig{file=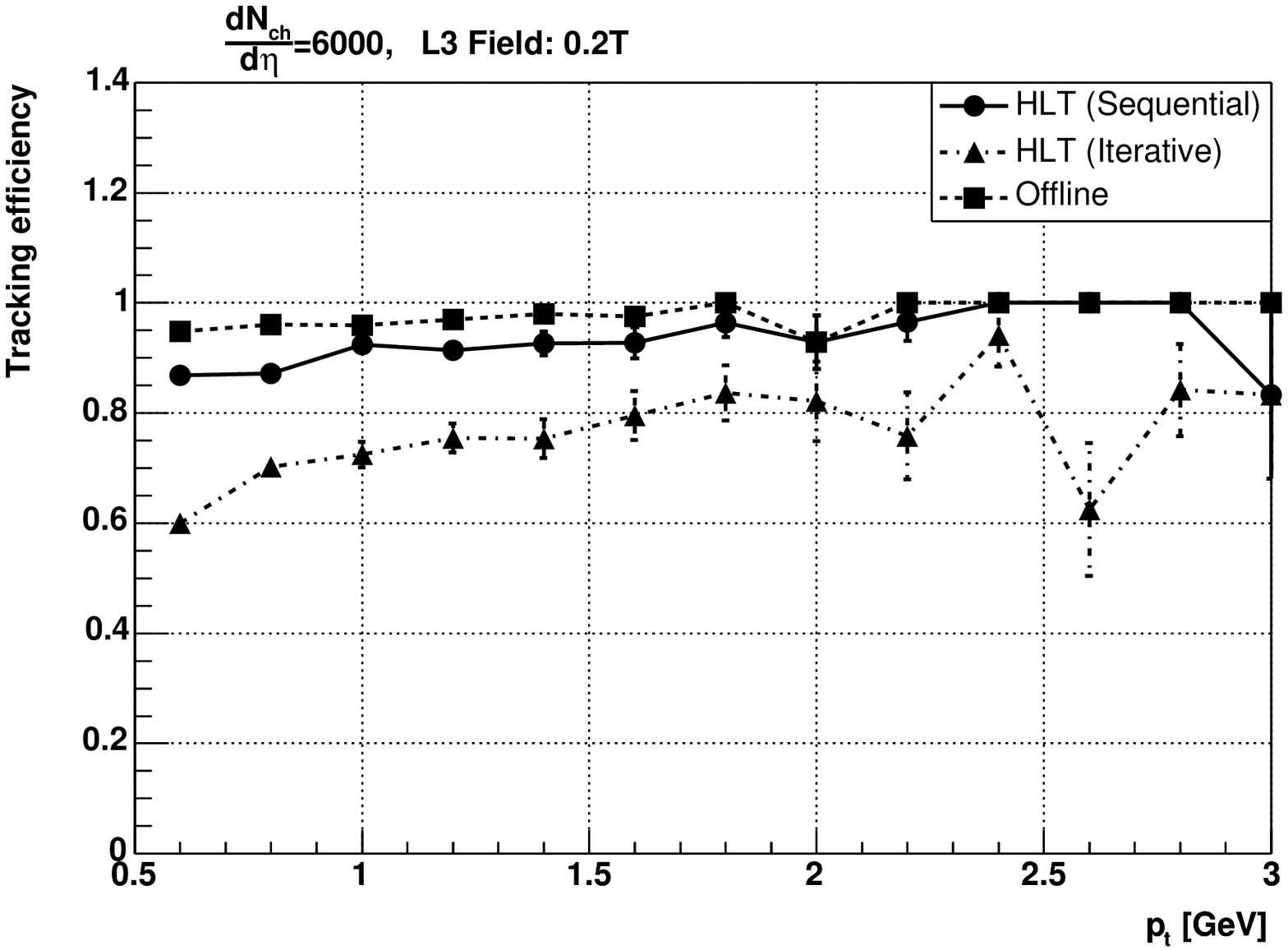,width=8cm}
\epsfig{file=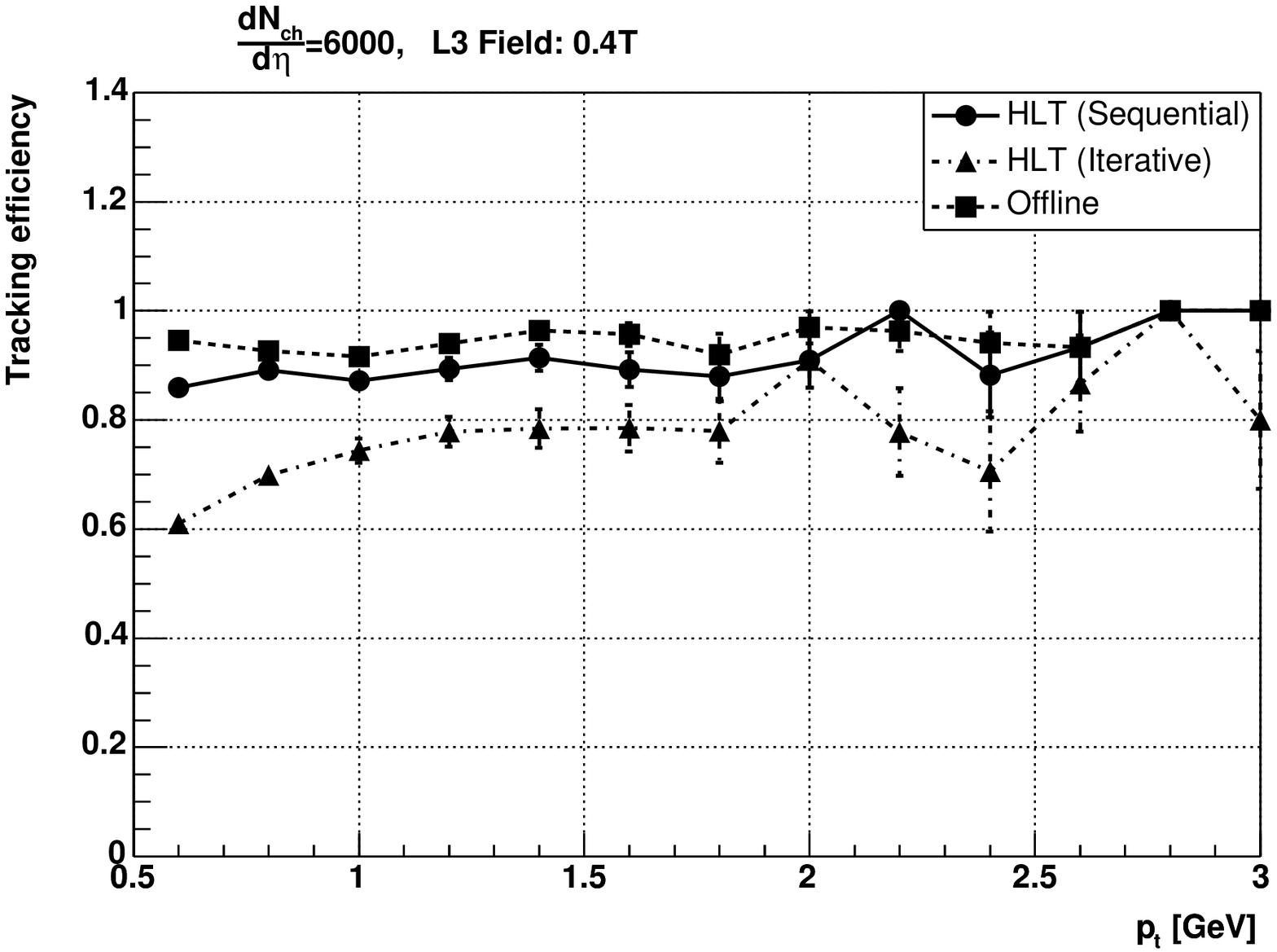,width=8cm}}

\caption[Tracking efficiencies as a function of \pt for the HLT
iterative track reconstruction chain for \pt$\geq$\,0.5\,GeV.]
	{Tracking efficiencies as a function of \pt for four different
multiplicities at both magnetic
field settings of 0.2T (left) and 0.4T (right). Only tracks with
\pt$\geq$\,0.5\,GeV/c were included.}
\label{TRACK_hthighpteff}
\efig

\subsubsection{Momentum resolution}
Figure~\ref{TRACK_hthighptres} shows the transverse momentum
resolution for two of the event samples in
Figure~\ref{TRACK_hthighpteff}. At the lower (\dndy=\,1000)
multiplicity the iterative approach is slightly better than the
sequential approach. At higher multiplicities (\dndy=\,4000), however,
the iterative approach is lower than both sequential and Offline.

\bfig[htb]
\centerline{\epsfig{file=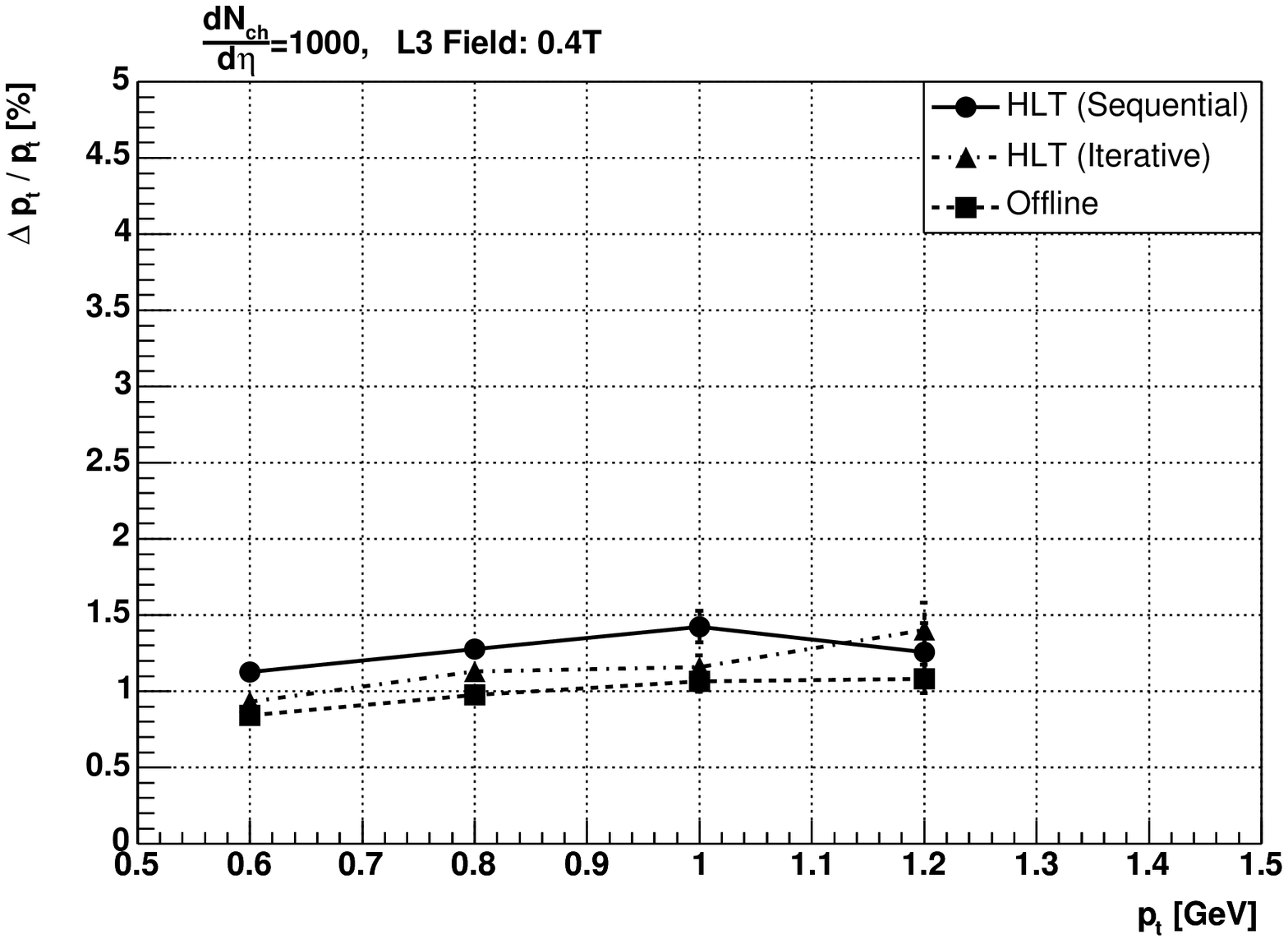,width=8cm}
\epsfig{file=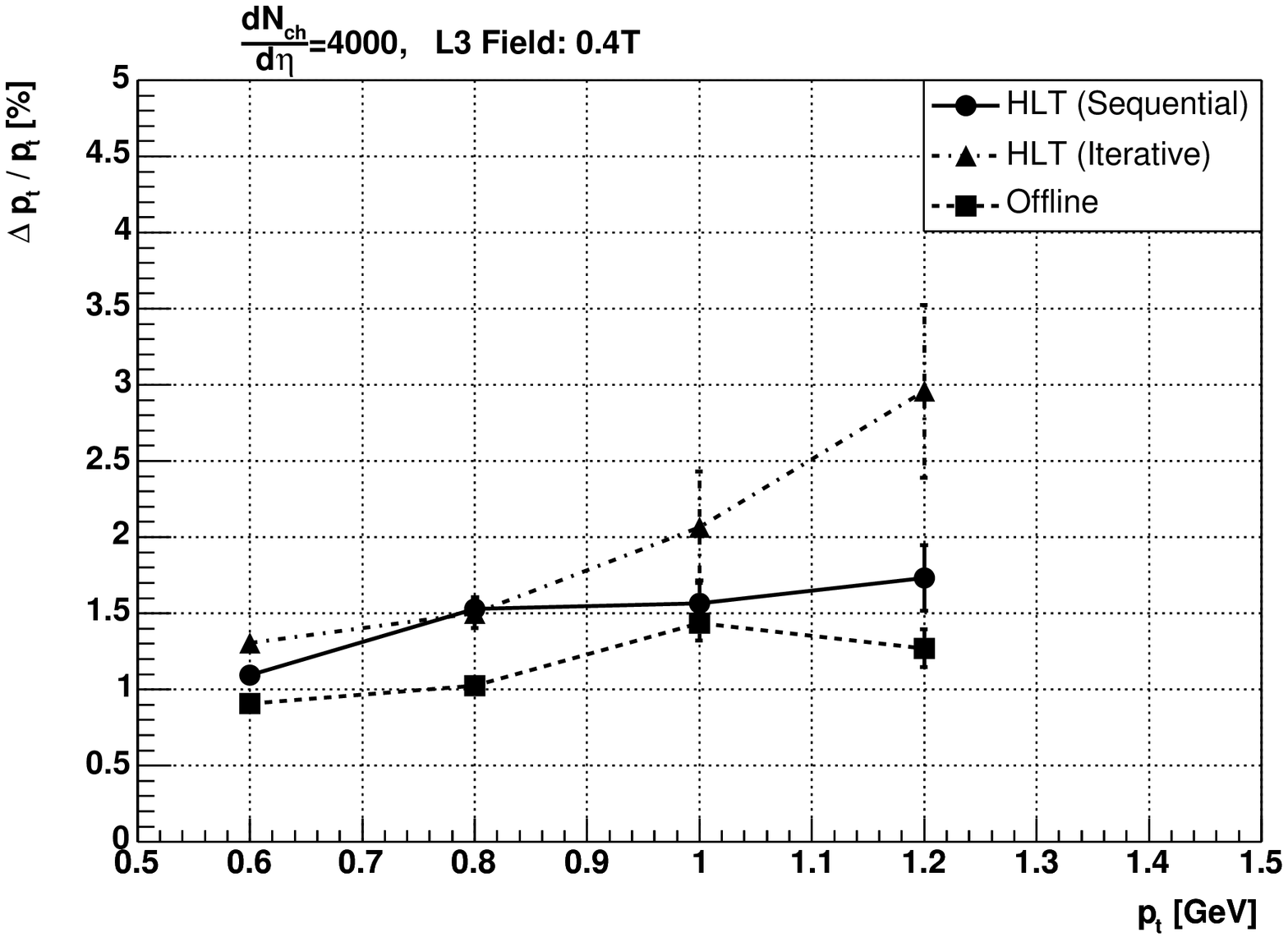,width=8cm}}
\caption[Transverse momentum resolution for the HLT iterative track
reconstruction chain for \dndy=\,1000 and 4000.]
	{Transverse momentum resolution for two multiplicities. Only
tracks with $p_t\geq$\,0.5\,GeV were included.}
\label{TRACK_hthighptres}
\efig

\subsubsection{Remarks on the tracking performance}
The results indicate that the implemented iterative reconstruction
chain has clear limitations when going to higher multiplicities. 
The main reason for this shortcoming is the relative high number of
falsely identified track
candidates from the HT. The subsequent Cluster Fitter has no way of
distinguishing these false track candidates from the valid ones. As a
consequence, the input to the fitting procedure of a given cluster distribution
might be wrong, and the fit procedure fails. This effect becomes
increasingly significant as the occupancy increases, which is
demonstrated by the fact that both the tracking efficiency and
transverse momentum resolution deteriorates at higher multiplicities. 

The relatively high number of fake tracks is mainly due to
the complexity of the formed structures in the parameter space. In
general, the peak formation depends on both the location of the
points on a track segment in space and the errors of the points.
In addition, energy loss and multiple scattering will cause the
trajectories to deviate from their circular pattern, which leads to an
additional smearing of the peaks. All of these effects complicates the peak finding,
since the structured backgrounds is difficult the distinguish from the
real peak formations. As a consequence, the criteria for selecting
a valid peak needs to be relaxed, and background formations
will be wrongly identified as track candidates.



\subsubsection{Computing requirements}
The required processing time of the various steps of the HT has been
measured on a standard
PC of 800\,MHz Twin Pentium III, 256\,KB L2 cache, running a Linux
kernel v2.4. Table~\ref{TRACK_httiming} lists the CPU-time for a
multiplicity of \dndy=\,4000. The measurements are resolved with respect
to the different computing steps, and the values thus corresponds to
the time needed by a single processing module when implemented
according to the
scheme shown in Figure~\ref{TRACK_houghtopology}. The HT needs
about 3.5\,s to process the data within a single sub-sector, resulting
in a total HT processing time of 750\,s for the complete event.
It should be noted that the algorithm was already optimized
by utilizing look-up tables for the accumulation procedure~\cite{hltdaqtrig}.



\begin{table}
    \begin{center}
    \begin{tabular}{|c|c|c|}
	\hline
	{\bf Process} 	& {\bf Locality}  & {\bf CPU-time [s]} \\
    \hline
    \hline
        HT		& Sub-sector & 3.5 \\
	Histogram add	& Sector & 1.2 \\
	Peak finder	& Sector & 1.8 \\
	\hline	
    \end {tabular}
    \caption[Measured CPU-time for the different processing steps in
the HT.]
            {Measured CPU-time for the different processing steps in
the HT. The measurement was done for an event sample with multiplicity
of \dndy=\,4000. The {\it locality} refers to where the respective
process is running, according to the data flow shown in
Figure~\ref{TRACK_houghtopology}.}
    \label{TRACK_httiming}
    \end{center}
\end{table}

The results clearly indicate that the HT is very CPU-inefficient. The
reason is that the processing is particularly I/O-bound as is needs extensive
and repetitive access to large accumulator arrays. Due to
their relative large sizes and numbers, the memory requirements are
too large to fit in the cache of the CPU, and thus the timing numbers
will scale poorly with the CPU clock frequency. However, the
inherit degree of locality and parallelism of the HT makes it suited
for implementation in FPGA co-processors~\cite{hltdaqtrig}.

\section{Summary}
In this chapter two different approaches for pattern recognition in
the ALICE TPC for the High Level Trigger system have been presented.
The two approaches have both their advantages and limitations with
respect to complexity, implementation and performance. 


Sequential tracking has shown to be efficient for lower multiplicities,
both with respect to the quality of the reconstructed tracks and the
required processing time. The
achieved tracking performance is at the same level as the Offline
reconstruction chain for multiplicities \dndy$\leq$\,2000. At higher
multiplicities, the tracking
efficiency drops, in particular in the low momentum regime. The
reasons for this are well understood. It is evident that the straight forward
Cluster Finder approach fails to reconstruct the cluster centroids once the
occupancy is too high. This is due to its lacking capabilities of
unfolding the overlapping charge distributions.

In the iterative approach, an attempt has been made to improve the
cluster reconstruction by providing an estimate of the track
parameters prior to a cluster fitting procedure. This was done by
applying an implementation of the Hough Transform on the raw ADC-data.
Here the located track candidates were used as an input for a
two-dimensional cluster fitting and deconvolution of the overlapping
clusters. The advantage of this approach is the simplicity and inherit
degree of parallelization of the Hough Transform, which makes it an
clear candidate to be
implemented very early on the readout chain within the HLT
system, possibly utilizing the FPGA co-processors planned for the
HLT-RORC. However, the HT produces a very high number of false positives, which
complicates the cluster fitting. As a result, the resulting tracking
performance is less than that of the sequential approach. 

Alternative solutions to improve the tracking efficiency for higher
multiplicities are discussed in Chapter~\ref{CONC}.

\chapter{TPC Data Compression}
\label{COMP}

The option to compress the detector data efficiently enables a
potential increase of the event rate to mass storage, even without
performing event selection. All physics observables will benefit from
such an application. Applicable data compression schemes and their
expected performance on ALICE TPC data are presented in this chapter.

\section{Introduction}
The ultimate goal of data compression is to reduce the number of bits
required to store information, without any significant information
loss. From a information theory point of view, data compression techniques
can be divided into two main categories; lossy and lossless. A lossy
data compression does not guarantee a bit-by-bit reconstruction of the
original data set, and may therefore concede a certain loss of
accuracy. In exchange however they may provide a greatly increased
compression factor. Lossy compression techniques are commonly applied to
graphics and digitized sound, as the digitized representations of such
analogue phenomena includes a certain degree of noise anyhow. Most lossy
algorithms can be adjusted to different quality levels, gaining higher
accuracy in exchange for less effective compression. 
Lossless compression techniques, on the other hand, guaranty that an
exact duplicate of the input data stream is generated after a
compress/expand cycle. These techniques are applied when the loss of
even a single bit can affect the information content significantly.

In general data compression consists of taking a stream of symbols and
transforming them into codes. If the compression is effective, the
resulting stream of codes will be significantly smaller than the original data
size. The decision to output a certain code for a certain symbol or
data subset is based on a model. The model is a collection of
data and rules to process the input data and determine which
codes to output. 

The most effective compression results can be
achieved if the model is well adapted to the underlying data. A
well-known example is the approach used by the MP3-format for audio
files, where the results achieved when sound is compressed adapted to
human hearing characteristics are much better than the results from
general purpose algorithms. In case of TPC data the underlying data
is the ADC-data, the clusters and tracks, while all the relevant information
is contained in the final reconstructed physical observables. Before
discussing the various compression algorithms applied, a brief
summary of the characteristics of TPC data with respect to the
various signal generation errors is given.

\section{TPC signal generation and models}
An effective data compression scheme has to be well adapted to the
underlying data model and the noise already present in the data. As
long as the allowed loss of accuracy does not exceed the inherit
accuracy already present in the data, the compression scheme will in general not alter
the information content.
Prior to discussing applicable data compression techniques for TPC
data, the
understanding of the signal generation and the resulting inherit error
sources within the original data stream is necessary.

\subsubsection{Error sources within the TPC signal generation}
The ultimate task of the TPC
is to measure the kinematics of the traversing particles and
contribute to the particle identification by energy loss
measurement. Before the data is readout out from the detector, several
factors contributes to alter the data. These can be summarized as follows:
\begin{itemize}
\item {\bf Elastic and inelastic scattering.}\\
Both prior to entering the sensitive volume of the TPC and within
the gas of the TPC volume, the charged
particles interact with the surrounding material, causing the
particle track to deviate from its original trajectory. These effects
include both elastic scattering from nuclei,
and inelastic collisions with the atomic electrons of the material.
The energy loss is described by the Bethe-Bloch formula, and typically
follows a Landau distribution.
\item {\bf Diffusion.}\\
The drifting electrons diffuse when drifting towards the
end-caps of the TPC, and hence influence the position resolution of the
reconstructed space points. The drifting cluster of electrons can be
described by e.g. a 3D Gaussian distribution, where the widths are
determined by the diffusion constants of the gas
(see Section~\ref{TRACK_detresp}, page~\pageref{TRACK_gaussdiff}).
\item {\bf Electron attachment.}\\
During the drift the electrons can be absorbed in the gas by formation
of negative ions. 
\item {\bf Gas amplification.}\\
Each of the liberated electrons is subject to gas multiplication
when entering the readout chamber. This amplification is described by
a exponential probability distribution. The gas gain properties of the
gas has also a strong dependence on the temperature.
\item {\bf E$\times$B-effects.}\\
The fact that the {\bf E} and {\bf B} fields are not parallel near the anode
wires, leads to a displacement of the drifting electrons when entering
the readout chamber. 
\item {\bf Front-end electronics.}\\
When the image charge induced on the readout pads is processed through
the front-end electronics several sources contribute to the noise and distortion of
the original signal. Tail cancellation, pedestal subtraction and zero
suppression are all lossy compression techniques in their manipulation
of the original raw data. 
\item {\bf 10-to-8 bit compression.}\\
The ADC conversion gain is typically chosen so that $\sigma_{\mathrm{noise}}$
corresponds to 1 ADC count. This means that the
relative accuracy increases with the ADC-values, and is not needed for
the upper part of the dynamic range. The ADC-values can therefore be
compressed non-linearly from 10 to 8 bits leading to a constant
relative accuracy over the whole dynamic range
(Figure~\ref{COMP_convtable}).
The conversion from 10 to 8 bit of the ADC-values is thus a data volume
reduction with some information loss, but keeping the relative
accuracy for the single ADC-value.
\end{itemize}
All of these effects modify the original data, preventing an
{\it exact} event reconstruction. Therefore, the complete readout
system must be considered in order to decide to which extend lossy or
lossless data compression should be applied.
Also, since the relevant
information in the data lies in the final outcome of the analysis, any
lossy compression should be evaluated from its impact on the
reconstructed physics observables.

\subsubsection{Local and global TPC data models}
In the context of TPC data compression, the applicable compression
techniques can be divided into two main categories; local and global
modeling. Local modeling techniques are applied on the scale of
raw ADC-data, i.e. after a compress/expand cycle the resulting data
will still consist of the same ADC-data format. In the latter case
the data is described within a more global model, the clusters and
tracks. In such a compression scheme, the final uncompressed data may not
consist of ADC-data, but rather cluster and track parameters
themselves. The difference between these two approaches is obviously the
elimination of the original raw data from the data stream. However,
as the vital information in the TPC data is contained in the
reconstructed particle trajectories, there is in principle no need to
keep the raw ADC-data.

\section{TPC data format and coding}

\subsubsection{TPC readout data format}
The ALICE TPC raw data format which will be used during the
running of the experiment is defined by the ALTRO chip in the TPC
readout-electronics chain, Section~\ref{ALICE_tpcreadout}. In
addition to the digitization of the input signals, the ALTRO performs
the {\it zero-suppression} of the digitized data. This is
basically a data compression technique that consists of
detecting the hits in the time-direction and discarding the noise in
between by replacing it with zeros. It is implemented using a
sequence detection scheme, which is based on the rejection of isolated samples
with a value smaller than a threshold level. Both the
thresholding and hit-finding are lossy data compression
techniques, as small clusters or tails of the clusters might be
discarded from the data.

The zero-suppression results in long
sequences of zeros between the hits. Run-Length
Encoding (RLE) subsequently compresses these zeros. The principle of
RLE is to replace a sequence of
identical symbols by a certain character or tag, and the length of the
sequence. In case of the TPC the RLE is performed on the zeros. This
is thus equivalent to storing only the hits and their
positions. As the RLE does not modify or discard any data, it is a
lossless data compression technique.

The ALTRO data format consists of a list of sequences written in a
back-linked structure~\cite{alicetdr}, where each sequence is described by three
fields: Temporal information (temporal position of the last time-bin
in the sequence), sequence length (number of time-bins in the
sequence) and the ADC-values themselves.


\subsubsection{Simulation and coding}
Simulated data were generated within the AliROOT framework
Section~\ref{TRACK_aliroot}. All events were
generated using the HIJING parameterization for different
multiplicities, and the standard TPC slow simulator including the
complete detector response simulation package. The output from the
simulation thus correspond to zero-suppressed 10-bit ADC-data, whose
format is defined within special AliROOT data containers.
Before any of the data compression techniques presented in this
chapter were applied, the data was further modified in three steps:

\begin{enumerate}
\item A 45-degree (relative to the beam axis) cone was cut out of the
data in order to remove data
from particles which are not in the geometrical acceptance of the
outer detectors. This will be done also during readout of the TPC
detector, as the low \pt tracks crossing the TPC under small angles
are too problematic to resolve. This cutout reduces the original data
volume by $\sim$40\%.
\item The simulated 10-bit ADC-data were transformed into
8-bit data by a 10-to-8 bit table shown in
Figure~\ref{COMP_convtable}.
\item The final 8-bit ADC-data were then RLE and written to binary
files for subsequent comparison to the compressed data. The adapted
data format uses a coding scheme whose size resemble the
ALTRO-format. Further details about the format used can be
found in Appendix~\ref{APP2_rle}.
\end{enumerate}
All compression ratios in the following are calculated as
\begin{eqnarray*}
\frac{\mathrm{Compressed\ size}}{\mathrm{Original\ size}}\times
100\hspace{0.5cm} [\%]
\end{eqnarray*}
where the original size refers to the RLE 8 bit ADC-data.

\bfig
\insertplot{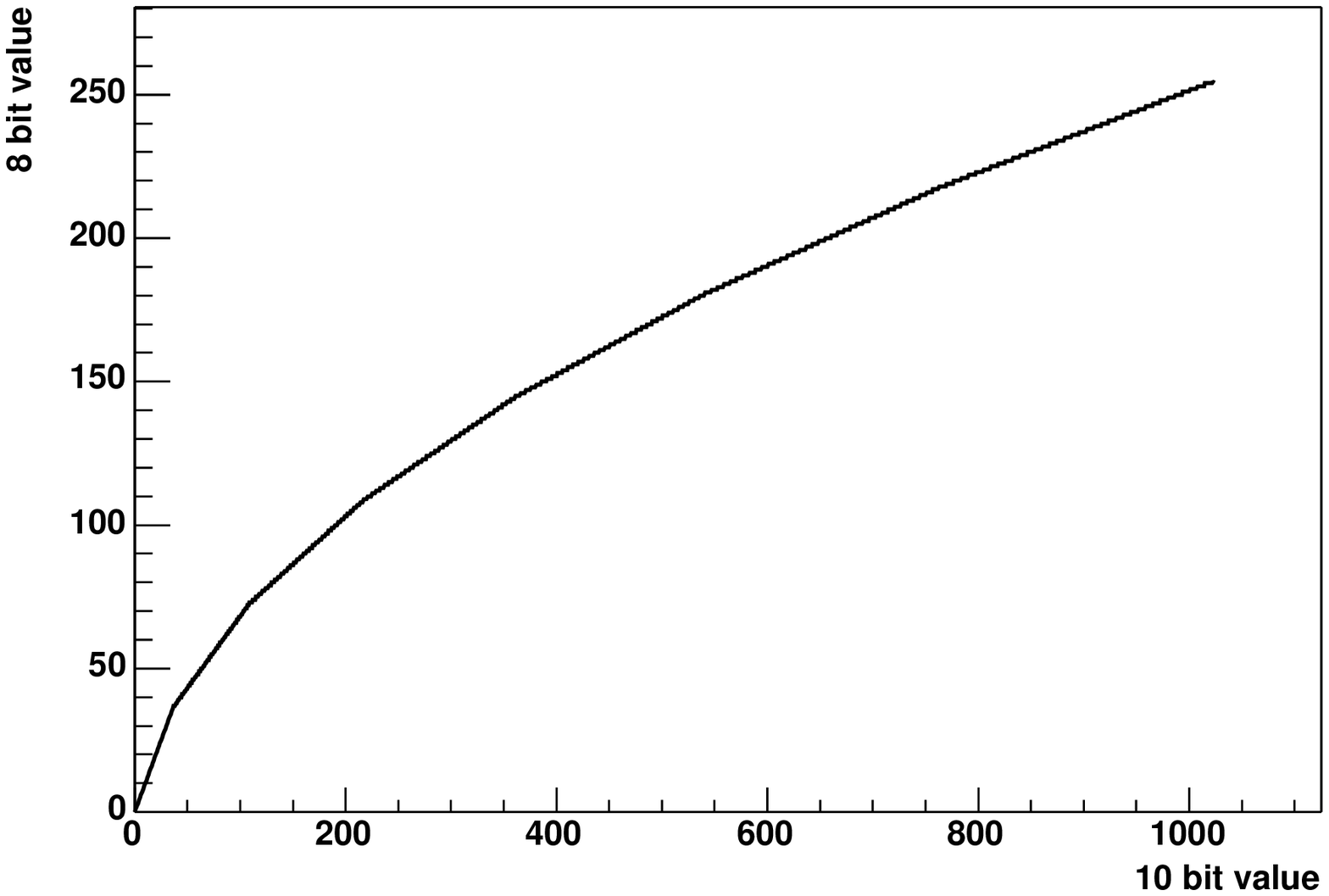}{8cm}
	{Plot of the 10-to-8 bit conversion table used.}
	{Plot of the 10-to-8 bit conversion table used~\cite{marian8bit}.}
\label{COMP_convtable}
\efig

\section{Local modeling techniques}
In~\cite{berger02} several data compression
techniques, based on local data modeling, were applied on TPC data. Both
data from the NA49
experiment, and simulated data from the ALICE TPC, were studied. A brief
introduction to the different algorithms applied and the achieved result
is given in the following.

\subsection{Lossless TPC data compression}
In general, most lossless compression techniques are based on {\it
entropy coding}. Such algorithms exploit any possible redundancy within the
information message, and the fact that the different symbols within a message
are not equally probable. A symbol, which has a very high probability
of occurrence, will be coded using very few bits, while symbols with a low
probability are coded with a larger number of bits. For TPC data, such
a scheme can exploit the fact that different ADC-samples are not
equally probable: Small ADC-values occur more often than larger
ones, Figure~\ref{COMP_adcdist}. The theoretical limit on the
average word size that can be
achieved with such a strategy is given by the {\it entropy} of the
data source. For TPC data the entropy can be computed as
\beq
E = -\sum_{A\in\Omega}p(A)\log_2 p(A).
\label{COMP_entropy}
\eeq
Here $p(A)$ is the probability of having a ADC-value, $A$, in the data,
and $\Omega$ is the set of all possible words that are contained in
the data source. The difference between the number of bits used to
represent a single character and its information entropy is the
potential for entropy coding techniques.

\bfig
\insertplot{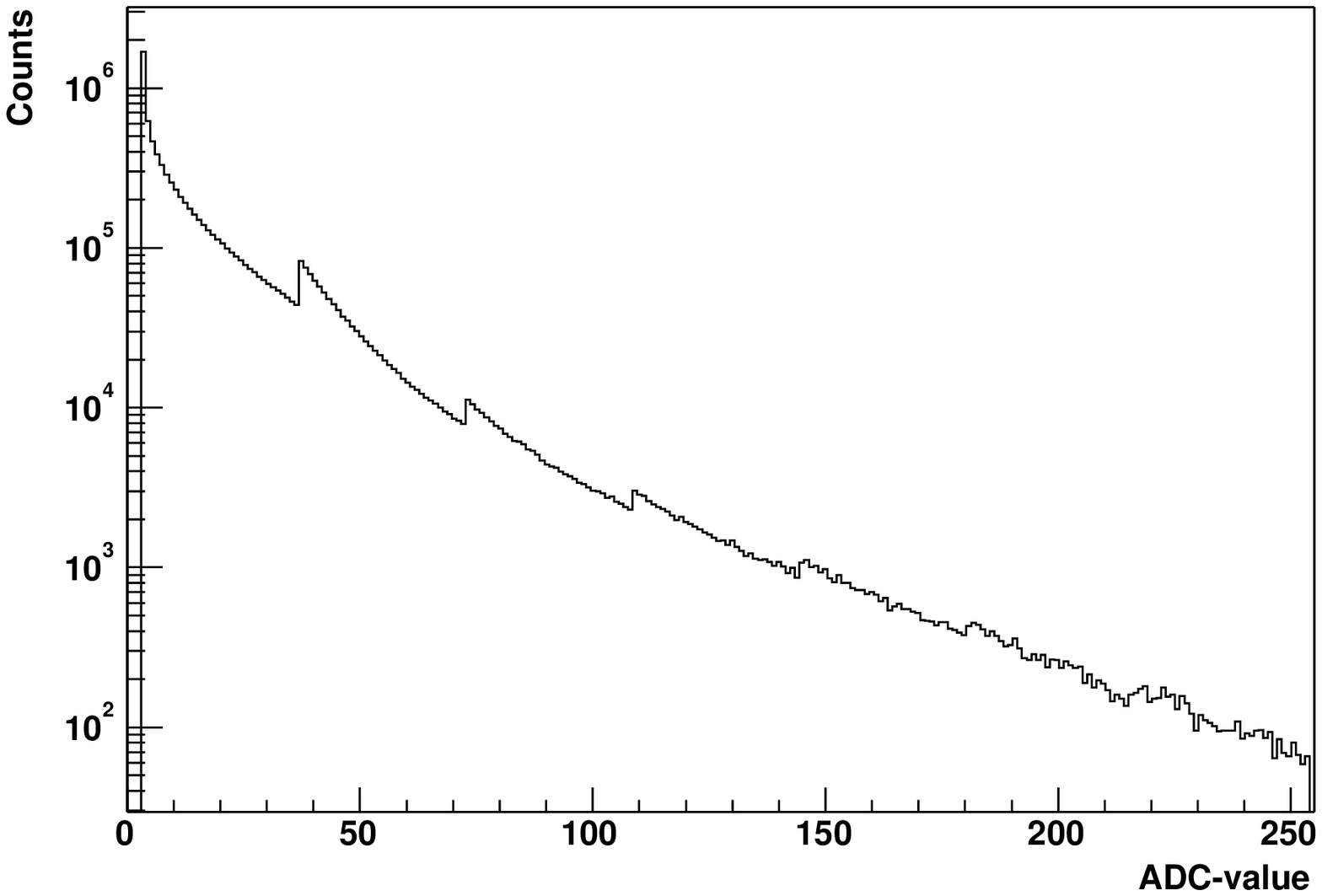}{8cm}
	{Distribution of ADC-values in simulated ALICE TPC-data.}
	{Distribution of ADC-values in simulated ALICE TPC-data.}
\label{COMP_adcdist}
\efig

{\bf Huffman Coding}
One of the most common entropy coding techniques is the 
{\it Huffman Coding}~\cite{huffman}. This algorithm has proved
to be easily implemented and achieve good compression
results without extensive processing power. The basic idea is to
assign each input signal to a leaf of a binary tree, the so-called
Huffman Tree. Each branch on this tree is either assigned the 0 or 1 bit,
and the path from the root node to the leaf defines the code used for
this symbol. By adapting the tree to the probability distribution
within the data sample, symbols with higher probabilities get shorter
codes and vice versa. 

{\bf Arithmetic Coding}
A somewhat more complicated but potentially more effective compression
technique is {\it Arithmetic Coding}. In contrast to Huffman coding,
this approach does not produce a code for each symbol, but rather a
code for an entire message. It is therefore not restricted to using
integral number of bits per code, and is thus potentially more
effective than the Huffman Coding. This compression technique can
theoretically approach the lower limit given by the
entropy~\cite{compbook}. The main idea is to assign to every symbol
an interval between 0 and 1, with a size according to the occurrence
probability of that symbol. For each input symbol the interval is
shrunk to the range assigned to the new symbol. In this way the interval
gets smaller and smaller and the input stream of symbols is replaced
by a single floating point number representing the encoded message.
The main drawback with Arithmetic Coding is that a relatively large
number of operations are needed to encode a single symbol.

{\bf Differentiation}
A common approach for lossless graphics compression is
{\it differentiation}, which exploits the fact that adjacent symbols
may be similar. This is particularly true for TPC-data, where adjacent
ADC-values are highly correlated. In this case the distribution of the
derivative of the input has lower entropy than the data itself. 

{\bf Code table coding}
A similar approach to differentiation is predictive encoding. In this
case the next value in the data
stream is guessed based on the previous sent value, and only the
difference between the guess and the actual value is sent. Thus, if the
guess is close to the actual value, the entropy of the difference is
small.

\subsection{Lossy TPC data compression}
The best compression results can always be achieved if small noise-like
changes of the data can be tolerated. The term lossy originates from
the fact that these methods do not allow a bit-by-bit reconstruction
of the original data set. Lossy compression techniques
usually introduce some kind of quantization of the data, thereby
lowering the required number of bits needed to store the
information. In the simplest form, the data samples can be quantized
into intervals corresponding to the required resolution.

{\bf Vector Quantization}
One of the more sophisticated quantization techniques is Vector
Quantization~\cite{vectorq}. Here statistical dependencies between
successive data samples are exploited. Instead of quantizing data
samples independently, several samples are grouped together to form a
vector of data samples. In the TPC data, such a vector can typically
be a sequence of ADC-values. The vector is compared to entries in a
codebook of vectors, and the index of the best matching vector in the
codebook is stored. In order to achieve effective compression factors
the codebook has to be {\it trained} on the statistical properties of
the data. 
Since such a codebook is pre-produced, only the given vectors are
available to represent the data. Since this can lead to rather large
quantization errors, also the difference between the input data and
the selected codebook entry, the residuals, can be stored. These
residuals can further be quantized and entropy encoded. 


\subsection{Results}
The resulting compression factors obtained with the algorithms
described above are summarized in Table~\ref{COMP_compfactors}. For
the Huffman Coding, separate trees were built for the ADC-values and
the header information, i.e. position and length. The Code table
coding implemented a table that included a best guess for each
combination of the preceding values. The results did not improve
if the table was dependent on two of the preceding values. 
For the Vector Quantization it was shown that the impact on the
physics observables is measurable but small, with the space point
resolution being the most sensitive quantity.

\begin{table}
    \begin{center}
    \begin{tabular}{|c|c|c|}
    \hline
	{\bf Type of encoder} & {\bf Entropy} & {\bf
Relative event size [\%]}\\
    \hline \hline
	Zero suppressed raw event size & 8 & 100\\
	Arithmetic Coding & 6.4 & 80\\
	Code table coding & 5.7 & 71\\
	Huffman Coding    & 5.2 & 65\\
	Vector Quantization & 5.1-3.8 & 64-48\\
    \hline
    \end {tabular}
    \caption[Compression performance for local data modeling
techniques on simulated ALICE TPC data.]
	{Compression performance for local data modeling
techniques on simulated ALICE TPC data~\cite{berger02}. The entropy is
given as the average number of bits used to encode a sample. For the
Vector Quantization the entropies depends on the treatment of the residuals.}
    \label{COMP_compfactors}
    \end{center}
\end{table}

\section{Global modeling techniques}
The standard compression techniques listed in
Table~\ref{COMP_compfactors} are all applied on the scale of ADC
samples and sequences. However, the TPC data can be described within a more
global model, the tracks and their corresponding clusters. Such
models can be exploited by compression methods by using it to
transform the data into an more
efficient representation, and thereby reducing the redundancy information (entropy) in
the input data. Instead of storing the TPC data on the ADC-level, the
information can be stored on a higher extraction level such as cluster
and track parameters. The main difference of such models from the
local techniques is that they require some form of pattern
recognition to be done prior to the compression scheme. 
In this context the purpose of the pattern recognition is different
from the normal case since the aim is not to extract physics
information about the particle kinematics, but to build a data model. 

In general the TPC pattern recognition scheme is a two-folded process,
which corresponds to reconstructing the space points from the clusters
and combining the space points into tracks,
Chapter~\ref{TRACK}. From this point of view, there are two levels in
which TPC data can be represented:
\begin{itemize}
\item Space point data.
\item Space point data relative to their tracks.
\end{itemize}
The first case corresponds to storing only the space points and their
properties. Thus, only a cluster finding procedure needs to be done
prior to the compression. In the latter case, it is assumed that a
zeroth-order tracking step is performed in which the space points
are encoded with respect to their distance to the tracks they
belong to. Both of these methods will eliminate the original raw data from
the data stream, as the final data set after decompression in either
case will consist of a list of clusters.


\subsection{Storing cluster data}
\label{COMP_cldata}
The TPC clusters represent the three dimensional space points of the
particle trajectories. The information which is needed
from an analysis point of view consists of the three dimensional coordinates,
the shape of the cluster and the total charge of the
corresponding cluster. To first order each of these values would
have to be stored with the required number of bits to ensure the
information content. However, from the way the data is organized in the detector,
some simplifications can be made. 

The space points are reconstructed
from the cluster centroids, which is calculated from the
two-dimensional digitized charge distribution is the
pad-row-plane. The coordinate in one dimension is thus given by the
radius of the pad-row plane, and is consequently the same for all the
clusters on the same pad-row. The remaining two coordinates can
therefore be encoded with respect to the pad-row they are located on.




\bfig[htb]
\centerline{\epsfig{file=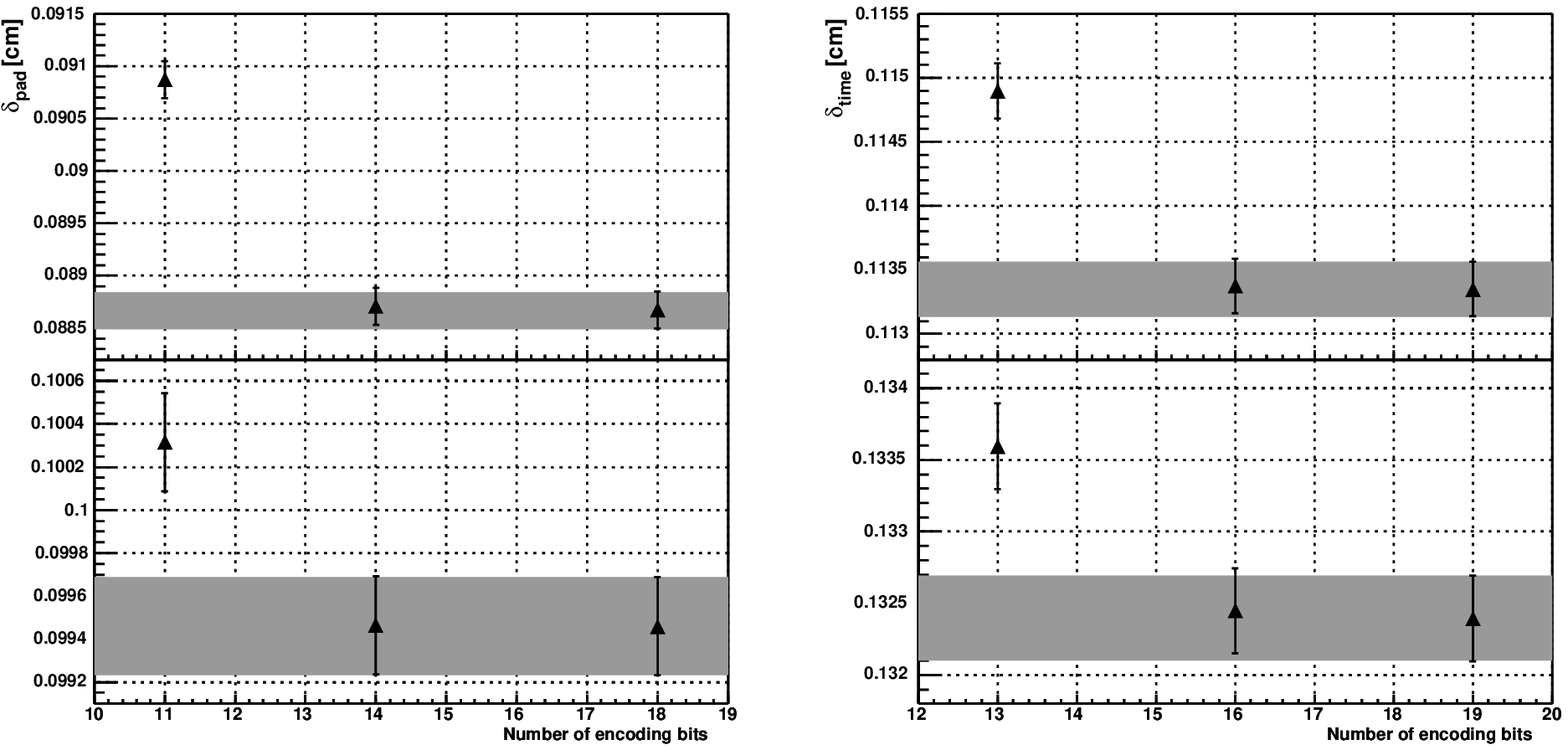,width=17cm}}
\caption[Space point resolution for different encoding sizes of the clusters.]
	{Space point resolution for different encoding sizes of the
cluster centroids. The upper plots show the result for the outermost TPC
chambers, while the inner plots shows the inner chambers. The shaded
area shows the resolution within the original data set.
The results were obtained by comparing the cluster centroids
and the position of the corresponding simulated particle
trajectory. The values correspond to the
standard deviation from a Gauss fit.}
\label{COMP_spres_vs_encsize}
\efig

The number of bits required to store the space point coordinates is
determined by the inherit resolution within the data. 
The typical space point resolution of the
ALICE TPC is at the order of 0.08\,cm and 0.1\,cm for the pad and time
direction respectively. With a pad-width of 0.4\,cm and 0.6\,cm
for the inner and outer TPC chambers respectively
(Table~\ref{ALICE_tpcparams1}), this
corresponds to a precision of $\sim$0.1 in pad-coordinates. In order
to avoid any deterioration of this resolution,
this coordinate should thus be stored with at least 2 decimal precisions.
Given that the maximum number of pads
on a single pad-row is 140, a total number of 14 bits\footnote{The
number is estimated from the assumption that one need to encode
numbers up to 140\,00, 2$^{14}$=163\,84.}
is needed to encode the
pad-coordinate. Similar arguments hold for the time-direction whose
absolute value can take up to 512 time-bins. With a 2 decimal precision
this would then correspond to a spatial resolution of
$\sim$0.006\,cm and thus a
total number of 16 bits for the encoding size.

Table~\ref{COMP_spacepoints} lists the number of bits required to
\begin{table}[htb]
\begin{center}
\begin{tabular}{|c|c|c|c|}
\hline
		  & & \multicolumn{2}{|c|}{\bf Precision [cm]}\\\cline{3-4}
{\bf Cluster centroid}   & {\bf Size (bits)} & {\bf Inner} & {\bf Outer}\\
\hline \hline
		         & 11	& 0.04   & 0.06  \\
	Pad centroid	 & 14   & 0.004  & 0.006 \\
			 & 18   & 0.0004 & 0.0006\\\hline
	  	         & 13	& \multicolumn{2}{|c|}{0.06}\\
	Time centroid	 & 16   & \multicolumn{2}{|c|}{0.006}\\
			 & 19   & \multicolumn{2}{|c|}{0.0006}\\
\hline
\end {tabular}
\caption[Space point data and their required encoding size.]
	{Space point data and their required encoding size.}
\label{COMP_spacepoints}
\end{center}
\end{table}
\noindent encode the positions using 1, 2 and 3 decimals for both pad and time
direction together with their respective spatial
resolutions\footnote{These numbers represent the maximum number of
bits required to store the coordinates. E.g. for the pad-direction,
the number of pads varies for each pad-row and the encoding size may be
reduced by a bit on the innermost pad-rows.}.
Figure~\ref{COMP_spres_vs_encsize} shows the impact on the space point
resolution of keeping the respective levels of precision within the data. 
In the plots also the resolution within the original
data, i.e. keeping the full floating point precision, is shown. As
expected there is some impact on the resolution by keeping only a
single decimal precision for both directions. No difference can
however be seen when using 2 and 3 decimals. This indicates that the a
total of 14 and 16 bits is needed to store the pad and time coordinate
respectively without any information loss.

\begin{table}[htb]
\begin{center}
\begin{minipage}[t]{0.4\linewidth}
\centering
\vglue0.2cm 
\begin{tabular}{|c|c|}
\hline
{\bf Cluster parameters}    & {\bf Size (Bit)} \\
\hline \hline
	Pad coordinate  & 14\\
	Time coordinate & 16\\
	$\sigma_{\mathrm{pad}}$  & 8\\
	$\sigma_{\mathrm{time}}$ & 8\\
	Total charge    & 12\\
\hline\hline
	Total 		& 58\\
\hline
\end {tabular}
\end{minipage}
\hspace{0.5cm}
\begin{minipage}[t]{0.4\linewidth}
\centering
\vglue0.2cm 
\begin{tabular}{|c|c|}
\hline
{\bf Raw-data parameters}    & {\bf Size (Bit)} \\
\hline \hline
	ADC-values	 & 3$\times$3$\times$8=72\\
	Pad-numbers	 & 3$\times$8=24\\
	Time information & 3$\times$32=96\\
	&\\
	&\\
\hline\hline
	Total 		& 192\\
\hline
\end {tabular}
\end{minipage}
\caption[Cluster and raw-data parameters and their respective encoding size.]
	{Data parameters and their encoding size. Left table: Cluster
parameters and the estimated number of bits required to store the
respective parameters within the intrinsic resolution. Right table:
Raw-data parameters and their estimated encoding size assuming a single
isolated cluster of size 3$\times$3 bins.}
\label{COMP_rawdatasizes}
\end{center}
\end{table}

In Table~\ref{COMP_rawdatasizes} the cluster parameters together
with their required encoding size are listed.
In addition to the two-dimensional centroid, also the cluster shape
and the total charge are needed. In order to store the cluster widths
with the same level of precision as for the centroids, each
direction is encoded with 8 bits. The total cluster
charge is stored using 12 bits. In total 58 bits are needed to store
the complete cluster using this representation.

The above storage considerations can now be compared to the
required size of storing the
raw-data itself. However, the required encoding size for the raw-data depends
in general on the size of the clusters in terms of number of bins in
the pad-row-plane, and also on the multiplicity. For instance,
assuming a single 'ideal' cluster whose size is 3$\times$3 bins, a total
of 3$\times$3$\times$8=72 bits are required to store the
ADC-values. Furthermore, header information is needed to encode the
position of the ADC-values in the pad-row-plane. This
typically contains the respective pad-numbers on which the individual
ADC-sequences are located and the relative time-information
(Appendix~\ref{APP2_rle}).
Each pad-number can be encoded using 8 bits, thus 3$\times$8=24
additional bits are required for the 3 different pad-numbers. For the
time-information 4$\times$8=32 bits are
needed for encoding of zero-sequences and RLE tags.
This results in a
total of 192 bits, Table~\ref{COMP_rawdatasizes}. Comparing with
the encoding size for the cluster parameters, a reduction of about
70\% may be achieved when storing the cluster parameters instead of
the raw-data. This estimate however assumes that only one isolated
cluster is present on a
single pad-row, which of course in general is not the case.
The relative size of the header information is strongly dependent on the
occupancy, and will decrease as a function of multiplicity due to the
fact that the sequences on a given pad are encoded according to their
internal distance and the pad-number is stored only once per pad. 
This is illustrated in Figure~\ref{COMP_clratio_vs_mult} which shows the
relative size of the cluster data for different multiplicities. Here, the
HLT Cluster Finder algorithm was used to reconstruct the clusters,
Section~\ref{TRACK_clfinder}. For
low multiplicities the ratio is $\sim$25\%, while it increases to
$\sim$30\% for higher multiplicities. The fact that the ratio for very
low multiplicity is $\sim$5\% lower than the value estimated above,
can be explained by the fact that the clusters are in general larger
than the assumed 3$\times$3 bins.

\bfig[htb]
\insertplot{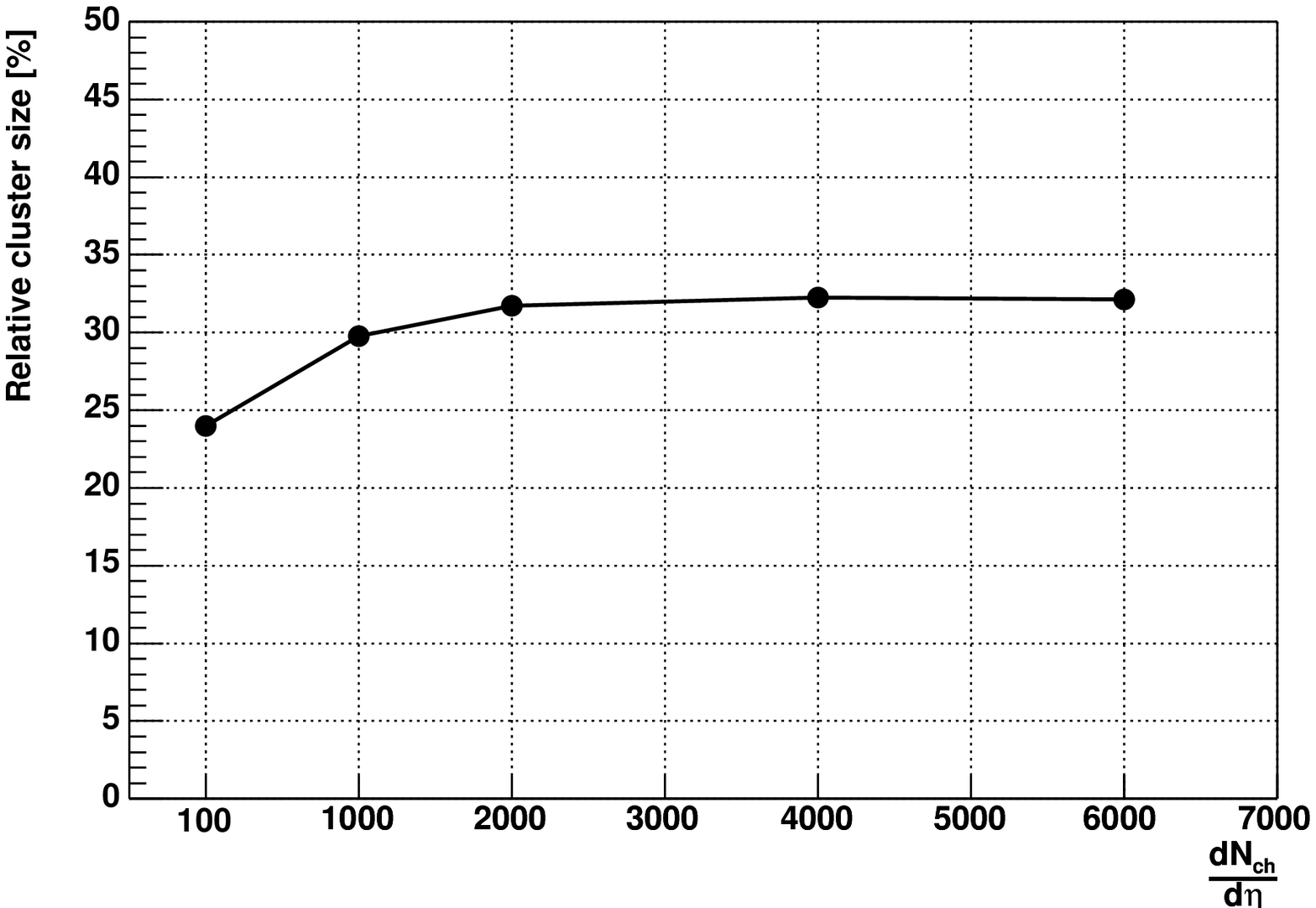}{8cm}
	{Ratio between cluster data and raw-data as a function of
multiplicity.}
	{Ratio between cluster data and raw-data as a function of
multiplicity. The cluster data was stored using the representation
listed in Table~\ref{COMP_rawdatasizes}.}
\label{COMP_clratio_vs_mult}
\efig

\subsection{Track and cluster modeling}
\label{COMP_trackmodeling}
Given the resolution of the space points and the size of the detector
volume, the space points will lead to a rather large encoding size of
the clusters. However, if the track reconstruction has been done the
clusters are assigned to their respective tracks. Within this data
model, the cluster positions can be encoded relative to the track
parameters and thereby lowering the required number of bits needed to
store the information.

The basic idea of this compression technique is to encode the data within
the context of the reconstructed tracks and their corresponding
clusters. Thus, the pattern recognition has to be solved prior to
the data compression. Once the track reconstruction is completed,
the tracks can be represented by helix parameters, and
the clusters by their two-dimensional crossing point with the
pad-row-plane. Let $(\Lambda_{\mathrm{pad}},\Lambda_{\mathrm{time}})$ denote a
intersection between a track and the pad-row plane, and
$(\lambda_{\mathrm{pad}},\lambda_{\mathrm{time}})$ the centroid of the corresponding
cluster assigned to the track, Figure~\ref{COMP_illres}.
\bfig[htb]
\insertplot{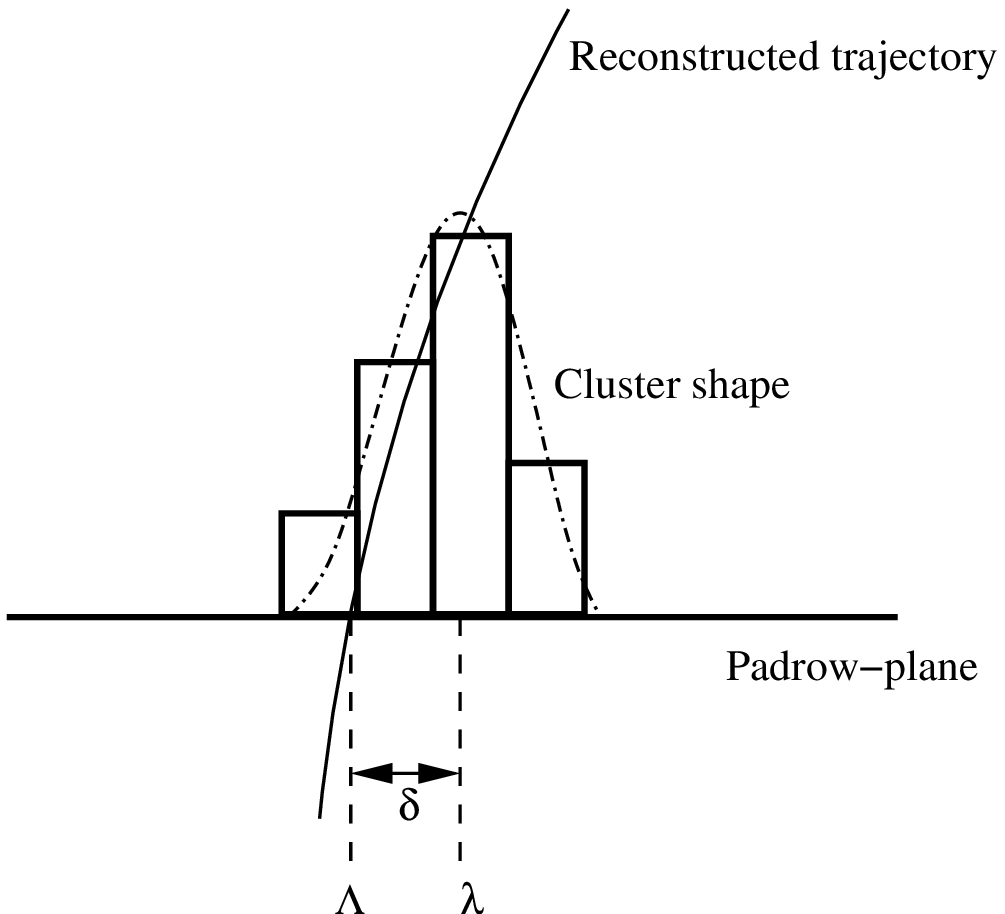}{8cm}
	{Definition of a residual.}
	{Definition of a residual. The residuals are defined as the
distance, $\delta$, between the track crossing point, $\Lambda$, and
the cluster centroid, $\lambda$, of the assigned clusters.}
\label{COMP_illres}
\efig
\noindent Further, let the distance between the track
crossing point and cluster position, the {\it residuals}, be
defined as 
\begin{eqnarray}
\delta_{\mathrm{pad}} = \Lambda_{\mathrm{pad}} - \lambda_{\mathrm{pad}}, \hspace{1cm}
\delta_{\mathrm{time}} = \Lambda_{\mathrm{time}} - \lambda_{\mathrm{time}}.
\label{COMP_residuals}
\end{eqnarray}
The residuals represent small deviations of the cluster position from
the track model, and are subject to the intrinsic detector resolution. 
This offers the possibility to quantize the residuals with a
transfer function whose resolution is adapted to the detector noise and
resolution,
\begin{eqnarray}
\delta_{\mathrm{pad}}\rightarrow\delta_{\mathrm{pad,Q}} =
\frac{\delta_{\mathrm{pad}}}{\epsilon_{\mathrm{pad}}},\hspace{1cm} 
\delta_{\mathrm{time}}\rightarrow\delta_{\mathrm{time,Q}} =
\frac{\delta_{\mathrm{time}}}{\epsilon_{\mathrm{time}}},
\label{COMP_qresiduals}
\end{eqnarray}
where $\epsilon$\ is the respective quantization intervals chosen for
the two directions. The resulting
numbers can now be stored with a minimum number of bits required to
encode the quantized residuals. 

In addition to the cluster positions, also the cluster shape and total
cluster charge is needed for the later offline analysis. As discussed
in Section~\ref{ALICE_tpcprinc} the shape of a cluster is determined by
the detector specific variables and the track inclination with the
pad-row-plane. The cluster shape can thus be parameterized as a function
of the track parameters, which allows the cluster shape to be restored
without storing the actual shape. Optionally, the deviation of the
cluster widths from the model can be stored. 

Regarding the cluster charge, its value is determined by the number
of primary collisions by the traversing particle in the gas which is a
random variable described by a Poisson distribution. Due to secondary
ionization and gas gain fluctuations the total charge is described by
a Landau distribution with a long tail. Because of this, the
representation of the cluster charge can not easily be reduced and is
thus stored as the original value as calculated during cluster
reconstruction. 





\subsubsection{Encoding scheme}
A compression scheme which stores the track and cluster information in a
compressed format has been implemented, and will be described in the
following.

The input to the compression scheme are the reconstructed tracks and
their corresponding assigned clusters. For every pad-row-plane the
residuals are calculated and quantized according to
Equations~\ref{COMP_residuals} and~\ref{COMP_qresiduals}.
The encoding scheme assumes the tracks have the following
properties\footnote{This definition only serves to optimize the
compression scheme, and has nothing to do with the actual tracking
procedure being used.}:
\begin{itemize}
\item The first and last point of the trajectory correspond to the
innermost and outermost pad-row of the TPC respectively.
\item A track may contain clusters from different TPC sectors.
\end{itemize}
In this scheme there is no need to store the pad-row number for every
cluster, as the initial assumption is that there is a corresponding
cluster on every successive pad-row throughout the TPC radius. Instead,
only 1 bit is required to
flag if this is not the case, i.e. to denote whether a cluster is present or
not on a particular pad-row. Furthermore, an additional bit is also
required to denote a possible change of the TPC sector. This may happen
if the trajectory crosses the boundary between two neighboring TPC
sectors.

A given data sample is encoded as illustrated in
Figure~\ref{COMP_encoding}.
\bfig[htb]
\insertplot{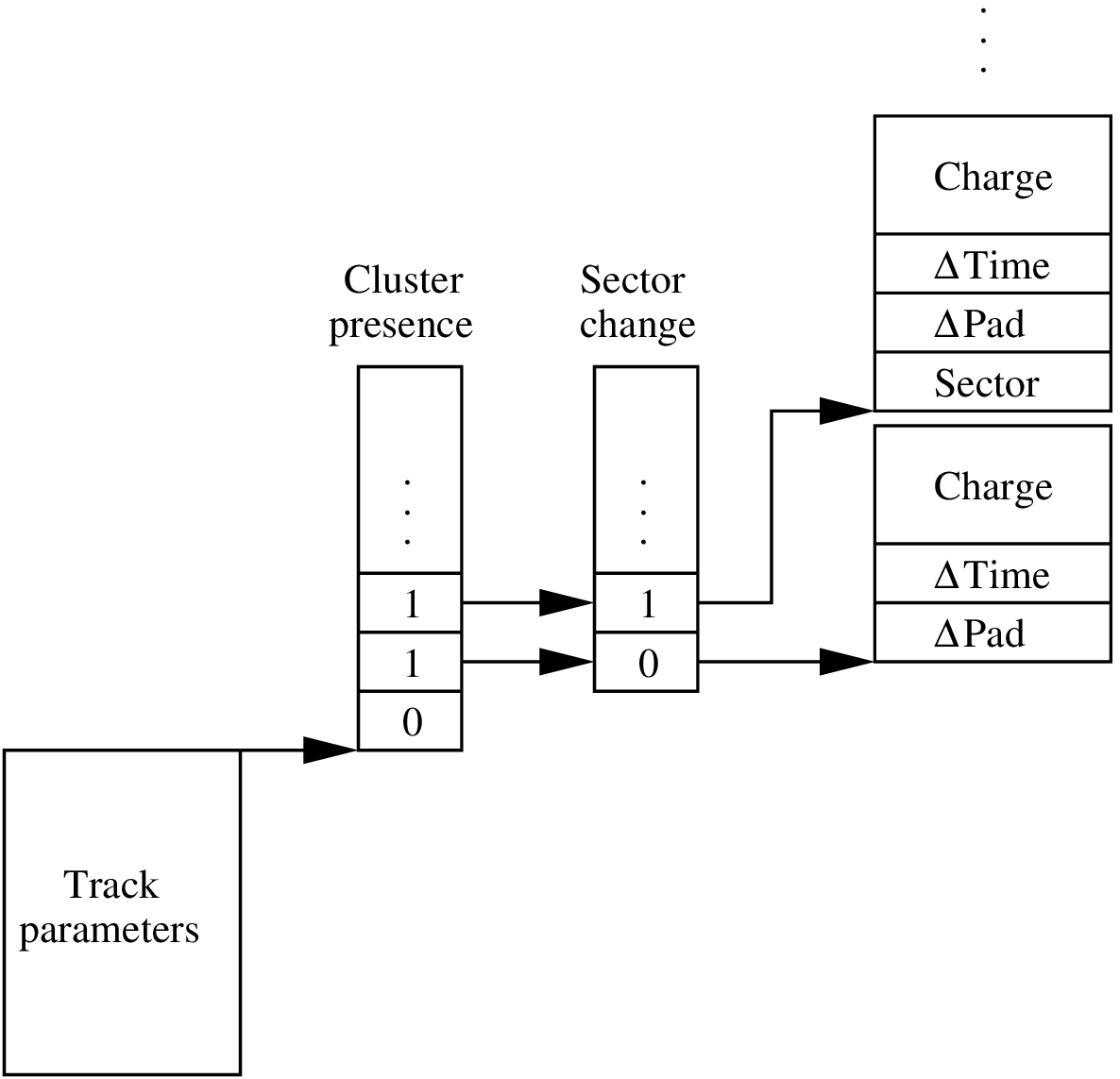}{8cm}
	{Data compression encoding scheme.}
	{Data compression encoding scheme.}
\label{COMP_encoding}
\efig
\noindent Each track contains a bitmap which encodes the {\it cluster
presence}. For every pad-row 1 bit is used to flag whether a cluster
has been assigned or not on that pad-row. If no cluster is present,
only the '0' bit is written, otherwise the cluster information is
also stored. A second bitmap is used to store any change in TPC sector
number. If a cluster in the list belongs to a different sector than the
previous, this is stored by setting this bit to '1'. In that case,
the new sector number is stored in addition to the cluster information.


The compressed data sample consists of the track
parameters of every track together with the coded cluster information
for the corresponding clusters. The track parameters correspond to
a minimum set of required parameters to describe a helix,
Appendix~\ref{APP_helix}. These are the curvature of the circular
motion, $\kappa$, the coordinates of the point at the distance of closest approach
(DCAO), $(r_{\mathrm{DCAO}},\phi_{\mathrm{DCAO}},z_{\mathrm{DCAO}})$,
and the dip-angle of the track, $\lambda$. In
Table~\ref{COMP_clusterparams} these parameters are
listed together with their respective encoding sizes.
The size of the cluster parameters depends on the choice of quantization
intervals, $\epsilon$, in Equation~\ref{COMP_qresiduals}. The minimum
number of bits required to store a quantized residual $\delta_Q$ is given by
\beq
n_{\mathrm{bits}} =
\mathrm{max}[\frac{\log(|\delta_Q|)}{\log(2)}]+1\equiv n_{\mathrm{bits}}(\epsilon)
\label{COMP_nbits}
\eeq
where max$[x]$ denotes the closest integer larger than $x$. The extra bit
is needed to encode the sign. Thus for a given data
sample the number of bits used to encode the residuals are calculated
from the maximum value in the sample using
Equation~\ref{COMP_nbits}. The track--cluster parameters and their
respective sizes are listed in
Table~\ref{COMP_clusterparams}. 
\begin{table}[htb]
\begin{center}
\begin{minipage}[t]{0.4\linewidth}
\centering
\vglue0.2cm
\begin{tabular}{|c|c|}
\hline
{\bf Track parameters}      & {\bf Size (Byte)} \\
\hline \hline
   $\kappa$                 &       4 (float)\\
   $\phi_{\mathrm{DCAO}}$   &       4 (float)\\
   r$_{\mathrm{DCAO}}$      &       4 (float)\\
   z$_{\mathrm{DCAO}}$      &       4 (float)\\
   $\lambda$                &       4 (float)\\
\hline 
\end {tabular}
\end{minipage}
\hspace{0.5cm}
\begin{minipage}[t]{0.5\linewidth}
\centering
\vglue0.2cm
\begin{tabular}{|c|c|}
\hline
{\bf Track--cluster parameters} & {\bf Size (Bit)} \\
\hline \hline
   Cluster presence             & 1\\
   Sector / change              & 6 / 1\\
   Cluster charge               & 12\\	
   $\delta_{\mathrm{pad,Q}}$    & $n_{\mathrm{bits}}(\epsilon)$\\
   $\delta_{\mathrm{time,Q}}$   & $n_{\mathrm{bits}}(\epsilon)$\\

\hline
\end {tabular}
\end{minipage}
\caption[Cluster and track parameters and their respective size used
in the data compression scheme.]
	{Cluster and track parameters and their respective size used in the
compression scheme. The track parameters are defined at the distance
of closest approach to origin (DCAO). The number of bits used to encode the
residuals is determined by the maximum value of the respective
residuals in the sample, and are calculated from Equation~\ref{COMP_nbits}.}
\label{COMP_clusterparams}
\end{center}
\end{table}

\bfig[htb]
\insertplot{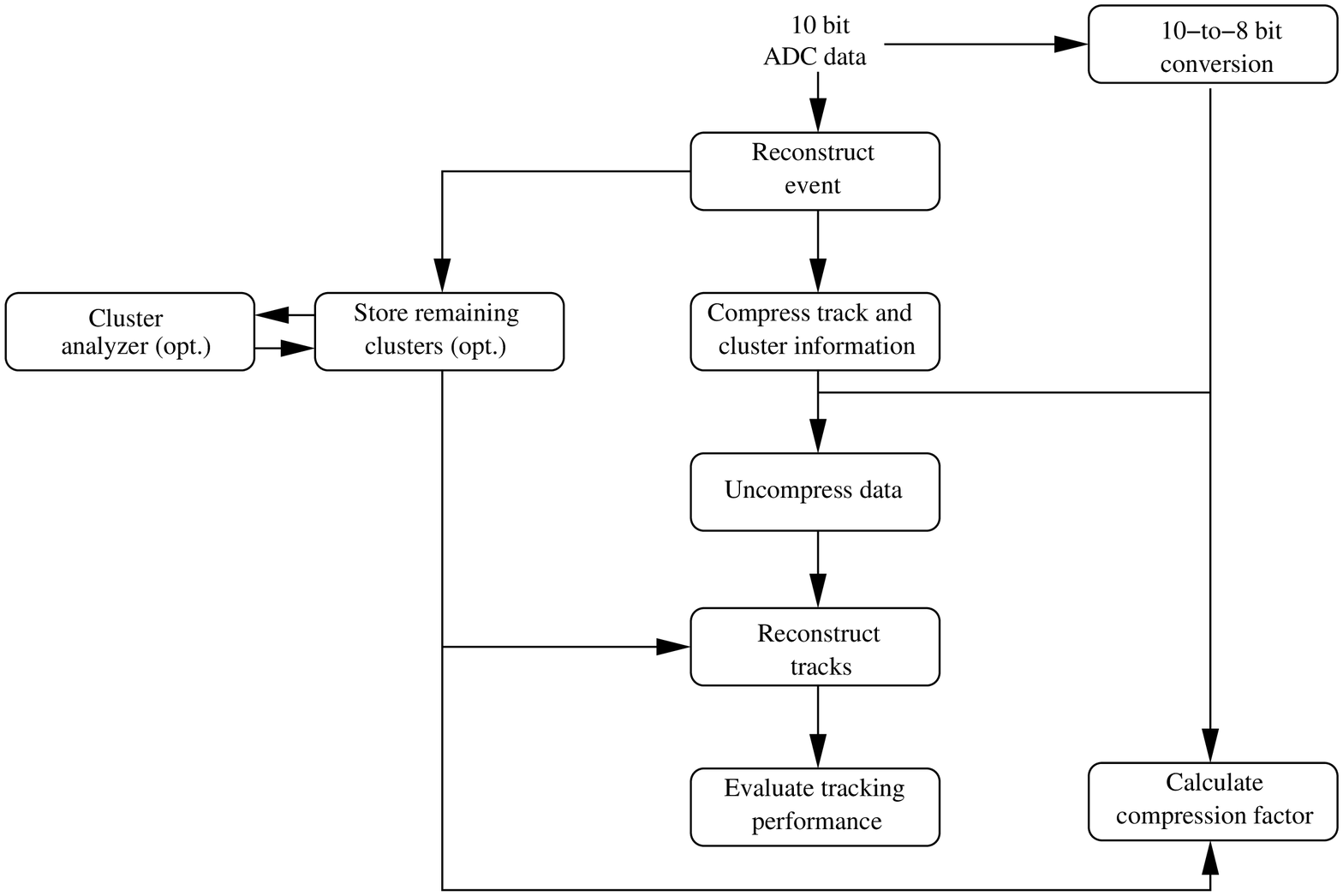}{12cm}
	{Flow diagram of the implemented data compress/expand cycle.}
	{Flow diagram of the implemented data compress/expand cycle.}
\label{COMP_cycle}
\efig

\subsubsection{The compress/expand cycle}
In general, any implementation of a data compression scheme consists
of two separate steps: compression and expansion. In the first step
the given input data sample is encoded following the implemented
encoding scheme. The compressed data sample can then be compared to
the original in order to evaluate the compression ratio. Then, during
the expansion, the compressed data sample is decompressed and restored
into the same data format as the original data. Finally, the
uncompressed data can be compared to the original data set in
order to evaluate any impact on the relevant information content.

Figure~\ref{COMP_cycle} show a flow diagram of the complete
compress/expand cycle which has been applied. 
In a first step the
clusters and tracks are reconstructed from the 10 bit ADC-data.
The resulting clusters and tracks are then encoded according to the
scheme illustrated in
Figure~\ref{COMP_encoding}. The remaining clusters in the event which
were not assigned to any tracks may optionally be
written in addition to the compressed data, or they
may be discarded all together. Alternatively a special {\it selection} of the
remaining clusters can be done
using a {\it cluster analyzer} (discussed on page~\pageref{COMP_remcl}).
The data is then uncompressed by restoring the cluster centroids and
shape from the compressed data, and optionally merging them with the
remaining cluster data. The final list of clusters is then given to
the offline track finder
program. Once the tracks have been reconstructed, the resulting
tracking performance can be evaluated in order to check the impact the
compression has on the tracking performance.
Furthermore, the compressed data size is compared to the
original raw-data size, represented by the 8 bit ADC-data, to obtain
the relative compression ratios.

\subsubsection{On choosing the quantization intervals}
The quantization interval, $\epsilon$, in Equation~\ref{COMP_qresiduals}
determines the accuracy of the compression scheme. This is an essential
parameter in the algorithm as it limits both the accuracy and the
achievable compression factor of the algorithm.
Selecting a very coarse quantization interval
will in general allow for a greater compression factor as the number of bits
required to encode the residuals will be small. Any such gain in
compression factor will however come at the expense of a
correspondingly lower space point resolution in the final uncompressed
data sample.

\bfig[]
\centerline{\epsfig{file=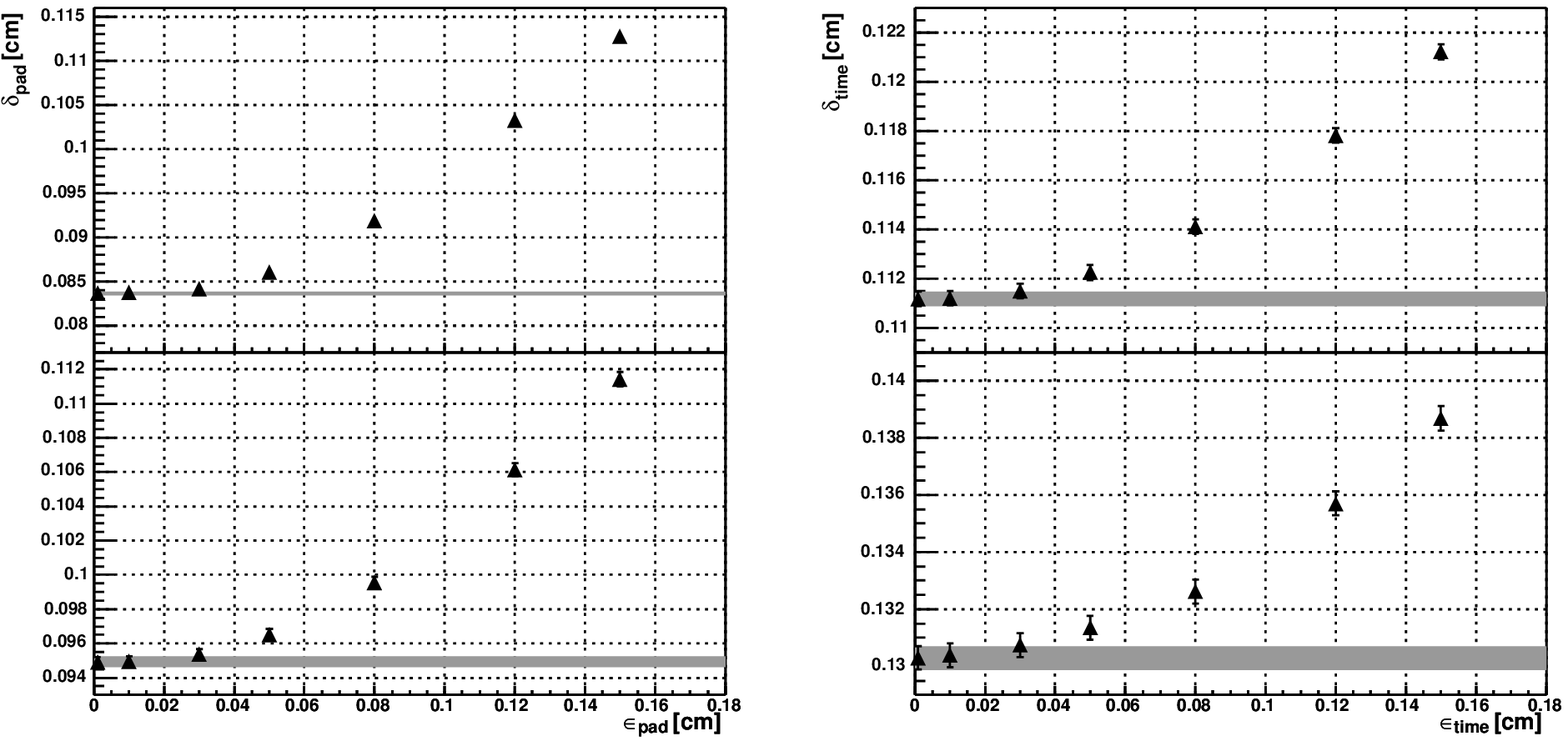,width=17cm}}
\caption[Impact on the space point resolution.]
	{Impact of the quantization scheme on the space point
resolution. The upper plots show the result for the outer TPC
chambers, while the inner plots shows the inner chambers.
The shaded area corresponds to the resolution within the
original data sample. The resolutions were obtained from
isolated clusters from all primary tracks with $p_t\geq$\,0.1\,GeV. The
values correspond to the standard deviation from a Gauss fit of the
respective distributions.}
\label{COMP_residuals_vs_epsilon}
\vspace{2cm}
\centerline{\epsfig{file=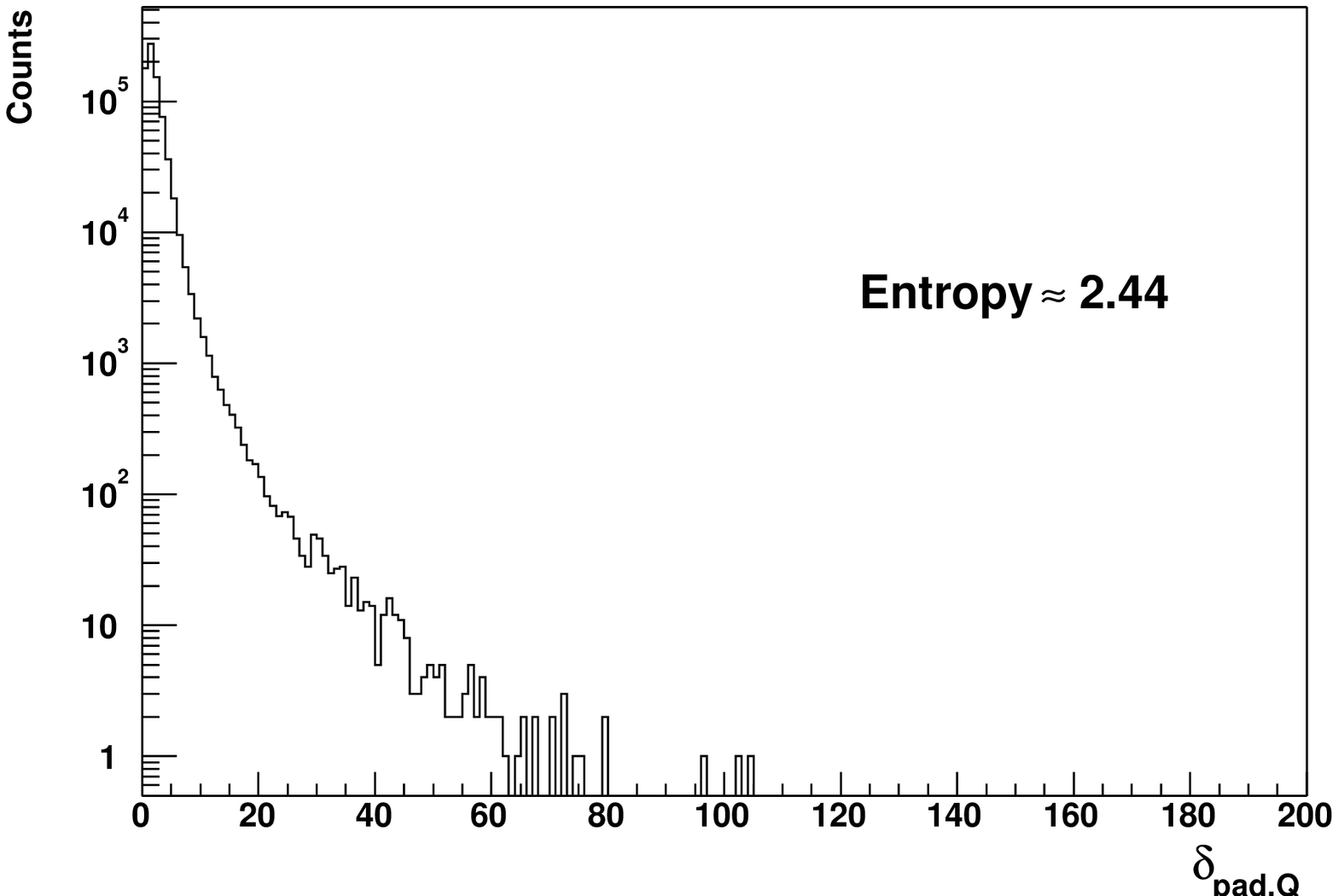,width=8cm}
\epsfig{file=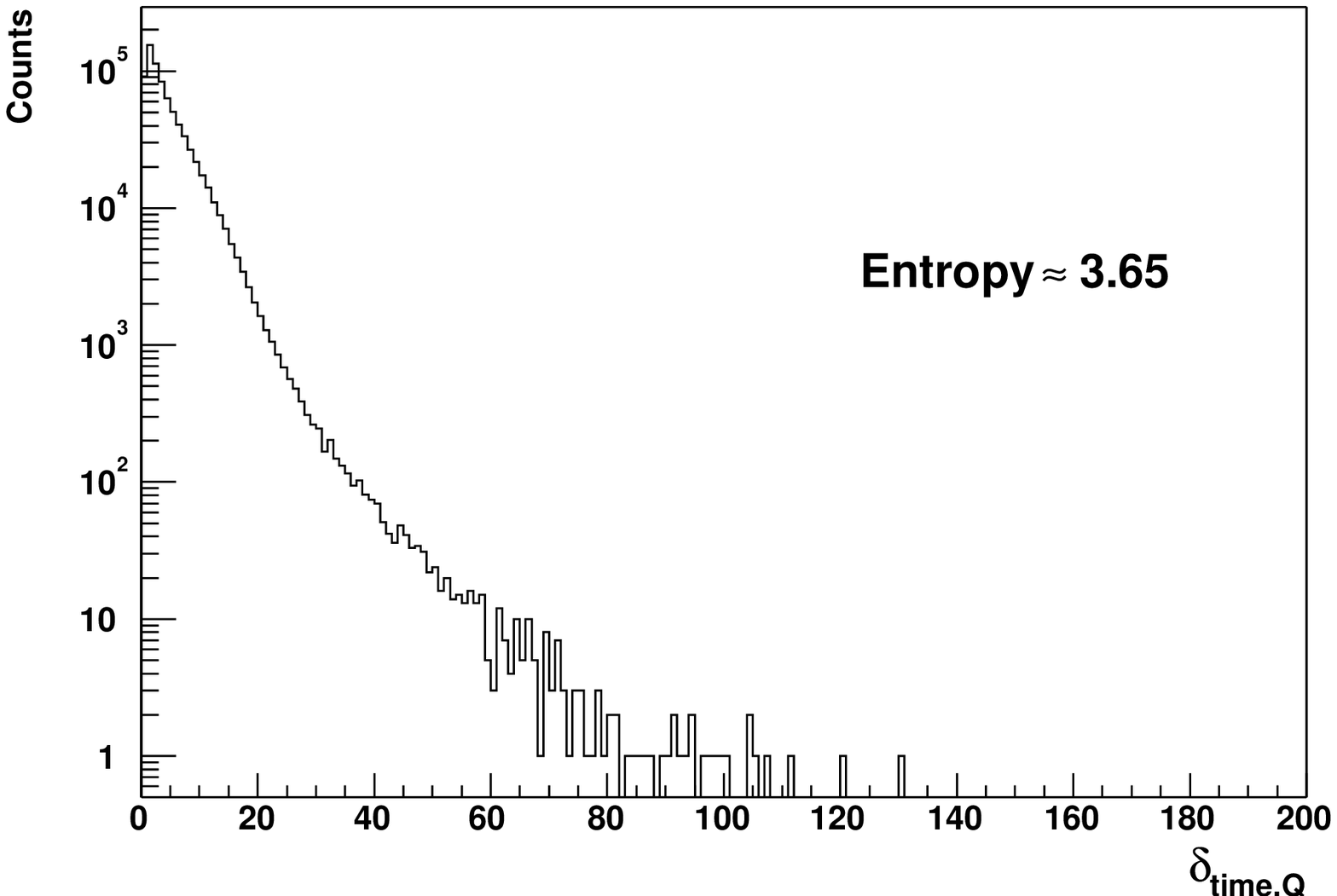,width=8cm}}
\caption[Distribution of quantized residuals.]
	{Distribution of the quantized residuals for a 
quantization interval of $\epsilon$=0.05\,cm. The data sample used are
the same as used in Figure~\ref{COMP_residuals_vs_epsilon}.}
\label{COMP_nepsilons}
\efig

Figure~\ref{COMP_residuals_vs_epsilon} illustrates the impact of the
quantization steps on the space point resolution. The plots show the
final space point resolution after a compress/expand cycle as a
function of the choice of quantization steps in the pad and time
direction, respectively. The event sample correspond to a low
multiplicity event (\dndy=\,100), i.e. well separated clusters and tracks,
and the events were reconstructed by the HLT
sequential track reconstruction chain, Section~\ref{TRACK_seqtracking}.
As one would expect the space point resolution deteriorates as the
quantization intervals increases. For $\epsilon$=0.05\,cm, the loss
is about 2\% and 1\% for the pad and time direction, respectively.

Figure~\ref{COMP_nepsilons} shows the distribution of the quantized
residuals for $\epsilon$=0.05\,cm. The distributions are approximately
exponential. The maximum value of the distributions are 104 and 130
for the pad and time distribution respectively, which according to
Equation~\ref{COMP_nbits} means that
8 and 9 bits are required to encode the residuals with a fixed bit-rate.
However, given that values are not equally probable, further
compression is possible by entropy coding. The entropy of the samples
has been calculated according to Equation~\ref{COMP_entropy}, and suggest
a theoretical lower limit on the average word size needed to encode
these residuals of 2.4 and 3.6 for the pad and time direction respectively.

\subsubsection{Cluster widths}
In addition to the cluster centroids, also the space point errors
are needed for the track reconstruction. 
For the track finding algorithms, these
errors are for practical purposes calculated from the
cluster widths using suitable proportional factors optimized from
the simulated data, Section~\ref{TRACK_clerrors}. As the cluster widths can be
parameterized according to the track parameters
(Equation~\ref{TRACK_clwidths}), this can be exploited by the
compression scheme by removing the cluster shape information from the
data stream and restoring it from parameterization during the
decompression step. 

However, since the cluster generation in the TPC is a result of
stochastic processes, the cluster shape and also the space point errors
are subject to fluctuations.
As a consequence, the parameterization of the cluster shape
may lead to deviations of the space point
errors from the ``true'' errors. This can furthermore be a
potential source for loss in tracking performance as e.g. the track
reconstruction algorithm may make wrong decisions based on
$\chi^2$-criteria if the provided error estimates differ
significantly from the true ones. Such loss would in particular be
expected to occur if the occupancy is large, in which the space point
assignment critically depends on a good estimate of the individual
space point errors. 

\bfig[htb]
\insertplot{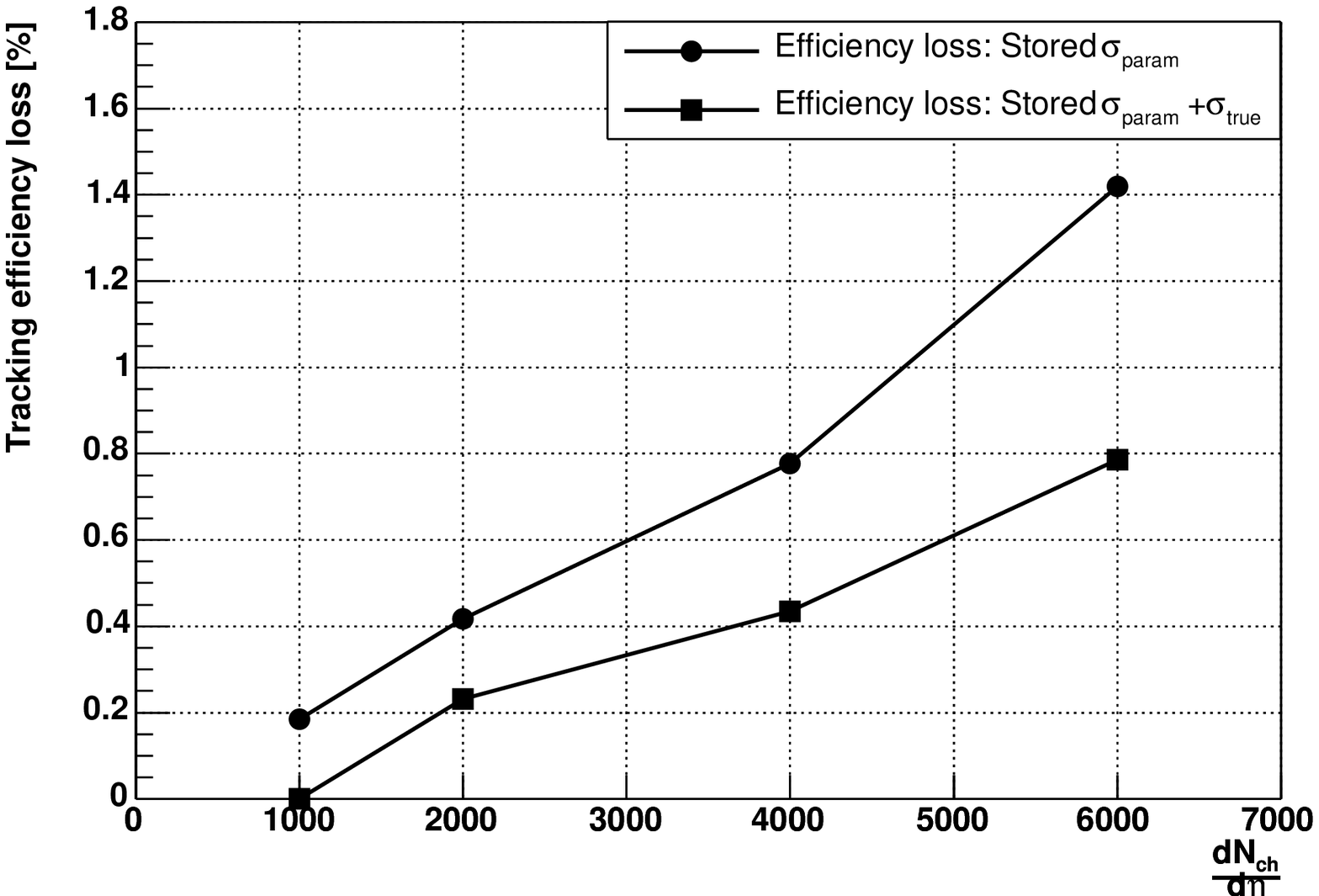}{9cm}
	{Efficiency loss due to removing the cluster shape information
from the data stream.}
	{Integrated tracking efficiency loss due to removing the
cluster shape information from the data stream
($\sigma_{\mathrm{param}}$). Instead of storing the
cluster widths for each cluster, the widths are parameterized by
Equation~\ref{TRACK_clwidths}. The
efficiency loss when the original cluster widths are stored for
clusters where the measured widths deviate substantially from the 
parameterized ones are also shown,
Equation~\ref{COMP_widthsel} ($\sigma_{\mathrm{param}}+\sigma_{\mathrm{true}}$).}
\label{COMP_compvswidth}
\efig

Figure~\ref{COMP_compvswidth} shows the loss in tracking
efficiency due to removing the cluster shape information from the data
stream as a function of multiplicity. In this case clusters
reconstructed by the Offline cluster finder was
used as input in order to avoid any effects resulting from difference
in performance between the HLT and Offline cluster finder
algorithms. The loss was estimated by taking the difference between
the integrated Offline tracking efficiency in the case where the
cluster shape was parameterized according to the track parameters and
the original data set in which the cluster shape was stored. The
results indicate that the loss increases slightly as a function of
multiplicity, and reaches the maximum of $\sim$1.4\% for \dndy=\,6000. 

In order to minimize such a loss, the original shape information can
be kept by storing the {\it deviation} of the parameterized widths from
the true ones. Alternatively, the widths can be stored only for
clusters whose widths differs from the parameterization more than a
certain threshold value. The scenario is also shown in
Figure~\ref{COMP_compvswidth}. In this case the cluster widths has
been stored for clusters where
\beq
|\sigma_{\mathrm{param}}-\sigma_{\mathrm{true}}| > \frac{1}{2}\sigma_{\mathrm{true}}.
\label{COMP_widthsel}
\eeq
This selection criteria includes $\sim$10-15\% of all the clusters in the
samples, and results in no observable efficiency loss at
\dndy=\,1000. However, at higher multiplicities, \dndy=\,6000, a
efficiency loss of $\sim$0.8\% is observed.

The results indicates that the Offline track
finder algorithm is rather sensitive to the estimation of the cluster
shape. 
This may be explained by the fact that the Offline track finder has
been highly tuned to the space point errors obtained from the
simulated cluster shape~\cite{belikov}.
In order to further optimize the compression scheme in terms of
cluster shape and space point resolutions, a detailed understanding of
the TPC response and the impact on the final Offline reconstruction
chain is needed. Such a study can however only be done using real
TPC data, as these data will contain other effects as well. 
The results presented in the following are obtained using only a 
parameterization of the cluster widths.

\subsubsection{Remaining clusters}
\label{COMP_remcl}
After the track finding procedure, a certain amount of {\it
remaining} clusters will be present. These are clusters which were not
assigned to any tracks during the track reconstruction procedure. The
are several possible sources of these clusters,
and they depend foremost on the efficiency of the track finding
algorithm at hand. In general the most common source is particles with a
very low \pt whose tracks could not be reconstructed due to large
multiple scattering and energy loss. This is in particular true for the so-called
$\delta$-electrons, which are produced when particles transfer
a relatively large amount of their energy into a single electron when
traversing through the gas. The typical signal from such particles forms
a large continuous cluster area because of their large inclination with
the pad-row-plane. Since these particles are in general not of any
interest, they will not be included in any Offline analysis and can
therefore be considered as noise. Thus from a data compression point of
view, such clusters should be removed from the data stream as they
do not contain any relevant information.

The list of remaining clusters may however also include
clusters that are not originating from ``noise'' particle tracks, but
are merely a result of an inefficient track reconstruction chain. In order
to minimize any loss of accuracy and efficiency in the output data,
the remaining clusters may be written in addition to the compressed data. 
These clusters can however not be encoded with respect to any track
model, and thus needs to be stored with the cluster parameters as
proposed in Table~\ref{COMP_rawdatasizes}. This will then contribute
with an
additional overhead to the compressed data size. On the other hand, if
they are completely disregarded from the data, vital information may
be lost as the list may contain ``valid'' clusters which were not
assigned to any tracks by the track reconstruction algorithm.
Such a loss can be significant if the tracking performance of
the applied tracking algorithm is lower than the one
that will be used for the later offline analysis. 

An option to store the remaining clusters, or a certain selection of them, has
been included in the compression scheme
(Figure~\ref{COMP_cycle}). Here, a {\it cluster analyzer}
can classify clusters based on a selection scheme, and only store a
fraction of the cluster list. The
analyzer selects remaining clusters located within the
detector region being used for the seed-finding procedure in
the Offline track finder algorithm,
Section~\ref{TRACK_offlinechain}.

The impact on the tracking efficiency from removing the remaining
clusters is illustrated for a given data sample in
Figure~\ref{COMP_trackeff_vs_remaining}. 
\bfig[htb]
\centerline{\epsfig{file=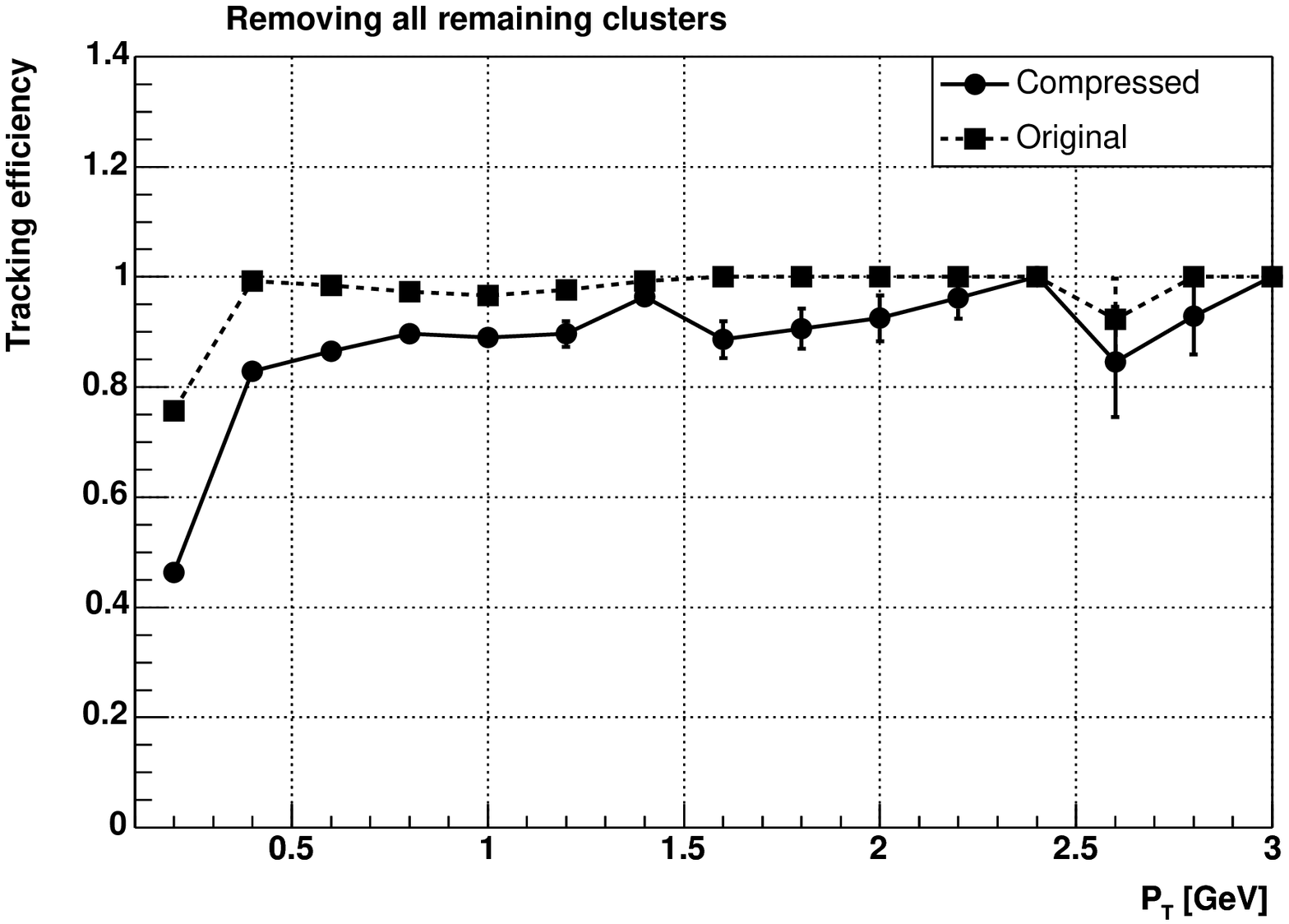,width=8cm}
\epsfig{file=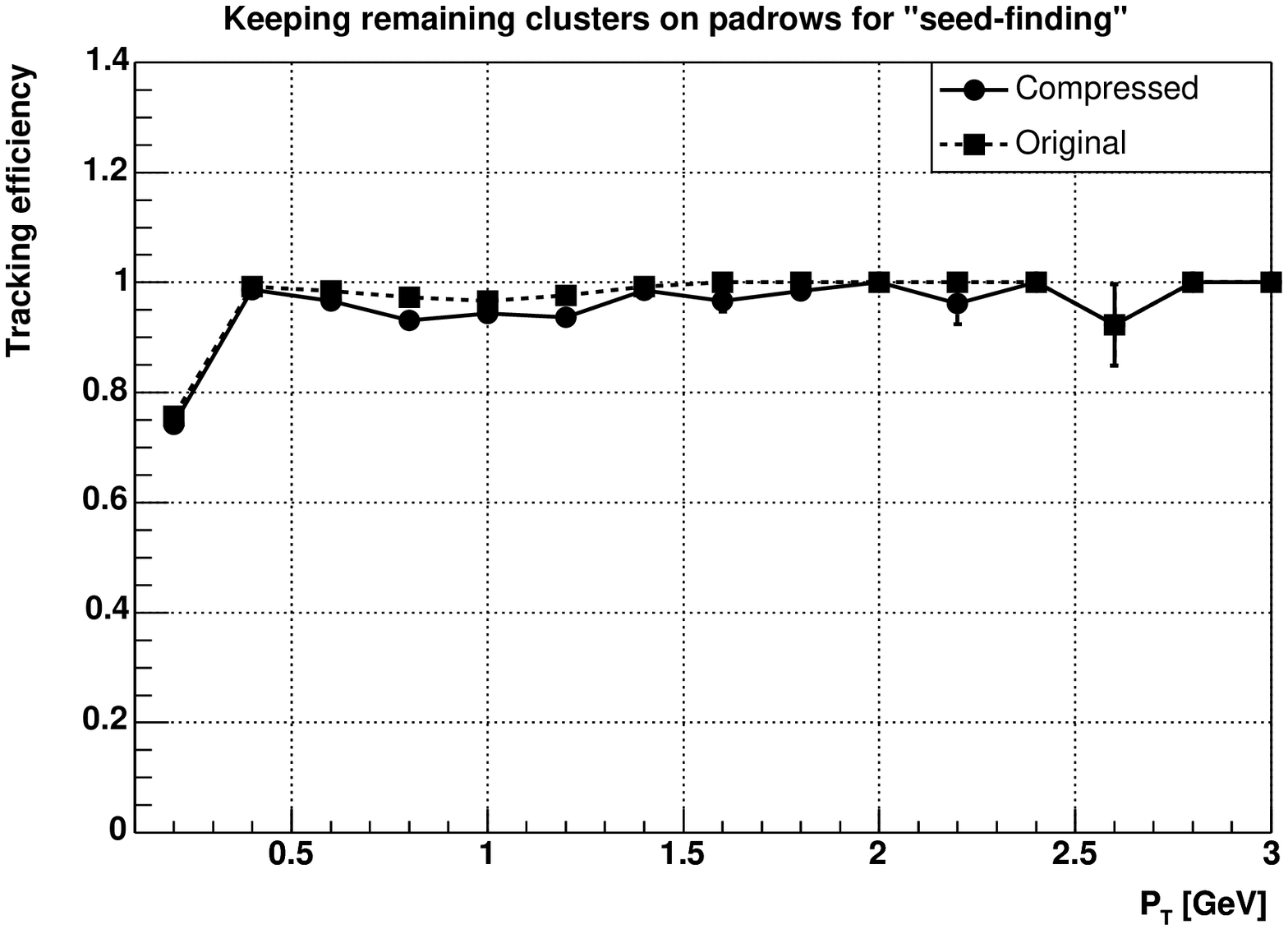,width=8cm}}
\caption[Impact on tracking efficiency from disregarding the remaining
clusters in the compression scheme.]
	{Impact on tracking efficiency from disregarding the remaining
clusters in the compression scheme. The data sample corresponds to an
event with multiplicity of \dndy=\,1000. The plots show the tracking
efficiency as a function of \pt before (original) and after
(compressed) the compress/expand cycle has been applied. Left: All
remaining clusters are removed from the data stream. Right:
Only remaining clusters which are located in the region where
track seed-finding is performed are stored.}
\label{COMP_trackeff_vs_remaining}
\efig
\noindent The two plots display
the Offline tracking efficiency as a function of \pt both before and after a
compress/expand cycle. In both cases, the input clusters and tracks to
the compression scheme were reconstructed by the HLT sequential track
reconstruction chain, Section~\ref{TRACK_seqtracking}.
In the left plot, all the remaining clusters
have been removed from the data sample. A significant
loss of tracking performance is observed, resulting in an integrated loss
of about 16\%. In the
right plot, the cluster analyzer was applied to the remaining clusters
in which the resulting selection were
stored while the rest is removed. In this case, the integrated efficiency loss is
reduced to about 1.5\%.

\subsection{Results}
In order to investigate the performance of the compression scheme
outlined in the previous sections, the compress/expand cycle was
applied to various simulated data samples. The input to the
compression scheme presented in the following are clusters and tracks
which were reconstructed using the implemented HLT sequential
tracking approach presented in Section~\ref{TRACK_seqtracking}. 
All tracking performance results and impact thereon refer to the
standard Offline track finder.

\subsubsection{Impact on the momentum and dip-angle resolution}
One of the observables, which should be sensitive to the
compression, is the momentum resolution. According
to Equation~\ref{TRACK_ptresformula} the relative transverse
momentum resolution depends on both the space point resolution and the
number of space points assigned to the track. In
Figure~\ref{COMP_compvsptres} the impact on the relative momentum
resolution is shown for an event with multiplicity \dndy=\,1000. 
\bfig[htb]
\centerline{\epsfig{file=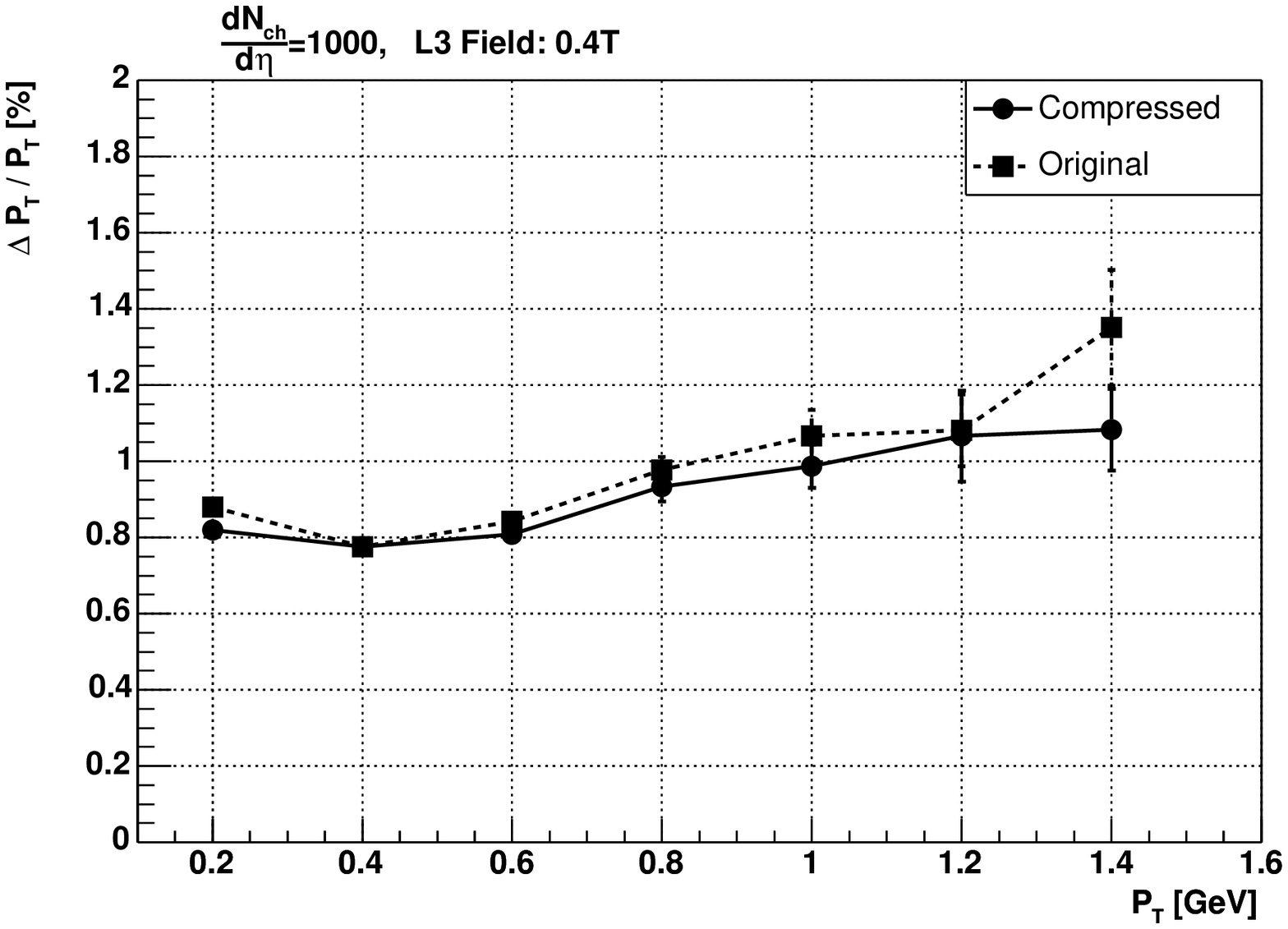,width=8cm}
\epsfig{file=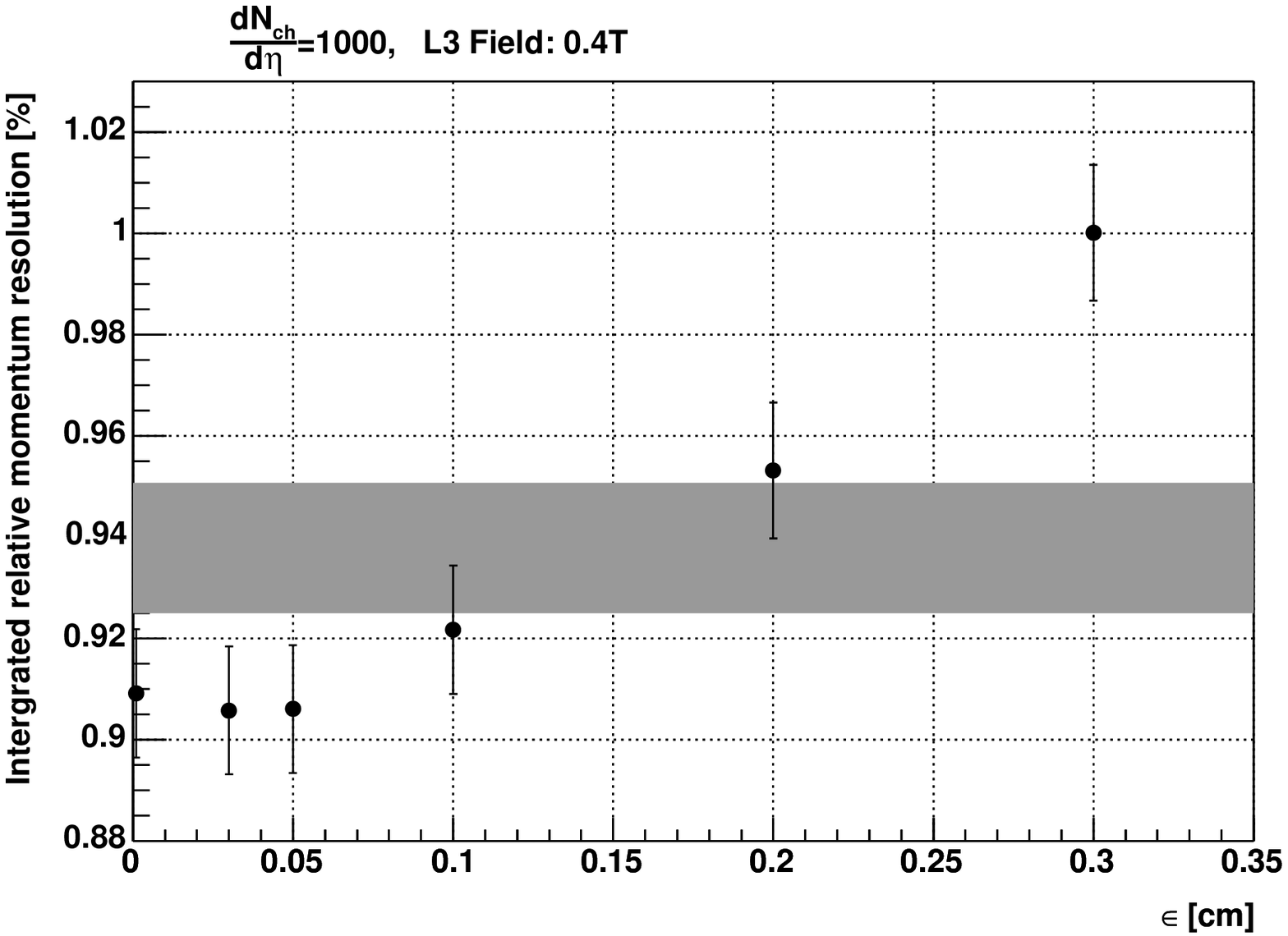,width=8cm}}
\caption[Impact on the relative momentum resolution.]
	{Impact on the relative momentum
resolution from the compression. Left: Relative momentum resolution as
a function of \pt
for both original data set and the compressed data sample. In the
compression scheme the quantization intervals were
$\epsilon$=0.05\,cm. Right:
Integrated relative momentum resolution as a function of the
quantization intervals in the compression scheme. The shaded area
corresponds to the resolution within the original data set.}
\label{COMP_compvsptres}
\efig
\noindent In the left plot the
resolution is shown as a function of \pt for a quantization interval
of $\epsilon$=0.05\,cm for both the original and the compressed data
sample. In the compression the remaining clusters selected by the
cluster analyzer
(Figure~\ref{COMP_trackeff_vs_remaining}) were stored in addition to
the compressed data. The plot indicates a slight improvement of the momentum
resolution in the compressed data compared to the original for all
\pt-bins shown. In the right plot the integrated resolution is shown
as a function of the quantization intervals used. For
$\epsilon\leq$\,0.1\,cm the momentum resolution is $\sim$3\% better
relative to the original data.

\bfig[]
\centerline{\epsfig{file=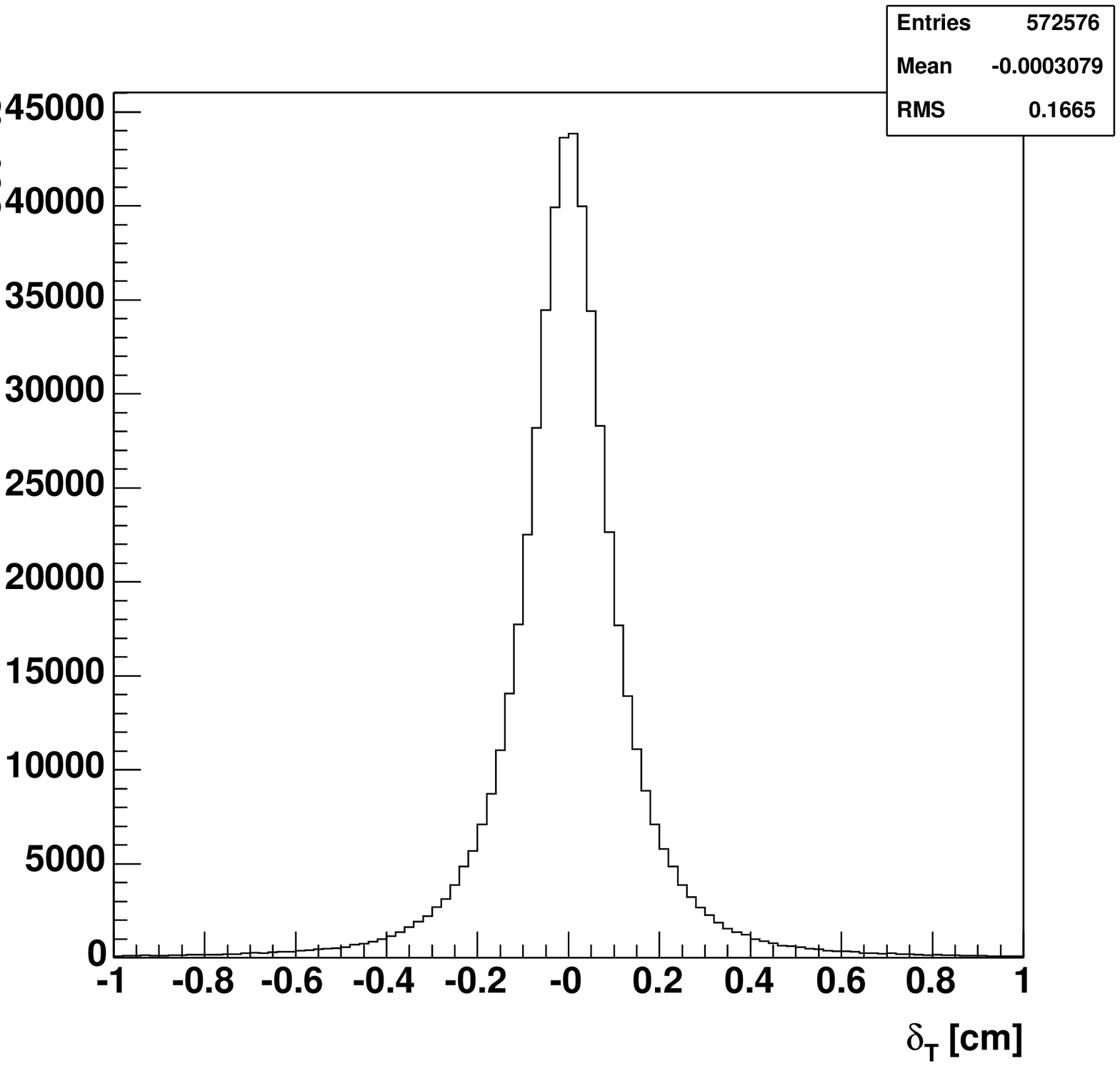,width=7cm}
\hspace{0.5cm}\epsfig{file=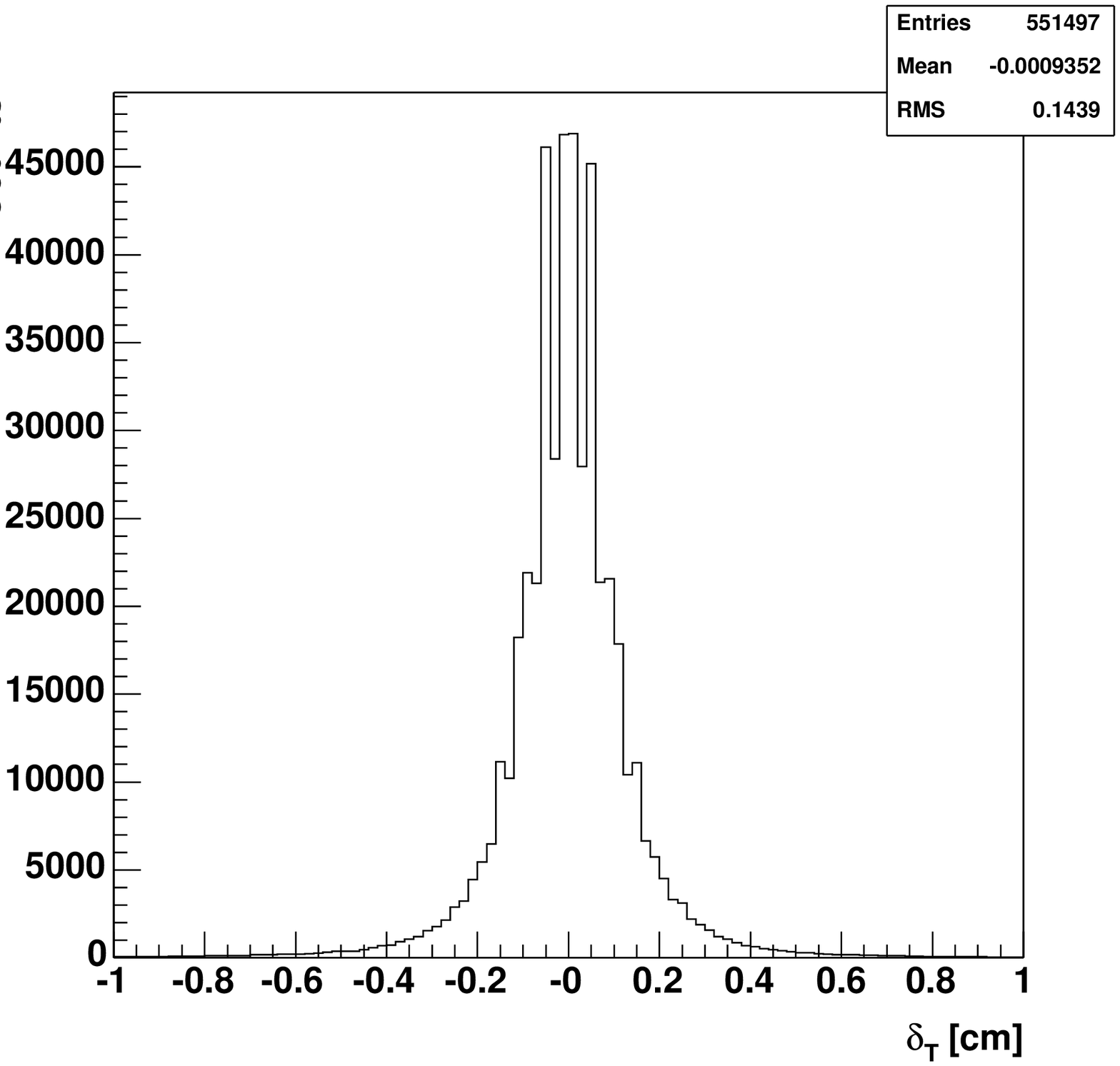,width=7cm}}
\caption[Transverse residual distribution before and after data
compression.]
	{Transverse residual distribution before and after data
compression. Left: Transverse residual distribution within the
original data set. Right: Transverse residual distribution in the
post-compression data set. The distribution has been obtained from
clusters in the outer TPC chambers, and is averaged over all tracks
in the event sample (\dndy=\,1000). Note the steps in the distribution which is due
to the quantization.}
\label{COMP_residuals_beforeandafter}
\vspace{2cm}
\insertplot{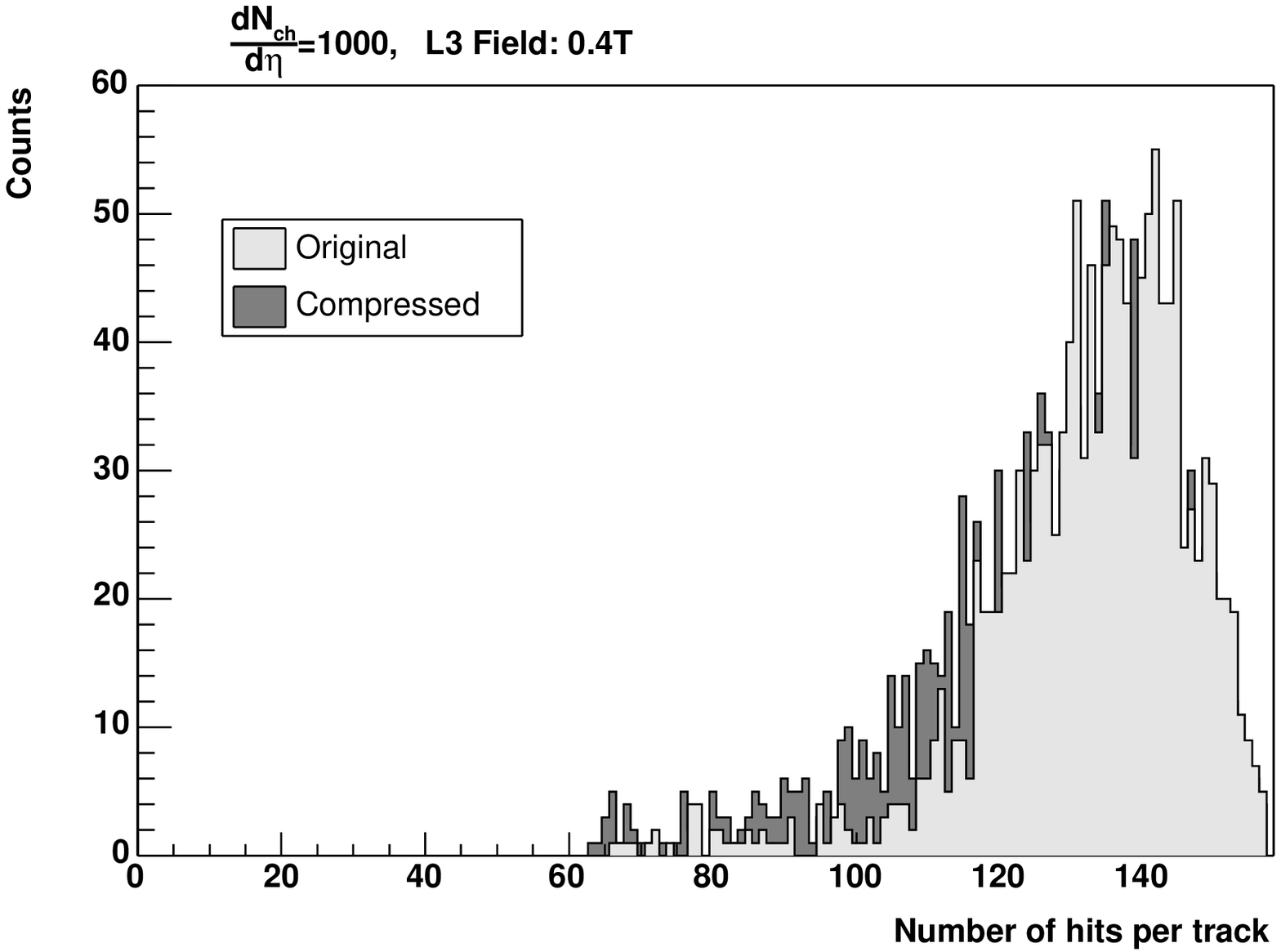}{9cm}
	{Impact on the number of assigned clusters per track.}
	{Impact on the number of assigned clusters per track from the compression. The
distributions corresponds to the number of assigned clusters per track
before (original) and after (compressed) data compression has been applied.}
\label{COMP_nhitspertrack}
\efig

The improvement in the resolution is an effect of the quantization of
the residuals. As a consequence of this quantization, the resulting
residual distributions are slightly {\it narrowed} as the residuals which are below
the quantization interval are mapped to zero. This narrowing is illustrated in
Figure~\ref{COMP_residuals_beforeandafter}. 
Here the transverse
residual distribution before and after the applied compression scheme
is presented, showing that the width of the distribution (represented
by the RMS-value) is smaller for the post-compression distribution
compared to the original data set. This effect leads to
a better fit in the post-compression track finding and thus a more precise
estimate of the curvature and correspondingly the momentum
resolution. Furthermore, the potential loss of clusters 
will have a less significant impact on the momentum
resolution as the number of space points has a 1/$\sqrt{\mathrm{N}}$
dependence compared to the linear dependence on the space point
resolution. Figure~\ref{COMP_nhitspertrack} shows the number of
clusters per track before and after the compression. A slight shift
towards smaller values is observed for the compressed data
sample. The mean value of the distribution changes by less than two
clusters per track for the quantization intervals investigated.

\bfig[htb]
\insertplot{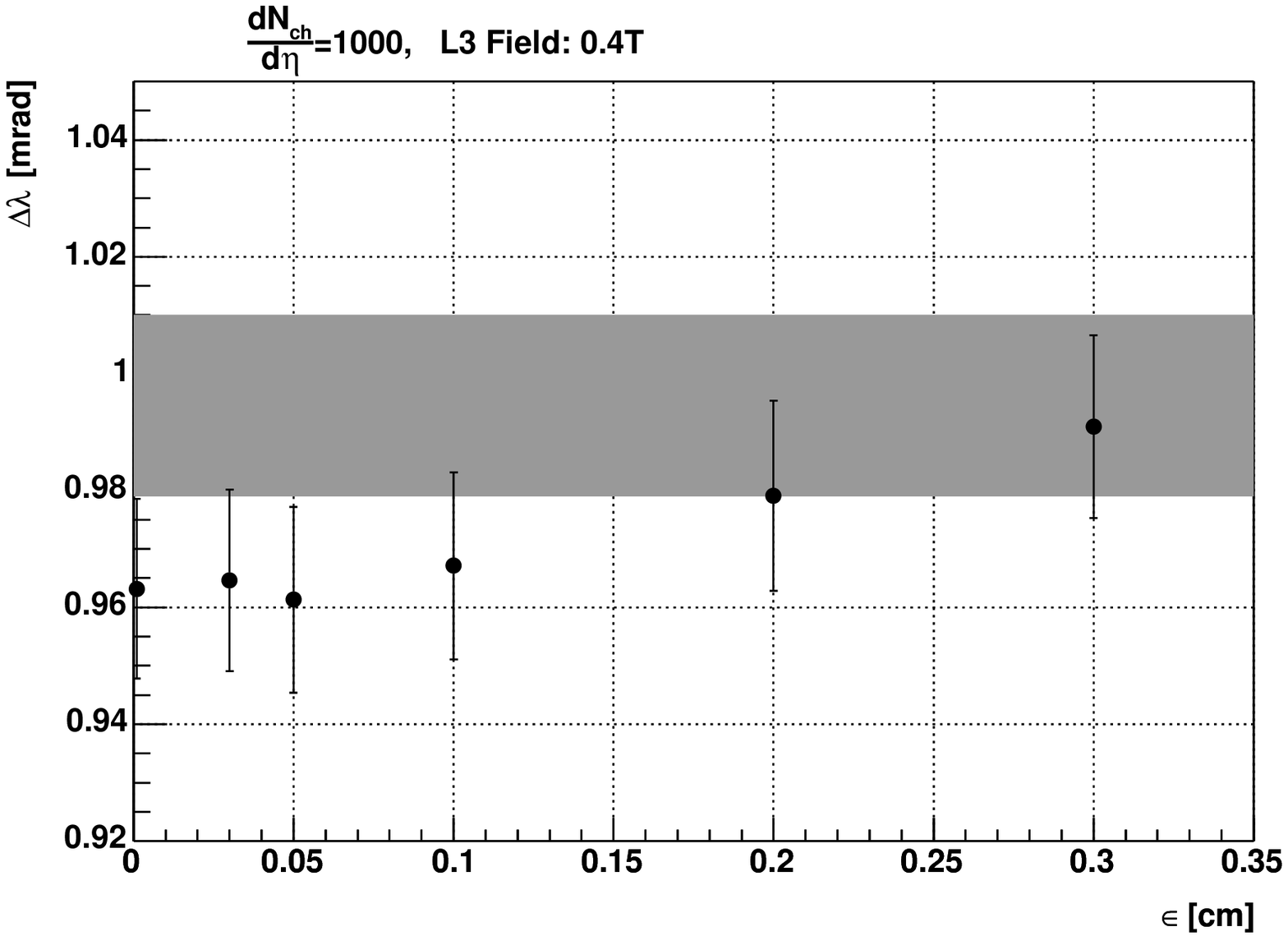}{9cm}
	{Impact on the dip-angle resolution.}
	{Integrated track dip-angle resolution, $\Delta\lambda$, as a
function of the quantization intervals in the compression scheme.}
\label{COMP_compvsdipangle}
\efig

Similar improvement of resolution can also be seen for the
longitudinal part of the reconstructed
tracks, which is illustrated by the impact on dip-angle resolution in
Figure~\ref{COMP_compvsdipangle}. Also here the
resolution is shown as a function of the quantization intervals which
have been used during the compression scheme. For $\epsilon\leq$\,0.1\,cm
the resolution is $\sim$4\% better than within the original data.

\subsubsection{Compression ratios versus tracking performance}
In Figure~\ref{COMP_compandeff_vs_mult} the achieved compression ratios
are shown together with the corresponding loss in efficiency for 5
different multiplicities, with \dndy ranging from 100 to 6000. The
quantization intervals was set to $\epsilon$=0.05\,cm for both
pad and time direction. For every data set the compress/expand
cycle was applied using three different options:
\begin{itemize}
\item Keeping all the remaining clusters.
\item Keeping a selection of remaining clusters.
\item Disregarding all the remaining clusters. 
\end{itemize}
The second option applies the selection option discussed above which
includes the clusters located in the region being used for
seed-finding by the Offline track finder.
The compression results are also summarized in
Table~\ref{COMP_compresults1}.

\bfig[htb]
\insertplot{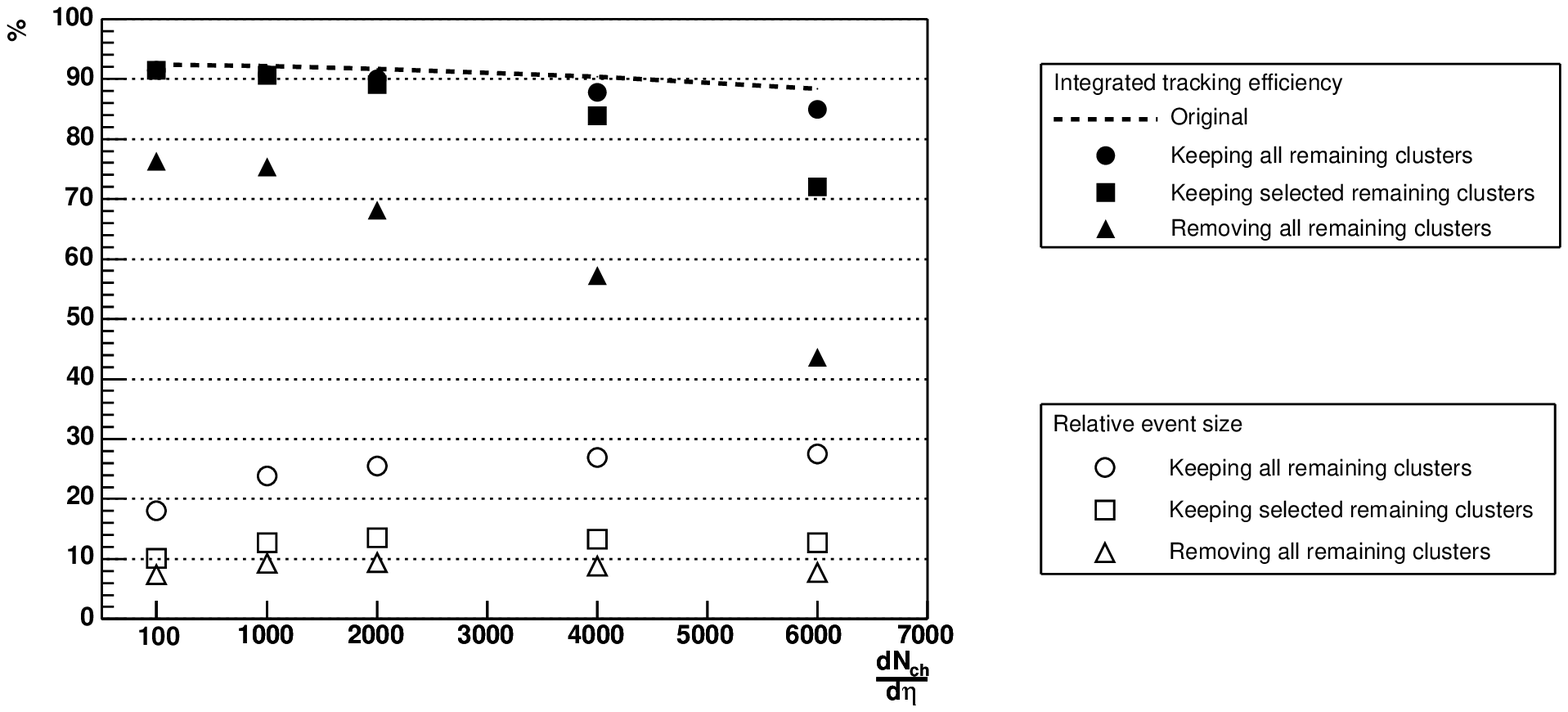}{16cm}
	{Achieved compression ratios and the corresponding efficiency
loss.}
	{Achieved compression ratios and the corresponding efficiency
loss for the different event samples investigated.}
\label{COMP_compandeff_vs_mult}
\efig
\begin{table}[htb]
    \begin{center}
    \begin{tabular}{|c|c|c|c|}
    \hline
	      & \multicolumn{3}{|c|}{{\bf Relative event size [\%]}}\\\cline{2-4}
	\dndy & All remaining & Selected remaining & No remaining\\
    \hline\hline
	100  & 18.0 & 10.1 & 7.3\\
	1000 & 23.8 & 12.7 & 9.3\\
	2000 & 25.4 & 13.5 & 9.3\\
	4000 & 26.9 & 13.3 & 8.8\\
	6000 & 27.5 & 12.8 & 7.7\\\hline
    \end {tabular}
    \caption[Data compression ratios for the different event samples.]
            {Data compression ratios for the different event samples
investigated. The three columns correspond to the various options of
storing the remaining clusters.}
    \label{COMP_compresults1}
    \end{center}
\end{table}

It is generally observed that both the efficiency loss and the
relative event size increases as a function of multiplicity.
If all remaining clusters are kept in the data stream
the resulting compression ratios range from 18-28\%. In
this case, a very small loss in efficiency can be observed for all the event
samples with a maximum of 3.5\% for \dndy=\,6000. In the case where only
a selection of remaining clusters is stored, compression
ratios of 10-14\% are achieved. In this case, however, some loss of tracking
efficiency can be seen, particularly for the higher multiplicity
events. For \dndy=\,1000 the loss is on average $\sim$1.5\%, while for
\dndy=\,6000 it goes up to $\sim$16\%. The third scenario, which
disregards all remaining clusters in the data stream, indicates a
significant impact for all the data samples investigated. In this
case, the efficiency loss is on average $\geq$16\% even for the lowest
multiplicity of \dndy=\,100.

The observed loss in tracking efficiency is mainly caused by the
shortcomings of the track reconstruction chain used to model the data,
and not the compression scheme itself. This is also supported by the
fact that the loss is insignificant at lower multiplicities, both with
respect to the tracking efficiency and the impact on resolutions.
For higher multiplicities, the applied tracking scheme has clear limitations
compared to the approach used by Offline, which was demonstrated in
Section~\ref{TRACK_seqperformance}.

In Table~\ref{COMP_compresults2} the properties of the compressed
data samples are listed. 
\begin{table}[htb]
    \begin{center}
    \begin{tabular}{|c|c|c|c|c|c|c|}
    \hline
	 & \multicolumn{2}{|c|}{{\bf Bits used}} &
\multicolumn{2}{|c|}{{\bf Entropy}}
& \multicolumn{2}{|c|}{{\bf Relative size [\%]}}\\\cline{2-7}
	\dndy & Pad & Time & Pad & Time & Track parameters & Cluster parameters \\
    \hline\hline
	100  & 7 & 8 & 2.4 & 3.6 & 6.0 & 94.0 \\
	1000 & 9 & 9 &2.7  & 3.9 & 5.9 & 94.1\\
	2000 & 8 & 9 &2.7  &3.9  & 6.6 & 93.4  \\
	4000 & 9 & 9 &2.9  &4.1  & 6.9 & 93.1 \\
	6000 & 9 & 9 &3.0  &4.3  & 7.3 & 92.7 \\ \hline
    \end {tabular}
    \caption[Properties of the compressed data samples.]
            {Properties of the compressed data samples. The {\it bits
used} refer to the fixed bit-rate used to encode the residuals, while
the {\it entropy} is the calculated entropy of the respective
quantized residual distribution in the sample. The {\it relative size}
gives the relative
size of the track and cluster data in the compressed format.}
    \label{COMP_compresults2}
    \end{center}
\end{table}
\noindent In particular, the difference between the
fixed bit-rate used to encode the residuals and their entropies
indicate that the samples can be compressed even further by
e.g. Arithmetic Coding. By simply taking the ratio between the calculated
entropies and the number of bits used, an additional factor of
2-3 is achieved. The potential gain in the total compression ratio will however be
slightly less than this factor, as the compressed
data also contain the cluster charges, header information and the
track parameters (Figure~\ref{COMP_encoding}) which are difficult to
compress any further.

\section{Summary}
In this chapter various options for compressing the TPC data as an
application for the High Level Trigger System are presented. 

Extensive studies applying local data modeling techniques on both real NA49
TPC data and simulated ALICE TPC data are published
in~\cite{berger02}. The results show that lossless compression techniques
such as Huffman and Arithmetic Coding can achieve compression ratios
of $\geq$60\%. The lossy Vector Quantizer may achieve ratios of
$\sim$50\% with a measurable, but small, impact on the space point
resolution. 

By introducing online pattern recognition, the TPC data can be
modeled more efficiently by representing the data using cluster and
track information. A data compression scheme which utilize the
redundant information content by representing the cluster data
relative to the track model has been implemented. Such a scheme will
lower the bit-rate needed to encode the cluster parameters as the
cluster model critically depends on the track parameters. Additional
options to compensate for the potential loss of clusters in the
compression scheme have been implemented in the cluster
analyzer. These options include clusters that should be stored in
addition to the compressed data sample.
The results using simulated ALICE TPC data
indicate that compression ratios of 10-15\% are achievable. Even lower
compressing ratios are possible by utilizing the entropy factor of the cluster
residuals, e.g. using Arithmetic Coding. The entropies of the data samples
indicate an additional compression factor of 2, depending on the overhead from the
remaining clusters. 

The relative loss in tracking efficiency is small for \dndy$\leq$\,2000,
and increases for
higher multiplicities. The impact on the relative \pt resolution and
dip-angle resolutions was shown to be insignificant. 
The efficiency loss at higher multiplicities is solely due
to the inefficiency of the HLT reconstruction chain and not
the compression scheme itself.

\chapter{Conclusions and Outlook}
\label{CONC}

In summary, two main topics are addressed in this work:
\begin{itemize}
\item Online TPC pattern recognition.
\item Online TPC data compression.
\end{itemize}
The pattern recognition forms the basis of the HLT system as all the
foreseen applications, both event selection and efficient data compression, relies on
a full or partial online event reconstruction. The latter may be
considered as the ultimate HLT-application
as an effective online compression of the TPC data would potentially enable
higher statistics for all the physics observables.

\section{Online TPC pattern recognition}
The critical performance issue both in terms of tracking efficiencies
and computing requirements is the particle multiplicity. Present
predictions for the multiplicity in central Pb--Pb collisions at LHC
range from 2000 to 6000 charged particles per unit rapidity, and
extrapolations from RHIC suggesting multiplicity values of about 2000-3000.

\subsubsection{Tracking performance}
Figure~\ref{CONC_inteff} shows the integrated
tracking efficiency of primary particles as a function of multiplicity
obtained with the HLT sequential tracking scheme. Two theoretical predictions,
\dndy$\approx$\,2200~\cite{eskola} and
\dndy$\approx$\,3200~\cite{amelin}, are marked with the shaded area.
Within this multiplicity range, the simulations
indicate that tracking efficiencies of 85--90\% are feasible.
\bfig[htb]
\insertplot{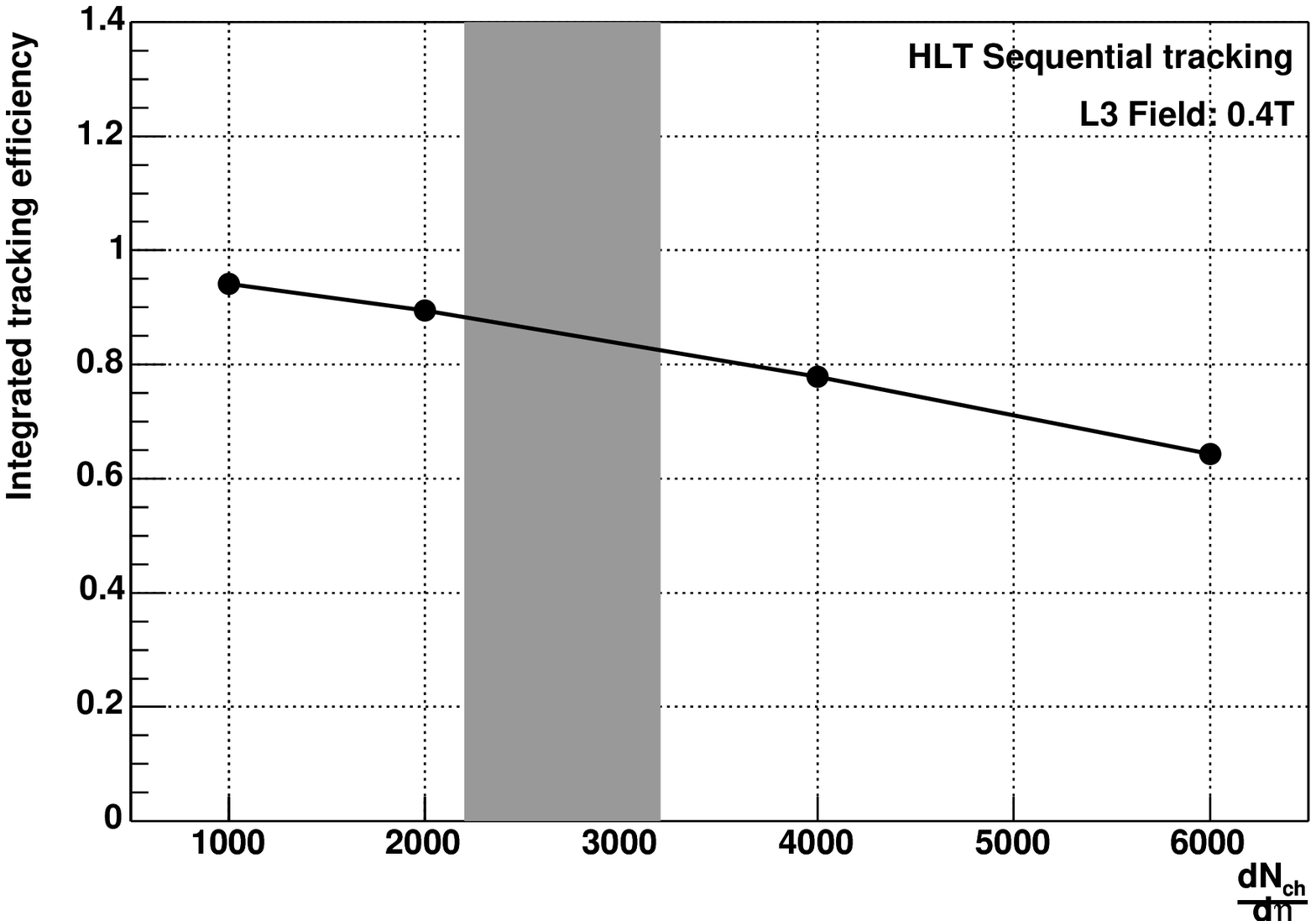}{8cm}
	{Integrated tracking efficiency and predicted multiplicities.}
	{Integrated tracking efficiency obtained with the HLT
sequential tracking approach. The shaded area indicates the present
theoretical predictions based on extrapolations from
RHIC~\cite{eskola,amelin}.}
\label{CONC_inteff}
\efig
For the secondary particles, simulations indicate that more
than 60\% of the decay products from $K$ and $\Lambda$ particles
within the TPC acceptance can be reconstructed with the current
tracking scheme. With further optimizations of the
tracking parameters this number is likely to improve. 

At these relatively low multiplicities the resulting occupancy might
be handled satisfactory by the implemented sequential tracking scheme.
For higher multiplicities it is obvious that an iterative tracking approach is
required in order to achieve the desired online tracking efficiencies. 
The implemented algorithms, the Hough Transform and a Cluster
Fitter, however do not show any improved performance compared to the
sequential chain. The main reason being the relatively high number of
falsely identified peaks in the HT.

\subsubsection{Impact on physics cases}
The performance of the online track reconstruction algorithms must
eventually be evaluated with respect to the impact on the various
trigger applications. Such evaluation can only be done by extensive
studies using the reconstructed tracks as input to the trigger
algorithms. Some preliminary conclusion may however be drawn based on the 
current outlined trigger algorithms.

Most of the physics trigger applications such as the jet and
open charm trigger relies on a good tracking efficiency for high
transverse momentum particles. The jet trigger will mainly focus on particles
with $p_t\geq$\,2\,GeV, and the open charm trigger will utilize a online
momentum filter where only tracks with $p_t\sim$\,0.5--1\,GeV will be
included in the search for $D$-candidates. In this region the tracking
efficiency of the sequential approach is comparable to Offline for
all multiplicities, as the relative efficiency loss is mainly
restricted to lower momentum particle tracks. Also
the iterative tracking approach shows promising performance
in this momentum regime. This suggest that both
tracking approaches may be utilized in such trigger
applications. Further refinement of the open charm trigger in terms of
secondary vertex analysis can only be done by including information by
the ITS detector.

\subsubsection{Suggestions for further work}
Based on the current understanding of the ALICE TPC tracking
performance, there are several optimizations and alternative solutions
that should be pursued. 

The fact that the sequential tracking scheme demonstrates a good performance at lower
occupancies suggests the possibility of using it also for higher
multiplicities in regions where the occupancy is sufficiently low,
i.e. at large TPC radius. Reconstruction the tracks in the outermost
pad-rows of the TPC with the Cluster Finder and Track Finder, these
track segments can be extended towards the inner pad-rows by utilizing
the Cluster Fitter. In such a scenario, the tracks reconstructed by
the sequential tracking scheme serve as input to the Cluster Fitter
where the charge distributions along the trajectory at the inner
pad-rows are fitted and deconvoluted in order to properly reconstruct
and collect the cluster centroids. The final list of assigned clusters
may then be fitted to a helix in order to obtain the track parameters.
Since the occupancy is low, the tracking parameters of the
Track Finder should be relaxed in order to recover as many of the
tracks as possible. 


The cluster fitting procedure may also be further optimized.
One option is to compare the estimated width of the cluster to be
fitted to the calculated RMS-value of the actual charge
distribution. Such a comparison can be used to
determine the validity
of the input track, and in particular reject input tracks whose
estimated widths deviate substantially from that of the charge
distribution. This may consequently reduce the number of
fake tracks in the fitting procedure. Furthermore, global information
about a track may be utilized in order to reject fake tracks at an
early stage in the cluster fitting. For instance, a simple check
may be performed on the input track on whether it points to valid
clusters on a certain number of successive pad-rows. If the track is
valid, there should be corresponding charge distributions along its
trajectory on all the pad-rows. 

Final optimization and tuning of the cluster fitting procedure can
only be obtained using real ALICE TPC data, as a detailed understanding of the
TPC response functions is needed to optimize the parameterization of
the cluster shapes.

\subsubsection{Computing requirements}
The required computing power of the HLT system is directly influenced
by the CPU-time needed by the individual processing modules in the
event reconstruction chain. The largest amount of computing is by far
required by the track reconstruction in the TPC, and the measured
processing times of the TPC reconstruction chain may serve as a
indicator on the amount of processing power needed for the complete
system. Such estimates will however ignore any overhead from the
interprocess communication and synchronization in order to operate
the HLT system, and can only be indicative.

Given the rapid rate of increasing CPU performance, the number of
single processors at the time of purchasing the HLT components will
scale down relative to current available CPU performance. 
According to Moore's law the processing power of a single processor will
approximately double for every 18-24 months. 
Such a scaling is however restricted to
applications which have a very low memory access requirements, i.e. it
is only approximately applicable if all the data and code references
are in the internal memory-cache of the CPU. This means that if an
algorithm is heavily I/O bound, its processing time will not scale
accordingly. 

Table~\ref{CONC_computing} lists the CPU-time measured on
two different CPUs for the sequential reconstruction chain. The
benchmark CPUs consist of Pentium III 800\,MHz and Pentium 4 2800\,MHz,
both with 1\,GB of RAM and 256\,kB and 512\,kB L2 cache
respectively. Both architectures were running a Linux kernel v2.4.
The measured CPU-time is integrated over all the processing modules, and is thus
equivalent to the processing time needed to reconstruct one event on a
single CPU. Based on these numbers one can estimate the required number of
processors needed to process the data within the
time-budget of 5\,ms in central Pb--Pb rate of 200\,Hz. For instance,
assuming a multiplicity of \dndy=\,2000, about 6300\,ms/5\,ms=1260 CPUs (Pentium 4) is
required to reconstruct the tracks in the TPC, while approximately the
double amount is needed for \dndy=\,4000. Comparing the
performance on the two processor types there is a factor of
about 2.2 improvement in processing time, which illustrates that the
processing times does not scale with the CPU clock frequency (3.5). 

The Cluster Finder algorithm needs approximately 25\% of the measured CPU-time.
This algorithm has been synthesized on a FPGA, and will
most likely utilize the FPGA co-processor functionality planned
for the HLT-RORC. Assuming that the Cluster Finder can be processed in
a single FEP, this can potentially reduce the number of required
processors by a factor 4. 

Given that the HLT components will be purchased in 2006-2007, an
additional factor of increase of the CPU performance can be expected compared with
todays measurements. How much this will affect
the overall requirements of the HLT system, however, has to be
monitored and evaluated during the purchasing procedure.

\begin{table}
    \begin{center}
    \begin{tabular}{|c|c|c|c|c|}
	\hline
	      & \multicolumn{2}{|c|}{\bf Pentium III,
800\,MHz}&\multicolumn{2}{|c|}{\bf Pentium 4, 2800\,MHz}\\\cline{2-5}
	 \dndy & CPU-time [s] &  \#CPU& CPU-time [s] &  \#CPU\\
    \hline
    \hline
	1000  & 7.5  & 1500 & 3.4  & 680  \\
	2000  & 14.0 & 2800 & 6.3  & 1260 \\
	4000  & 29.5 & 5900 & 13.2 & 2650 \\
	6000  & 47.3 & 9460 & 21.2 & 4240 \\
    \hline
    \end{tabular}
    \caption[Computational demands on the HLT system.]
            {Computational demands on the HLT system from TPC
tracking. The CPU-time is integrated over all the processing modules
in the sequential tracking approach,
and is equivalent to the processing time needed to reconstruct the
event on a single CPU. The number of processors corresponds to the measured CPU-time
divided by the available time-budget of 5\,ms assuming event rate of 200\,Hz.}
    \label{CONC_computing}
    \end{center}
\end{table}

\section{Online data compression}
The modeling techniques implemented
indicate that one can compress the TPC data by a factor
6-10 with a low impact on the tracking performance. This however
assumes that the online pattern recognition scheme which are used to
model the data is comparable to that of which will be used during
the Offline analysis.

Based on the achieved compression ratios one can estimate the
potential online data rate reduction as far as the TPC is
concerned. In Table~\ref{CONC_datared}, the event sizes estimated
from RLE 8 bit ADC data and the
corresponding data rate assuming a central event rate of
200\,Hz are shown. These event sizes are scaled down using the compression
ratios obtained and the resulting data rate is given accordingly.
\begin{table}
    \begin{center}
    \begin{tabular}{|c|c|c|c|c|}
	\hline
	      &\multicolumn{2}{|c|}{\bf RLE 8 bit TPC
data}&\multicolumn{2}{|c|}{\bf Compressed track--cluster data}\\\cline{2-5}
	\dndy & Event size [MB]& Data rate [MB/s]& Event size [MB]& Data rate[MB/s]\\\hline\hline
	1000  & 13.8 & 2760  & 1.8& 360\\
	2000  & 24.1 & 4820  & 3.3& 660\\
	4000  & 44.1 & 8820  & 5.9& 1180\\
	6000  & 61.6 & 12320 & 7.8& 1560\\
    \hline
    \end{tabular}
    \caption[Estimated TPC data rate reduction based on obtained
TPC data compression ratios.]
            {Estimated TPC data rate reduction based on obtained
TPC data compression ratios. The compression ratios used corresponds
to the results where a selection of the remaining clusters has been stored.}
    \label{CONC_datared}
    \end{center}
\end{table}
Given the foreseen bandwidth to mass storage of $\sim$1200\,MB/s, the
results indicates that up to a multiplicity of
dN$_{\mathrm{ch}}$/d$\eta$=4000 the data
rate can sufficiently be reduced to write all the measured data to mass
storage. 

Entropy encoding of the quantized residuals by e.g. Arithmetic Coding may
further improve the compression ratio by a factor of 2. However, 
also clusters belonging to secondary tracks must be considered,
resulting in an increase of the compressed data size.
With a good tracking efficiency they can be modeled
and compressed similarly to what is done for the
primaries. Including data from these tracks is not likely to have a
significant impact on the compression ration, as their relative size
is small. Their presence will therefore mainly concern the additional
computing power needed for their reconstruction.

The difference between the modeling scheme, the general readout
scheme and/or using e.g. entropy coding, is the elimination of the original
raw-data from the data stream. However, the raw-data itself contains a
significant amount of redundant information as its characteristics
are governed by the detector specific constants such as the
diffusion constants of the gas, electronic response functions etc.
Therefore, once the Offline reconstruction and analysis chain has been
set up correctly and fully understood, there will in principle be no
reason to return to the raw-data anymore.
During the first years of running the experiment, however,
zero-suppressed raw-data will
have to be recorded without any further processing in order to ensure
a complete understanding of the TPC detector response. 
Any data compression scheme exploiting the global characteristics of
the data is therefore likely not to be used during these first years
of operation.



\section{Outlook}
Although recent results from RHIC indicate a particle
multiplicity which is well below the original ALICE design
value of dN$_{\mathrm{ch}}$/dy=8000, the uncertainties are still very
large. More effort should therefore be invested in further optimizing the
algorithms for the higher multiplicity regime, possibly utilizing
alternative combined tracking approaches as outlined
above. Extensive studies are also needed to investigate the impact of
the tracking performance on the various trigger applications, and
adapt the tracking parameters accordingly. Furthermore, the other
detectors, in particular the ITS and TRD, need to be incorporated
into the HLT track reconstruction framework in order to enable full
event reconstruction functionality. An obvious approach would be to
implement an online version of the Kalman filter as
used within the Offline framework. 

Further optimization of the TPC data modeling and compression scheme
relies on a detailed understanding of the TPC response and its data
models, and should therefore primarily be done using real ALICE TPC
data. The main impact on its performance is however strongly correlated
with the efficiency of the preceding pattern recognition step, and the
achievable compression ratio versus information loss will therefore
benefit from any improvement thereon.

\appendix
\chapter{Track parameterizations}
\label{APP}

In general, the motion of a particle within a given detector
depends on both the magnetic and electric fields, and the
interactions with the surrounding material such as multiple scattering
and energy loss. For pattern recognition and track fitting
purposes, it is often practical to assume that the
trajectory of the particle is not affected by the material. In a
static uniform magnetic field the trajectory of a charged particle is
then described by a helix. In the following, the track model and its helix
parameterization commonly used in collider experiments
will be outlined. A detailed description of general track
models and fitting procedures is given by Bock et. al.~\cite{bock}.

\section{The equations of motion}
A charged particle moving in a static magnetic field, {\bf B}, is
subject to the Lorentz force, {\bf F}, which is given by,
\beq
{\bf F} = q{\bf v}\times{\bf B}
\eeq
where {\bf v} is the velocity of the particle and $q$ is the
charge. From this equation the equation of motion can be derived,
\beq
m\gamma\frac{d^2{\bf x}}{dt^2} = c^2 K q{\bf v}(t)\times{\bf
B}({\bf x}(t))
\label{APP_motion}
\eeq
where $K$ is a proportionality factor, {\bf B}({\bf x}) is the static
magnetic field as a function of particle position, {\bf x}, $m$ is the
rest mass, $c$ is the velocity of light and $t$ is the time in the
laboratory rest frame. The relativistic Lorentz factor is given by
$\gamma = (1-\beta^2)^{-\frac{1}{2}}$ and $\beta=v/c$.

Equation~\ref{APP_motion} 
can be rewritten in the form of geometrical quantities only,
\beq
\frac{d^2{\bf x}}{ds^2} = \frac{Kq}{P}\frac{d{\bf x}}{ds}\times{\bf
B}({\bf x}(s))
\label{APP_gmotion}
\eeq
where $s(t)$ is the distance along the trajectory and 
$P=m\gamma v$
in the laboratory frame. The following units are commonly used:
\begin{itemize}
\item $q$ in multiples of the positive elementary charge
(dimensionless).
\item {\bf x} and s in cm.
\item P in GeV/c.
\item {\bf B} in Tesla.
\item $K$ is proportional to the velocity of light and is defined as
0.00299792458\,T$^{-1}$\,cm$^{-1}$.
\end{itemize}

\section{Helix parameterizations}
\label{APP_helix}
By integrating formula~\ref{APP_gmotion} one can obtain a parameterization for
the particle trajectory. There are in total six integration
constant in addition to the unknown momentum $P$. However, taking the identity
\begin{eqnarray*}
\left(\frac{dx}{ds}\right)^2+\left(\frac{dy}{ds}\right)^2+\left(\frac{dz}{ds}\right)^2\equiv1
\end{eqnarray*}
and by choosing a ``reference surface'' there will be
only five independent parameters defining the trajectory, three defining
the circular projection in the plane orthogonal to the magnetic field
direction and two for the linear motion along the direction of the field.
By further assuming that the magnetic field is parallel to the
$z$-axis, equation~\ref{APP_gmotion} reduces to,
\begin{eqnarray}
\frac{d^2x}{ds^2} &=& \frac{Kq}{P}\frac{dy}{ds}\nonumber\\
\frac{d^2y}{ds^2} &=&-\frac{Kq}{P}\frac{dx}{ds}B\nonumber\\
\frac{d^2z}{ds^2} &=& 0
\end{eqnarray}
and the solution is a helix with an axis parallel to $z$,
\begin{eqnarray}
x(s) &=& x_{0} + \frac{1}{\kappa}[\cos(\Phi_{0} + h s
\kappa\cos\lambda)-\cos\Phi_{0}]\nonumber\\
y(s) &=& y_{0} + \frac{1}{\kappa}[\sin(\Phi_{0} + h s \kappa\cos\lambda)-\sin\Phi_{0}]\nonumber\\
z(s) &=& z_{0} + s\sin\lambda.
\label{APP_helixparam}
\end{eqnarray}
The helix parameters are illustrated in Figure~\ref{APP_illhelixparam} and are
defined as follows:
\begin{itemize}
\item $s$ is the path length along the helix. It increases when
moving in the direction of the particle's momentum vector.
\item $(x_0,y_0,z_0)$ are the coordinates of the starting point of
the helix, where $s=s_0$=0.
\item $\lambda = \sin^{-1}(dz/ds)$ ($-\pi/2<\lambda\leq\pi/2$)
is the slope of the helix, commonly referred to as the dip-angle.
\item $\kappa=1/R$ is the curvature of the circular projection in the $xy$-plane.
\item $q$ is the charge of the particle in units of positron charge.
\item $h$ is the sense of rotation of the projected helix in the
$xy$-plane, where $h$=-sign(qB)=$\pm$1 (=sign($d\phi/ds$), where $\phi$ is
the track direction).
\item $\psi_0$ is the azimuthal angle of the track direction at the
starting point.
\item $\Phi_0$ is the azimuth angle of the
starting point (polar coordinates) with respect to the helix axis
($\Phi_0=\psi_0-h\pi/2$).
\end{itemize}

\bfig[htb]
\centerline{\epsfig{file=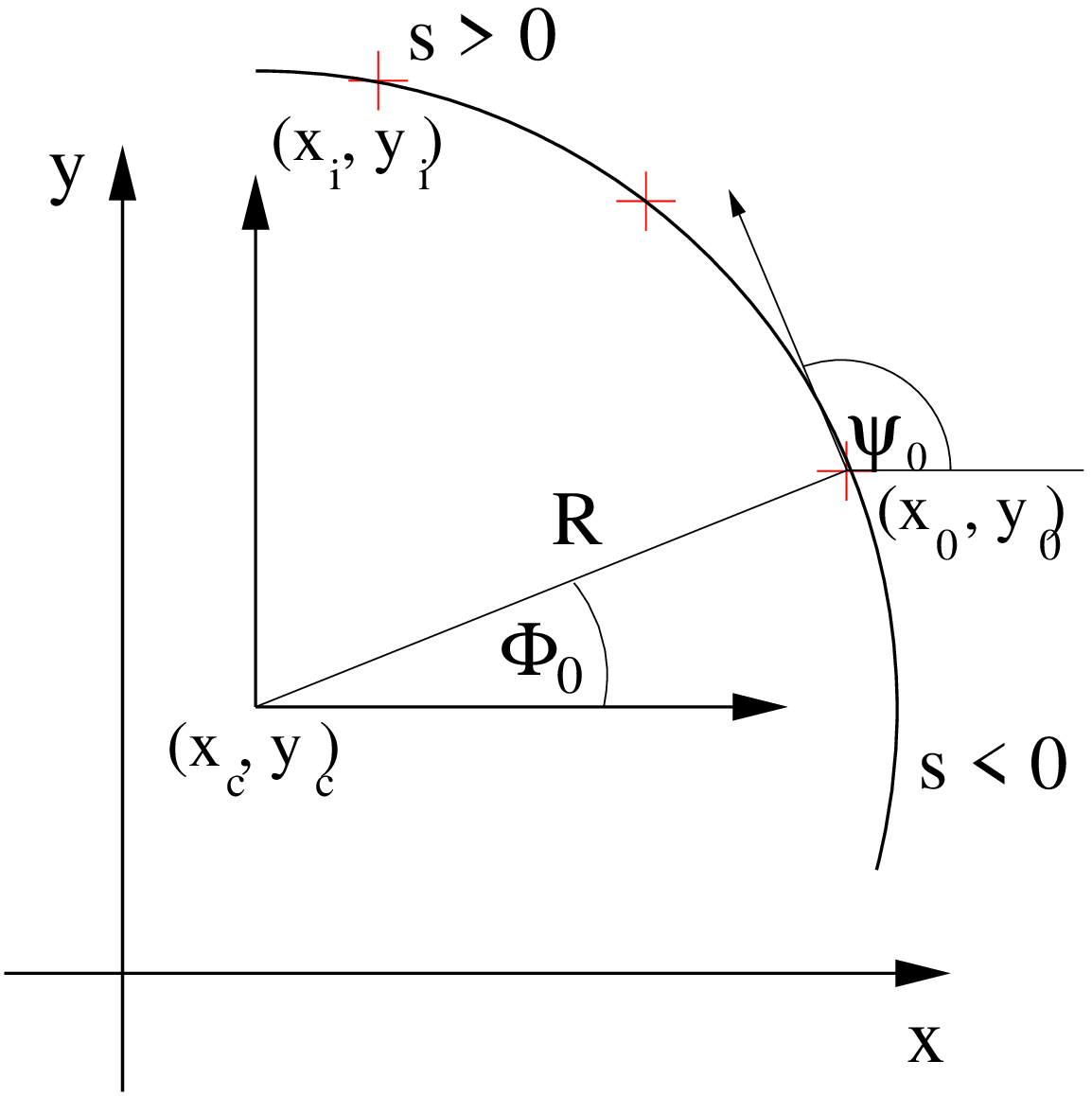,width=6cm}
\hspace{1cm}\epsfig{file=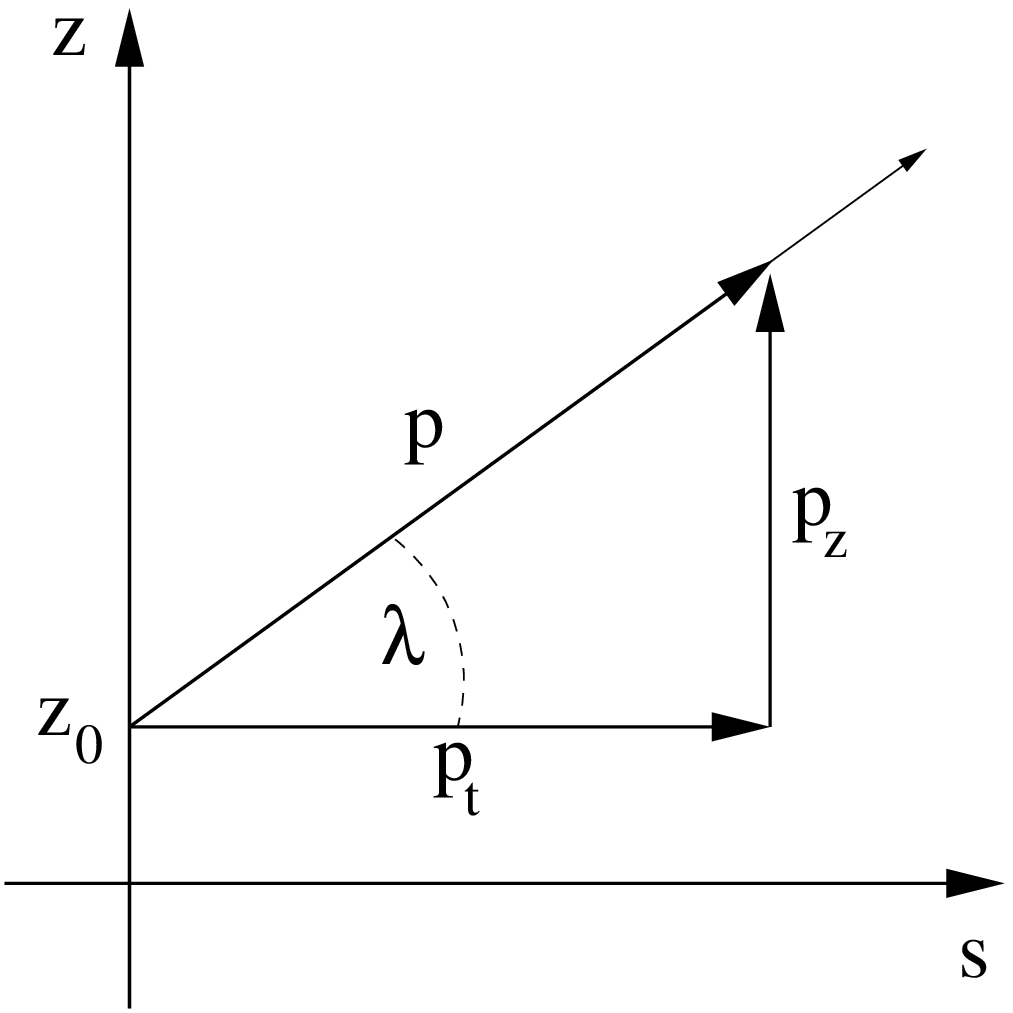,width=6cm}}
\caption[Schematic view of the helix parameters.]
	{Schematic view of the helix parameters. Projection of
the helix in the $xy$-plane (left). Projection of the helix in the
$sz$-plane (right).}
\label{APP_illhelixparam}
\efig

\subsubsection{Track fit parameters}
The track fit in the transverse plane determines the center of
curvature $(x_c,y_x)$ and the radius of curvature $R$, while the linear
fit in $(s,z)$-plane returns the $z_0$ and $\tan\lambda$. In order to
calculate the various track parameters, one need to determine the
reference/starting point of the helix. This is typically chosen as a
point closest to the innermost layer of the detector, i.e.
relative to the innermost assigned space point of the track. If the
innermost assigned space point on the track is ($x_1,y_1$), the
azimuthal angle can be defined as
\beq
\Phi_0 = \tan^{-1}\frac{y_1-y_c}{x_1-x_c}.
\label{APP_phase}
\eeq
The closest point $(x_0,y_0)$ on the trajectory and the azimuthal
angle of the track direction at that point are then calculated
according to
\begin{eqnarray}
x_0 &=& x_c + \frac{\cos\Phi_0}{\kappa}\nonumber\\
y_0 &=& y_c + \frac{\sin\Phi_0}{\kappa}\nonumber\\
\psi_0 &=& \Phi_0 + h\frac{\pi}{2}
\end{eqnarray}
The remaining track parameters can be calculated as
\begin{eqnarray}
p_t &=& \frac{KB}{\kappa}\nonumber\\
p_z &=& p_t\tan\lambda\nonumber\\
p &=& \sqrt{p_t^2+p_z^2}.
\end{eqnarray}

\subsubsection{Helix going through origin}
If the helix is going through the origin (0,0,0), the helix
parameterization in equation~\ref{APP_helixparam} is reduced to
\begin{eqnarray}
x(s) &=& \frac{1}{\kappa}[\cos(\Phi_0+hs\kappa\cos\lambda)-\cos\Phi_0]\nonumber\\
y(s) &=&
\frac{1}{\kappa}[\sin(\Phi_0+hs\kappa\cos\lambda)-\sin\Phi_0]\nonumber\\
z(s) &=& s\sin\lambda
\end{eqnarray}
Defining $t=s\kappa\cos\lambda$, and utilizing the relations
\begin{eqnarray}
\cos(\Phi_0)&=&\cos(\psi_0-h\frac{\pi}{2})=h\sin\psi_0\nonumber\\
\sin(\Phi_0)&=&\sin(\psi_0-h\frac{\pi}{2})=-h\cos\psi_0
\end{eqnarray}
gives
\begin{eqnarray}
x(t) &=& \frac{1}{\kappa}[\sin(\psi_0+ht)-\sin\psi_0]\nonumber\\
y(t) &=& \frac{1}{\kappa}[-\cos(\psi_0+ht)+\cos\psi_0]\nonumber\\
z(t) &=& \gamma t
\end{eqnarray}
where
\beq
\gamma = \frac{\tan\lambda}{\kappa}
\eeq
In this case the number of independent parameters describing the helix
is reduced from five to 3, i.e. two parameters for the circular projection in the
transverse plane ($\kappa$ and $\psi_0$), and one parameter for
describing the longitudinal motion ($\gamma$).

\subsubsection{Defining the track parameters at DCAO}
The reference point of the track can be defined as the point of 
distance of closest approach to the coordinate system origin (DCAO),
Figure~\ref{APP_dcao}. This is useful if e.g. all
tracks should be defined relative to the same point.
In this case, the
point $(x_1,y_1)$ in
equation~\ref{APP_phase} is (0,0) and $(x_0,y_0,z_0)$ is the point at
DCAO. The helix can now be defined by the parameter set,
\begin{eqnarray*}
(\kappa,r_{\mathrm{DCAO}},\phi_{\mathrm{DCAO}},z_{\mathrm{DCAO}},\lambda)
\end{eqnarray*}
where $r_{\mathrm{DCAO}}$, $\phi_{\mathrm{DCAO}}$ and
$z_{\mathrm{DCAO}}$ denotes the radius,
azimuthal angle and z-coordinate at the point,
respectively. In addition to these five parameters, two signs are needed to
denote the sense of rotation, $h$, and the relative distance of the point
with respect to the center of curvature
(sign($R-\sqrt{x_c^2+y_c^2}$)). The five parameters thus completely
define the helix, and the track parameters can be calculated as
\begin{eqnarray}
x_0 &=& |r_{\mathrm{DCAO}}|\cos\phi_{\mathrm{DCAO}}\nonumber\\
y_0 &=& |r_{\mathrm{DCAO}}|\sin\phi_{\mathrm{DCAO}}\nonumber\\
\Phi_0 &=& \left\{\begin{array}{ll}
			\phi_{\mathrm{DCAO}}               &\mbox{if $R>\sqrt{x_c^2+y_c^2}$}\\
			\phi_{\mathrm{DCAO}}+\frac{\pi}{2} &\mbox{if $R<\sqrt{x_c^2+y_c^2}$}\end{array}\right.\nonumber\\
\psi_0 &=& \Phi_0 + h\frac{\pi}{2}\nonumber\\
x_c &=& x_0-\frac{1}{\kappa}\cos\Phi_0\nonumber\\
y_c &=& y_0-\frac{1}{\kappa}\sin\Phi_0
\end{eqnarray}

\bfig[t]
\insertplot{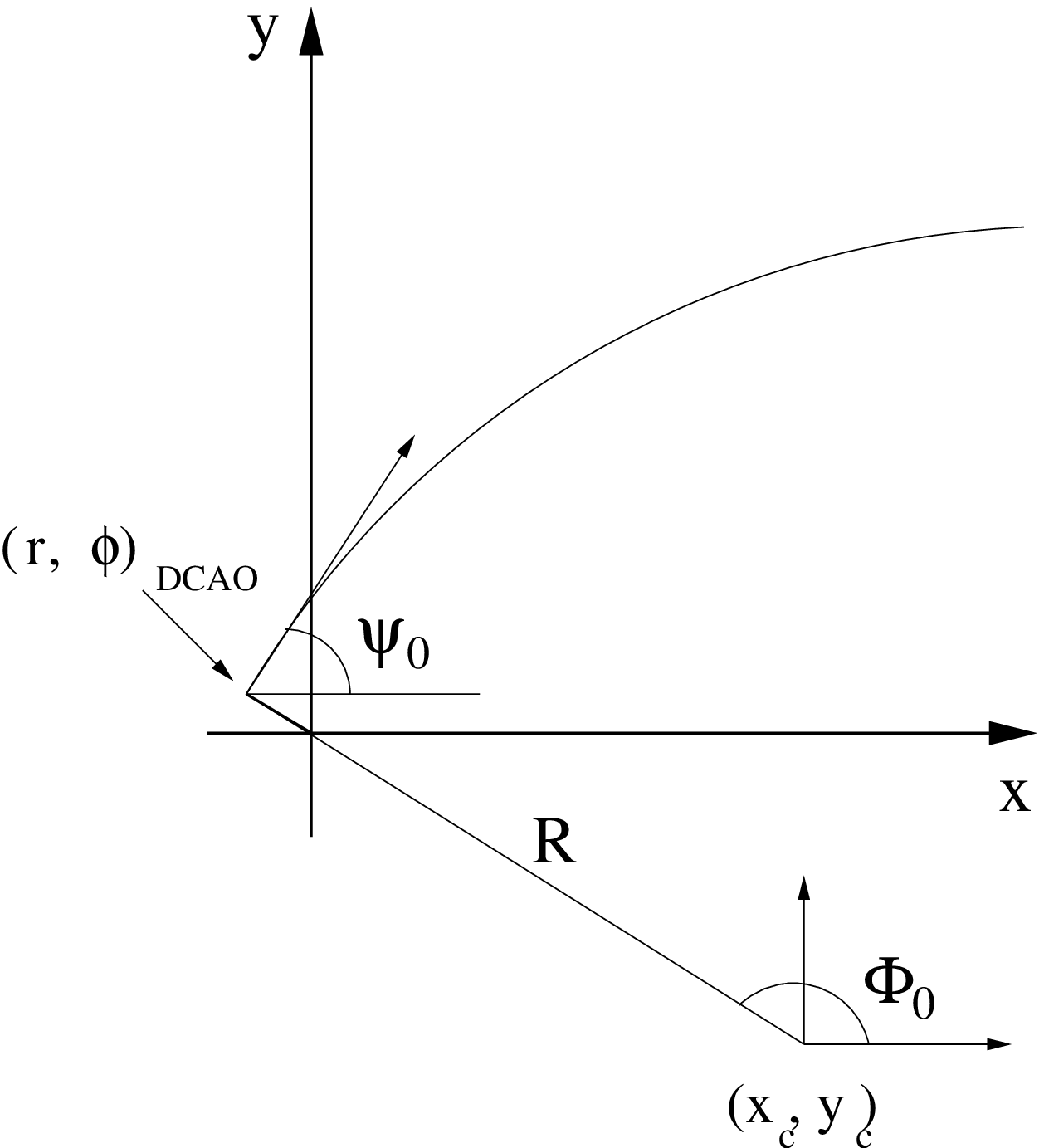}{6cm}
	{Definition of the track parameters at the point of DCAO.}
	{Defining the track parameters at the point of DCAO.}
\label{APP_dcao}
\efig




\chapter{Software and data formats}
\label{APP2}

\section{Analysis software structure}
All the HLT analysis software has been written in C++.
The various processing steps are implemented in individual modules,
allowing various processing topologies to be implemented.
The data structures used internally in the framework are simple
C-structures which has been adapted to the format which is likely to be used
in the readout chain. In order to be
compatible with the modular communication framework to be used in the
HLT system (page~\pageref{HLT_pubsub}), the data
payloads between the processing modules are communicated by data
references to the actual data in memory. This means that each module takes
a data pointer as input, and reads the input data from corresponding
location in memory. Similarly, the output data is written back to memory and
the corresponding data pointer is communicated to the next module in
the processing chain. This concept is illustrated in
Figure~\ref{APP2_overview}. 
\bfig[htb]
\insertplot{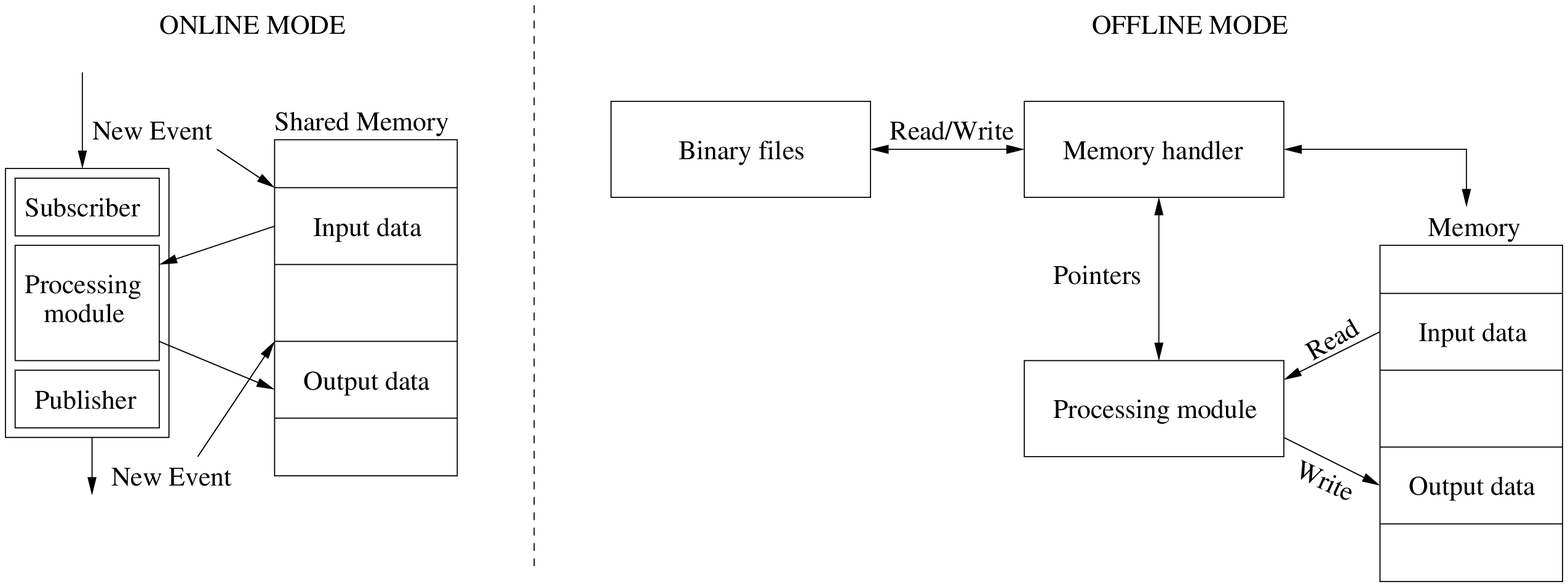}{12cm}
	{Schematic overview of the data payload
communication in the HLT reconstruction chain.}
	{Schematic overview of the data payload
communication in the HLT reconstruction chain. In the {\it online
mode} the communication is handled by the Publisher/Subscriber
framework (left), while in {\it offline mode} the memory handling is
handled transparently by a dedicated memory handler class (right).}
\label{APP2_overview}
\efig

When running the chain on a parallel prototype system ({\it online
mode}), the inter-process communication is completely handled within the
Publisher/Subscriber framework~\cite{timm}. This framework is
responsible for connecting all the various modules in the chain together,
and the data payloads are handled transparent of the
underlying network interface. For code development and tracking performance
evaluation purposes/debugging etc, i.e. when running the chain in
{\it offline mode}, the memory handling is done by a dedicated memory
handler class. In this case, the output data is written to binary files for
subsequent evaluation. In this way there is no difference between the two
running modes as far as the processing modules are concerned. This
makes it possible to use the same analysis code both for prototype
testing of the HLT system and offline evaluation within the AliROOT
framework.

\subsection*{Interface to AliROOT}
The interface between the HLT and AliROOT framework is provided via special functions
incorporated into a derived class of the HLT memory handling
class. This functionality
enables reading of the AliROOT data containers and transforming the
data into the format used within the HLT framework,
Figure~\ref{APP2_memhandler}.
\bfig[htb]
\centerline{\epsfig{file=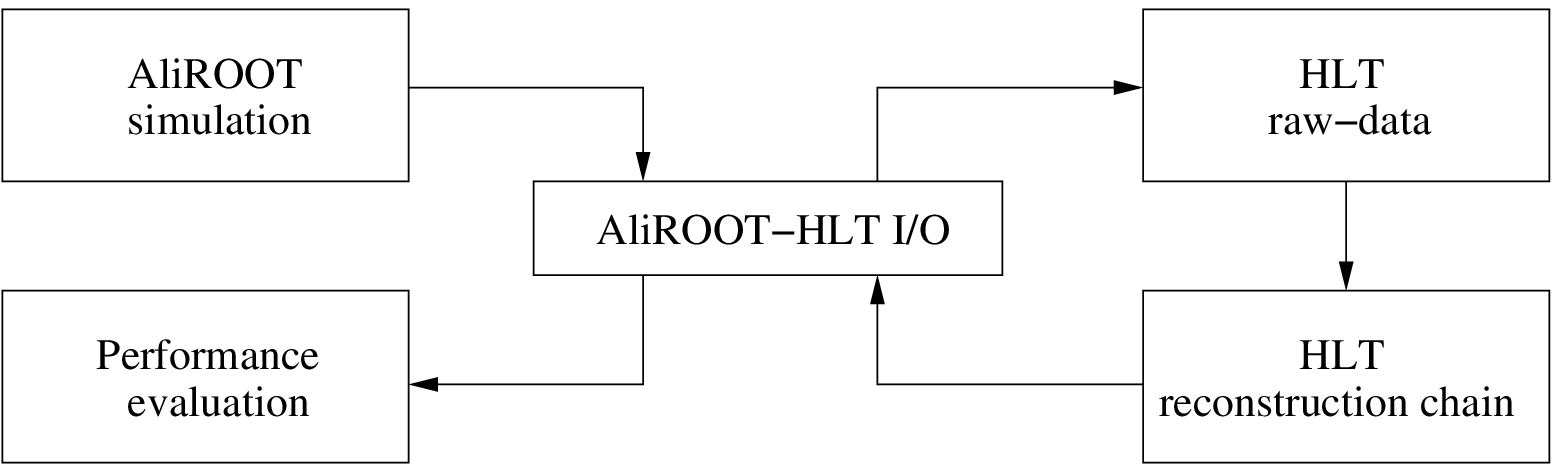,width=6cm}}
\caption[Schematic overview of the interface between HLT analysis code
and AliROOT.]
	{Schematic overview of the interface between HLT analysis code
and AliROOT.}
\label{APP2_memhandler}
\efig
\noindent This furthermore enables
storing and sending the Monte Carlo information from the simulation
through the HLT reconstruction chain in order to evaluate the tracking
performance. 

The HLT code can be compiled into shared libraries which can be loaded into an AliROOT
interactive session where it can run and evaluated using standard
ROOT-macros. The AliROOT support is
controlled via preprocessor options during compiling.

\section{Compressed data formats}

\subsubsection{8-bit RLE ADC-data format}
\label{APP2_rle}
In order to enable a simple approach to estimate the raw-data event
sizes, a RLE ADC-data formatting scheme has been implemented. The format
basically consists of replacing the zero-intervals between the
sequences by a tag and the length of the intervals, and will therefore
be comparable in size with the ALTRO-format.

Initially the 10 bit ADC-data from the simulation is converted to 8
bit using a 10-to-8 bit conversion table,
Figure~\ref{COMP_convtable}. The data is
then compressed by the RLE scheme and written to binary files. 
The data is organized by its inherit granularity from the detector readout
scheme, i.e. 1 file per (sub-)sector.
Each file thus contains a certain number of pad-rows which is written
in the beginning of the file. The remaining data stream consists of
the data on the successive pad-rows, 
\begin{verbatim}
NROWS PADROW_0 PADROW_1 ...
\end{verbatim}
For every pad-row, {\tt PADROW\_\#}, the corresponding row number is written together
with the number of pads containing data on that row. (pads which
has at least 1 sequence of time-bins above threshold)
Both number are
written using a 8 bit word, as the number of rows and pads on
a given pad-row is 159 and 140, respectively. For every pad containing
data the pad number is written, and then the ADC-values on that
pad. When a series of zeros occur, a zero is written followed by the
number of zeros. Example:
\begin{verbatim}
PAD 0 NZEROS ADC ADC ADC ADC 0 NZEROS ADC ADC ADC 0 0
\end{verbatim}
This pad with number {\tt PAD} contains two sequences with 4 and 3
consecutive time-bins
respectively. The two zeros at the end is used to mark the end of the
data stream on that pad.
Each entry consists of a 8 bit word.
If the number of zeros in a sequence is more than 255, an additional 8
bit word is written. 

\subsubsection{Bitwise I/O handling}
For the implemented data compression schemes, bitwise handling of the data is
necessary. The standard C I/O libraries only accommodates I/O on even byte
boundaries, and all bitwise I/O has to be done using bitwise operators
on integer values. In order to enable a more conventional way of
handling bitwise I/O to files, special routines has been adapted
from~\cite{compbook}. 
All bitwise I/O operations to file is done via the structure:
\begin{verbatim}
#include <stdio.h>
typedef struct bit_file {
    FILE *file;
    unsigned char mask;
    int rack;
} BIT_FILE;
\end{verbatim}
All data is read/written to file via a pointer to the normal {\tt FILE}
structure. The bitwise handling of the data is done with the
additional {\tt mask} and {\tt rack} elements.
The {\tt rack} contains the current byte of data either read in from
the file or waiting to be written out to the file. {\tt mask} contains
a single bit mask used either to set or clear the current output bit
or to mask in the current input bit. The mask element is initialized
to {\tt 0x80}, and during the output the first write to the
{\tt BIT\_FILE} will set or clear that bit and the mask element shifts to the
next. Once the mask has shifted to the point at which all the bits in
the output rack have been set or cleared, the rack is written out to
the file, and a new rack byte is started. Performing input from a
{\tt BIT\_FILE} is done in a similar fashion.

Four types of I/O routines are defined, which read or write a single
bit or multiple bits at a time:
\begin{verbatim}
void          OutputBit( BIT_FILE *bit_file, int bit );
void          OutputBits( BIT_FILE *bit_file,
                          unsigned long code, int bit_count );
int           InputBit( BIT_FILE *bit_file );
unsigned long InputBits( BIT_FILE *bit_file, int bit_count );
\end{verbatim}
{\tt code} denotes the value which is read/written using {\tt
bit\_count} number of bits.

\subsubsection{Compressed cluster data format}
The compressed cluster data format has been used for the estimation of
the data size needed to store the clusters as raw-data-arrays,
Section~\ref{COMP_cldata}. 
For every pad-row the corresponding row-number is written with a 8 bit
word. Next the number of clusters present on the row is encoded with
10 bits. The data stream for a single pad-row then becomes:
\begin{verbatim}
PADROW NCLUSTERS CLUSTER CLUSTER CLUSTER ... 
\end{verbatim}
E.g. for a given pad-row the two first is written to the output file by:
\begin{verbatim}
OutputBits(bitfile,padrow,8);        //Padrow number
OutputBits(bitfile,n_clusters,10);   //Number of clusters on padrow
\end{verbatim}
where
\begin{verbatim}
BIT_FILE *bitfile;     //Pointer set to the to the output file.
\end{verbatim}
A cluster on the current pad-row is now written as
\begin{verbatim}
OutputBits(bitfile,pad_centroid,n_pad_bits);    //Pad centroid
OutputBits(bitfile,time_centroid,n_time_bits);  //Time centroid
OutputBits(bitfile,pad_width,n_padwidth_bits);  //Pad width
OutputBits(bitfile,time_width,n_timewidth_bits);//Time width
OutputBits(bitfile,tot_charge,n_charge_bits);   //Cluster charge
\end{verbatim}

\subsubsection{Compressed cluster--track data format}
This format refers to the format used for the compressed track and
cluster information, Section~\ref{COMP_trackmodeling}. It
basically consists of the track parameters and
the clusters assigned to the given track. The overall data stream
has the following format
\begin{verbatim}
TRACK CLUSTER CLUSTER CLUSTER ... CLUSTER _CLEAR_
\end{verbatim}
where {\tt TRACK} and {\tt CLUSTER} denotes the track and relative
cluster parameters respectively. {\tt \_CLEAR\_} marks that the
current bit-racks in {\tt BIT\_FILE} is written to file and the mask
element is cleared.

The track parameters are defined within a standard C-structure as:
\begin{verbatim}
struct AliL3TrackModel {
  Float_t fKappa; //curvature
  Float_t fPhi;   //azimuthal angle of DCAO 
  Float_t fD;     //radius of DCA0
  Float_t fZ0;    //z-coordinate of DCA0
  Float_t fTgl;   //tan of dipangle
};
typedef struct AliL3TrackModel AliL3TrackModel;
\end{verbatim}
For a given track the parameters are written to the output file:
\begin{verbatim}
fwrite(&track,sizeof(AliL3TrackModel),1,bitfile->file);
\end{verbatim}
where
\begin{verbatim}
BIT_FILE *bitfile;     //Pointer set to the to the output file.
AliL3TrackModel track; //Structure filled with the track parameters
\end{verbatim}
The first cluster in the stream is encoded as:
\begin{verbatim}
OutputBit(bitfile,1);        //A single bit to flag cluster presence
OutputBits(bitfile,sector,6); //TPC sector number of the first cluster
OutputBit(output,0);                              //Sign of time residual 
OutputBits(output,abs(delta_time_q),n_time_bits); //Absolute timeresidual value
OutputBit(output,0);                              //Sign of pad residual 
OutputBits(output,abs(delta_pad_q),n_pad_bits);   //Absolute padresidual value
OutputBits(output,tot_charge,n_charge_bits);      //Cluster charge
\end{verbatim}
The following clusters in the track list are encoded in the same fashion,
with the only difference that instead of writing the TPC sector number
for every cluster, a single bit is used to denote a
possible change of sector. E.g. if there is no change:
\begin{verbatim}
OutputBit(bitfile,1); //Cluster present
OutputBit(bitfile,0); //No change of sector
OutputBit(output,0);                              //Sign of time residual 
...
\end{verbatim}
and if there was a change of sector
\begin{verbatim}
OutputBit(bitfile,1); //Cluster present
OutputBit(bitfile,1); //Change of sector
OutputBits(bitfile,sector,6); //New TPC sector number
OutputBit(output,0);                              //Sign of time residual 
...
\end{verbatim}
Reading/uncompressing the data is done similarly, using the
corresponding input routines.

\end{document}